\documentclass[twocolumn,tighten]{aastex63}

\usepackage{natbib}
\usepackage{epsfig}
\usepackage{amsmath}
\usepackage{xcolor}
\usepackage{hyperref}
\usepackage{booktabs}
\usepackage[encapsulated]{CJK}
\usepackage{ucs}
\usepackage[utf8x]{inputenc}
\usepackage{url}

\graphicspath{{./Figures/}{./Figures/Atlas/Combined/}}

\newcommand{\hi}{\mbox{\rm \ion{H}{1}}}
\newcommand{\hii}{\mbox{\rm \ion{H}{2}}}

\newcommand{\coone}{\mbox{\rm CO($1\text{--}0$)}} 
\newcommand{\cotwo}{\mbox{\rm CO($2\text{--}1$)}} 
\newcommand{\cothree}{\mbox{\rm CO($3\text{--}2$)}}

\newcommand{\acounits}{\mbox{M$_\odot$~pc$^{-2}$ (K~km~s$^{-1}$)$^{-1}$}}

\newcommand{\mtolwise}{\mbox{$\Upsilon_\star^{3.4}$}}
\newcommand{\mtolunits}{\mbox{${\rm M}_\odot~{\rm L}_\odot^{-1}$}}

\newcommand{\ntarget}{\mbox{$90$}}

\newcommand{\OSU}{\affil{Department of Astronomy, The Ohio State University, 140 West 18th Avenue, Columbus, Ohio 43210, USA}}

\newcommand{\Alberta}{\affil{Department of Physics, University of Alberta, Edmonton, AB T6G 2E1, Canada}}

\newcommand{\ANU}{\affil{Research School of Astronomy and Astrophysics, Australian National University, Canberra, ACT 2611, Australia}}

\newcommand{\IPAC}{\affil{Caltech-IPAC, 1200 E. California Blvd. Pasadena, CA 91125, USA}}

\newcommand{\Carnegie}{\affil{Observatories of the Carnegie Institution for Science, 813 Santa Barbara Street, Pasadena, CA 91101, USA}}

\newcommand{\CCAPP}{\affil{Center for Cosmology and Astroparticle Physics, 191 West Woodruff Avenue, Columbus, OH 43210, USA}}

\newcommand{\CfA}{\affil{Harvard-Smithsonian Center for Astrophysics, 60 Garden Street, Cambridge, MA 02138, USA}}

\newcommand{\CNRS}{\affil{CNRS, IRAP, 9 Av. du Colonel Roche, BP 44346, F-31028 Toulouse cedex 4, France}}

\newcommand{\ESO}{\affil{European Southern Observatory, Karl-Schwarzschild Stra{\ss}e 2, D-85748 Garching bei M\"{u}nchen, Germany}}

\newcommand{\GEMINI}{\affil{Gemini Observatory/NSF’s NOIRLab, 950 N. Cherry Avenue, Tucson, AZ, 85719, USA}}

\newcommand{\Heidelberg}{\affil{Astronomisches Rechen-Institut, Zentrum f\"{u}r Astronomie der Universit\"{a}t Heidelberg, M\"{o}nchhofstra\ss e 12-14, D-69120 Heidelberg, Germany}}

\newcommand{\IRAM}{\affil{Institut de Radioastronomie Millim\'{e}trique (IRAM), 300 Rue de la Piscine, F-38406 Saint Martin d'H\`{e}res, France}}

\newcommand{\ITA}{\affil{Universit\"{a}t Heidelberg, Zentrum f\"{u}r Astronomie, Institut f\"{u}r Theoretische Astrophysik, Albert-Ueberle-Str 2, D-69120 Heidelberg, Germany}}

\newcommand{\IWR}{\affil{Universit\"{a}t Heidelberg, Interdisziplin\"{a}res Zentrum f\"{u}r Wissenschaftliches Rechnen, Im Neuenheimer Feld 205, D-69120 Heidelberg, Germany}}

\newcommand{\JHU}{\affil{Department of Physics and Astronomy, The Johns Hopkins University, Baltimore, MD 21218, USA}}

\newcommand{\Maryland}{\affil{Department of Astronomy, University of Maryland, College Park, MD 20742, USA}}

\newcommand{\MPE}{\affil{Max-Planck-Institut f\"{u}r extraterrestrische Physik, Giessenbachstra{\ss}e 1, D-85748 Garching, Germany}}

\newcommand{\MPIA}{\affil{Max-Planck-Institut f\"{u}r Astronomie, K\"{o}nigstuhl 17, D-69117, Heidelberg, Germany}}

\newcommand{\NRAO}{\affil{National Radio Astronomy Observatory, 520 Edgemont Road, Charlottesville, VA 22903-2475, USA}}

\newcommand{\OAN}{\affil{Observatorio Astron\'{o}mico Nacional (IGN), C/Alfonso XII, 3, E-28014 Madrid, Spain}}

\newcommand{\ObsParis}{\affil{Sorbonne Universit\'{e}, Observatoire de Paris, Universit\'{e} PSL, CNRS, LERMA, F-75014, Paris, France}}

\newcommand{\Princeton}{\affil{Department of Astrophysical Sciences, Princeton University, Princeton, NJ 08544 USA}}

\newcommand{\UToledo}{\affil{Ritter Astrophysical Center, University of Toledo, 2801 W. Bancroft St., Toledo, OH, 43606}}

\newcommand{\Toulouse}{\affil{Universit\'{e} de Toulouse, UPS-OMP, IRAP, F-31028 Toulouse cedex 4, France}}

\newcommand{\UBonn}{\affil{Argelander-Institut f\"ur Astronomie, Universit\"at Bonn, Auf dem H\"ugel 71, 53121 Bonn, Germany}}

\newcommand{\UChile}{\affil{Departamento de Astronom\'{i}a, Universidad de Chile, Camino del Observatorio 1515, Las Condes, Santiago, Chile}}

\newcommand{\UCSD}{\affil{Center for Astrophysics and Space Sciences, Department of Physics,  University of California,\\ San Diego, 9500 Gilman Drive, La Jolla, CA 92093, USA}}

\newcommand{\UGent}{\affil{Sterrenkundig Observatorium, Universiteit Gent, Krijgslaan 281 S9, B-9000 Gent, Belgium}}

\newcommand{\ULyon}{\affil{Univ Lyon, Univ Lyon 1, ENS de Lyon, CNRS, Centre de Recherche Astrophysique de Lyon UMR5574,\\ F-69230 Saint-Genis-Laval, France}}

\newcommand{\UMass}{\affil{University of Massachusetts—Amherst, 710 N. Pleasant Street, Amherst, MA 01003, USA}}

\newcommand{\UVA}{\affil{University of Virginia, 530 McCormick Rd, Charlottesville, VA 22904, USA}}

\newcommand{\UWyoming}{\affil{Department of Physics and Astronomy, University of Wyoming, Laramie, WY 82071, USA}}

\newcommand{\LAM}{\affil{
Aix Marseille Universit\'{e}, CNRS, CNES, LAM (Laboratoire d’Astrophysique de Marseille), F-13388 Marseille,
France}}

\newcommand{\UHawaii}{\affil{Institute for Astronomy, University of Hawaii, 2680 Woodlawn Drive, Honolulu, HI 96822, USA}}

\newcommand{\UCM}{\affil{Departamento de F\'{\i}sica de la Tierra y Astrof\'{\i}sica, Universidad Complutense de Madrid, E-28040, Spain}}

\newcommand{\STScI}{\affil{Space Telescope Science Institute, 3700 San Martin Drive, Baltimore, MD 21218, USA}}

\newcommand{\INAF}{\affil{INAF -- Osservatorio Astrofisico di Arcetri, Largo E. Fermi 5, I-50157, Firenze, Italy}}

\newcommand{\Sydney}{\affil{Sydney Institute for Astronomy, School of Physics A28, The University of Sydney, NSW 2006, Australia}}

\newcommand{\SAI}{\affil{Sternberg Astronomical Institute, Lomonosov Moscow State University, Universitetsky pr. 13, Moscow, Russia}}

\shorttitle{PHANGS--ALMA: Cloud Scale Imaging of Galaxies}
\shortauthors{Leroy, Schinnerer et al.}

\begin{document}

\title{PHANGS--ALMA: Arcsecond CO(2--1) Imaging of Nearby Star-Forming Galaxies}

\begin{abstract}
We present PHANGS--ALMA, the first survey to map CO~$J=2\rightarrow1$ line emission at $\sim 1\arcsec \sim 100$~pc spatial resolution from a representative sample of $\ntarget$ nearby ($d \lesssim 20$~Mpc) galaxies that lie on or near the $z=0$ ``main sequence'' of star-forming galaxies. CO line emission traces the bulk distribution of molecular gas, which is the cold, star-forming phase of the interstellar medium. At the resolution achieved by PHANGS--ALMA, each beam reaches the size of a typical individual giant molecular cloud (GMC), so that these data can be used to measure the demographics, life-cycle, and physical state of molecular clouds across the population of galaxies where the majority of stars form at $z=0$. This paper describes the scientific motivation and background for the survey, sample selection, global properties of the targets, ALMA observations, and characteristics of the delivered ALMA data and derived data products. As the ALMA sample serves as the parent sample for parallel surveys with VLT/MUSE, HST, AstroSat, VLA, and other facilities, we include a detailed discussion of the sample selection. We detail the estimation of galaxy mass, size, star formation rate, CO luminosity, and other properties, compare estimates using different systems and provide best-estimate integrated measurements for each target. We also report the design and execution of the ALMA observations, which combine a Cycle~5 Large Program, a series of smaller programs, and archival observations. Finally, we present the first $1\arcsec$ resolution atlas of CO emission from nearby galaxies and describe the properties and contents of the first PHANGS--ALMA public data release.
\end{abstract}


\correspondingauthor{Adam K. Leroy}
\email{leroy.42@osu.edu}

\author[0000-0002-2545-1700]{Adam~K.~Leroy}
\footnote{This paper represents a collective effort by the PHANGS-ALMA team. Please see a description of contributions of individual team members in Appendix~\ref{sec:contrib}.}
\OSU \CCAPP

\author[0000-0002-3933-7677]{Eva~Schinnerer}
\MPIA

\author[0000-0002-9181-1161]{Annie~Hughes}
\CNRS
\Toulouse

\author[0000-0002-5204-2259]{Erik~Rosolowsky}
\Alberta

\author[0000-0003-3061-6546]{J\'er\^ome~Pety}
\IRAM
\ObsParis

\author{Andreas~Schruba}
\MPE

\author[0000-0003-1242-505X]{Antonio~Usero}
\OAN

\author[0000-0003-4218-3944]{Guillermo A. Blanc}
\Carnegie
\UChile

\author[0000-0002-5635-5180]{M\'elanie Chevance}
\Heidelberg

\author[0000-0002-6155-7166]{Eric Emsellem}
\ESO
\ULyon

\author[0000-0001-5310-467X]{Christopher M. Faesi}
\UMass

\author[0000-0001-6405-0785]{Cinthya~N.~Herrera}
\IRAM

\author[0000-0001-9773-7479]{Daizhong~Liu}
\MPIA

\author[0000-0002-6118-4048]{Sharon E. Meidt}
\UGent

\author[0000-0002-0472-1011]{Miguel~Querejeta}
\OAN

\author[0000-0002-2501-9328]{Toshiki~Saito}
\MPIA

\author[0000-0002-4378-8534]{Karin M. Sandstrom}
\UCSD

\author[0000-0003-0378-4667]{Jiayi~Sun \begin{CJK*}{UTF8}{gbsn}(孙嘉懿)\end{CJK*}}
\OSU

\author[0000-0002-0012-2142]{Thomas G. Williams}
\MPIA

\author[0000-0002-5259-2314]{Gagandeep S. Anand}
\UHawaii

\author[0000-0003-0410-4504]{Ashley~T.~Barnes}
\UBonn

\author[0000-0002-2333-5474]{Erica A. Behrens}
\UVA
\OSU

\author[0000-0002-2545-5752]{Francesco Belfiore}
\INAF

\author[0000-0003-4826-9079]{Samantha M. Benincasa}
\OSU \CCAPP

\author[0000-0003-0583-7363]{Ivana Be\v{s}li\'c}
\UBonn

\author[0000-0003-0166-9745]{Frank Bigiel}
\UBonn

\author[0000-0002-5480-5686]{Alberto D. Bolatto}
\Maryland

\author[0000-0002-8760-6157]{Jakob S. den Brok}
\UBonn

\author[0000-0001-5301-1326]{Yixian Cao}
\LAM

\author[0000-0003-0085-4623]{Rupali Chandar}
\UToledo

\author[0000-0002-5235-5589]{J\'er\'emy Chastenet}
\UGent
\UCSD

\author[0000-0003-2551-7148]{I-Da Chiang \begin{CJK*}{UTF8}{bkai}(江宜達)\end{CJK*}}
\UCSD

\author[0000-0002-8549-4083]{Enrico Congiu}
\UChile

\author[0000-0002-5782-9093]{Daniel A. Dale}
\UWyoming

\author[0000-0003-1943-723X]{Sinan Deger}
\IPAC

\author[0000-0002-1185-2810]{Cosima Eibensteiner}
\UBonn

\author[0000-0002-4755-118X]{Oleg V. Egorov}
\Heidelberg
\SAI

\author[0000-0002-0697-0177]{Axel Garc\'ia-Rodr\'iguez}
\OAN

\author[0000-0001-6708-1317]{Simon C. O. Glover}
\ITA

\author[0000-0002-3247-5321]{Kathryn Grasha}
\ANU

\author[0000-0001-9656-7682]{Jonathan~D.~Henshaw}
\MPIA 

\author[0000-0002-0757-9559]{I-Ting Ho}
\MPIA

\author[0000-0002-3227-4917]{Amanda A.~Kepley}
\NRAO

\author[0000-0002-0432-6847]{Jaeyeon Kim}
\Heidelberg

\author[0000-0002-0560-3172]{Ralf S.\ Klessen}
\ITA
\IWR

\author[0000-0001-6551-3091]{Kathryn Kreckel}
\Heidelberg

\author[0000-0001-9605-780X]{Eric~W.~Koch}
\CfA
\Alberta

\author[0000-0002-8804-0212]{J.~M.~Diederik Kruijssen}
\Heidelberg

\author[0000-0003-3917-6460]{Kirsten L. Larson}
\IPAC

\author[0000-0002-2278-9407]{Janice C. Lee}
\GEMINI 
\IPAC

\author[0000-0002-1790-3148]{Laura A. Lopez}
\OSU \CCAPP

\author[0000-0001-8895-6784]{Josh Machado}
\OSU

\author[0000-0002-5993-6685]{Ness Mayker}
\OSU \CCAPP

\author[0000-0003-2290-7060]{Rebecca McElroy}
\Sydney

\author[0000-0001-7089-7325]{Eric J. Murphy}
\NRAO

\author[0000-0002-0509-9113]{Eve C.~Ostriker}
\Princeton

\author[0000-0002-1370-6964]{Hsi-An Pan}
\MPIA

\author{Ismael Pessa}
\MPIA

\author[0000-0003-1111-3951]{Johannes~Puschnig}
\UBonn

\author[0000-0001-7876-1713]{Alessandro Razza}
\UChile

\author[0000-0003-0651-0098]{Patricia S\'anchez-Bl\'azquez}
\UCM

\author[0000-0002-6363-9851]{Francesco Santoro}
\MPIA

\author[0000-0002-5783-145X]{Amy Sardone}
\OSU \CCAPP

\author[0000-0003-2707-4678]{Fabian Scheuermann}
\Heidelberg

\author{Kazimierz Sliwa}
\MPIA

\author[0000-0001-6113-6241]{Mattia C.\ Sormani}
\ITA

\author[0000-0002-9333-387X]{Sophia K. Stuber}
\MPIA

\author[0000-0002-8528-7340]{David A. Thilker}
\JHU

\author[0000-0003-2261-5746]{Jordan A. Turner}
\UWyoming

\author[0000-0003-4161-2639]{Dyas~Utomo}
\NRAO

\author[0000-0002-7365-5791]{Elizabeth J. Watkins}
\Heidelberg

\author[0000-0002-3784-7032]{Bradley Whitmore}
\STScI

\section{Introduction}
\label{sec:intro}

This paper presents PHANGS--ALMA, an Atacama Large Millimeter/\linebreak[0]{}submillimeter Array (ALMA) survey aimed at studying the physics of molecular gas across the nearby galaxy population. PHANGS--ALMA is a key component of the multiwavelength observational campaign conducted by the Physics at High Angular resolution in Nearby Galaxies (PHANGS) project\footnote{\url{http://phangs.org/}}. Combining a Cycle~5 ALMA Large Program, a suite of smaller programs, and data from the ALMA archive, PHANGS--ALMA mapped the CO $J=2\rightarrow1$ emission, hereafter \cotwo , from a cleanly-selected sample of \ntarget\ of the nearest, ALMA-accessible, massive, star-forming galaxies. The resulting \cotwo\ data have high spatial and spectral resolution, good surface brightness sensitivity, full flux recovery, and good coverage of the area of active star formation in each target.

These characteristics make PHANGS--ALMA the first ``cloud scale,'' $\sim100$ pc, survey of molecular gas across a local galaxy sample that is representative of where stars form in the $z=0$ universe. Though the data are suitable for many scientific applications, the survey was designed with the broad goals of quantifying the physics of star formation and feedback at the scale of individual giant molecular clouds (GMCs) and connecting these measurements to galaxy-scale properties and processes. 

With these goals in mind, the PHANGS team has followed up PHANGS--ALMA with a suite of multi-wavelength programs that span the spectrum from far-UV to radio, aiming to sample all stages of the star formation and feedback cycle. ``PHANGS--MUSE'' is obtaining optical integral field spectroscopy using the VLT/MUSE instrument to measure the properties of ionized gas and stellar populations at resolution matched to ALMA (PI: E. Schinnerer; E. Emsellem et al. in preparation). ``PHANGS--HST'' is using HST/WFC3 five-filter broad-band imaging to find and characterize stellar clusters and associations \citep[PI: J. Lee;][]{LEE21}. Other programs include new, high resolution far-UV mapping by AstroSAT (PI: E. Rosolowsky), new ground-based narrow-band H$\alpha$ imaging using the MPG 2.2m/WFI and Du Pont/DirectCCD instruments (PIs: G. Blanc, I-T. Ho; A. Razza et al. in preparation), and new \textsc{Hi} imaging using the VLA and MeerKAT (PI: D. Utomo; D. Utomo et al. in preparation).

This paper begins with an overview of the background, design, and goals of PHANGS--ALMA (\S\ref{sec:motivation}). Because PHANGS--ALMA served as the parent sample for many of the multi-wavelength efforts descibed above, we discuss the sample selection in some detail in \S\ref{sec:sample}. We also present our best estimates for the integrated properties of our target galaxies in \S\ref{sec:sampleprops}. In \S\ref{sec:observations}, we describe the ALMA observations.  A full description of our data processing pipeline is presented in a companion paper \citep{LEROY21a}, and we summarize our approach in \S\ref{sec:processing}. In \S\ref{sec:products}, we describe the properties of the science-ready data products. Then we present an atlas of the PHANGS--ALMA data in \S \ref{sec:atlas}. We give a brief summary in \S\ref{sec:summary}.

\section{Scientific Motivation}
\label{sec:motivation}

\subsection{Background}

\subsubsection{Previous Surveys of Molecular Gas in Galaxies}

\begin{figure*}[t!]
\gridline{
\fig{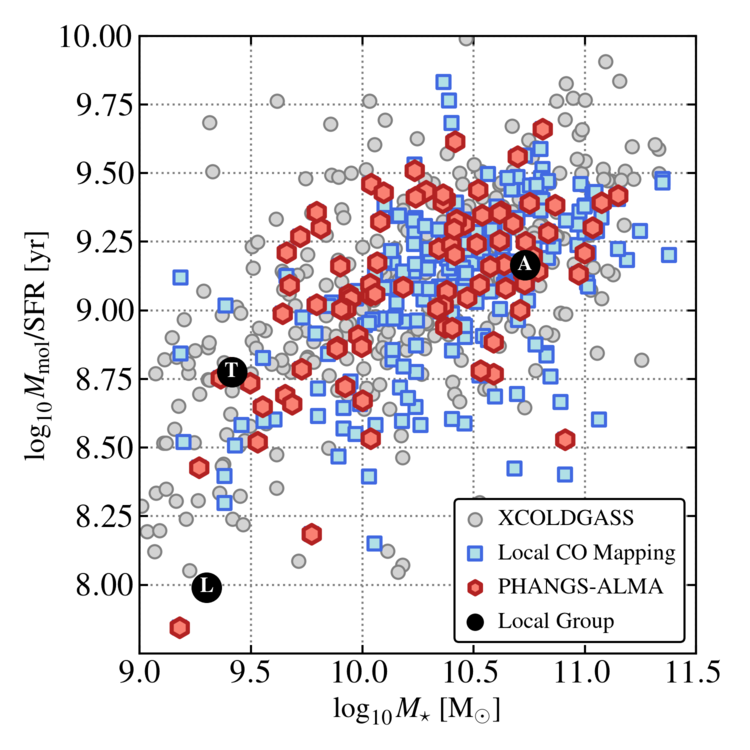}{0.475\textwidth}{}
\fig{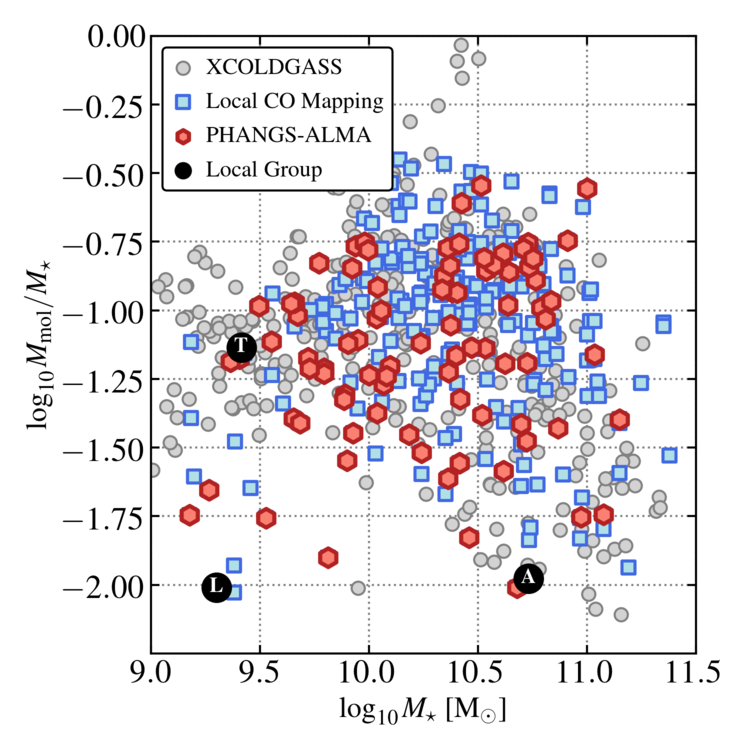}{0.475\textwidth}{}
}
\vspace{-1cm}
\gridline{
\fig{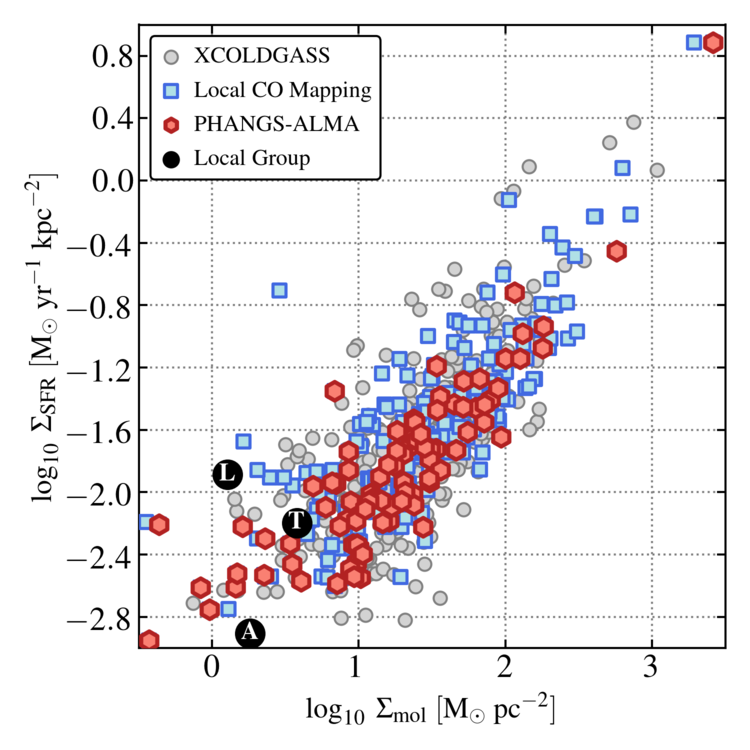}{0.475\textwidth}{}
\fig{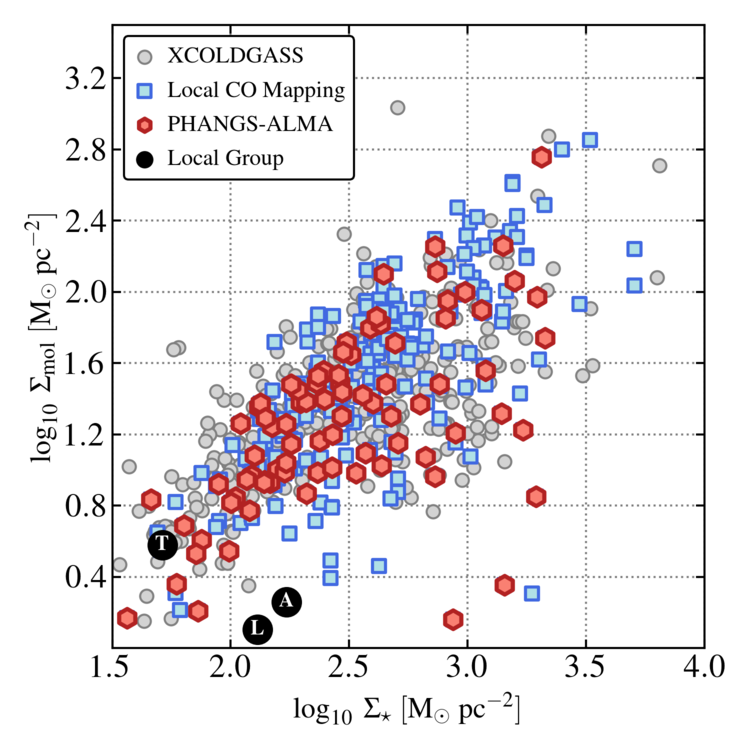}{0.475\textwidth}{}
}
\vspace{-5mm}
\caption{\textbf{Global trends in the molecular gas content of galaxies relative  to their stellar mass and star formation rates. PHANGS--ALMA aims to link these global trends to local properties and local physics in the molecular gas.} Each point in each panel shows an individual galaxy from PHANGS--ALMA (red), xCOLD~GASS \citep[gray;][]{SAINTONGE17}, or previous-generation CO mapping of local galaxies \citep[blue;][A.~Schruba et al.\ in preparation]{KUNO07,LEROY09,SORAI19}. The \textit{top left} panel shows the ratio of molecular gas mass, $M_\mathrm{mol}$, to SFR as a function of stellar mass, $M_\star$. The \textit{top right} panel shows the ratio of $M_{\rm mol}$ to $M_\star$ as a function of $M_\star$. The \textit{bottom left} panel shows the average SFR surface density, $\Sigma_{\rm SFR}$, within the half-light radius as a function of the molecular gas mass surface density, $\Sigma_{\rm mol}$. The \textit{bottom right} panel shows $\Sigma_{\rm mol}$ as a function of the average stellar surface density inside the half-light radius, $\Sigma_\star$. All four panels assume a fixed, Galactic CO-to-H$_2$ conversion factor of $\alpha_{\rm CO} = 4.35$~\acounits \citep{BOLATTO13A} and adopt a single CO~(2--1)-to-CO~(1--0) ratio, $R_{21} = 0.65$ \citep{DENBROK21,LEROY13} when needed. For local galaxies, we estimate SFR and $M_\star$ using WISE mid-IR and GALEX UV following \citet{LEROY19}. In the bottom two panels, we assume that the effective radius for stellar mass also represents the half-light radius for CO emission and SFR. For more details see \S\ref{sec:sampleprops}. For reference we show three Local Group galaxies with previous GMC-scale CO mapping: the LMC~(L), M31~(A), and M33~(T) in each panel \citep[values from][]{NIETEN06,FUKUI08,DRUARD14,JAMESON16,LEROY19}.
\label{fig:intscaling}}
\end{figure*}

Much of our knowledge about the behavior of the molecular interstellar medium (ISM) in $z=0$ galaxies has been established by CO surveys that either integrate over whole galaxies (e.g., the FCRAO survey, \citealt{YOUNG95}; AMIGA, \citealt{LISENFELD11}; COLD~GASS, \citealt{SAINTONGE11}; ALLSMOG, \citealt{BOTHWELL14}; xCOLD~GASS \citealt{SAINTONGE17}, and JINGLE, \citealt{SAINTONGE18}) or resolve the large-scale structure of galaxy disks but do not distinguish individual molecular clouds (e.g., BIMA~SONG, \citealt{HELFER03}; the Nobeyama CO Atlas, \citealt{KUNO07}; HERACLES, \citealt{LEROY09}, the JCMT NGLS, \citealt{WILSON12}; CARMA STING, \citealt{RAHMAN12}; ATLAS--3D CO, \citealt{ALATALO13, DAVIS14}; CARMA~EDGE~\citealt{BOLATTO17}; NRO~COMING, \citealt{SORAI19}; and ALMAQUEST, \citealt{LIN19}).

These surveys have demonstrated a close link between molecular gas and star formation, showing that the location and rate of recent star formation in a galaxy tracks the distribution of molecular gas \citep[e.g.,][]{WONG02,KENNICUTT07,BIGIEL08,LEROY08,SCHRUBA11}. Yet despite this good overall correspondence, observations reveal important variations in the amount of star formation per unit molecular gas both among different types of galaxies \citep[e.g.,][and see Figure~\ref{fig:intscaling}]{SAINTONGE11,LEROY13,DAVIS14,HUANG15} and within different regions of the same galaxy \citep[e.g.,][]{LONGMORE13,LEROY13,MEIDT13,MOMOSE13,LEROY17A,UTOMO17,BROWNSON20}. The normalized CO emission of galaxies also varies, with CO emission appearing fainter relative to starlight or tracers of recent star formation in low mass and early-type galaxies \citep[e.g.,][]{YOUNG91,YOUNG96,SCHRUBA12,HUNT15,SAINTONGE16,SAINTONGE17}. This change in brightness arises from both changes in molecular gas content and changes in CO emissivity per unit of molecular gas mass, but the relative magnitude of these effects is uncertain \citep[for a review see][]{BOLATTO13A}. 

Figure~\ref{fig:intscaling} illustrates some of these global trends using data from PHANGS--ALMA (red, see \S \ref{sec:sampleprops} for details), xCOLD~GASS \citep{SAINTONGE17} as the largest homogeneous unresolved survey, and a compilation of local CO mapping surveys \citep[COMING, the Nobeyama CO Atlas, HERACLES, and a HERA follow up survey;][and A. Schruba et al. in preparation]{SORAI19,KUNO07,LEROY09}. The top panels show how the ratios of molecular gas mass to star formation rate and molecular gas mass to stellar mass change across the local galaxy population. The lower panels show the relationship between the average surface densities of molecular gas, stars, and star formation. Together, the four panels of Figure~\ref{fig:intscaling} demonstrate the good overall correspondence between star formation, stellar mass, and molecular gas in galaxies, but also illustrate important variations in the molecular content of galaxies normalized by size, star formation rate, or stellar mass. The abundance, structure, and ability of molecular gas to form stars varies across the $z=0$ galaxy population.

\begin{figure*}[t!]
\gridline{
\fig{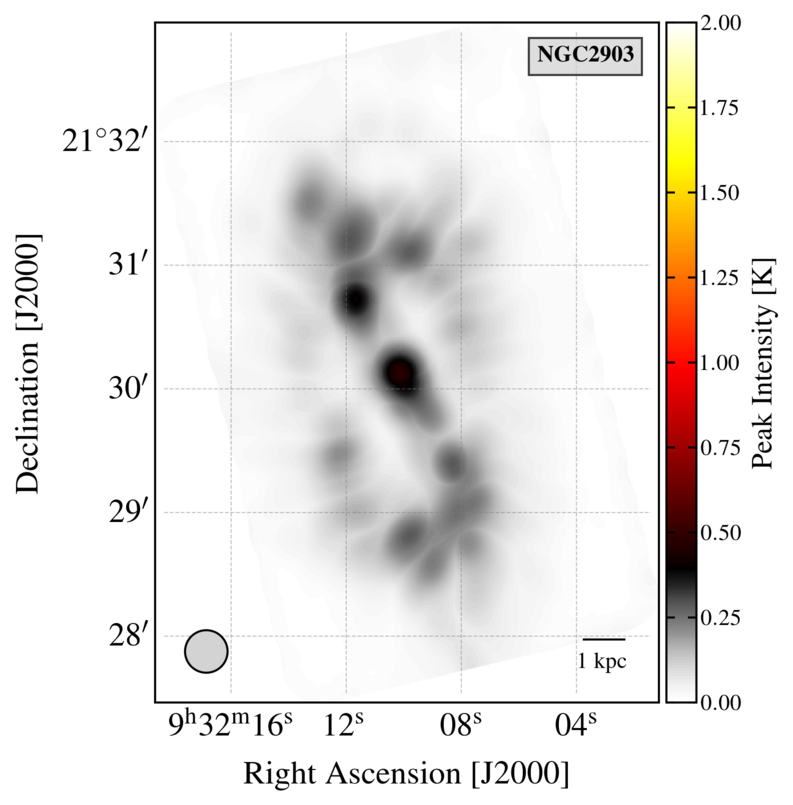}{0.4\textwidth}{}
\hspace{-3cm}
\fig{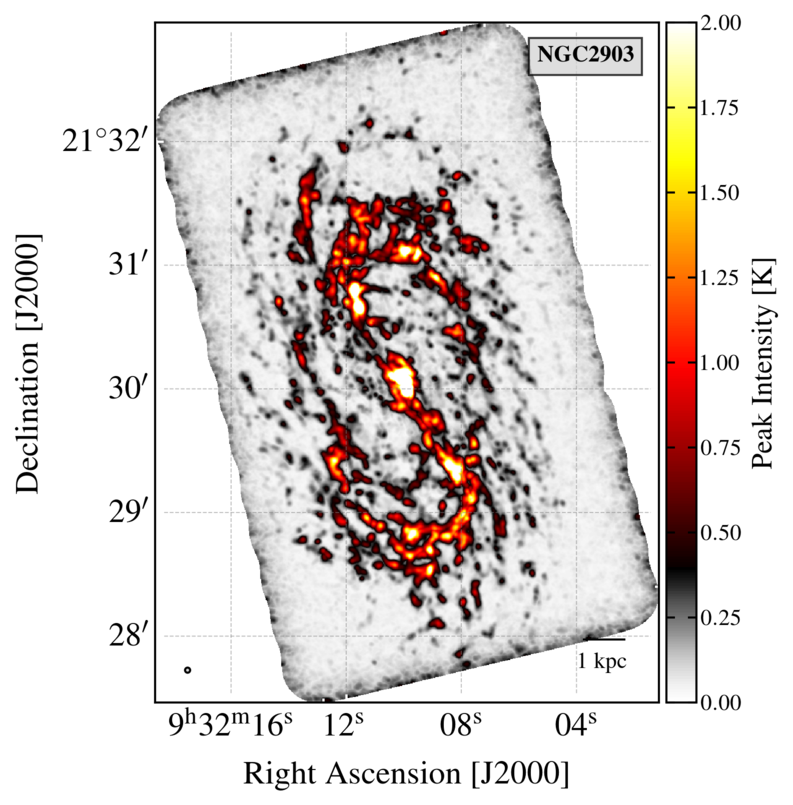}{0.4\textwidth}{}
}
\vspace{-1cm}
\gridline{
\fig{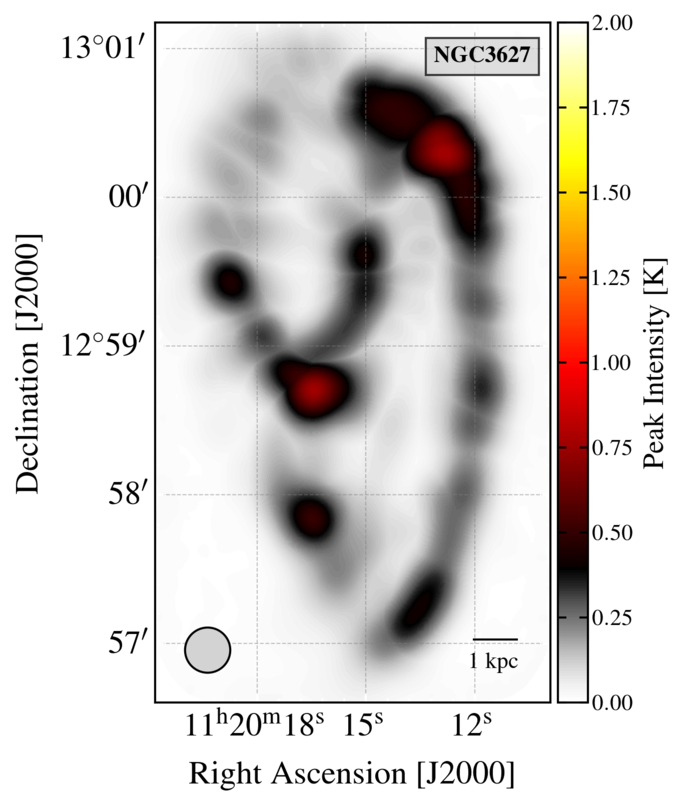}{0.4\textwidth}{}
\hspace{-3cm}
\fig{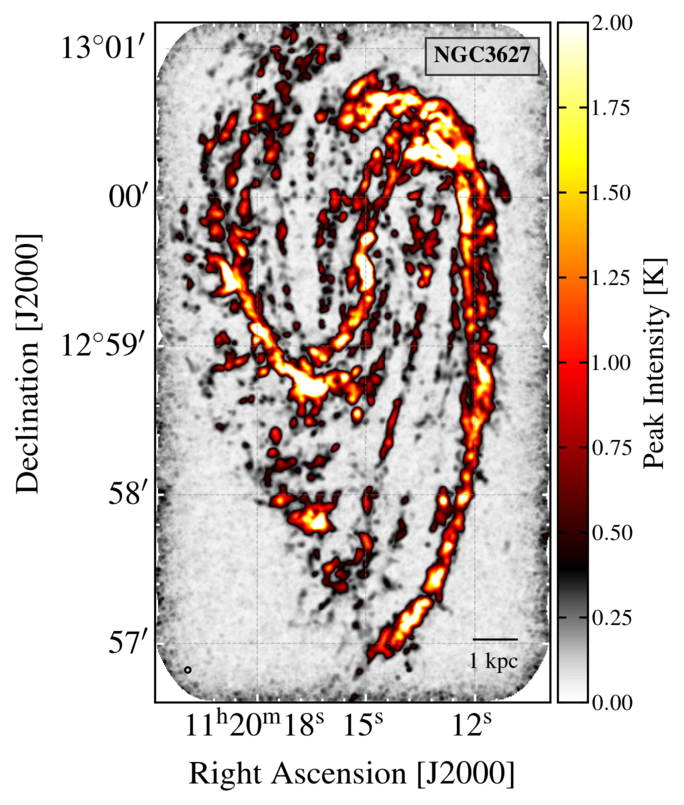}{0.4\textwidth}{}
}
\caption{\textbf{CO emission at the kiloparsec resolution of previous surveys and the ``cloud scale'' resolution of PHANGS--ALMA.} The \textit{left} panels shows \cotwo\ emission measured by PHANGS--ALMA for NGC~2903 and NGC~3627 convolved to $1$~kpc resolution, roughly corresponding to the resolution of previous large CO mapping surveys. The \textit{right} panels show the \cotwo\ emission from the same galaxies at the typical $120$~pc resolution of PHANGS--ALMA. Circles show the beam in each panel, and each map shows the maximum intensity of emission along a line of sight at 12.5 km~s$^{-1}$ velocity resolution. The high resolution view shows clumpy structures corresponding to individual massive molecular clouds. The high resolution images also show a strong influence of dynamical features; both galaxies have strong stellar bars and spiral arms.
\label{fig:kpc2cloud}}
\end{figure*}

\subsubsection{Key Physics at or Near Cloud Scales}

Unfortunately, low resolution observations offer limited insight into the physical state of molecular gas. In the Milky Way and its Local Group neighbors, most of the molecular gas resides in GMCs with masses ${\sim}10^4{-}10^{7}$~M$_\odot$. These clouds have sizes of tens of parsecs and appear dominated by supersonic turbulence \citep[e.g.,][]{SOLOMON87,BLITZ07,FUKUI10,ROMANDUVAL10,GRATIER12,HEYER15,RICE16,MIVILLE17,SCHRUBA19}. 

GMCs do not fill the galaxy disk. In low resolution extragalactic observations like those mentioned in the previous section, the CO emission from GMCs is diluted with nearby non-CO emitting regions. That is, the intrinsic distribution of CO emission in galaxies is strongly clumped on scales much smaller than the $\gtrsim$\,kiloparsec resolution of the previous generation of large CO mapping surveys \citep[e.g.,][]{LEROY13B}. Figure~\ref{fig:kpc2cloud} illustrates this phenomenon by showing CO emission from two PHANGS--ALMA targets at two resolutions: $1$~kpc and $120$~pc resolution. The sharp, clumpy structure that is striking in maps at high resolution (right panels), blurs into faint, low-contrast structures when observed at kiloparsec resolution (left panels).

Current models of star formation predict a link between star formation, feedback, and gas properties on the scale of individual GMCs, which can be inferred using high resolution observations. For example, the mean density and density distribution within a cloud may set the characteristic timescale for star formation \citep[e.g.,][]{PADOAN02,HENNEBELLE11,KRUMHOLZ05,FEDERRATH12,KRUMHOLZ12}. The strength of self-gravity in the cloud relative to turbulence and magnetic fields may affect the efficiency with which gas is converted into stars \citep[e.g.,][]{PADOAN12,PADOAN17,BURKHART18,JGKIM21}. The density, turbulence, and self-gravity may also determine how the new-born stars cluster \citep[e.g.,][]{KRUIJSSEN12,HOPKINS13d,GRUDIC20,KRUMHOLZ20}. The local gas (column) density distribution may also interact with sources of stellar feedback to determine whether a cloud is disrupted or not and how much gas and radiation leaves the system  \citep[e.g.,][]{THOMPSON05,WALCH15,THOMPSON16,GEEN16,RASKUTTI16,RASKUTTI17,REISSL18,JGKIM18,JGKIM19,GEEN21}. Because the timescale, efficiency, and spatial clustering of star formation and feedback have a qualitative impact on the GMC-scale gas properties \citep[e.g.,][]{HOPKINS13c,GENTRY17,KELLER20}, star formation, stellar feedback, and GMC properties form a complex, multi-scale system with many types of physics at play.

Dynamical processes acting on ${\sim}10{-}1000$~pc scales also play a key role in setting the abundance, structure, and ability of molecular gas to form stars \citep[for recent reviews see][]{DOBBS14,KRUMHOLZ14b,CHEVANCE20b}. As a concrete example, the high resolution images in Figure~\ref{fig:kpc2cloud} show the unmistakable imprint of a stellar bar and spiral arms in both galaxies, which is far less obvious in the low resolution maps of these targets. Spiral arm passage may collect individual quiescent molecular clouds into large star-forming associations \citep[e.g.,][]{KODA09,MEIDT15}, or trigger phase changes from the atomic to molecular medium \citep[e.g.,][]{DOBBS14}. This can trigger star formation \citep[e.g.,][]{EGUSA17} and/or organize the star-forming structures \citep[e.g.,][]{SCHINNERER17,ELMEGREEN18,TRESS20,WTKIM20}. Meanwhile, gas flows along stellar bars and arms may prevent collapse of the streaming gas, suppressing star formation \citep[e.g.,][]{MEIDT13}. These same streaming motions can also redistribute the gas, fuel star formation in the inner parts of the galaxy, or even trigger nuclear starbursts \citep[e.g,][]{KENNEY92,SAKAMOTO99A,SHETH05,SCHMIDT16}. Collisions between gas clouds may also trigger star formation \citep[e.g.,][]{TAN00,INOUE13,FUKUI20}.

\begin{figure*}[t!]
\gridline{
\fig{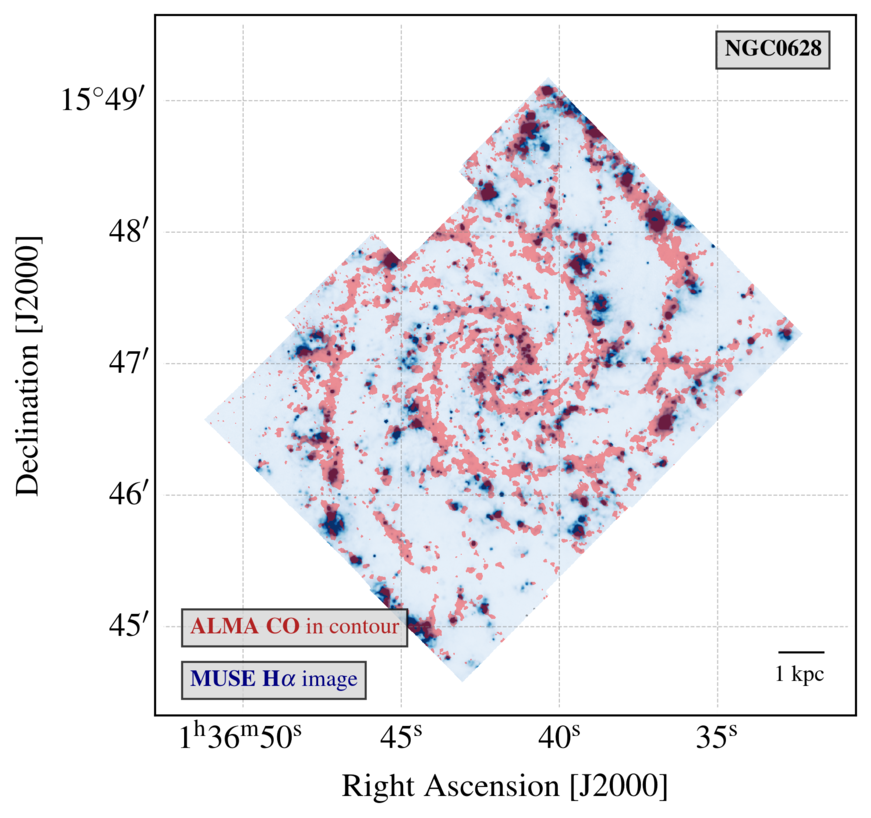}{0.4\textwidth}{}
\hspace{-3cm}
\fig{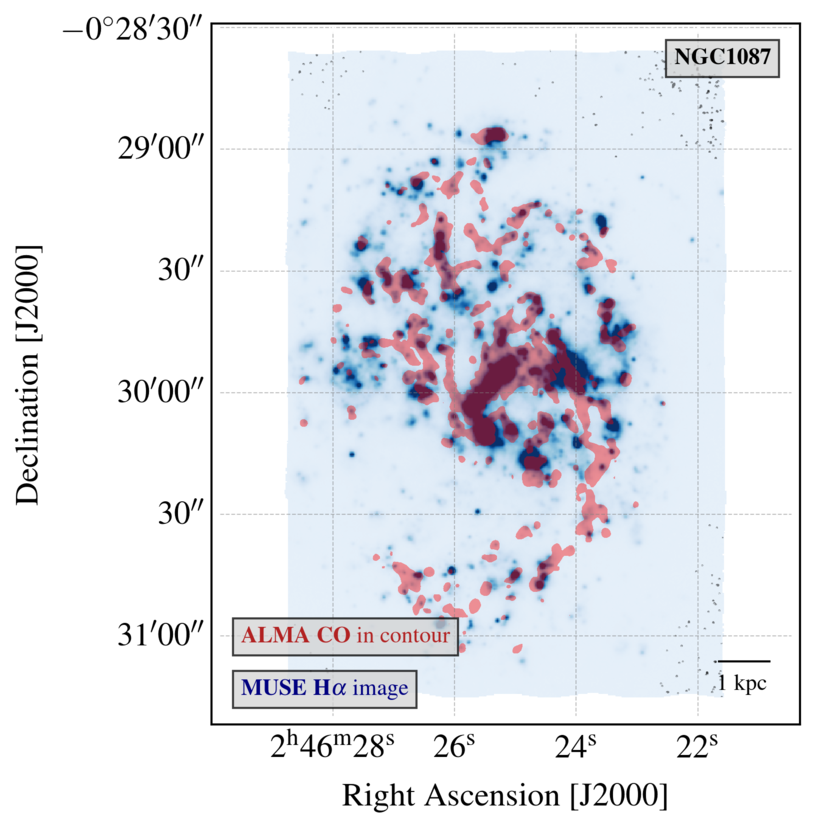}{0.4\textwidth}{}
}
\vspace{-5mm}
\gridline{
\fig{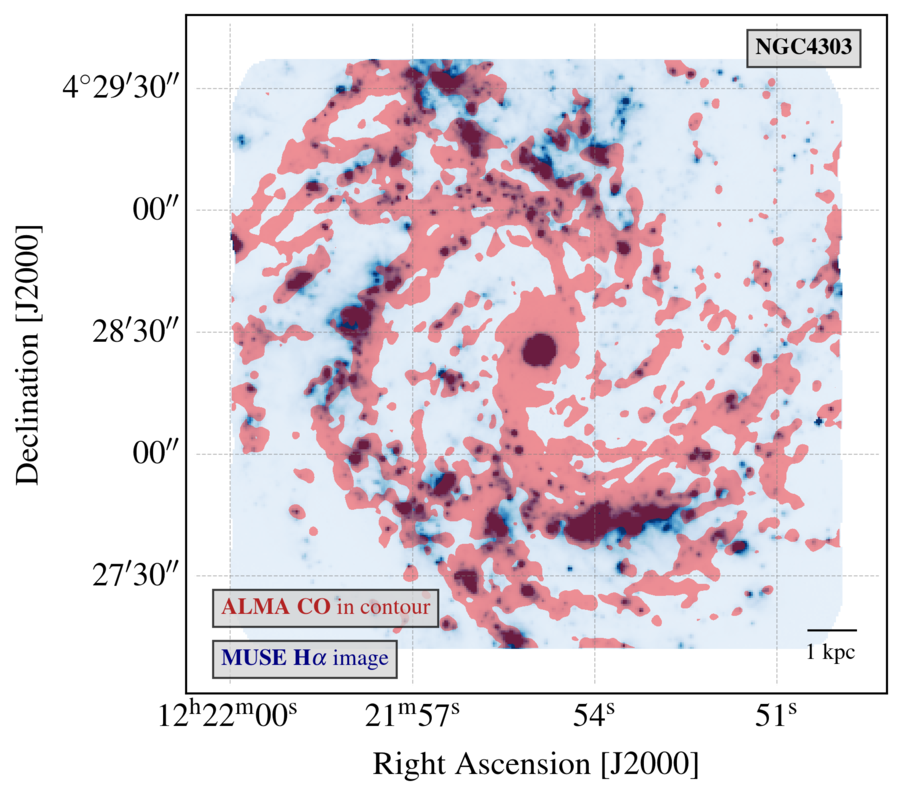}{0.4\textwidth}{}
\hspace{-3cm}
\fig{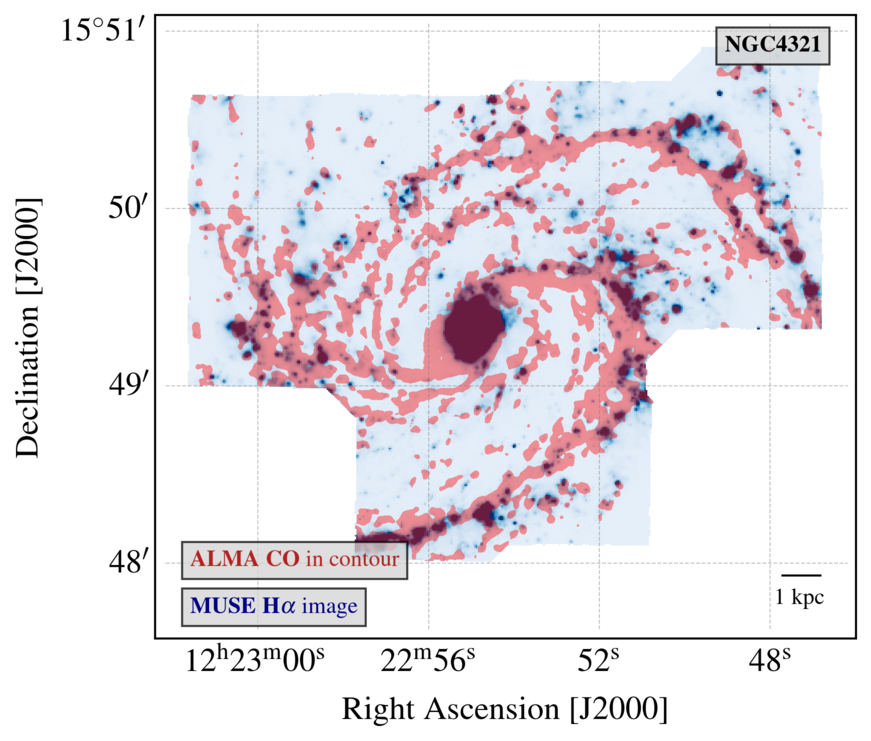}{0.4\textwidth}{}
}
\vspace{-5mm}
\caption{\textbf{At high spatial resolution, tracers of massive young stars, here H$\alpha$, and cold star-forming gas, here CO, visibly separate, providing statistical constraints on timescales for star formation and feedback.} Each figure shows the \cotwo\ peak intensity maps from PHANGS--ALMA in a red contour plotted over H$\alpha$ emission measured from VLT-MUSE integral field unit observations in blue \citep[PHANGS-MUSE, E.~Emsellem et al.\ in preparation; see also][]{KRECKEL16,KRECKEL18}. The \cotwo\ traces cold, often star-forming gas. The H$\alpha$ traces ionizing photons produced by young stars. Though the two track one another on large scales, they show distinct distributions on small scales, reflecting that they trace different phases of the star formation process \citep[e.g.,][]{KAWAMURA09,SCHRUBA10,KRUIJSSEN14}. The joint distributions of these two tracers, as well as stellar clusters identified by \textit{Hubble}, constrain the timescales associated with different phases of the star formation process. In turn, these inferred timescales offer observational constraints on topics from stellar feedback to the interplay of gravity and turbulence \citep[e.g.,][]{CHEVANCE20b}.
\label{fig:hacodemo}}
\end{figure*}

A further advantage of high resolution observations is that they give access to the temporal domain of interstellar processes. Star formation is a dynamic process, with clouds evolving rapidly under the influence of both gravity \citep[e.g.,][]{ELMEGREEN00} and ``stellar feedback'' \citep[e.g.,][]{LEE16,KRUIJSSEN19,CHEVANCE20}, a term used to describe the combined influence of ionizing photons, direct and indirect radiation pressure, gas heating, stellar winds, and supernova explosions \citep[e.g.,][]{LOPEZ11,LOPEZ14,DALE15,RAHNER17,RAHNER19}. Over the last decade it has been recognized that the details of the various stellar feedback mechanisms  have a large impact on molecular gas properties, star formation rates, and galaxy evolution \citep[e.g.,][]{HOPKINS12,AGERTZ13,WALCH15,KLESSEN16,KIMOSTRIKER17}. But many such details remain poorly constrained by observations. The interplay between turbulence and gravity also remains imperfectly understood, with models variously positing short-lived clouds in near free-fall \citep[e.g.,][]{VAZQUEZSEMADENI94,ELMEGREEN00,BALLESTEROS01,BALLESTEROS01,IBANEZMEJIA16}, star formation proceeding at a steady pace \citep[e.g.,][]{KRUMHOLZ05}, or steadily accelerating star formation within clouds \citep[e.g.,][]{MURPHY11,LEE16}.

When observed at sufficient resolution, molecular gas, \hii\ regions, stellar clusters, and other tracers of star formation and feedback visibly separate \citep[e.g.,][]{KAWAMURA09,ONODERA10,SCHRUBA10,GRATIER12,SCHINNERER13,CORBELLI17}. To illustrate this effect, we overplot the distribution of H$\alpha$ and CO emission for four PHANGS--ALMA targets in Figure~\ref{fig:hacodemo}. The distributions of CO, tracing GMCs, and of H$\alpha$, tracing recent star formation, mostly track each other at large scales, but are clearly distinct at spatial resolution of a few times $10$~pc. Comparing GMCs to stellar tracers with well-understood ages or lifespans allows one to infer the timescales for star formation and feedback from high resolution imaging \citep[e.g.,][]{KAWAMURA09,KRUIJSSEN14,CORBELLI17,KRUIJSSEN18}. This offers the prospect to build a picture of the evolutionary sequence of star formation \citep[e.g.,][]{MURRAY11,LEE16,KRUIJSSEN19}, to constrain the feedback mechanisms responsible for cloud destruction \citep[e.g.,][]{CHEVANCE21}, and to assess the fraction of non-star-forming molecular material \citep[e.g.,][]{SCHINNERER19,CHEVANCE20,KIM21}.

\subsubsection{Cloud Scale Surveys Before ALMA}

High resolution CO observations measure the physical state of the gas, probe crucial dynamical processes, and constrain the timescales for star formation and feedback. So far, most cloud scale studies of normal, non-starburst galaxies have targeted members of the Local Group \citep[see review by][]{FUKUI10}. During the past two decades, there have been high spatial resolution, wide-field mapping surveys of the CO emission in the Magellanic Clouds \citep{FUKUI99,MIZUNO01,WONG11}, M31 \citep[][]{NIETEN06,ROSOLOWSKY07,SCHRUBA21}, M33 \citep[][]{ENGARGIOLA03,ROSOLOWSKY07B,ONODERA12,DRUARD14}, and various Local Group dwarf galaxies \citep[e.g.,][]{LEROY06} as well as small-area mapping of Local Group targets with ALMA \citep[e.g.,][]{RUBIO15,SCHRUBA17,WONG19}. 

Unfortunately, the range of galaxy types and dynamical environments in the Local Group is limited. With one massive early-type spiral and a modest number of dwarf galaxies, the Local Group is not representative of the galactic environments where most star formation occurs at $z=0$. Local Group galaxies also do not harbor the environmental extremes found in more distant galaxies. While their proximity offers significant advantages in terms of resolution and surface brightness sensitivity, observations in LMC, M33, and M31 cannot capture the full range of behavior seen in Figure~\ref{fig:intscaling}. 

The lack of diversity in the Local Group is problematic because we know from the low resolution surveys discussed above that the amount and behavior of molecular gas is closely linked to properties of the host galaxy. The balance between atomic gas and molecular gas depends sensitively on the interstellar pressure and local dust content \citep[e.g.,][]{WONG02,BLITZ06,LEROY08,WONG13,SCHRUBA18,SUN20}. The distribution of molecular gas strongly reflects the structure of the stellar disk \citep[e.g.,][]{YOUNG95,REGAN01,LEROY08,SCHRUBA11}. These trends also hold across the whole galaxy population \citep[e.g.,][]{YOUNG91,YOUNG95,SAINTONGE11,SAINTONGE17}, since the molecular gas content, or at least the CO emission, of a galaxy depends strongly on its mass and metallicity \citep[e.g.,][]{SCHRUBA12,BOTHWELL14,HUNT15,SAINTONGE17}.

Mapping CO emission at the scale of individual GMCs across a diverse, representative sample of star-forming galaxies is thus the logical step forward in this field. Before ALMA, however, mapping GMC-scale CO emission from a single normal star-forming galaxy required a major time investment. As a result, mapping studies --- especially with the PdBI and OVRO interferometers --- typically focused on bright, compact starburst galaxies \citep[e.g.,][]{DOWNES98} and nuclear regions hosting starbursts and active galactic nuclei \citep[e.g., NUGA and MAIN;][]{GARCIABURILLO03,JOGEE05}. After a number of studies targeting individual galaxies or dwarf galaxies \citep[e.g.,][]{ROSOLOWSKY05,BOLATTO08,RAHMAN11}, the CANON survey took an important first step toward synthetic cloud-scale imaging of a sample of normal galaxies beyond the Local Group. CANON mapped \coone\ emission at ${\sim}2$\arcsec\ resolution over the inner regions of a sample of spiral galaxies \citep[][]{KODA09,DONOVANMEYER12,DONOVANMEYER13,MOMOSE13}. This survey provided important evidence for variations in molecular gas properties as a function of galactic environment, and for the role of spiral arms in GMC formation and evolution.

\begin{figure*}[t!]
\begin{center}
\includegraphics[width=0.40\textwidth]{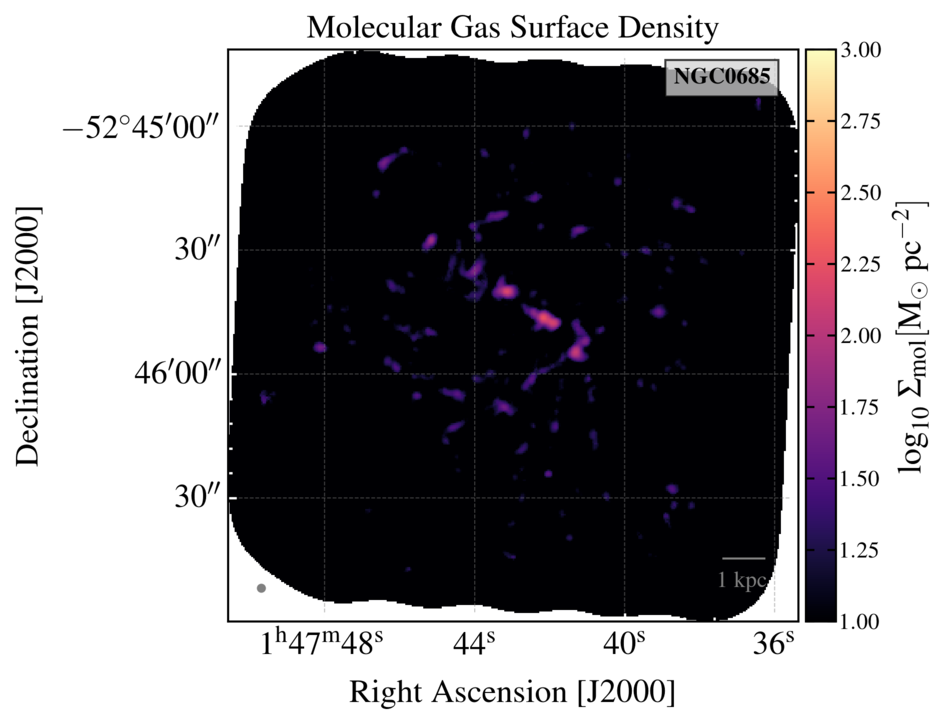}
\hspace{1cm}
\includegraphics[width=0.40\textwidth]{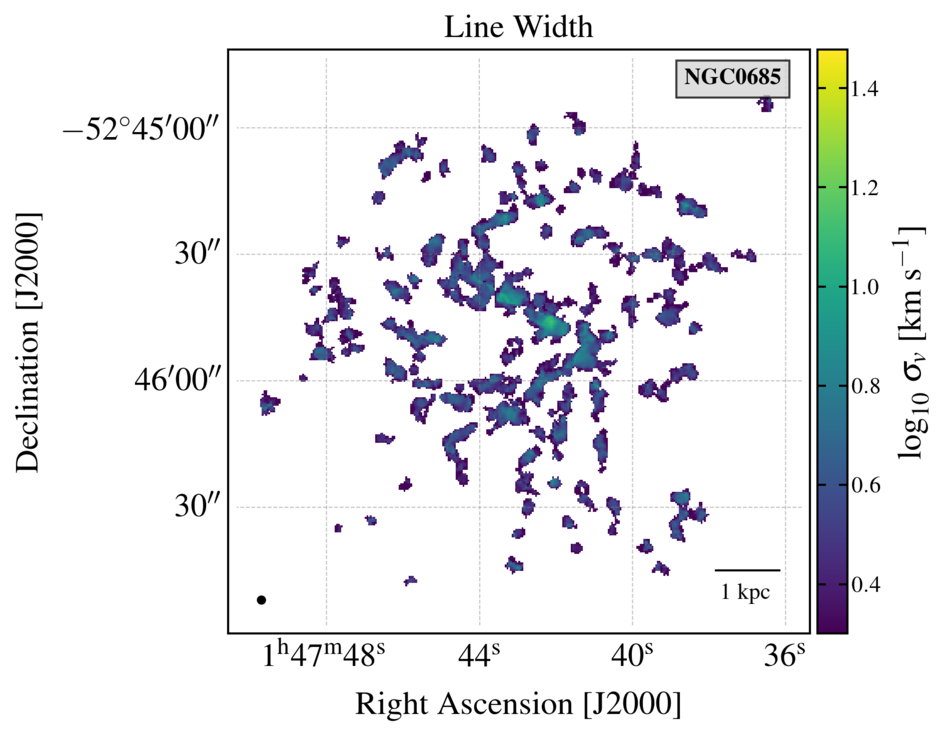}

\includegraphics[width=0.40\textwidth]{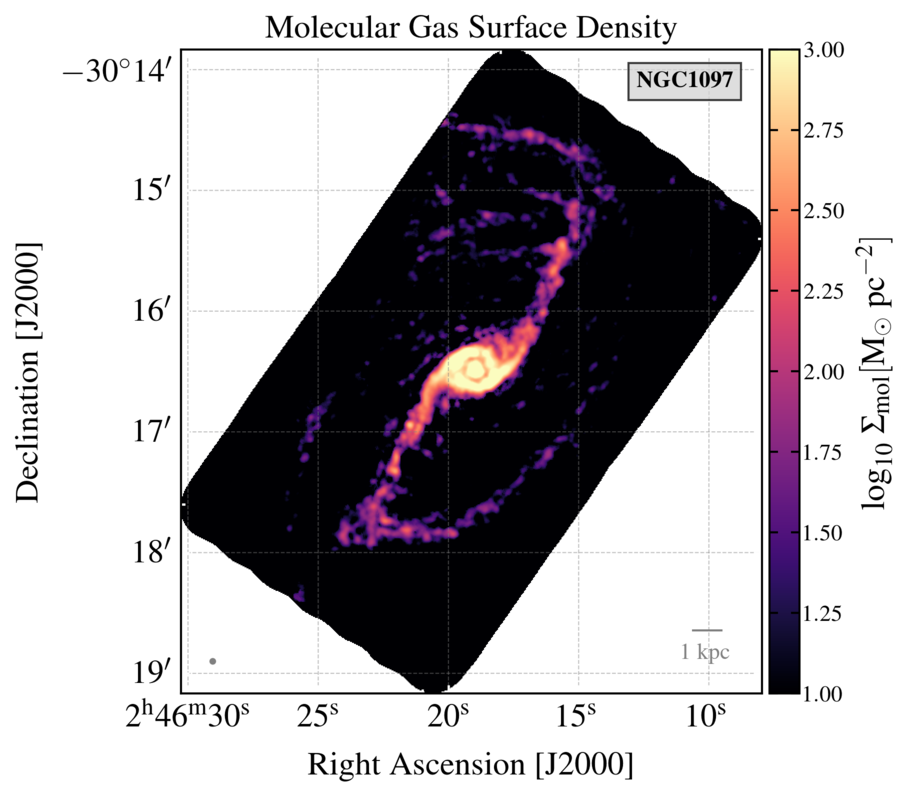}
\hspace{1cm}
\includegraphics[width=0.40\textwidth]{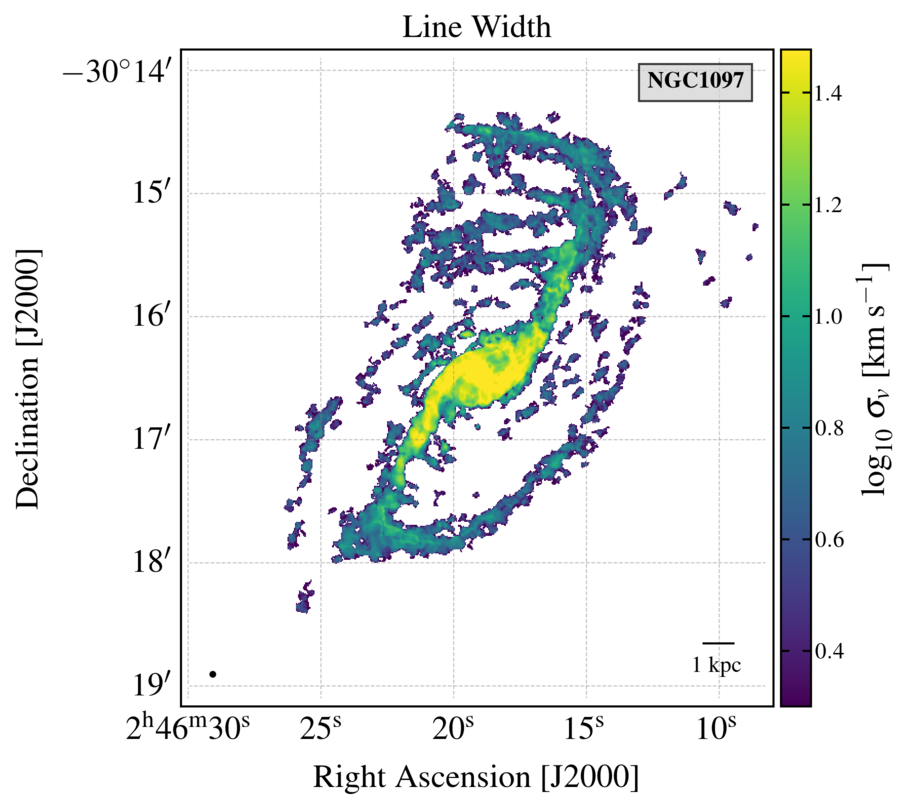}

\includegraphics[width=0.38\textwidth]{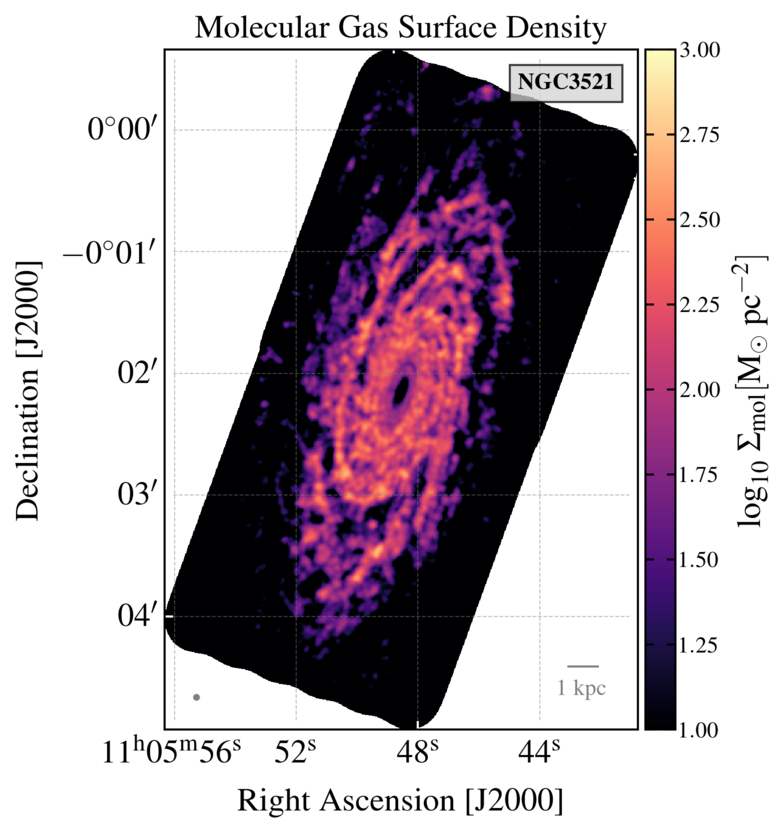}
\hspace{1.5cm}
\includegraphics[width=0.38\textwidth]{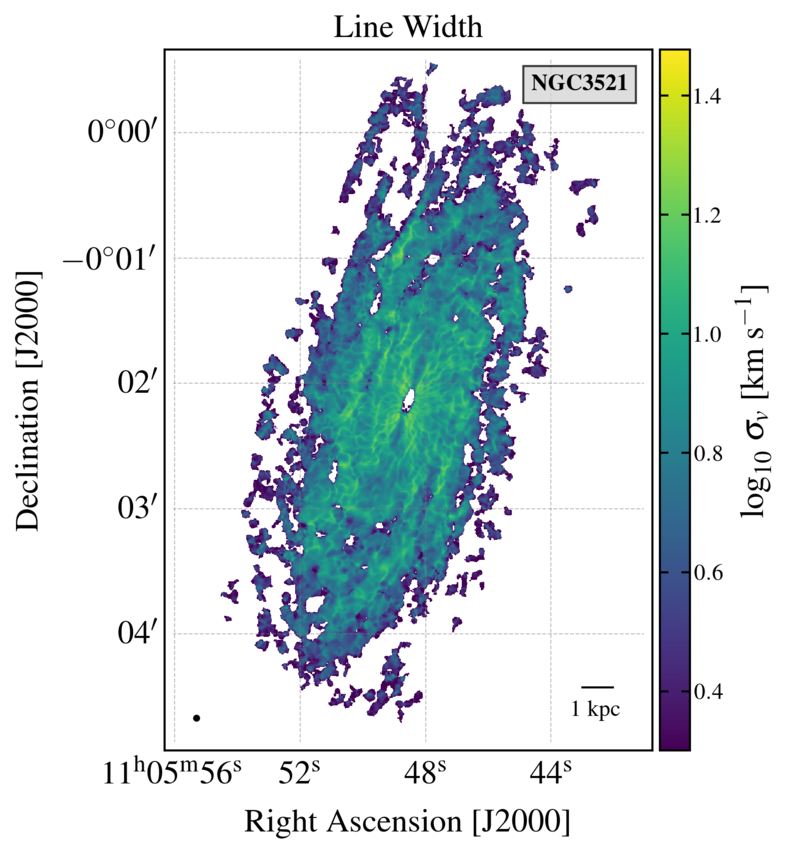}
\end{center}
\caption{\textbf{Cloud scale molecular gas properties depend on host galaxy and location in the galaxy.} PHANGS--ALMA \cotwo-based estimates of (\textit{left}) molecular gas surface density, $\Sigma_{\rm mol}$, and (\textit{right}) effective line width, $\sigma_v$, for three galaxies. All three galaxies are shown at the same physical resolution of $150$~pc and using the same logarithmic color stretch. The galaxies show striking differences in their overall surface density and line width, the morphology of their gas distribution at $150$~pc scales, and in the distributions of gas and line width within each galaxy. Note that the line width is only measured within pixels detected at high confidence, while the integrated intensity is estimated over an area selected for completeness. See \S \ref{sec:products} for details on the calculation individual maps and \S \ref{sec:atlas} for similar images of the whole sample.
\label{fig:varyingclouds}}
\end{figure*}

Subsequently, the PdBI Arcsecond Whirlpool Survey \citep[PAWS;][]{SCHINNERER13,PETY13} mapped M51 at $1\arcsec \approx 40$~pc resolution \citep[improving on a similar effort at $3\arcsec$ resolution using CARMA by][]{KODA09}. At this high resolution, the contrast between the molecular gas in M51 and that of Local Group galaxies proved striking \citep[e.g.,][]{HUGHES13A, HUGHES13B}. Figure \ref{fig:varyingclouds} illustrates a similar contrast. It shows that at fixed $150$~pc resolution, the surface density and line width of molecular gas vary significantly and systematically as a function of location in the galaxy and among host galaxies. Analysis of the PAWS data helped establish that the cloud-scale structure of the molecular ISM depends on dynamical environment and host galaxy properties \citep[][]{HUGHES13A, HUGHES13B, COLOMBO14B, LEROY16} and showed that the local star formation activity in M51 depends on cloud-scale ISM structure \citep[][]{MEIDT13, LEROY17A}. PAWS studies also demonstrated how high resolution imaging yields insight into the evolution and timescales of individual star-forming regions \citep[][]{SCHINNERER13,MEIDT15,SCHINNERER17}.

In spite of these important first efforts, the sample size, field of view, and sensitivity of high physical resolution CO observations targeting normal, star-forming galaxies remained limited before ALMA, preventing 
%Simply put, there has been no 
a synthetic view of molecular gas properties across the full local galaxy population.

\subsection{PHANGS--ALMA}

ALMA has transformed our ability to observe molecular line emission from nearby galaxies. In ${\sim}1{-}2$~hours of on-source main array time, ALMA can map \cotwo\ emission at $1\arcsec$ resolution across a ${\sim}2\arcmin \times 2\arcmin$ field with two times better sensitivity than achieved by PAWS. For comparison, PAWS required almost $130$~hours on-source to map the \coone\ emission from the inner $4.5\arcmin \times 3\arcmin$ of M51 with the IRAM PdBI \citep[][]{PETY13}. The dramatically faster survey speed of ALMA provides the first opportunity of surveying a large, representative sample of local galaxies at ${\sim}1\arcsec$ resolution.

PHANGS--ALMA applies these capabilities to image \cotwo\ emission across almost all massive, nearby, southern, star-forming galaxies (see \S \ref{sec:sample}). The key elements of the survey are:

\begin{enumerate}
\item Observations of \cotwo\ emission across the region of active star formation of each galaxy with sensitivity to detect individual GMCs ($M_{\rm mol} > 10^5$~M$_\odot$) along each line of sight.
\item \ntarget\ targets with a simple selection function (\S\ref{sec:sample}) that selects most local massive galaxies on the star-forming main sequence (\S\ref{sec:sampleprops}).
\item $1'' \sim 100$~pc angular and physical resolution, $\approx 2.5$~km~s$^{-1}$ velocity resolution (\S\ref{sec:observations}), and inclusion of short-spacing data to ensure complete flux recovery.
\end{enumerate}

The core of the survey is a Cycle~5 ALMA Large Program (PI: E.~Schinnerer) that mapped \cotwo\ emission at ${\sim}1\arcsec$ resolution from $58$ galaxies (\S\ref{sec:observations}). The Large Program built on several smaller pilot programs, and was supplemented by observations to complete the sample and extend the range of parameter space studied. Wherever feasible, archival ALMA \cotwo\ observations that match the PHANGS--ALMA observing strategy have been incorporated into the PHANGS sample for processing and analysis (see \S\ref{sec:sample} and~\S\ref{sec:observations}).

Because our selection strategy (\S\ref{sec:sample}) is simple, our targets include almost every massive, star-forming galaxy visible to ALMA within $12$~Mpc. Our coverage of more distant targets, with $12 \lesssim d \lesssim 17$~Mpc, is also good, but less complete due to the combination of distance uncertainties (where targets have a true distance $<17$~Mpc but a measurement of $d>17$~Mpc) and volume effects (the number of galaxies with distances $<d$ grows $\propto d^3$ precluding complete coverage in a reasonable amount of time).

This simple selection strategy yields a diverse sample of galaxy types (\S\ref{sec:sampleprops}). Our targets span more than a decade in stellar mass, star formation rate, and specific star formation rate. They include strongly barred galaxies, grand design spirals, flocculent galaxies, and even some early-type galaxies. In the ${\rm SFR}{-}M_\star$ space commonly used to discuss galaxy evolution, the PHANGS--ALMA targets provide good sampling of the local ``main sequence'' of star-forming galaxies \citep[][]{NOESKE07}.

We refer to PHANGS--ALMA as a ``cloud scale'' spectroscopic imaging survey. This means that our resolution and sensitivity are well matched to the scale of an individual GMC. The ${\sim}100$~pc resolution of PHANGS--ALMA matches the thickness of the molecular disk in the Milky Way and other galaxies \citep[][]{HEYER15,YIM20}. Our beam also has roughly the same diameter as massive GMCs, which are often found to have radii of ${\sim}30{-}60$~pc \citep[e.g.,][]{SOLOMON87,BOLATTO08,COLOMBO14A,FREEMAN17,MIVILLE17,ROSOLOWSKY21}. The point source sensitivity of PHANGS--ALMA is also well matched to detecting individual GMCs, with a characteristic mass scale of $M_{\rm mol} \approx 10^5$~M$_\odot$ and power-law mass distribution \citep[e.g.,][]{FUKUI10}. These characteristics make PHANGS--ALMA ideally suited to measure the demographics, motions, and organization of molecular gas (clouds) in galaxies.

The PHANGS--ALMA imaging is not designed to heavily resolve individual GMCs. Instead, we target cloud scale resolution across entire galaxies for a large sample. Although ALMA can achieve resolutions much better than $1\arcsec$ at the $\nu = 230$~GHz of \cotwo, such observations have poor surface brightness sensitivity and require a prohibitive amount of time to detect CO emission from molecular clouds. Specifically, the integration time ($t$) required to reach a fixed surface brightness sensitivity at a resolution ($\theta$) scales as $t \propto \theta^{-4}$. Thus, even targeting an order of magnitude poorer sensitivity, ALMA could only survey one or two nearby galaxies at ${\sim}10$~pc resolution during the time required to map all \ntarget\ PHANGS--ALMA targets at ${\sim}100$~pc resolution.

\subsection{Science goals}

The science goals driving the PHANGS--ALMA survey design motivate this ``cloud scale imaging'' philosophy. We constructed the survey to address major open questions about the demographics of GMCs, the life cycle of star-forming regions, and the link between cloud scale physics, galactic scale processes and host galaxy properties.

The sample selection (\S\ref{sec:sample}) and observing strategy (\S\ref{sec:observations}) for PHANGS--ALMA were designed to address five core science goals:

\begin{enumerate}
\setcounter{enumi}{0}
\item \textit{Measure the demographics of molecular clouds, and measure how GMC populations depend on host galaxy and location in a galaxy.}
\end{enumerate}

Despite more than three decades studying GMCs in other galaxies, we lack a quantitative, observationally-grounded understanding of their demographics. Put another way, we still lack an answer to the question: ``For a given set of local conditions inside a given host galaxy, what population of GMCs should be present?'' 

As described above, this mostly reflects the technical obstacles to observing entire GMC populations before ALMA. These limitations induced GMC studies to focus on a handful of nearby galaxies, e.g., the LMC, M33, M31, and M51.

PHANGS--ALMA aims to change this situation by measuring the distributions of GMC mass, line width, surface density, internal pressure, and virial parameter\footnote{We adopt a simple virial parameter definition $\alpha_\mathrm{vir} = 2K/U_{\rm g}$, where $K$ is the kinetic energy and $U_g$ is the gravitational potential energy.} in each region of each galaxy. Because we target a diverse sample of galaxies that represents where stars are forming at $z=0$, we expect PHANGS--ALMA to provide a solid empirical foundation to understand the link between GMCs, host galaxy, and dynamical environment. These measurements will provide important constraints on GMC formation, destruction, and evolution \citep[e.g.,][]{JEFFRESON18}.

This will quantitatively connect GMC studies to models of galaxy evolution \citep[e.g.,][]{SOMERVILLE15} and provide key benchmarks for numerical simulations aiming to ``get the cold gas right'' \citep[e.g.,][]{DOBBS19,JEFFRESON20,TRESS20b}. First work on this topic using PHANGS--ALMA appears in \citet{SUN18,SUN20,SUN20B}, \citet{HERRERA20}, and \citet{ROSOLOWSKY21}.

This science goal drives us to observe a galaxy sample that spans the star-forming main sequence, to reach a resolution that approaches the scale of individual GMCs, and to achieve sensitivity to individual GMCs.

\begin{enumerate}
\setcounter{enumi}{1}
\item \textit{Measure the star formation efficiency per free fall time, $\epsilon_{\rm ff}$, at cloud scales. Measure how $\epsilon_{\rm ff}$ depends on the density, dynamical state, and turbulence in molecular clouds.} 
\end{enumerate}

Star formation is inefficient: only a small fraction of the mass of a cloud is converted to stars over the time it takes for the cloud to gravitationally collapse \citep[e.g.,][]{ZUCKERMAN74, MCKEE07}. Over the last two decades, many analytic and numerical models have considered star formation in turbulent molecular clouds (e.g., following \citealt{PADOAN02} and \citealt{KRUMHOLZ05}). These models often treat the efficiency of star formation relative to direct collapse, i.e., the ``star formation efficiency per free fall time,'' as a crucial prediction \citep[e.g., see a synthesis in][]{FEDERRATH12,FEDERRATH13}. 

Put more simply, much work over the last two decades views either the gravitational free fall time at the scale of an individual cloud ($\tau_{\rm ff} \propto \rho^{-0.5}$, with $\rho$ the gas volume density) or the turbulent crossing time ($\tau_{\rm cross} \propto l / \sigma$, with $\sigma$ the turbulent velocity dispersion) as the relevant timescale for star formation \citep[e.g.,][]{ELMEGREEN00,HARTMANN01,MACLOW04,KRUMHOLZ12,PADOAN16}. In this view, the relevant efficiency for star formation is the fraction of gas converted to stars over the relevant timescale, e.g., $\epsilon_{\rm ff} \equiv M_{\rm mol}/{\rm SFR}/\tau_{\rm ff}$ or  $\epsilon_{\rm cross} \equiv M_{\rm mol}/{\rm SFR}/\tau_{\rm cross}$.

These models are increasingly central to how we understand star formation in galaxies \citep[see the review by][]{KRUMHOLZ19}. Testing them requires estimating the key timescales, $\tau_\mathrm{ff}$ and $\tau_\mathrm{cross}$, on the scales of interest. In turn, estimating these timescales requires measuring the density and velocity dispersion of cold gas at the scale of an individual GMC. This requires at least ``cloud scale'' resolution. Because such observations have been scarce, direct measurements of $\epsilon_\mathrm{ff}$ and tests of turbulent models have been mostly confined to studies of the Milky Way \citep[][]{EVANS14,VUTISALCHAVUAKUL16} and a handful of the nearest galaxies \citep[e.g.,][]{LEROY17A,OCHSENDORF17,SCHRUBA19}. 

By making measurements of the mass surface density and line width of cold gas at cloud scales, PHANGS--ALMA yields estimates of $\tau_\mathrm{ff}$ and $\tau_\mathrm{cross}$. Combining these with measurements of SFR and the total molecular gas reservoir, $M_{\rm mol}$, allows us to make resolved estimates of $\epsilon_\mathrm{ff}$ across the whole local galaxy population. 

This is the second core science goal of PHANGS--ALMA: to measure $\epsilon_{\rm ff}$ across the local galaxy population and quantify how $\epsilon_{\rm ff}$ depends on host galaxy properties and local conditions in the cold gas. Doing so, we aim to provide a benchmark and test for current and future models of star formation in molecular clouds. First work on this topic using the pilot PHANGS--ALMA data appears in \citet{KRECKEL18} and \citet{UTOMO18}. Similar to the first goal, this science goal drives us to observe a diverse galaxy sample, to reach a resolution that approaches the scale of individual GMCs, and to achieve sensitivity to individual GMCs.

Combining these first two goals, we aim to link the observed global trends in molecular gas content and star formation within the molecular gas to local physics. We will measure how molecular cloud populations depend on local and global environment, and we will also measure how the properties of molecular clouds affect the star formation and feedback process. This will allow us to understand if global trends in the gas depletion time stem from underlying changes in the GMC population.

\begin{enumerate}
\setcounter{enumi}{2}
\item \textit{Quantify the ``violent cycling'' between phases of the star formation process. Use this to constrain the life cycle of clouds and feedback.} 
\end{enumerate}

Several lines of evidence suggest that GMCs experience dramatic evolution and violent disruption on timescales of a few Myr to a few tens of Myr \citep[e.g.,][among many others]{KAWAMURA09,MEIDT15,LEE16,CORBELLI17,KRUIJSSEN19}. The details of stellar feedback, its interaction with the ISM, the preconditions for star formation on cloud scales, and the dominant mechanism for cloud disruption all remain highly uncertain and areas of active theoretical research \citep[e.g.,][]{GATTO15,JEFFRESON18,SEMENOV18}.

A main way to constrain these physics is to measure the relative distributions of emission tracing different phases of the star formation process. At ${\sim}100$~pc resolution, GMCs, \hii\ regions, and stellar clusters appear distinct from one another \citep[e.g.,][among many others, including a first illustration of PHANGS--ALMA data in \citealt{KRECKEL18}]{KAWAMURA09,SCHRUBA10}.  Figure~\ref{fig:hacodemo} illustrates the dissimilar spatial distributions of H$\alpha$ and CO, which is thought to reflect the evolution of star-forming regions \citep[e.g.,][]{KAWAMURA09,SCHRUBA10, KRUIJSSEN14}, e.g., from quiescent clouds to star-forming clouds to disrupted clouds. In the simplest terms, the fraction of clouds in different states maps to the timescales for a cloud to evolve through that state, though more sophisticated modeling techniques have been developed \citep[e.g.,][]{KRUIJSSEN18}, including treatment of gas flows along streamlines \citep[e.g.,][]{MEIDT15,EGUSA17}.

Despite many observations of \hii\ regions and stellar clusters at $\lesssim 100$~pc resolution, sensitive, wide area CO observations that isolate individual clouds have been scarcer and mostly focused on the Local Group. PHANGS--ALMA aims to change this situation, producing CO maps suitable to combine with H$\alpha$ maps, HST-based cluster catalogs, and integral field unit data to constrain the timescales and evolutionary sequence of GMCs across many environments. First work on this topic using the PHANGS--ALMA data appears in \citet{KRECKEL18}, \citet{SCHINNERER19}, and \citet{CHEVANCE20,CHEVANCE21}.

This science goal requires PHANGS--ALMA to observe CO with high enough resolution to resolve the discrete distributions of molecular gas for comparison to high resolution maps of ionized gas and young stars. The required resolution varies, but is usually better than a few hundred parsecs \citep[e.g.,][]{CHEVANCE20}. As for the first and second goal, the great diversity of galaxies observed in PHANGS--ALMA is instrumental for quantifying how the lifecycle of GMC evolution, star formation and feedback may vary with the galactic environment.

\begin{enumerate}
\setcounter{enumi}{3}
\item \textit{Measure how the self-regulated, large scale structure of galaxy disks emerges from a medium made of individual clouds and star-forming regions.} 
\end{enumerate}

The disks of normal, star-forming galaxies at $z=0$ are often viewed as quasi-equilibrium systems. With that framework, vertical force balance is described by hydrostatic equilibrium with a dynamical pressure term balancing gravity \citep[e.g.,][among many others]{ELMEGREEN89,BLITZ06,OSTRIKER10}. Radial equilibrium is often considered in terms of Toomre stability \citep[e.g.,][]{KENNICUTT89,SILK97,THOMPSON05}. 

Past observations testing these models mostly had $\sim$kpc resolution, i.e., resolving galaxy disks but not breaking emission into individual star-forming regions \citep[e.g.,][]{MARTIN01,WONG02,BOISSIER03,LEROY08,COLOMBO18}. Though simulations have explored how self-regulation emerges from a chaotic, high resolution view of the ISM spanning molecular clouds to galactic disks \citep[e.g.,][]{KIM13,ORR18}, few observations had a comparable dynamic range. As a result, we lack a clear measurement of how the physical effects thought to be essential for self regulation --- turbulent motions \citep[e.g.,][]{FEDERRATH10,PADOAN16}, supported by dynamical forces due to the galactic potential \citep[e.g.,][]{MEIDT18,MEIDT20}, and gravitational collapse \citep[e.g.,][]{VAZQUEZSEMADENI94,IBANEZMEJIA16} --- relate to one another as a function of scale.

Concretely, PHANGS--ALMA aims to assess force (pressure) balance in the radial and vertical directions as a function of scale. This will place strong observational constraints on the dynamical state of the ISM and the scales on which the self-regulation of galactic disks sets in. First work on this topic using the PHANGS--ALMA data appears in \citet{SUN20}.

This goal requires high physical resolution to break the molecular gas into individual clouds, high spectral resolution to assess the kinetic energy and other motions, and the inclusion of short-spacing data to allow studies that span a broad range of spatial scales.

\begin{enumerate}
\setcounter{enumi}{4}
\item \textit{Measure the motions, flows, and organization of cold gas in galaxies at ${\sim}$100$-$1,000~pc scales.} 
\end{enumerate}

At ${\sim}100$~pc resolution, CO maps of massive disk galaxies reveal a highly structured medium \citep[e.g.,][]{SCHINNERER13,HIROTA18,SUN18,SCHRUBA21}. Many galaxies show strikingly well-defined features associated with gas flows along bars, gas in spiral arms, and ``feathers'' associated with spiral arms \citep[e.g.,][]{LYNDS70,LAVIGNE06,CORDER08,SCHINNERER17}. 

This structure is not captured either in low resolution studies of galaxy disks or GMC studies, which treat the gas as individual units. The last main goal of PHANGS--ALMA is to quantitatively characterize the structure and motions of the gas on scales bigger than a cloud but below the $\sim$kpc resolution at which disks appear relatively smooth. 

Among metrics, we aim to quantify gas clumping \citep{LEROY13B}, concentration into spiral arms \citep[e.g.,][]{FOYLE10}, and organization into filamentary structures \citep[e.g.,][]{JACKSON10,KOCH15,ZUCKER18}. Our goal is to approach these measurements in a quantitative, reproducible manner, similar to the techniques used to characterize density fields in studies of large scale structure \citep[for a recent applications to CO maps see][]{GRASHA18}. These measurements will represent sophisticated benchmarks for simulations aiming to reproduce realistic cold gas structure or simulated CO emission.

The high-resolution kinematic information in PHANGS--ALMA also allows qualitatively new measurements related to these same phenomena. With high signal to noise, velocity resolution, and spatial resolution, we aim to measure streaming motions along spiral arms and bars, search for colliding gas flows, signatures of gas inflow, and to identify when---and if---self-gravitating gas structures decouple from the global velocity field \citep[e.g.,][]{ROSOLOWSKY03,BRAINE18,MEIDT18,HERRERA20}.  The first application of the PHANGS data to measure detailed kinematic structure appears in \citet{HENSHAW20} and \citet{LANG20}.

Together, these five science goals inform the sample selection (\S\ref{sec:sample}) and observing strategy (\S\ref{sec:observations}) of PHANGS--ALMA. All can be met by a sensitive, wide area CO survey of a representative sample of galaxies that reaches cloud scale resolution. The rest of this paper describes our sample selection (\S\ref{sec:sample}), current best-estimate properties of the selected galaxies (\S \ref{sec:sampleprops}), observation design and execution (\S\ref{sec:observations}), processing pipeline (\S\ref{sec:processing}), and the resulting data (\S\ref{sec:products}).

\section{Sample Selection}
\label{sec:sample}

PHANGS--ALMA aims to obtain cloud-scale CO maps for all ALMA-visible, massive, star-forming disk galaxies out to the near side of the Virgo Cluster. In this section, we discuss the motivation (\S\ref{sec:selectmotivation}), implementation (\S\ref{sec:selectimplement}), and uncertainties associated with our selection strategy (\S\ref{sec:selectaccuracy}).  Further, we present several extensions to the main sample in \S\ref{sec:selectextensions}.

\subsection{Requirements}
\label{sec:selectmotivation}

\begin{deluxetable}{ll}
\tabletypesize{\small}
\tablecaption{PHANGS--ALMA Selection \label{tab:selection}}
\tablewidth{0pt}
\tablehead{
\colhead{Quantity} & 
\colhead{Value}
}
\startdata
\hline
\multicolumn{2}{c}{Selection criteria for main sample\tablenotemark{a}} \\
\hline
Declination & $-75\degr< \delta < +25\degr$ \\
Inclination & $i < 75\degr$ \\
Distance & $d < 17$~Mpc \\
$\log_{10} M_\star [{\rm M}_\odot]$ & $>9.75$ \\
$\log_{10} {\rm SFR/M}_\star [{\rm yr}^{-1}]$ & $>-11$ \\
Main sample selection\tablenotemark{b} & $75$ galaxies \\
Extensions\tablenotemark{c} & $15$ galaxies \\
\hline
\multicolumn{2}{c}{Monte Carlo results (\S \ref{sec:selectaccuracy}) \tablenotemark{d}} \\
\hline
Without distance uncertainties & \\
$\ldots$ expected sample size & $82 \pm 4$\\
$\ldots$ false positive rate & $13\% \pm 3\%$\\
$\ldots$ false negative rate & $16\% \pm 3\%$ \\
$\ldots$ correct selection rate & $87\% \pm 3\%$ \\
With distance uncertainties & \\
$\ldots$ expected sample size & $76 \pm 6$ \\
$\ldots$ false positive rate & $42\% \pm 4\%$ \\
$\ldots$ false negative rate & $47\% \pm 5\%$ \\
$\ldots$ correct selection rate & $58\% \pm 4\%$ \\
\enddata
\tablenotetext{a}{These are the selection criteria used to design the sample. As discussed in \S \ref{sec:sample} and \S \ref{sec:sampleprops}, some selected sample members no longer meet the selection criteria because we have improved our estimates of their properties. See the text for more details.}
\tablenotetext{b}{The main sample is quoted as $74$ galaxies in a number of our earlier works, because NGC~1068 met our selection criteria but was excluded due to previous archival mapping. As of the writing of this paper, we refer to the main sample as having $75$ galaxies and we do include NGC~1068, which has new CO~(2-1) 7{-}m+TP mapping (PI: M. Querejeta).}
\tablenotetext{c}{These are members of PHANGS--ALMA known to not meet the selection criteria. Their scientific focus is contrasting with the properties and resolution of the main sample.}
\tablenotetext{d}{Galaxies rejected ``by eye'' are entirely excluded from this calculation.}
\end{deluxetable}

Our science goals (\S\ref{sec:motivation}) require us to observe molecular gas at ``cloud scales'' and associate it with multi-wavelength signatures of star formation, feedback, and galactic structure. With that in mind, we selected our main sample according to the following criteria, summarized in Table~\ref{tab:selection}.

\begin{enumerate}

\item \textbf{Close enough that 1\arcsec\ \boldmath$\leq$ 100~pc.} We targeted galaxies with an estimated distance $d \leq 17$~Mpc. Our core science goals require resolving molecular gas into individual cloud-sized resolution elements, a requirement associated with a fixed physical resolution of ${\sim}100$~pc. 

\item \textbf{Not highly inclined.} We selected galaxies with inclination $i < 75^\circ$. This allows us to distinguish individual gas clouds and dynamical features. It also allows for clean association between CO and emission at other wavelengths, e.g., H$\alpha$ and the near-IR continuum.

\item \textbf{Visible to ALMA.} We considered targets with declination between $\delta = -75\degr$ and $\delta = +25\degr$.

\item \textbf{Relatively massive.} We targeted galaxies with stellar mass $\log_{10} M_\star [{\rm M}_\odot] \gtrsim 9.75$. Our adopted mass cutoff translates to ${\sim}2$ times the mass of the LMC or M33 and lies about one order of magnitude below the knee in the local galaxy mass function, $M^*$ where $\log_{10} M^* \left[{\rm M}_\odot\right] \approx 10.65$ at $z=0$ \citep[e.g.,][]{BALDRY08}. In star-forming galaxies, stellar mass correlates with star formation activity, gas fraction, molecular-to-atomic gas ratio, and metallicity \citep[e.g., see review by][]{BLANTON09}. By adopting this mass threshold, we aimed to capture a wide range of galaxy properties but to avoid focusing on low metallicity, low mass galaxies where detecting CO can be a major challenge \citep[e.g.,][]{BOLATTO13A,HUNT15,SCHRUBA17}. 

At both low and high redshift, the shape of the galaxy mass function and the star-forming main sequence implies that most stars form in galaxies within one dex of $M^*$ \citep[ e.g.,][]{KARIM11,LESLIE20}, where $M^*$ refers to the characteristic mass scale in the galaxy mass function \citep[$\log_{10} M^* \approx 10.6$ at $z=0$, e.g.,][]{WEIGEL16}. We thus expect that our target mass range  captures conditions representative of much of the secular build-up of galaxies.

\item \textbf{Actively star-forming.} We targeted galaxies with specific star formation rate ${\rm SFR}/M_\star > 10^{-11}$~yr$^{-1}$. This selects galaxies close to the $z=0$ star-forming main sequence \citep[e.g.,][]{BLANTON09} and removes passive, non-star-forming galaxies that are less likely to have massive cold gas reservoirs. Our selection includes starburst galaxies with high ${\rm SFR}/M_\star$. However such systems are rare in the $z=0$ universe \citep[e.g.,][]{SANDERS96} and they are mostly excluded by our distance cut.

\end{enumerate}

\subsection{Implementation}
\label{sec:selectimplement}

We worked on the selection of targets for the main PHANGS--ALMA sample from 2015 to 2017. The quantities that we used in the sample selection process were extracted from public databases or derived from public images. We caution that the property estimates that we used while making the sample selection may no longer represent our best estimates of certain galaxy properties for some targets. In particular, the distances to nearby galaxies can have large uncertainties and our best estimates for our targets' distances have significantly evolved since the original selection process. We have also revised our approaches to estimate stellar mass and SFR since the original sample selection. We describe our current best estimates of galaxy properties in \S\ref{sec:sampleprops}. 

For selection, we implemented our sample criteria in the following way:

\begin{enumerate}
\item \textbf{Super-sample:} We considered objects classified as galaxies in LEDA, and required that they have either a deprojected rotation velocity $> 120$~km~s$^{-1}$ or an absolute $B$ magnitude $M_B < -18$~mag. These represent less stringent cuts than those imposed on distance, mass, or specific star formation rate below. We do not expect that these criteria had a significant impact on the final sample selection.

\item \textbf{Distance:} Distance represents the dominant uncertainty in our selection. For the original selection, we used the median redshift-independent distance from NED\footnote{The NASA/IPAC Extragalactic Database (NED) is operated by the Jet Propulsion Laboratory, California Institute of Technology, under contract with the National Aeronautics and Space Administration.}. Our requirement of $1\arcsec \approx 100$~pc entails that our targets are too close for accurate Hubble flow distances, and very few of these nearby galaxies have high-quality, redshift-independent distances (see \S\ref{sec:distance}).

\item \textbf{Orientation:} We adopted positions and photometric inclination from HyperLEDA \citep{MAKAROV14}. These inclinations can be uncertain, e.g., due to the uncertain handling of disk thickness in high-inclination cases or ambiguity in the geometry of the galaxy. They introduce a modest uncertainty into the selection.

\item \textbf{SFR and M$_\star$ from WISE:} Our selection depended on stellar mass, $M_\star$, and specific star formation rate, ${\rm SFR}/M_\star$. We estimated $M_\star$ by carrying out photometry on WISE band~1 ($3.4$~\micron) images from the unWISE reprocessing \citep{LANG14} of the WISE all-sky survey \citep{WRIGHT10}. We estimated the SFR using  unWISE band~4 ($22$~\micron) images. 

During our original sample selection, we translated the WISE band~1 luminosity to $M_\star$ assuming a fixed mass-to-light ratio of $\mtolwise \approx 0.53$~\mtolunits . This is roughly consistent with \citet{MEIDT14}, \citet{MCGAUGH14}, \citet{QUEREJETA15}, and other results from \textit{Spitzer}'s S$^4$G survey \citep{SHETH10}. We converted from WISE band~4 to SFR using a factor $C \approx 10^{-42.7}$ to convert from $\nu L_\nu$ in erg~s$^{-1}$ to M$_\odot$~yr$^{-1}$. This agrees well with \citet{KENNICUTT12} and \citet{JARRETT13}; for more details on both notation and appropriateness of this value see \citet{LEROY19}. We verified during sample selection that our WISE-based approach yielded SFRs consistent with estimates based on the IRAS Revised Bright Galaxy Survey \citep{SANDERS03} and other common approaches to estimate the SFR \citep[e.g.,][]{KENNICUTT12}. Since our adopted conversions between WISE luminosity and SFR or $M_\star$ were linear, our original selection can  be stated as a WISE band~1 luminosity cut combined with a WISE band~4 to WISE band~1 color cut.

The WISE band~1 photometry becomes uncertain at low Galactic latitude, $b$, due to the presence of foreground stars and the limited angular resolution of WISE. Our selection is therefore less accurate at low $\lvert b\rvert$. The adopted conversion factors also affected our sample selection. We did not include a UV or another ``unobscured'' term in the SFR estimate when selecting our target sample. This introduced some bias against dwarf galaxies and other galaxies with little or no dust. We do not expect this to be a significant concern for relatively high mass ``main sequence'' galaxies. We also adopted a single $\mtolwise$, which likely led us to overestimate the stellar mass of low mass, high SFR/$M_\star$ galaxies (see \S\ref{sec:selectaccuracy}). Most importantly, because we use stellar mass to select the sample, uncertainty in distance also affected this step of the selection.

\item \textbf{Rejection of incorrect selections:} After applying the above criteria, we identified a few cases where a target's true inclination appeared to be nearly edge-on based on visual inspection of WISE and optical images. Visual inspection likewise revealed a few other targets with highly concentrated nuclear IR emission that is likely to be dominated by an AGN or a compact starburst. We rejected these targets as being unsuitable for wide-area CO mapping. Another handful of potential targets, usually at low Galactic latitude, are located directly behind bright Milky Way stars, making multi-wavelength analysis impractical. The overall list of manually excluded galaxies is ESO~138-010, ESO~494-026, IC~5201, NGC~1055, NGC~4802, NGC~6221, NGC~6875, PGC~18855, and PGC~54411. 

\end{enumerate}

\begin{figure*}
\plottwo{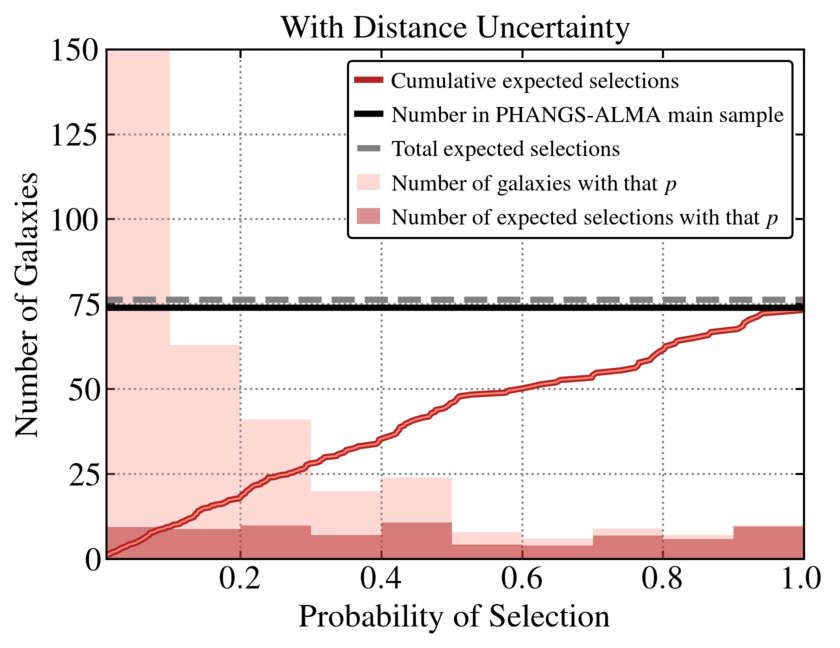}{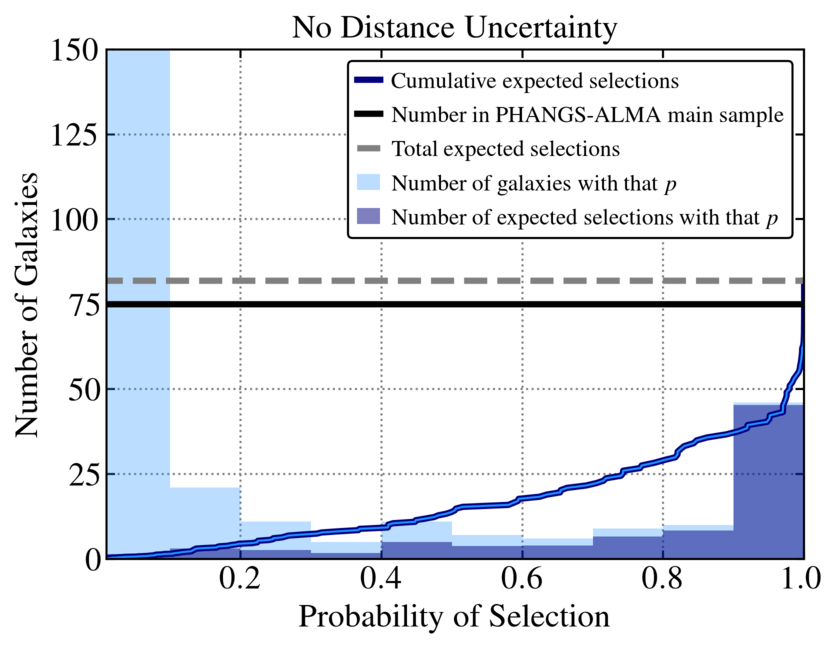}
\caption{\textbf{Sample selection expectations and uncertainties based on a Monte Carlo calculation.} Histograms show the number of galaxies as a function of the probability that they meet the main PHANGS--ALMA sample selection criteria. We take the physical property estimates and associated uncertainties for ${\sim}15{,}000$ local galaxies from \citet{LEROY19} as ``true''. Then we repeatedly realize random versions of galaxy properties to estimate the probability of selecting each galaxy. The {\em left} panel includes uncertainties on distance, the {\em right} panel considers only uncertainties on stellar mass, star formation rate, and inclination. Light histograms show the total number of galaxies with that probability of selection. Dark histograms show the number of galaxies that we would expect to select from that bin based on these probabilities. The (red/blue) solid and gray dashed lines show the resulting cumulative distribution function and total expected number of selected galaxies. The solid black line shows the actual number of galaxies ($75$) selected for the PHANGS--ALMA main sample. Our sample contains roughly the expected number of galaxies, but once distance uncertainties are accounted for there is substantial uncertainty associated with the exact selection (see also Figure~\ref{fig:sample_location}).
\label{fig:sample_monte}}
\end{figure*}

\begin{deluxetable*}{cllcl}
\tabletypesize{\small}
\tablecaption{ALMA Projects Processed as Part of PHANGS--ALMA \label{tab:projects}}
\tablewidth{0pt}
\tablehead{
\colhead{Cycle} &
\colhead{Project Code} & 
\colhead{P.I.} &
\colhead{Galaxies} &
\colhead{Notes}
}
\startdata
\hline
\multicolumn{5}{c}{PHANGS--ALMA Projects} \\
\hline \hline
1 & 2013.1.00650.S & E. Schinnerer & 1 & NGC~0628, pilot project \\
3 & 2015.1.00925.S & G. Blanc & 9\tablenotemark{a} & Pilot project \\
3 & 2015.1.00956.S & A. K. Leroy & 8 & Pilot project \\
5 & 2017.1.00886.L & E. Schinnerer & 54 & Large Program \\
& & \multicolumn{3}{l}{co-P.I.s A.K. Leroy, G. Blanc, A. Hughes, E. Rosolowsky, A. Schruba} \\
5 & 2017.1.00392.S & G. Blanc & 9\tablenotemark{a} & Pilot project completion \\
5 & 2017.1.00766.S & M. Chevance & 7\tablenotemark{b} & Early-type extension \\
6 & 2018.1.00484.S & M. Chevance & 7\tablenotemark{b} & Early-type extension completion\\
6 & 2018.1.01651.S & A. K. Leroy & 9\tablenotemark{a} & Pilot project completion \\
6 & 2018.1.01321.S & C. Faesi & 3 & 7-m and total power, very close galaxies \\
6 & 2018.A.00062.S & C. Faesi & 5\tablenotemark{c,d} & 7-m and total power, very close galaxies \\
7 & 2019.1.01235.S & C. Faesi & 5\tablenotemark{c,d} & 7-m and total power, very close galaxies completion\\
7 & 2019.2.00129.S & M. Querejeta & 1 & 7{-}m and total power, NGC~1068 \\
\hline \hline
\multicolumn{5}{c}{Archival \cotwo\ Data Processed with PHANGS--ALMA} \\
\hline \hline
1 & 2013.1.01161.S & K. Sakamoto & 2 & NGC~1365 and NGC~5236 (M83) 12-m, 7-m, and total power \\
1 & 2013.1.00803.S & D. Espada & 1 & NGC~5128\tablenotemark{c} 12-m, 7-m, and total power \\
3 & 2015.1.00782.S & K. Johnson & 1 & NGC~7793\tablenotemark{d} 12-m; see \citet{GRASHA18} \\
5 & 2015.1.00121.S & K. Sakamoto & 1 & NGC~5236 (M83) 12-m, 7-m, and total power \\
6 & 2016.1.00386.S & K. Sakamoto & 1 & NGC~5236 (M83) 12-m, 7-m, and total power 
\enddata
\tablenotetext{a}{These three programs targeted the same set of $9$ total galaxies.}
\tablenotetext{b}{These two programs targeted the same set of $7$ total galaxies.}
\tablenotetext{c}{NGC~5128 is Centaurus~A. Projects 2018.A.0062.S and 2019.1.01235.S targeted NGC~5128 using the \mbox{7-m} and total power antennas. Archival project 2013.1.00803.S targeted the galaxy with \mbox{12-m}, \mbox{7-m}, and total power observations, but the \cotwo\ total power observations were not usable. See the closely related \coone\ observations in \citet{ESPADA19}.}
\tablenotetext{d}{Projects 2018.A.0062.S and 2019.1.01235.S targeted NGC~7793 using the \mbox{7-m} and total power antennas. Archival project 2015.1.00782.S observed this galaxy using only the \mbox{12-m} antennas.}
\end{deluxetable*}

This implementation yielded $75$ primary PHANGS--ALMA targets. Of these, $18$ were observed as part of several pilot programs, which we list in Table~\ref{tab:projects}. Another $54$ were observed as part of the ALMA Large Program ``\mbox{100,000} Molecular Clouds Across the Main Sequence: GMCs as the Drivers of Galaxy Evolution'' (2017.1.00886.L, P.I.\ E.~Schinnerer), which was carried out during Cycles~5 and~6. Two further galaxies, NGC~1365 and NGC~5236 (M83), meet our selection criteria and have been targeted for wide-area CO mapping by other programs (see Table~\ref{tab:projects}). We include these galaxies in our sample for most science analysis, using a version of the archival data that we reprocessed using the PHANGS--ALMA pipeline. A final galaxy, NGC~1068, meets our criteria but was excluded from the Large Progrma due to the presence of archival CO~(3-2) mapping \citep{GARCIABURILLO14}. New PHANGS CO~(2-1) 7{-}m+TP mapping (PI: M. Querejeta) appear in this paper. We now formally include NGC~1068 in the main sample, bringing the main sample size to $75$ galaxies.

We supplement this main sample with several extensions that relax one or more of the selection criteria, which we describe in \S\ref{sec:selectextensions}, where we also note several public data sets with similar properties to PHANGS--ALMA. These extensions also leverage archival ALMA data, including the \cotwo\ observations of NGC~7793 presented in \citet{GRASHA18} and \cotwo\ observations of Centaurus~A (NGC~5128) closely related to the \coone\ observations presented by \citet{ESPADA19}.

Combining these extensions with the main sample, PHANGS--ALMA currently consists of \ntarget\ nearby galaxies observed at similar physical resolution in \cotwo. We present the list of all targets, along with best estimates of their global properties in \S\ref{sec:sampleprops}.

\subsection{Accuracy and Uncertainty in Selection}
\label{sec:selectaccuracy}

\begin{figure}
\includegraphics[width=0.5\textwidth]{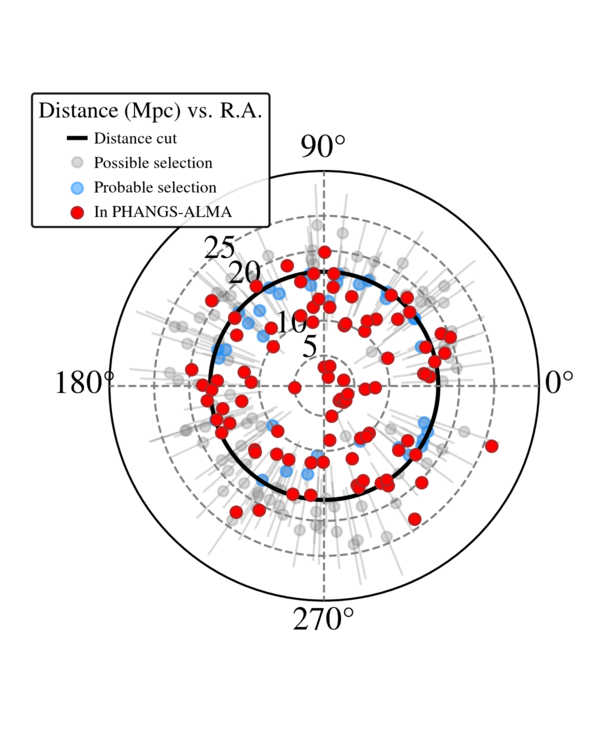}
\caption{\textbf{Sample selection and location.} Right ascension and current best-estimate distance of PHANGS--ALMA targets (red points) and candidate targets that we did not select (blue and gray points). The overdensity of points towards $180\degr$ reflects the presence of the Virgo cluster. Blue points show targets close to meeting our selection criteria at their present distance; given the uncertainty in their stellar mass, SFR, and inclination we estimate that they are $\gtrsim 50$\% likely to meet our selection criteria. Gray points show galaxies where, given the uncertainty in distance, the targets are between $50$\% and $10$\% likely to have true values that meet our selection criteria. Red points show the actual PHANGS--ALMA sample. The dashed gray concentric circles indicate distances of 5, 10, 20 and 25~Mpc; our distance cutoff at $17$~Mpc is indicated with a thick black line. Our selection did a good job of picking out all good candidates within $\lesssim 10$~Mpc. The well-known uncertainties in distances to nearby galaxies leads to uncertainty in selection near our distance cutoff. Figure~\ref{fig:sample_monte} gives another view of the uncertainty in selection.
\label{fig:sample_location}}
\end{figure}

Although the selection criteria for PHANGS--ALMA are quite simple, uncertainties in parameter estimation lead to uncertainty in the sample selection. Distance remains the primary driver of uncertainty for PHANGS--ALMA, since the Hubble flow does not yield high quality distances to galaxies closer than $\sim 50$~Mpc \citep[e.g., see figure~1 in][]{LEROY19}. Any uncertainty in distance also affects the inferred stellar mass, which is another of our key selection criteria. Secondary uncertainties stem from how we estimate stellar mass and the star formation rate.

To assess the uncertainty associated with our sample selection, we carry out a Monte Carlo exercise. We begin with estimates of stellar mass, star formation rate, inclination, and distance to ${\sim}15{,}000$ nearby galaxies drawn from \citet{LEROY19}. These property estimates leverage GALEX and WISE photometry, with calibrations pinned to the properties of SDSS galaxies estimated by \citet{SALIM18}. The distances are drawn from the \citet{TULLY09} Extragalactic Distance Database. We note that the stellar masses, SFRs, and distances in \citet{LEROY19} should all be superior to those that we used for selection (\S\ref{sec:selectimplement}). 

The Monte Carlo exercise proceeds as follows:

\begin{enumerate}
    \item We adopt the catalog values as the ``true'' values. We exclude the galaxies removed from the target list by hand from any calculations.
    
    \item We randomly perturb the inclination, mass, and star formation rate of each galaxy according to their uncertainties. We adopted a $\pm 5\degr$ uncertainty for the inclination, consistent with \citet{LANG20}, and we cap the value at $90^\circ$. For the SFR and stellar mass, we draw the uncertainties from \citet{LEROY19}. These are typically $0.1$~dex for $M_\star$ and $0.15$~dex for the SFR; this primarily reflects uncertainty in the stellar mass-to-light ratio and conversion from IR and UV luminosity to SFR.
    
    \item We randomly shift the distance, with the magnitude of the shift set by the uncertainty in the distance to each galaxy, which depends on the quality of the distance indicator\footnote{See section~2.2 in \citet{LEROY19}. Roughly, we adopt $0.03$~dex uncertainty for TRGB distances, $0.06$~dex for other quality distances, and $0.125$~dex for other distances.}. We adjust the stellar mass and star formation rate to reflect the new distance.
    
    \item For each galaxy and each realization, we check whether the galaxy's new properties would qualify for our selection.
    
    \item For each realization of the full sample, we check how many false positives and false negatives have been created by perturbing our best estimates of the galaxy properties. 
    
    To do this, we first record if each set of true galaxy properties meets our selection criteria. Then we note whether each true selection would meet our criteria after perturbing the galaxy properties. The number of galaxies not selected but that have ``true'' properties that meet our selection criteria establishes the false negative rate. 
    
    To establish the false positive rate, we note how many selected sample members in the random realization would not have been selected if we used their ``true'' properties.
\end{enumerate}

\noindent We repeat the exercise twice, each time using $10{,}000$ realizations. In the first case, we impose distance uncertainties. In the second case we skip step \#3, and only consider uncertainties unrelated to distance.

\begin{figure}
\gridline{\fig{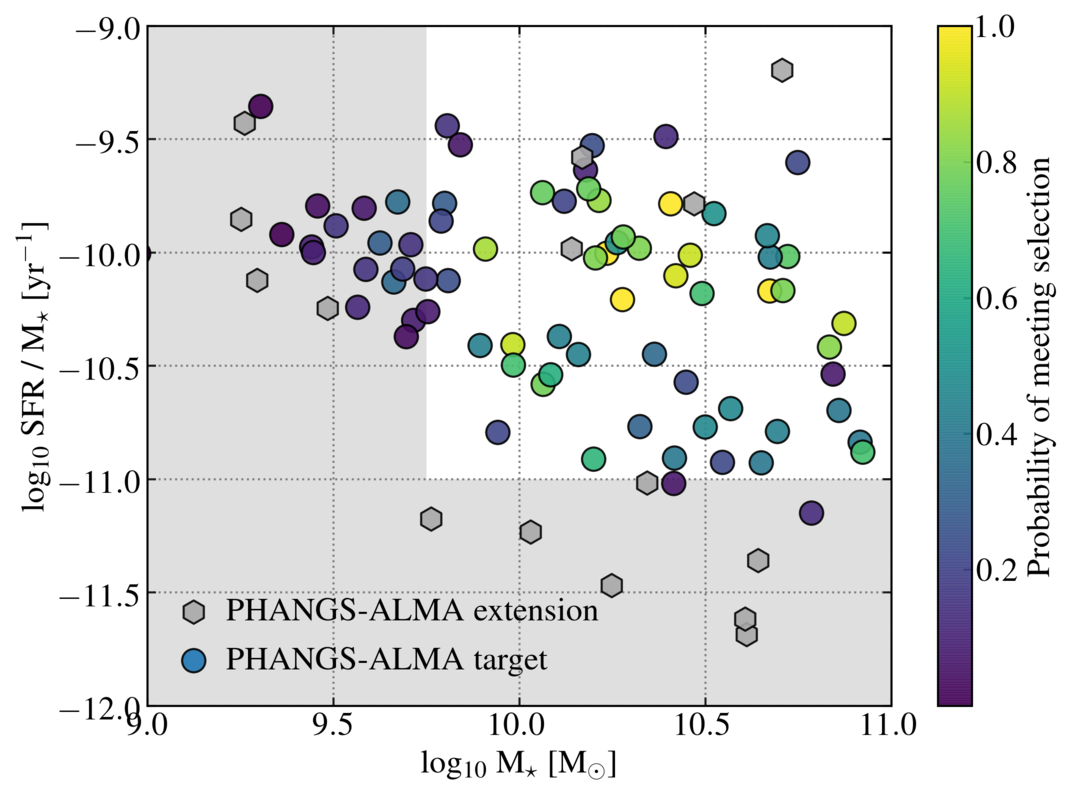}{0.48\textwidth}{}}
\vspace{-12pt}
\caption{\textbf{Probability of re-selection for PHANGS--ALMA targets.} Probability that actual PHANGS--ALMA targets would be re-selected, based on our Monte Carlo calculations and galaxy property estimates by \citet{LEROY19}. The gray region illustrates values excluded by our nominal selection criteria. The color of each point shows the fraction of realizations it is re-selected during the Monte Carlo calculations based on distance, inclination, mass, and specific star formation rate. Our high mass, high SFR targets usually meet our selection criteria. The variable WISE band~1 mass-to-light ratio used in \citet{LEROY19} and adopted in \S \ref{sec:sampleprops} leads us to reassess some of our lower mass targets as likely having $\log_{10} M_\star [M_\odot] \approx 9.5{-}9.75$. Otherwise, most of deviation from $p=1$ reflects the persistent uncertainty in the distances to nearby galaxies.
\label{fig:sample_mainseq}}
\end{figure}

Based on the number of times a galaxy is selected over all $10{,}000$ realizations, we assign each target a probability, $p$, of meeting our selection criteria. Figures~\ref{fig:sample_monte}, \ref{fig:sample_location}, and~\ref{fig:sample_mainseq} visualize the results of this calculation, which we also summarize in the second part of Table~\ref{tab:selection}. 

Figure~\ref{fig:sample_monte} shows histograms of the number of local galaxies that have probability $p$ of matching our selection. The total number of galaxies with $p$ appears as a light shaded region. The expected number of selections in that bin, which is the sum of all $p$ in that bin, appears as a dark shaded region. For example, $20$ galaxies with a $5$\% chance of selection yield $1$ expected selection. The cumulative distribution function (CDF) of selections appears as a solid line. This line shows the total number of galaxies with probability of selection $< p$ expected to be selected. The total number of predicted selections, i.e., the last point in the CDF, appears as a dashed gray line. For comparison, the solid black line shows the total number of PHANGS--ALMA ``main sample'' selections, i.e., our total number of targets less the $15$ extension galaxies that we know do not meet our original selection function.

There are several results in Figure~\ref{fig:sample_monte} and Table~\ref{tab:selection} that are worth noting.

\medskip

\textbf{PHANGS--ALMA selects about the expected number of galaxies:} First, the agreement between the dashed gray and black lines shows that our selected sample has about the expected size. Based on our Monte Carlo calculations, there should be ${\sim}75$ galaxies that meet our selection criteria. Specifically, as listed in Table~\ref{tab:selection}, we select $76$ and $82$ galaxies on average, depending on whether we randomize the distance. We selected $75$ targets for PHANGS--ALMA. In good agreement with this, without any Monte Carlo calculation, the \citet{LEROY19} catalog yields $79$ galaxies that meet our selection criteria.

\medskip

\textbf{Distance represents the dominant source of uncertainty in sample selection:} Distance uncertainties are included in the left panel but not the right one of Figure~\ref{fig:sample_monte}. When including distance uncertainties, far fewer galaxies have high probabilities. We also expect many of our selections to be uncertain, which is reflected by their modest~$p$. This demonstrates that distance uncertainties can easily shift galaxies into or out of our sample. By contrast, in the right panel, with no distance uncertainties, the sample selection appears clean, with most selected targets having high probabilities and relatively few ambiguous cases. Table~\ref{tab:selection} shows the same result. When distance uncertainties are included, the false positive and false negative rates are both much higher than in the case without distance uncertainties.

\medskip

\textbf{Distance uncertainties lead to high false positive and false negative rates:} The uncertainty in distance leads to both a high false positive rate and a high false negative rate. Based on the Monte Carlo exercise, we estimate a false positive rate of ${\sim}40\%$ and a false negative rate of ${\sim} 50\%$ (Table~\ref{tab:selection}) when the distance uncertainty is fold into the model. This appears in the left panel of Figure~\ref{fig:sample_monte} as a substantial contribution to the sample from bins with low~$p$. A large fraction of our sample consists of galaxies with moderate $p$ values, i.e., relatively uncertain selections. This is an unavoidable result of selecting local galaxies on mass and distance. It implies that as distance estimates improve, some of our targets will no longer meet our selection criteria, while other targets, not originally selected, will meet our criteria.

Figure~\ref{fig:sample_location} shows the location of our selected galaxies along with ``possible'' and ``probable'' selections. Here a ``possible'' selection means $0.5 \gtrsim p > 0.1$ (gray dots) with distance uncertainties. A~``probable'' selection refers to a galaxy with $p > 0.5$ (blue dots) without distance uncertainties. We show $106$ possible candidates and $31$ probable candidates. As expected, the figure shows that almost all of these probable and possible selections hover near the distance cutoff of $17$~Mpc. Put another way, there are a significant ($>100$) number of galaxies in the local Universe that have at least a moderate probability of having true properties that fulfil the PHANGS--ALMA selection criteria but were not included in our original main sample due to uncertainties in the estimation of nearby galaxy properties.

\medskip

\textbf{Probability of re-selection and mass-to-light ratio for low mass galaxies:} In Figure~\ref{fig:sample_mainseq}, we examine the probability that the actual PHANGS--ALMA targets would be re-selected based on the \citet{LEROY19} physical parameter estimates. We plot all PHANGS--ALMA targets in SFR/$M_\star$ versus $M_\star$ space and highlight our selection criteria using a shaded gray region. The color of each point indicates $p$ for that galaxy from the Monte Carlo exercise above. Targets from the extension programs are also plotted in the figure as gray points.

Overall, the figure shows that our high mass, high SFR targets have a reasonably high chance of re-selection. However, the figure shows ${\sim} 20$ low $M_\star$ galaxies with low re-selection probability in the upper left part of the plot. The low M$_\star$ and low re-selection probability for these targets reflect differences in how we estimated stellar mass during selection and the method used in \citet{LEROY19}, which is similar to what we use in \S\ref{sec:sampleprops}. Our selection (\S\ref{sec:selectimplement}) adopted a fixed WISE band~1 mass-to-light ratio. The \citet{LEROY19} values used in this plot and the Monte Carlo analysis adopt a variable mass-to-light ratio. Their calibration yields moderately lower masses for the same WISE band~1 luminosity for low mass, high SFR/$M_\star$ galaxies. Concretely, the median WISE1 mass-to-light ratio drops from $\mtolwise \approx 0.5$~\mtolunits\ during selection to $\mtolwise \approx 0.35$~\mtolunits\ on average in our current estimates. In practice, this highlights a modest systematic uncertainty in our selection by stellar mass, above and beyond the distance uncertainty. With our current best approach for mass estimation, PHANGS--ALMA actually extends down to $\log_{10} M_\star [M_\odot] \approx 9.5$.

\medskip

\textbf{Notable Omissions:} This exercise revealed two clear omissions that appear almost certain to meet our criteria, NGC~3344 and NGC~3368. Another galaxy, NGC~4984, also appears likely to meet our criteria but the distance is quite uncertain. These three galaxies appear as blue dots near the $10$~Mpc line in Figure~\ref{fig:sample_location}.

\medskip

\textbf{How to use this information:} For most PHANGS--ALMA users, the uncertainty in sample selection will have little impact. We adopt a simple selection function and implement it in a reasonable way. Even if the selections are uncertain, we do not expect any important biases due to this uncertainty. For most applications, using all available data will represent a satisfactory approach. In these cases, the most important implication of this section is that one should always adopt the latest integrated galaxy parameter estimates. As our knowledge of distances to local galaxies improves, our estimates of their properties can change dramatically.

For those interested in a more rigorous sample selection, the uncertainty built into the sample selection implies that \textit{one should re-construct the sample used for any science project using current best parameter estimates.} That is, best practice is to not consider the main sample as a fixed entity but instead to treat PHANGS--ALMA as a supersample from which rigorous subsamples using the most up-to-date parameter estimates can be drawn.

\subsection{Extensions to the Main Sample}
\label{sec:selectextensions}

In addition to observations of our main sample, we have pursued several PHANGS--ALMA extension programs, which we also list in Table~\ref{tab:projects}. These extensions adopt a similar observational setup as the core PHANGS--ALMA program but target galaxies that were missed or excluded by our initial selection. 

The first major extension targets $7$ early-type galaxies that have signs of active star formation (2017.1.00766.S, 2018.100484.S, P.I.\ M.~Chevance) but specific star formation rate, SFR/M$_\star$, too low to qualify for our main sample. Molecular gas and star formation are present in a significant fraction of early-type galaxies \citep[e.g.,][]{YOUNG11} and these systems represent a distinct environment for molecular cloud formation and evolution. This program pursues many of the same science goals as our main program with the goal of illuminating how the environments of early-type galaxies affect cold gas and star formation at cloud scales.

The other current major extension uses the ACA's \mbox{7-m} and total power facilities to target $8$ galaxies with $d \lesssim 4$~Mpc  (2018.1.01321.S, 2018.A.00062.S, 2019.1.01235.S, P.I.\ C.~Faesi). This sample serves two key goals. First, these local targets include lower mass galaxies. Galaxy mass and metallicity represent key drivers of ISM physics. However, the low surface brightness and small structure size of CO in low mass galaxies \citep[e.g., see][]{HUGHES13A,DRUARD14,FAESI18,SUN18,SCHRUBA19} means that only observations of the nearest such systems are practical. 

These very local targets also include several highly inclined, more massive galaxies, NGC~0253, NGC~4945, and the Circinus galaxy. These galaxies resemble other members of the main sample in their mass and star formation rate but were not selected due to inclination. The proximity of these targets means that telescopes with limited angular resolution still achieve high physical resolution, allowing cloud scale comparisons between molecular gas, far infrared emission, \hi\ $21$~cm emission, and radio continuum data. Furthermore, these targets are ideal for future ALMA follow-ups that achieve very high physical resolution over a wide area, building a bridge between PHANGS--ALMA and Milky Way and Local Group observations. Given the small set of such very nearby galaxies, we relaxed inclination cuts and added $3$ galaxies. At present, we map these using only the \mbox{7-m} and total power telescopes, producing data that effectively matches the properties of the other PHANGS--ALMA CO observations.

\textbf{Closely matched archival data:} The IRAM \mbox{30-m} \cotwo\ map of M33 by \citet{DRUARD14} \citep[see also][]{GRATIER10,BRAINE18}  closely resembles a PHANGS--ALMA map in terms of resolution and coverage of the whole area of active star formation. Though we do not treat it as a formal member of the PHANGS--ALMA sample, it serves as a valuable low-mass comparison galaxy for many PHANGS--ALMA analyses.

Several similar data sets exist targeting other CO transitions in nearby galaxies. For example, PAWS mapped \coone\ at $1''$ resolution in M51 \citep[][]{SCHINNERER13}, NOEMA has surveyed \coone\ emission from IC~342 at cloud scales (A.~Schruba et al., in preparation), and CARMA (\citealt{SCHRUBA21}, see also \citealt{CALDUPRIMO16}) and the IRAM 30-m telescope \citep{NIETEN06} have surveyed \coone\ in M31 at comparable resolution. The NANTEN \coone\ surveys of the Magellanic Clouds \citep[][]{FUKUI99,MIZUNO01} also achieve similar physical resolution as the PHANGS--ALMA maps. More recently, \citet{KRUIJSSEN19} presented an ALMA \coone\ map of NGC~300 at sufficient resolution to resolve GMCs (PI: A. Schruba). Similar to M33, we treat these as valuable complementary data sets but not members of the PHANGS--ALMA main sample.

\section{Properties of the Observed Sample}
\label{sec:sampleprops}

\startlongtable
 \begin{deluxetable*}{lcccccc} 
 \tablecaption{PHANGS--ALMA Positions, Orientations, and Distances  \label{tab:sample_orient}} 
 \tablewidth{0pt} 
 \tabletypesize{\footnotesize} 
 \tablehead{ 
 \colhead{Galaxy} & 
 \colhead{$\alpha_{J2000}$} & 
 \colhead{$\delta_{J2000}$} & 
 \colhead{$v_{\rm LSR}$} & 
 \colhead{P.A.} & 
 \colhead{$i$} & 
 \colhead{$d$} 
 \\ 
 \colhead{} &  
 \colhead{} &  
 \colhead{} &  
 \colhead{(km s$^{-1}$)} &  
 \colhead{($^\circ$)} &  
 \colhead{($^\circ$)} &  
 \colhead{(Mpc)}  
 }  
\startdata 
ESO097-013$^{\rm X}$ & $14^{\rm h}13^{\rm m}09.9^{\rm s}$ & $-65^{\circ}20^{\prime}21^{\prime\prime}$ & $430.3 \pm   5.2$ & $ 36.7 \pm  10.0$ & $ 64.3 \pm   5.0$ & $4.20 \pm 0.77$ (1)\\ 
IC 1954 & $3^{\rm h}31^{\rm m}31.1^{\rm s}$ & $-51^{\circ}54^{\prime}18^{\prime\prime}$ & $1039.1 \pm   5.0$ & $ 63.4 \pm   0.2$ & $ 57.1 \pm   0.7$ & $12.80 \pm 2.17$ (2,3,4)\\ 
IC 5273 & $22^{\rm h}59^{\rm m}26.7^{\rm s}$ & $-37^{\circ}42^{\prime}10^{\prime\prime}$ & $1286.0 \pm   5.0$ & $234.1 \pm   2.0$ & $ 52.0 \pm   2.1$ & $14.18 \pm 2.14$ (3,4)\\ 
IC 5332 & $23^{\rm h}34^{\rm m}27.5^{\rm s}$ & $-36^{\circ}06^{\prime}04^{\prime\prime}$ & $699.3 \pm   2.3$ & $ 74.4 \pm  10.0$ & $ 26.9 \pm   5.0$ & $9.01 \pm 0.40$ (5)\\ 
NGC 0247$^{\rm X}$ & $0^{\rm h}47^{\rm m}08.6^{\rm s}$ & $-20^{\circ}45^{\prime}38^{\prime\prime}$ & $148.8 \pm   1.6$ & $167.4 \pm  10.0$ & $ 76.4 \pm   5.0$ & $3.71 \pm 0.13$ (6)\\ 
NGC 0253$^{\rm X}$ & $0^{\rm h}47^{\rm m}33.1^{\rm s}$ & $-25^{\circ}17^{\prime}18^{\prime\prime}$ & $235.4 \pm   2.4$ & $ 52.5 \pm  10.0$ & $ 75.0 \pm   5.0$ & $3.70 \pm 0.12$ (6)\\ 
NGC 0300$^{\rm X}$ & $0^{\rm h}54^{\rm m}53.5^{\rm s}$ & $-37^{\circ}41^{\prime}04^{\prime\prime}$ & $155.5 \pm   7.3$ & $114.3 \pm  10.0$ & $ 39.8 \pm   5.0$ & $2.09 \pm 0.09$ (6)\\ 
NGC 0628 & $1^{\rm h}36^{\rm m}41.7^{\rm s}$ & $+15^{\circ}47^{\prime}01^{\prime\prime}$ & $650.8 \pm   5.0$ & $ 20.7 \pm   1.0$ & $  8.9 \pm  12.2$ & $9.84 \pm 0.63$ (6)\\ 
NGC 0685 & $1^{\rm h}47^{\rm m}42.8^{\rm s}$ & $-52^{\circ}45^{\prime}43^{\prime\prime}$ & $1346.6 \pm   5.0$ & $100.9 \pm   2.8$ & $ 23.0 \pm  43.4$ & $19.94 \pm 3.01$ (3,4)\\ 
NGC 1068$^{\rm X}$ & $2^{\rm h}42^{\rm m}40.7^{\rm s}$ & $-0^{\circ}00^{\prime}48^{\prime\prime}$ & $1130.1 \pm   6.4$ & $ 72.7 \pm  10.0$ & $ 34.7 \pm   5.0$ & $13.97 \pm 2.11$ (3,4)\\ 
NGC 1087 & $2^{\rm h}46^{\rm m}25.2^{\rm s}$ & $-0^{\circ}29^{\prime}55^{\prime\prime}$ & $1501.5 \pm   5.0$ & $359.1 \pm   1.2$ & $ 42.9 \pm   3.9$ & $15.85 \pm 2.22$ (7)\\ 
NGC 1097 & $2^{\rm h}46^{\rm m}18.9^{\rm s}$ & $-30^{\circ}16^{\prime}29^{\prime\prime}$ & $1257.5 \pm   5.0$ & $122.4 \pm   3.6$ & $ 48.6 \pm   6.0$ & $13.58 \pm 2.05$ (3,4)\\ 
NGC 1313$^{\rm X}$ & $3^{\rm h}18^{\rm m}15.7^{\rm s}$ & $-66^{\circ}29^{\prime}52^{\prime\prime}$ & $451.2 \pm   7.2$ & $ 23.4 \pm  10.0$ & $ 34.8 \pm   5.0$ & $4.32 \pm 0.17$ (6)\\ 
NGC 1300 & $3^{\rm h}19^{\rm m}41.0^{\rm s}$ & $-19^{\circ}24^{\prime}40^{\prime\prime}$ & $1545.4 \pm   5.0$ & $278.0 \pm   1.0$ & $ 31.8 \pm   6.0$ & $18.99 \pm 2.86$ (3,4)\\ 
NGC 1317 & $3^{\rm h}22^{\rm m}44.3^{\rm s}$ & $-37^{\circ}06^{\prime}14^{\prime\prime}$ & $1930.5 \pm   5.0$ & $221.5 \pm   2.9$ & $ 23.2 \pm   7.5$ & $19.11 \pm 0.85$ (6)\\ 
NGC 1365 & $3^{\rm h}33^{\rm m}36.4^{\rm s}$ & $-36^{\circ}08^{\prime}25^{\prime\prime}$ & $1613.3 \pm   5.0$ & $201.1 \pm   7.5$ & $ 55.4 \pm   6.0$ & $19.57 \pm 0.78$ (6)\\ 
NGC 1385 & $3^{\rm h}37^{\rm m}28.6^{\rm s}$ & $-24^{\circ}30^{\prime}04^{\prime\prime}$ & $1476.8 \pm   5.0$ & $181.3 \pm   4.8$ & $ 44.0 \pm   7.6$ & $17.22 \pm 2.60$ (3,4)\\ 
NGC 1433 & $3^{\rm h}42^{\rm m}01.5^{\rm s}$ & $-47^{\circ}13^{\prime}19^{\prime\prime}$ & $1057.4 \pm   5.0$ & $199.7 \pm   0.3$ & $ 28.6 \pm   6.0$ & $18.63 \pm 1.84$ (8)\\ 
NGC 1511 & $3^{\rm h}59^{\rm m}36.6^{\rm s}$ & $-67^{\circ}38^{\prime}02^{\prime\prime}$ & $1331.0 \pm   5.0$ & $297.0 \pm   2.1$ & $ 72.7 \pm   1.2$ & $15.28 \pm 2.26$ (2)\\ 
NGC 1512 & $4^{\rm h}03^{\rm m}54.1^{\rm s}$ & $-43^{\circ}20^{\prime}55^{\prime\prime}$ & $871.4 \pm   5.0$ & $261.9 \pm   4.2$ & $ 42.5 \pm   6.0$ & $18.83 \pm 1.86$ (8)\\ 
NGC 1546 & $4^{\rm h}14^{\rm m}36.3^{\rm s}$ & $-56^{\circ}03^{\prime}39^{\prime\prime}$ & $1243.8 \pm   5.0$ & $147.8 \pm   0.4$ & $ 70.3 \pm   0.6$ & $17.69 \pm 2.02$ (7)\\ 
NGC 1559 & $4^{\rm h}17^{\rm m}36.6^{\rm s}$ & $-62^{\circ}47^{\prime}00^{\prime\prime}$ & $1275.2 \pm   5.0$ & $244.5 \pm   3.0$ & $ 65.4 \pm   8.4$ & $19.44 \pm 0.45$ (9)\\ 
NGC 1566 & $4^{\rm h}20^{\rm m}00.4^{\rm s}$ & $-54^{\circ}56^{\prime}17^{\prime\prime}$ & $1483.3 \pm   5.0$ & $214.7 \pm   4.1$ & $ 29.5 \pm  10.6$ & $17.69 \pm 2.02$ (7)\\ 
NGC 1637 & $4^{\rm h}41^{\rm m}28.2^{\rm s}$ & $-2^{\circ}51^{\prime}29^{\prime\prime}$ & $698.9 \pm   1.6$ & $ 20.6 \pm  10.0$ & $ 31.1 \pm   5.0$ & $11.70 \pm 1.01$ (10)\\ 
NGC 1672 & $4^{\rm h}45^{\rm m}42.5^{\rm s}$ & $-59^{\circ}14^{\prime}50^{\prime\prime}$ & $1318.3 \pm   5.0$ & $134.3 \pm   0.4$ & $ 42.6 \pm  12.9$ & $19.40 \pm 2.93$ (3,4)\\ 
NGC 1809 & $5^{\rm h}02^{\rm m}05.0^{\rm s}$ & $-69^{\circ}34^{\prime}05^{\prime\prime}$ & $1290.4 \pm   5.0$ & $138.2 \pm   8.9$ & $ 57.6 \pm  23.6$ & $19.95 \pm 5.63$ (2)\\ 
NGC 1792 & $5^{\rm h}05^{\rm m}14.3^{\rm s}$ & $-37^{\circ}58^{\prime}50^{\prime\prime}$ & $1175.9 \pm   5.0$ & $318.9 \pm   0.9$ & $ 65.1 \pm   1.1$ & $16.20 \pm 2.44$ (3,4)\\ 
NGC 2090 & $5^{\rm h}47^{\rm m}01.9^{\rm s}$ & $-34^{\circ}15^{\prime}02^{\prime\prime}$ & $898.2 \pm   5.0$ & $192.5 \pm   0.6$ & $ 64.5 \pm   0.2$ & $11.75 \pm 0.84$ (11)\\ 
NGC 2283 & $6^{\rm h}45^{\rm m}52.8^{\rm s}$ & $-18^{\circ}12^{\prime}39^{\prime\prime}$ & $821.9 \pm   5.0$ & $ -4.1 \pm   1.0$ & $ 43.7 \pm   3.6$ & $13.68 \pm 2.06$ (3,4)\\ 
NGC 2566 & $8^{\rm h}18^{\rm m}45.6^{\rm s}$ & $-25^{\circ}29^{\prime}58^{\prime\prime}$ & $1609.6 \pm   5.0$ & $312.0 \pm   2.0$ & $ 48.5 \pm   6.0$ & $23.44 \pm 3.53$ (7)\\ 
NGC 2775 & $9^{\rm h}10^{\rm m}20.1^{\rm s}$ & $+7^{\circ}02^{\prime}17^{\prime\prime}$ & $1339.2 \pm   5.0$ & $156.5 \pm   0.1$ & $ 41.2 \pm   0.6$ & $23.15 \pm 3.49$ (3,4)\\ 
NGC 2835 & $9^{\rm h}17^{\rm m}52.9^{\rm s}$ & $-22^{\circ}21^{\prime}17^{\prime\prime}$ & $867.3 \pm   5.0$ & $  1.0 \pm   1.0$ & $ 41.3 \pm   5.3$ & $12.22 \pm 0.93$ (5)\\ 
NGC 2903 & $9^{\rm h}32^{\rm m}10.1^{\rm s}$ & $+21^{\circ}30^{\prime}03^{\prime\prime}$ & $547.0 \pm   5.0$ & $203.7 \pm   2.0$ & $ 66.8 \pm   3.1$ & $10.00 \pm 1.99$ (2,3,4)\\ 
NGC 2997 & $9^{\rm h}45^{\rm m}38.8^{\rm s}$ & $-31^{\circ}11^{\prime}28^{\prime\prime}$ & $1076.9 \pm   5.0$ & $108.1 \pm   0.7$ & $ 33.0 \pm   9.0$ & $14.06 \pm 2.80$ (7)\\ 
NGC 3059 & $9^{\rm h}50^{\rm m}08.2^{\rm s}$ & $-73^{\circ}55^{\prime}20^{\prime\prime}$ & $1236.5 \pm   5.0$ & $-14.8 \pm   2.9$ & $ 29.4 \pm  11.0$ & $20.23 \pm 4.04$ (7)\\ 
NGC 3137 & $10^{\rm h}09^{\rm m}07.5^{\rm s}$ & $-29^{\circ}03^{\prime}51^{\prime\prime}$ & $1086.6 \pm   5.0$ & $ -0.3 \pm   0.5$ & $ 70.3 \pm   1.2$ & $16.37 \pm 2.34$ (7)\\ 
NGC 3239 & $10^{\rm h}25^{\rm m}04.9^{\rm s}$ & $+17^{\circ}09^{\prime}49^{\prime\prime}$ & $748.3 \pm   3.2$ & $ 72.9 \pm  10.0$ & $ 60.3 \pm   5.0$ & $10.86 \pm 1.05$ (12)\\ 
NGC 3351 & $10^{\rm h}43^{\rm m}57.8^{\rm s}$ & $+11^{\circ}42^{\prime}13^{\prime\prime}$ & $774.7 \pm   5.0$ & $193.2 \pm   2.0$ & $ 45.1 \pm   6.0$ & $9.96 \pm 0.33$ (6)\\ 
NGC 3489$^{\rm X}$ & $11^{\rm h}00^{\rm m}18.6^{\rm s}$ & $+13^{\circ}54^{\prime}04^{\prime\prime}$ & $692.1 \pm   3.1$ & $ 70.0 \pm  10.0$ & $ 63.7 \pm   5.0$ & $11.86 \pm 1.63$ (2,13)\\ 
NGC 3511 & $11^{\rm h}03^{\rm m}23.8^{\rm s}$ & $-23^{\circ}05^{\prime}12^{\prime\prime}$ & $1096.7 \pm   5.0$ & $256.8 \pm   0.8$ & $ 75.1 \pm   2.2$ & $13.94 \pm 2.10$ (3,4)\\ 
NGC 3507 & $11^{\rm h}03^{\rm m}25.4^{\rm s}$ & $+18^{\circ}08^{\prime}08^{\prime\prime}$ & $969.4 \pm   5.0$ & $ 55.8 \pm   1.3$ & $ 21.7 \pm  11.3$ & $23.55 \pm 3.99$ (2)\\ 
NGC 3521 & $11^{\rm h}05^{\rm m}48.6^{\rm s}$ & $-0^{\circ}02^{\prime}09^{\prime\prime}$ & $798.0 \pm   5.0$ & $343.0 \pm   0.6$ & $ 68.8 \pm   0.3$ & $13.24 \pm 1.96$ (2)\\ 
NGC 3596 & $11^{\rm h}15^{\rm m}06.2^{\rm s}$ & $+14^{\circ}47^{\prime}13^{\prime\prime}$ & $1187.9 \pm   5.0$ & $ 78.4 \pm   1.0$ & $ 25.1 \pm  11.0$ & $11.30 \pm 1.03$ (6)\\ 
NGC 3599$^{\rm X}$ & $11^{\rm h}15^{\rm m}26.9^{\rm s}$ & $+18^{\circ}06^{\prime}37^{\prime\prime}$ & $836.8 \pm  20.2$ & $ 41.9 \pm  10.0$ & $ 23.0 \pm   5.0$ & $19.86 \pm 2.73$ (2,13)\\ 
NGC 3621 & $11^{\rm h}18^{\rm m}16.3^{\rm s}$ & $-32^{\circ}48^{\prime}45^{\prime\prime}$ & $724.3 \pm   5.0$ & $343.8 \pm   0.3$ & $ 65.8 \pm   1.8$ & $7.06 \pm 0.28$ (5)\\ 
NGC 3626 & $11^{\rm h}20^{\rm m}03.8^{\rm s}$ & $+18^{\circ}21^{\prime}25^{\prime\prime}$ & $1470.7 \pm   5.0$ & $165.2 \pm   2.0$ & $ 46.6 \pm   6.0$ & $20.05 \pm 2.34$ (2,13)\\ 
NGC 3627 & $11^{\rm h}20^{\rm m}15.0^{\rm s}$ & $+12^{\circ}59^{\prime}29^{\prime\prime}$ & $715.4 \pm   5.0$ & $173.1 \pm   3.6$ & $ 57.3 \pm   1.0$ & $11.32 \pm 0.48$ (6)\\ 
NGC 4207 & $12^{\rm h}15^{\rm m}30.4^{\rm s}$ & $+9^{\circ}35^{\prime}06^{\prime\prime}$ & $606.6 \pm   5.0$ & $121.9 \pm   2.0$ & $ 64.5 \pm   6.0$ & $15.78 \pm 2.34$ (2)\\ 
NGC 4254 & $12^{\rm h}18^{\rm m}49.6^{\rm s}$ & $+14^{\circ}25^{\prime}00^{\prime\prime}$ & $2388.2 \pm   5.0$ & $ 68.1 \pm   0.5$ & $ 34.4 \pm   1.0$ & $13.10 \pm 2.01$ (14)\\ 
NGC 4293 & $12^{\rm h}21^{\rm m}12.8^{\rm s}$ & $+18^{\circ}22^{\prime}57^{\prime\prime}$ & $926.2 \pm   5.0$ & $ 48.3 \pm   2.0$ & $ 65.0 \pm   6.0$ & $15.76 \pm 2.38$ (7)\\ 
NGC 4298 & $12^{\rm h}21^{\rm m}32.8^{\rm s}$ & $+14^{\circ}36^{\prime}22^{\prime\prime}$ & $1138.1 \pm   5.0$ & $313.9 \pm   0.7$ & $ 59.2 \pm   0.8$ & $14.92 \pm 1.36$ (5)\\ 
NGC 4303 & $12^{\rm h}21^{\rm m}54.9^{\rm s}$ & $+4^{\circ}28^{\prime}25^{\prime\prime}$ & $1559.8 \pm   5.0$ & $312.4 \pm   2.5$ & $ 23.5 \pm   9.2$ & $16.99 \pm 3.02$ (7)\\ 
NGC 4321 & $12^{\rm h}22^{\rm m}54.9^{\rm s}$ & $+15^{\circ}49^{\prime}20^{\prime\prime}$ & $1572.3 \pm   5.0$ & $156.2 \pm   1.7$ & $ 38.5 \pm   2.4$ & $15.21 \pm 0.50$ (11)\\ 
NGC 4424 & $12^{\rm h}27^{\rm m}11.6^{\rm s}$ & $+9^{\circ}25^{\prime}14^{\prime\prime}$ & $447.4 \pm   5.0$ & $ 88.3 \pm   2.0$ & $ 58.2 \pm   6.0$ & $16.20 \pm 0.69$ (6)\\ 
NGC 4457 & $12^{\rm h}28^{\rm m}59.0^{\rm s}$ & $+3^{\circ}34^{\prime}14^{\prime\prime}$ & $886.0 \pm   5.0$ & $ 78.7 \pm   2.0$ & $ 17.4 \pm   6.0$ & $15.10 \pm 2.00$ (2,13)\\ 
NGC 4459$^{\rm X}$ & $12^{\rm h}29^{\rm m}00.0^{\rm s}$ & $+13^{\circ}58^{\prime}43^{\prime\prime}$ & $1190.1 \pm  11.0$ & $108.8 \pm  10.0$ & $ 47.0 \pm   5.0$ & $15.85 \pm 2.18$ (2,13)\\ 
NGC 4476$^{\rm X}$ & $12^{\rm h}29^{\rm m}59.1^{\rm s}$ & $+12^{\circ}20^{\prime}55^{\prime\prime}$ & $1962.7 \pm   1.3$ & $ 27.4 \pm  10.0$ & $ 60.1 \pm   5.0$ & $17.54 \pm 2.41$ (2,13)\\ 
NGC 4477$^{\rm X}$ & $12^{\rm h}30^{\rm m}02.2^{\rm s}$ & $+13^{\circ}38^{\prime}11^{\prime\prime}$ & $1362.2 \pm  33.1$ & $ 25.7 \pm  10.0$ & $ 33.5 \pm   5.0$ & $15.76 \pm 2.38$ (7)\\ 
NGC 4496A & $12^{\rm h}31^{\rm m}39.3^{\rm s}$ & $+3^{\circ}56^{\prime}23^{\prime\prime}$ & $1721.8 \pm   5.0$ & $ 51.1 \pm   4.1$ & $ 53.8 \pm   3.5$ & $14.86 \pm 1.06$ (11)\\ 
NGC 4535 & $12^{\rm h}34^{\rm m}20.3^{\rm s}$ & $+8^{\circ}11^{\prime}53^{\prime\prime}$ & $1953.6 \pm   5.0$ & $179.7 \pm   1.6$ & $ 44.7 \pm  10.8$ & $15.77 \pm 0.37$ (11)\\ 
NGC 4536 & $12^{\rm h}34^{\rm m}27.1^{\rm s}$ & $+2^{\circ}11^{\prime}18^{\prime\prime}$ & $1794.6 \pm   5.0$ & $305.6 \pm   2.3$ & $ 66.0 \pm   2.9$ & $16.25 \pm 1.12$ (6)\\ 
NGC 4540 & $12^{\rm h}34^{\rm m}50.9^{\rm s}$ & $+15^{\circ}33^{\prime}06^{\prime\prime}$ & $1286.5 \pm   5.0$ & $ 12.8 \pm   4.3$ & $ 28.7 \pm  28.7$ & $15.76 \pm 2.38$ (7)\\ 
NGC 4548 & $12^{\rm h}35^{\rm m}26.5^{\rm s}$ & $+14^{\circ}29^{\prime}47^{\prime\prime}$ & $482.7 \pm   5.0$ & $138.0 \pm   2.0$ & $ 38.3 \pm   6.0$ & $16.22 \pm 0.38$ (11)\\ 
NGC 4569 & $12^{\rm h}36^{\rm m}49.8^{\rm s}$ & $+13^{\circ}09^{\prime}46^{\prime\prime}$ & $-225.6 \pm   5.0$ & $ 18.0 \pm   2.0$ & $ 70.0 \pm   6.0$ & $15.76 \pm 2.38$ (7)\\ 
NGC 4571 & $12^{\rm h}36^{\rm m}56.4^{\rm s}$ & $+14^{\circ}13^{\prime}02^{\prime\prime}$ & $343.0 \pm   5.0$ & $217.5 \pm   0.6$ & $ 32.7 \pm   2.1$ & $14.90 \pm 1.07$ (15)\\ 
NGC 4579 & $12^{\rm h}37^{\rm m}43.5^{\rm s}$ & $+11^{\circ}49^{\prime}06^{\prime\prime}$ & $1516.7 \pm   5.0$ & $ 91.3 \pm   1.6$ & $ 40.2 \pm   5.6$ & $21.00 \pm 2.03$ (16)\\ 
NGC 4596$^{\rm X}$ & $12^{\rm h}39^{\rm m}55.9^{\rm s}$ & $+10^{\circ}10^{\prime}34^{\prime\prime}$ & $1883.3 \pm   7.5$ & $120.0 \pm  10.0$ & $ 36.6 \pm   5.0$ & $15.76 \pm 2.38$ (7)\\ 
NGC 4654 & $12^{\rm h}43^{\rm m}56.6^{\rm s}$ & $+13^{\circ}07^{\prime}36^{\prime\prime}$ & $1051.5 \pm   5.0$ & $123.2 \pm   1.0$ & $ 55.6 \pm   5.9$ & $21.98 \pm 1.14$ (11)\\ 
NGC 4689 & $12^{\rm h}47^{\rm m}45.6^{\rm s}$ & $+13^{\circ}45^{\prime}46^{\prime\prime}$ & $1614.2 \pm   5.0$ & $164.1 \pm   0.3$ & $ 38.7 \pm   2.7$ & $15.00 \pm 2.26$ (2,3,4)\\ 
NGC 4694 & $12^{\rm h}48^{\rm m}15.0^{\rm s}$ & $+10^{\circ}59^{\prime}01^{\prime\prime}$ & $1168.4 \pm   5.0$ & $143.3 \pm   2.0$ & $ 60.7 \pm   6.0$ & $15.76 \pm 2.38$ (7)\\ 
NGC 4731 & $12^{\rm h}51^{\rm m}01.2^{\rm s}$ & $-6^{\circ}23^{\prime}34^{\prime\prime}$ & $1483.6 \pm   5.0$ & $255.4 \pm   2.0$ & $ 64.0 \pm   6.0$ & $13.28 \pm 2.11$ (7)\\ 
NGC 4781 & $12^{\rm h}54^{\rm m}23.8^{\rm s}$ & $-10^{\circ}32^{\prime}14^{\prime\prime}$ & $1248.3 \pm   5.0$ & $290.0 \pm   1.3$ & $ 59.0 \pm   3.8$ & $11.31 \pm 1.18$ (7)\\ 
NGC 4826 & $12^{\rm h}56^{\rm m}43.6^{\rm s}$ & $+21^{\circ}41^{\prime}00^{\prime\prime}$ & $409.7 \pm   5.0$ & $293.6 \pm   1.2$ & $ 59.1 \pm   0.9$ & $4.41 \pm 0.19$ (5)\\ 
NGC 4941 & $13^{\rm h}04^{\rm m}13.1^{\rm s}$ & $-5^{\circ}33^{\prime}06^{\prime\prime}$ & $1116.0 \pm   5.0$ & $202.2 \pm   0.6$ & $ 53.4 \pm   1.1$ & $15.00 \pm 5.00$ (7)\\ 
NGC 4951 & $13^{\rm h}05^{\rm m}07.7^{\rm s}$ & $-6^{\circ}29^{\prime}38^{\prime\prime}$ & $1176.1 \pm   5.0$ & $ 91.2 \pm   0.5$ & $ 70.2 \pm   2.2$ & $15.00 \pm 4.19$ (2)\\ 
NGC 4945$^{\rm X}$ & $13^{\rm h}05^{\rm m}27.3^{\rm s}$ & $-49^{\circ}28^{\prime}04^{\prime\prime}$ & $559.3 \pm   2.7$ & $ 43.8 \pm  10.0$ & $ 90.0 \pm   5.0$ & $3.47 \pm 0.12$ (6)\\ 
NGC 5042 & $13^{\rm h}15^{\rm m}31.0^{\rm s}$ & $-23^{\circ}59^{\prime}02^{\prime\prime}$ & $1385.6 \pm   5.0$ & $190.6 \pm   0.8$ & $ 49.4 \pm   8.6$ & $16.78 \pm 2.53$ (3,4)\\ 
NGC 5068 & $13^{\rm h}18^{\rm m}54.7^{\rm s}$ & $-21^{\circ}02^{\prime}19^{\prime\prime}$ & $667.2 \pm   5.0$ & $342.4 \pm   3.2$ & $ 35.7 \pm  10.9$ & $5.20 \pm 0.22$ (5)\\ 
NGC 5128 & $13^{\rm h}25^{\rm m}27.6^{\rm s}$ & $-43^{\circ}01^{\prime}09^{\prime\prime}$ & $549.5 \pm   5.7$ & $ 32.2 \pm  10.0$ & $ 45.3 \pm   5.0$ & $3.69 \pm 0.13$ (6)\\ 
NGC 5134 & $13^{\rm h}25^{\rm m}18.5^{\rm s}$ & $-21^{\circ}08^{\prime}03^{\prime\prime}$ & $1749.1 \pm   5.0$ & $311.6 \pm   2.0$ & $ 22.7 \pm   6.0$ & $19.92 \pm 2.69$ (7)\\ 
NGC 5236 & $13^{\rm h}37^{\rm m}00.9^{\rm s}$ & $-29^{\circ}51^{\prime}56^{\prime\prime}$ & $509.4 \pm   2.1$ & $225.0 \pm  10.0$ & $ 24.0 \pm   5.0$ & $4.89 \pm 0.18$ (6)\\ 
NGC 5248 & $13^{\rm h}37^{\rm m}32.0^{\rm s}$ & $+8^{\circ}53^{\prime}07^{\prime\prime}$ & $1163.0 \pm   5.0$ & $109.2 \pm   3.5$ & $ 47.4 \pm  16.3$ & $14.87 \pm 1.32$ (7)\\ 
NGC 5530 & $14^{\rm h}18^{\rm m}27.3^{\rm s}$ & $-43^{\circ}23^{\prime}18^{\prime\prime}$ & $1183.2 \pm   5.0$ & $305.4 \pm   1.0$ & $ 61.9 \pm   2.6$ & $12.27 \pm 1.85$ (3,4)\\ 
NGC 5643 & $14^{\rm h}32^{\rm m}40.8^{\rm s}$ & $-44^{\circ}10^{\prime}29^{\prime\prime}$ & $1191.3 \pm   5.0$ & $318.7 \pm   2.0$ & $ 29.9 \pm   6.0$ & $12.68 \pm 0.54$ (6)\\ 
NGC 6300 & $17^{\rm h}16^{\rm m}59.5^{\rm s}$ & $-62^{\circ}49^{\prime}14^{\prime\prime}$ & $1102.1 \pm   5.0$ & $105.4 \pm   2.3$ & $ 49.6 \pm   5.8$ & $11.58 \pm 1.75$ (3,4)\\ 
NGC 6744 & $19^{\rm h}09^{\rm m}46.1^{\rm s}$ & $-63^{\circ}51^{\prime}27^{\prime\prime}$ & $832.3 \pm   5.0$ & $ 14.0 \pm   0.2$ & $ 52.7 \pm   2.2$ & $9.39 \pm 0.42$ (5)\\ 
NGC 7456 & $23^{\rm h}02^{\rm m}10.3^{\rm s}$ & $-39^{\circ}34^{\prime}10^{\prime\prime}$ & $1192.3 \pm   5.0$ & $ 16.0 \pm   2.9$ & $ 67.3 \pm   4.3$ & $15.70 \pm 2.33$ (2)\\ 
NGC 7496 & $23^{\rm h}09^{\rm m}47.3^{\rm s}$ & $-43^{\circ}25^{\prime}40^{\prime\prime}$ & $1639.2 \pm   5.0$ & $193.7 \pm   4.2$ & $ 35.9 \pm   6.0$ & $18.72 \pm 2.82$ (3,4)\\ 
NGC 7743$^{\rm X}$ & $23^{\rm h}44^{\rm m}21.1^{\rm s}$ & $+9^{\circ}56^{\prime}02^{\prime\prime}$ & $1687.3 \pm   5.4$ & $ 86.2 \pm  10.0$ & $ 37.1 \pm   5.0$ & $20.32 \pm 2.80$ (2,13)\\ 
NGC 7793$^{\rm X}$ & $23^{\rm h}57^{\rm m}49.8^{\rm s}$ & $-32^{\circ}35^{\prime}28^{\prime\prime}$ & $222.1 \pm   2.3$ & $290.0 \pm  10.0$ & $ 50.0 \pm   5.0$ & $3.62 \pm 0.15$ (6)\\ 
\enddata 
 \tablecomments{ 
 $X$ --- extension member;  
 centers from \citet{SALO15}, \citet{JARRETT03}, or LEDA \citep{PATUREL03,MAKAROV14};  
 orientations and velocities from \citet{LANG20} or \citet{SHETH10} or LEDA;  
 distance reference key:  
1---\citet{KARACHENTSEV04} 
2---\citet{TULLY16} 
3---\citet{SHAYA17} 
4---\citet{KOURKCHI20} 
5---\citet{ANAND21} 
6---\citet{TULLY09} 
7---\citet{KOURKCHI17} 
8---\citet{ANAND21} 
9---\citet{HUANG20} 
10---\citet{LEONARD03} 
11---\citet{FREEDMAN01} 
12---\citet{BARBARINO15} 
13---\citet{TONRY01} 
14---\citet{NUGENT06} 
15---\citet{PIERCE94} 
16---\citet{RUIZ96} 
 } 
\end{deluxetable*}

\startlongtable
 \begin{deluxetable*}{lccccccccc} 
 \tablecaption{PHANGS-ALMA Physical Properties  \label{tab:sample_phys}} 
 \tablewidth{0pt} 
 \tabletypesize{\footnotesize} 
 \tablehead{ 
 \colhead{Galaxy} & 
 \colhead{$\log_{10} M_\star$} & 
 \colhead{Src} & 
 \colhead{$R_e$} & 
 \colhead{$l_\star$} & 
 \colhead{$\log_{10}$~SFR} & 
 \colhead{Src} & 
 \colhead{$\log_{10}~L_{\rm CO}$} & 
 \colhead{Corr.} & 
 \colhead{$\log_{10}$~M$_{\rm HI}$} 
 \\ 
 \colhead{} &  
 \colhead{(M$_\odot$)} &  
 \colhead{} &  
 \colhead{(kpc)} &  
 \colhead{(kpc)} &  
 \colhead{(M$_\odot$~yr$^{-1}$)} &  
 \colhead{} &  
 \colhead{(K~km~s$^{-1}$~pc$^2$)} &  
 \colhead{} &  
 \colhead{(M$_\odot$)}  
 }  
\startdata 
NGC 0247   $^{\rm X}$ & $ 9.53$ & I & $  5.0$ & $  3.3$ & $-0.75$ & FUVW4  & $ 6.79$ & $ 1.42$  & $ 9.24$ \\ 
NGC 0253   $^{\rm X}$ & $10.64$ & I & $  4.7$ & $  2.8$ & $ 0.70$ & FUVW4  & $ 8.96$ & $ 1.00$  & $ 9.33$ \\ 
NGC 0300   $^{\rm X}$ & $ 9.27$ & I & $  2.0$ & $  1.3$ & $-0.82$ & FUVW4  & $ 6.61$ & $ 1.50$  & $ 9.32$ \\ 
NGC 0628    & $10.34$ & I & $  3.9$ & $  2.9$ & $ 0.24$ & FUVW4  & $ 8.41$ & $ 1.73$  & $ 9.70$ \\ 
NGC 0685    & $10.07$ & I & $  5.0$ & $  3.1$ & $-0.38$ & W4ONLY & $ 7.87$ & $ 1.25$  & $ 9.57$ \\ 
NGC 1068   $^{\rm X}$ & $10.91$ & I & $  0.9$ & $  7.3$ & $ 1.64$ & FUVW4  & $ 9.23$ & $ 1.30$  & $ 9.06$ \\ 
NGC 1097    & $10.76$ & I & $  2.6$ & $  4.3$ & $ 0.68$ & FUVW4  & $ 8.93$ & $ 1.31$  & $ 9.61$ \\ 
NGC 1087    & $ 9.94$ & I & $  3.2$ & $  2.1$ & $ 0.11$ & FUVW4  & $ 8.32$ & $ 1.06$  & $ 9.10$ \\ 
NGC 1313   $^{\rm X}$ & $ 9.26$ & I & $  2.5$ & $  2.1$ & $-0.14$ & FUVW4  & \nodata & \nodata  & $ 9.28$ \\ 
NGC 1300    & $10.62$ & I & $  6.5$ & $  3.7$ & $ 0.07$ & FUVW4  & $ 8.50$ & $ 1.28$  & $ 9.38$ \\ 
NGC 1317    & $10.62$ & W & $  1.8$ & $  2.4$ & $-0.32$ & FUVW4  & $ 8.10$ & $ 1.28$  & \nodata \\ 
IC 1954     & $ 9.67$ & I & $  2.4$ & $  1.5$ & $-0.44$ & FUVW4  & $ 7.78$ & $ 1.10$  & $ 8.85$ \\ 
NGC 1365    & $11.00$ & I & $  2.8$ & $ 13.1$ & $ 1.24$ & FUVW4  & $ 9.49$ & $ 1.36$  & $ 9.94$ \\ 
NGC 1385    & $ 9.98$ & I & $  3.4$ & $  2.6$ & $ 0.32$ & FUVW4  & $ 8.37$ & $ 1.09$  & $ 9.19$ \\ 
NGC 1433    & $10.87$ & I & $  4.3$ & $  6.9$ & $ 0.05$ & FUVW4  & $ 8.47$ & $ 1.38$  & $ 9.40$ \\ 
NGC 1511    & $ 9.92$ & I & $  2.4$ & $  1.7$ & $ 0.35$ & FUVW4  & $ 8.22$ & $ 1.09$  & $ 9.57$ \\ 
NGC 1512    & $10.72$ & I & $  4.8$ & $  6.2$ & $ 0.11$ & FUVW4  & $ 8.26$ & $ 1.45$  & $ 9.88$ \\ 
NGC 1546    & $10.37$ & I & $  2.2$ & $  2.1$ & $-0.08$ & FUVW4  & $ 8.44$ & $ 1.13$  & $ 8.68$ \\ 
NGC 1559    & $10.37$ & I & $  3.9$ & $  2.4$ & $ 0.60$ & NUVW4  & $ 8.66$ & $ 1.11$  & $ 9.52$ \\ 
NGC 1566    & $10.79$ & I & $  3.2$ & $  3.9$ & $ 0.66$ & FUVW4  & $ 8.89$ & $ 1.22$  & $ 9.80$ \\ 
NGC 1637    & $ 9.95$ & I & $  2.8$ & $  1.8$ & $-0.20$ & W4ONLY & $ 7.98$ & $ 1.10$  & $ 9.20$ \\ 
NGC 1672    & $10.73$ & I & $  3.4$ & $  5.8$ & $ 0.88$ & FUVW4  & $ 9.05$ & $ 1.25$  & $10.21$ \\ 
NGC 1809    & $ 9.77$ & I & $  4.5$ & $  2.4$ & $ 0.76$ & NUVW4  & $ 7.49$ & $ 4.24$  & $ 9.60$ \\ 
NGC 1792    & $10.62$ & I & $  4.1$ & $  2.4$ & $ 0.57$ & FUVW4  & $ 8.95$ & $ 1.11$  & $ 9.25$ \\ 
NGC 2090    & $10.04$ & W & $  1.9$ & $  1.7$ & $-0.39$ & FUVW4  & $ 7.67$ & $ 1.47$  & $ 9.37$ \\ 
NGC 2283    & $ 9.89$ & W & $  3.2$ & $  1.9$ & $-0.28$ & W4ONLY & $ 7.69$ & $ 1.16$  & $ 9.70$ \\ 
NGC 2566    & $10.71$ & W & $  5.1$ & $  4.0$ & $ 0.93$ & W4ONLY & $ 9.06$ & $ 1.13$  & $ 9.37$ \\ 
NGC 2775    & $11.07$ & I & $  4.6$ & $  4.1$ & $-0.06$ & FUVW4  & $ 8.40$ & $ 1.29$  & $ 8.65$ \\ 
NGC 2835    & $10.00$ & W & $  3.3$ & $  2.2$ & $ 0.10$ & FUVW4  & $ 7.71$ & $ 1.72$  & $ 9.48$ \\ 
NGC 2903    & $10.64$ & I & $  3.7$ & $  3.5$ & $ 0.49$ & FUVW4  & $ 8.76$ & $ 1.18$  & $ 9.54$ \\ 
NGC 2997    & $10.73$ & W & $  6.1$ & $  4.0$ & $ 0.64$ & FUVW4  & $ 8.97$ & $ 1.25$  & $ 9.86$ \\ 
NGC 3059    & $10.38$ & W & $  5.0$ & $  3.2$ & $ 0.38$ & W4ONLY & $ 8.59$ & $ 1.07$  & $ 9.75$ \\ 
NGC 3137    & $ 9.88$ & W & $  4.1$ & $  3.0$ & $-0.30$ & FUVW4  & $ 7.60$ & $ 1.35$  & $ 9.68$ \\ 
NGC 3239    & $ 9.18$ & I & $  3.1$ & $  2.0$ & $-0.41$ & FUVW4  & $< 6.62^\tablenotemark{ul}$ & $ 1.54$  & $ 9.16$ \\ 
NGC 3351    & $10.37$ & I & $  3.0$ & $  2.1$ & $ 0.12$ & FUVW4  & $ 8.13$ & $ 1.55$  & $ 8.93$ \\ 
NGC 3489   $^{\rm X}$ & $10.29$ & I & $  1.3$ & $  1.4$ & $-1.59$ & FUVW4  & $ 6.89$ & $ 1.37$  & $ 7.40$ \\ 
NGC 3511    & $10.03$ & I & $  4.4$ & $  2.4$ & $-0.09$ & FUVW4  & $ 8.15$ & $ 1.07$  & $ 9.37$ \\ 
NGC 3507    & $10.40$ & I & $  3.7$ & $  2.3$ & $-0.00$ & FUVW4  & $ 8.34$ & $ 1.17$  & $ 9.32$ \\ 
NGC 3521    & $11.03$ & I & $  3.9$ & $  4.9$ & $ 0.57$ & FUVW4  & $ 8.98$ & $ 1.18$  & $ 9.83$ \\ 
NGC 3596    & $ 9.66$ & I & $  1.6$ & $  2.0$ & $-0.52$ & NUVW4  & $ 7.81$ & $ 1.13$  & $ 8.85$ \\ 
NGC 3599   $^{\rm X}$ & $10.04$ & I & $  1.7$ & $  2.0$ & $-1.35$ & FUVW4  & $< 6.70^\tablenotemark{ul}$ & $ 1.35$  & \nodata \\ 
NGC 3621    & $10.06$ & W & $  2.7$ & $  2.0$ & $-0.00$ & FUVW4  & $ 8.13$ & $ 1.27$  & $ 9.66$ \\ 
NGC 3626    & $10.46$ & I & $  1.8$ & $  2.1$ & $-0.68$ & NUVW4  & $ 7.75$ & $ 1.14$  & $ 8.89$ \\ 
NGC 3627    & $10.84$ & I & $  3.6$ & $  3.7$ & $ 0.59$ & FUVW4  & $ 8.98$ & $ 1.16$  & $ 9.09$ \\ 
NGC 4207    & $ 9.72$ & I & $  1.4$ & $  0.7$ & $-0.72$ & FUVW4  & $ 7.71$ & $ 1.03$  & $ 8.58$ \\ 
NGC 4254    & $10.42$ & I & $  2.4$ & $  1.8$ & $ 0.49$ & FUVW4  & $ 8.93$ & $ 1.15$  & $ 9.48$ \\ 
NGC 4293    & $10.52$ & I & $  4.7$ & $  2.8$ & $-0.30$ & FUVW4  & $ 8.12$ & $ 1.57$  & $ 7.67$ \\ 
NGC 4298    & $10.04$ & I & $  3.0$ & $  1.6$ & $-0.34$ & FUVW4  & $ 8.26$ & $ 1.09$  & $ 8.87$ \\ 
NGC 4303    & $10.51$ & I & $  3.4$ & $  3.1$ & $ 0.73$ & FUVW4  & $ 9.00$ & $ 1.40$  & $ 9.67$ \\ 
NGC 4321    & $10.75$ & I & $  5.5$ & $  3.6$ & $ 0.55$ & FUVW4  & $ 9.02$ & $ 1.25$  & $ 9.43$ \\ 
NGC 4424    & $ 9.93$ & I & $  3.7$ & $  2.2$ & $-0.53$ & FUVW4  & $ 7.59$ & $ 1.16$  & $ 8.30$ \\ 
NGC 4457    & $10.42$ & I & $  1.5$ & $  2.2$ & $-0.52$ & FUVW4  & $ 8.21$ & $ 1.15$  & $ 8.36$ \\ 
NGC 4459   $^{\rm X}$ & $10.68$ & W & $  2.1$ & $  3.3$ & $-0.65$ & FUVW4  & $ 7.46$ & $ 2.41$  & \nodata \\ 
NGC 4476   $^{\rm X}$ & $ 9.81$ & W & $  1.2$ & $  1.2$ & $-1.39$ & FUVW4  & $ 7.05$ & $ 1.09$  & \nodata \\ 
NGC 4477   $^{\rm X}$ & $10.59$ & W & $  2.1$ & $  2.1$ & $-1.10$ & FUVW4  & $ 6.76$ & $ 1.58$  & \nodata \\ 
NGC 4496A   & $ 9.55$ & I & $  3.0$ & $  1.9$ & $-0.21$ & FUVW4  & $ 7.55$ & $ 1.15$  & $ 9.24$ \\ 
NGC 4535    & $10.54$ & I & $  6.3$ & $  3.8$ & $ 0.34$ & FUVW4  & $ 8.61$ & $ 1.78$  & $ 9.56$ \\ 
NGC 4536    & $10.40$ & I & $  4.4$ & $  2.7$ & $ 0.53$ & FUVW4  & $ 8.62$ & $ 1.06$  & $ 9.54$ \\ 
NGC 4540    & $ 9.79$ & I & $  2.0$ & $  1.4$ & $-0.78$ & FUVW4  & $ 7.69$ & $ 1.16$  & $ 8.44$ \\ 
NGC 4548    & $10.70$ & I & $  5.4$ & $  3.0$ & $-0.28$ & FUVW4  & $ 8.16$ & $ 2.00$  & $ 8.84$ \\ 
NGC 4569    & $10.81$ & I & $  5.9$ & $  4.3$ & $ 0.12$ & FUVW4  & $ 8.81$ & $ 1.40$  & $ 8.84$ \\ 
NGC 4571    & $10.10$ & I & $  3.8$ & $  2.0$ & $-0.54$ & FUVW4  & $ 7.88$ & $ 1.55$  & $ 8.70$ \\ 
NGC 4579    & $11.15$ & I & $  5.4$ & $  4.4$ & $ 0.33$ & FUVW4  & $ 8.79$ & $ 1.38$  & $ 9.02$ \\ 
NGC 4596   $^{\rm X}$ & $10.59$ & I & $  2.7$ & $  3.8$ & $-0.96$ & FUVW4  & $ 6.72$ & $ 1.83$  & \nodata \\ 
NGC 4654    & $10.57$ & I & $  5.6$ & $  4.0$ & $ 0.58$ & FUVW4  & $ 8.84$ & $ 1.18$  & $ 9.75$ \\ 
NGC 4689    & $10.24$ & I & $  4.7$ & $  3.0$ & $-0.39$ & W4ONLY & $ 8.22$ & $ 1.19$  & $ 8.54$ \\ 
NGC 4694    & $ 9.90$ & I & $  1.9$ & $  1.6$ & $-0.81$ & FUVW4  & $ 7.41$ & $ 1.30$  & $ 8.51$ \\ 
NGC 4731    & $ 9.50$ & I & $  7.3$ & $  3.0$ & $-0.22$ & FUVW4  & $ 7.29$ & $ 2.52$  & $ 9.44$ \\ 
NGC 4781    & $ 9.64$ & I & $  2.0$ & $  1.1$ & $-0.32$ & FUVW4  & $ 7.82$ & $ 1.05$  & $ 8.94$ \\ 
NGC 4826    & $10.24$ & I & $  1.5$ & $  1.1$ & $-0.69$ & FUVW4  & $ 7.79$ & $ 1.28$  & $ 8.26$ \\ 
NGC 4941    & $10.18$ & I & $  3.4$ & $  2.2$ & $-0.35$ & FUVW4  & $ 7.80$ & $ 1.27$  & $ 8.49$ \\ 
NGC 4951    & $ 9.79$ & I & $  1.9$ & $  1.9$ & $-0.46$ & FUVW4  & $ 7.65$ & $ 1.22$  & $ 9.21$ \\ 
NGC 4945   $^{\rm X}$ & $10.36$ & W & $  4.5$ & $  1.6$ & $ 0.19$ & W4ONLY & $ 8.77$ & $ 0.97$  & $ 8.92$ \\ 
NGC 5042    & $ 9.90$ & I & $  3.3$ & $  2.4$ & $-0.22$ & FUVW4  & $ 7.69$ & $ 1.84$  & $ 9.29$ \\ 
NGC 5068    & $ 9.41$ & I & $  2.0$ & $  1.3$ & $-0.56$ & FUVW4  & $ 7.26$ & $ 1.38$  & $ 8.82$ \\ 
NGC 5134    & $10.41$ & I & $  2.9$ & $  2.1$ & $-0.34$ & FUVW4  & $ 7.98$ & $ 1.14$  & $ 8.92$ \\ 
NGC 5128    & $10.97$ & W & $  4.7$ & $  4.1$ & $ 0.09$ & FUVW4  & $ 8.40$ & $ 0.98$  & $ 8.43$ \\ 
NGC 5236    & $10.53$ & I & $  3.5$ & $  2.4$ & $ 0.62$ & FUVW4  & $ 8.84$ & $ 1.14$  & $ 9.98$ \\ 
NGC 5248    & $10.41$ & I & $  3.2$ & $  2.0$ & $ 0.36$ & FUVW4  & $ 8.77$ & $ 1.14$  & $ 9.50$ \\ 
ESO097-013$^{\rm X}$ & $10.53$ & W & $  1.9$ & $  1.8$ & $ 0.61$ & W4ONLY & $ 8.42$ & $ 1.40$  & $ 9.81$ \\ 
NGC 5530    & $10.08$ & W & $  3.4$ & $  1.7$ & $-0.48$ & W4ONLY & $ 7.89$ & $ 1.34$  & $ 9.11$ \\ 
NGC 5643    & $10.34$ & W & $  3.5$ & $  1.6$ & $ 0.41$ & W4ONLY & $ 8.56$ & $ 1.06$  & $ 9.12$ \\ 
NGC 6300    & $10.47$ & W & $  3.6$ & $  2.1$ & $ 0.29$ & W4ONLY & $ 8.46$ & $ 1.12$  & $ 9.13$ \\ 
NGC 6744    & $10.72$ & W & $  7.0$ & $  4.8$ & $ 0.38$ & FUVW4  & $ 8.27$ & $ 2.75$  & $10.31$ \\ 
IC 5273     & $ 9.73$ & I & $  2.5$ & $  1.3$ & $-0.27$ & FUVW4  & $ 7.63$ & $ 1.14$  & $ 8.95$ \\ 
NGC 7456    & $ 9.65$ & I & $  4.4$ & $  2.9$ & $-0.43$ & FUVW4  & $ 7.13$ & $ 2.02$  & $ 9.28$ \\ 
NGC 7496    & $10.00$ & I & $  3.8$ & $  1.5$ & $ 0.35$ & FUVW4  & $ 8.33$ & $ 1.15$  & $ 9.07$ \\ 
IC 5332     & $ 9.68$ & I & $  3.6$ & $  2.8$ & $-0.39$ & FUVW4  & $ 7.09$ & $ 2.26$  & $ 9.30$ \\ 
NGC 7743   $^{\rm X}$ & $10.36$ & I & $  2.9$ & $  1.9$ & $-0.67$ & FUVW4  & $ 7.50$ & $ 2.65$  & $ 8.50$ \\ 
NGC 7793   $^{\rm X}$ & $ 9.36$ & I & $  1.9$ & $  1.1$ & $-0.57$ & FUVW4  & $ 7.23$ & $ 1.34$  & $ 8.70$ \\ 
\enddata 
 \tablenotetext{ul}{\,We quote a $5\sigma$ upper limit on $L_{\rm CO}$ constructed using a broad velocity window across the whole map.} 
 \tablecomments{ 
 $X$ --- extension member;  
 } 
\end{deluxetable*}

A main goal of PHANGS--ALMA is to relate cloud scale gas properties, star formation timescales, and kinematics to the properties of the host galaxy and location within the galaxy (\S\ref{sec:motivation}). To do this, we require estimates of galaxy properties. In this section, we report our current best estimate galaxy properties: orientation (\S\ref{sec:orient}), distance (\S\ref{sec:distance}), stellar mass (\S\ref{sec:stellarmass}), size (\S\ref{sec:size}), star formation rate (\S\ref{sec:sfr}), CO luminosity (\S\ref{sec:colum}), and \hi\ masses (\S\ref{sec:himass}). Then we summarize the properties of the sample and show PHANGS--ALMA targets on two common scaling relations, the main sequence of star-forming galaxies and the size--mass relation (\S\ref{sec:propsummary}). Readers who are only interested in the sample and not the provenance of the property estimates may wish to skip to \S\ref{sec:propsummary}.

We report the properties of the PHANGS--ALMA targets in Tables \ref{tab:sample_orient} and \ref{tab:sample_phys}. These properties represent current best estimates. In many cases, these estimates have been derived or refined \textit{after} sample selection, e.g., from rotation curve fitting using the PHANGS--ALMA data \citep{LANG20}. As a result, they do not perfectly agree with those used for sample selection. In other cases, PHANGS--ALMA papers use several estimates that sometimes have different zero points and scales. We note the translation between these systems whenever possible. In each case, this section presents our preferred values for scientific analysis.

For stellar mass, size, star formation rate, CO luminosity calculation, we also make these estimates for a larger sample of $261$ local galaxies that have CO maps. These include the PHANGS targets, the targets of the COMING survey \citep{SORAI19}, the targets of the Nobeyama nearby galaxy atlas \citep{KUNO07}, the targets of HERACLES \citep{LEROY09} and follow-up programs (A.~Schruba et al.\ in preparation), and the targets of the JCMT Nearby Galaxy Legacy Survey \citep{WILSON12}. In this section, we use this larger sample for methodology tests and a few comparisons. The measurements appear as blue dots in Figure~\ref{fig:intscaling}. These data are treated exactly the same as the PHANGS--ALMA data when estimating stellar mass, SFR, and other properties.

\subsection{Orientation and Galaxy Center}
\label{sec:orient}

We adopt position angles and inclinations from \citet{LANG20}. They use the CO kinematics derived from the PHANGS--ALMA data to constrain the position angle of each target. For $48$ targets, they also obtain a kinematic fit for the inclination. In cases where the CO kinematics do not sufficiently constrain the inclination, \citet{LANG20} identified preferred photometric estimates. These photometric orientation estimates come from the S$^4$G analysis of Spitzer/IRAC 3.6$\mu$m imaging by \citet{SALO15} when available, and from 2MASS NIR imaging work by \citet{JARRETT03} when not available. \citet{LANG20} also identify a preferred systemic velocity for each target. For targets not considered by \citet{LANG20}, we default to orientation parameters from S$^4$G \citep[][]{SHETH10,MUNOZMATEOS15,SALO15} when available and HyperLEDA \citep{PATUREL03,MAKAROV14} when not available. Whenever we become aware of kinematic-based estimates, we update our adopted orientations to reflect these. Based on comparison of the S$^4$G and HyperLEDA orientation parameters, we adopt a $\pm 5^\circ$ typical uncertainty for the inclination and a $\pm 10^\circ$ typical uncertainty for the position angle \citep[consistent with the uncertainty estimates by][in the cases with well-measured orientations]{LANG20}.

Following \citet{LANG20}, we adopt photometric centers from \citet{SALO15} when available. These centers leverage sensitive near-infrared imaging and should accurately reflect the center of stellar mass in the galaxy. When these are not available, we adopt central positions from the 2MASS Large Galaxy Atlas \citep{JARRETT03}. When neither are available, we use the optically defined central position from HyperLEDA or NED. The choice of photometric center generally matters at the level of a few arcseconds or less. We adopt a fiducial uncertainty of~1\arcsec.

\subsection{Distance}
\label{sec:distance}

We adopt distance estimates from \citet[][]{ANAND21}. Their work compiles a mixture of literature estimates and new tip of the red giant branch (TRGB) distances based on \textit{Hubble Space Telescope} (HST) observations. The new distances are derived from observations carried out as part of PHANGS--HST \citep{LEE21}. We reproduce the \citet[][]{ANAND21} distance estimates in Table \ref{tab:sample_orient}, where we also note the original references for the literature distances. We recommend citing the original reference when adopting these distances. 

In total, \citet{ANAND21} compile distance estimates and associated uncertainties for $117$ galaxies, including all \ntarget\ PHANGS--ALMA targets. They evaluate the available distance estimates for each target, select the highest quality estimate, and assign an associated uncertainty. Approximately $40\%$ of these distances come from high quality primary distance indicators, either TRGB- or Cepheid-based distances. Group-based distances, results from a numerical action method, and Tully--Fisher estimates account for most of the rest of the distances. A~handful of distances come from other direct techniques, mostly surface brightness fluctuations. The numerical action method \citep[e.g.,][]{SHAYA17} may be the least familiar of these. This method assigns distance based on a galaxy's position and velocity using a sophisticated three dimensional model of gravitationally induced flows in the local volume. It can be roughly thought of as a vastly-improved version of the Virgocentric flow-corrected Hubble flow distance.

The location of PHANGS--ALMA targets, combined with the distances from \citet{ANAND21}, are shown in Figure~\ref{fig:sample_location} and we show the distribution of distances in Figure~\ref{fig:hists_1}. Perhaps the main thing to see from these figures is that PHANGS--ALMA spans a large dynamic range in distances. This makes accounting for distance effects, e.g., by convolving data to a common physical scale before scientific analysis, crucial. The figures also show that, as expected, our targets cluster near our $17$~Mpc distance cutoff with several targets beyond the nominal cutoff. This simply reflects the uncertainty in distances and larger volume present at large radius.

The distances for PHANGS galaxies provided by \citet{ANAND21} build on many time- and effort-intensive studies, and we also refer to the initial/\linebreak[0]{}original studies when quoting the distances (Table~\ref{tab:sample_orient}). Also note PHANGS--ALMA has made heavy use of the distance compilations by the Extragalactic Distance Database \citep[EDD;][]{TULLY09}, the closely related CosmicFlows projects \citep{COURTOIS12,TULLY16}, HyperLEDA \citep{PATUREL03,MAKAROV14}, and NED. For non PHANGS--ALMA targets not considered by \citet{ANAND21}, we utilize these databases following the method described by \citet{LEROY19}.

\subsection{Stellar Mass}
\label{sec:stellarmass}

Stellar mass informs our selection and plays a wide-ranging role in the PHANGS--ALMA scientific analysis. The stellar plays a major role setting gravitational potential in the disk across most of the survey area. As a result, maps that trace stellar mass are crucial to define distinct dynamical environments in galaxies \citep[e.g.,][]{COLOMBO14B}. For galaxies without high-quality optical spectroscopy (e.g., from PHANGS--MUSE), e also use stellar mass to predict the metallicity of a galaxy and to place the galaxy in the context of the larger galaxy population.

We use near-infrared maps, combined with a radially varying mass-to-light ratio, \mtolwise , to estimate stellar mass and stellar disk size in our targets\footnote{We adopt the same mass-to-light ratio, \mtolwise\ in units of \mtolunits , for WISE1 at $3.4\mu$m and IRAC1 at $3.6\mu$m. See \citet{LEROY19} for details on the conversions.}. Whenever possible, we use the high quality IRAC $3.6$~\micron\ data from S$^4$G \citep{SHETH10}. When these are not available, we use WISE1 $3.4$~\micron\ maps from \citet{LEROY19}. The wavelength coverage of the two bands heavily overlap. We expect both to be dominated by light from old stars and minimally affected by dust extinction. Never the less dust emission can contaminate the band \citep[e.g.,][]{MEIDT12,SIMONIAN17} and the age of the stellar population still affects the mass-to-light ratio at these wavelengths \citep[e.g.,][]{BELL01}. We prefer to use the IRAC imaging for practical reasons. Although the WISE maps cover every galaxy, the versions produced by \citet{LEROY19} have $7.5\arcsec$ spatial resolution, which presents a limitation when attempting to mask foreground stars. They also have poorer signal-to-noise ratio than the IRAC data, though this is less problematic since telescopes easily detect most of the emission in most targets. These mass calculations assume a \citet{CHABRIER03} stellar initial mass function \citep[via][]{BRUZUAL03,SALIM16,SALIM18}.

In both cases, we apply the mass-to-light ratio prescription from \citet{LEROY19}, which we develop in more detail. This estimator combines UV, near-IR, and mid-IR emission to estimate the specific star formation rate, which is a strong driver of \mtolwise . The numerical estimate is calibrated to match the results of the
GALEX-SDSS-WISE Legacy Catalog \citep[GSWLC;][]{SALIM16,SALIM18} studying the full SDSS main galaxy sample. Adopting this mass-to-light ratio places our measurements on an approximately matched system with $>10{,}000$ other local galaxies and the full SDSS main galaxy sample studied by \citet{SALIM16,SALIM18}. To carry out this estimate we also constructed matched $15\arcsec$ resolution maps for WISE1, WISE3, WISE4, GALEX FUV, and GALEX NUV emission. In \S\ref{sec:sfr}, we use these to help estimate the star formation rate of our targets. We discuss the exact procedure and compare the implied \mtolwise\ to other stellar mass estimates used in PHANGS in the next subsection.

We process the S$^4$G IRAC images before using them to measure stellar masses or sizes. First, we apply the S$^4$G masks described by \citet{MUNOZMATEOS15} outside the central $10\arcsec$ of each target. These stars remove likely foreground stars and artifacts from the image. We do not mask the central part of each galaxy to avoid blanking bright nuclei; even when these regions are affected by saturation, we need them in our calculations. Then we mask some remaining visible artifacts and stars by hand. These are usually away from the galaxy but relevant to background subtraction. Then, we estimate and subtract a local background. This is usually calculated using the mode in a circular annulus $1.5{-}3~r_{25}$ in radius, but we adjusted the range based on visual inspection. Finally, we interpolated to fill in masked pixels with the median value at that galactocentric radius (i.e., the value suggested by the galaxy's radial profile). After that, we convolved the masked, background subtracted, interpolated IRAC images to $7.5\arcsec$ resolution for processing in a way that matched the WISE data.

\citet{LEROY19} describe the background subtraction and masking for the WISE images. For stellar mass estimates, we begin with the WISE1 sky-subtracted images. We apply the Gaia- and 2MASS-based star masks described in \citet{LEROY19}, but only outside $0.5~r_{25}$. Then, we manually inspected each image and blanked any remaining bright stars or image artifacts by hand. This caught some low-level background artifacts missed by the automatic masking in \citet{LEROY19} and also ensured that we only blanked objects that were foreground stars in the galaxy itself. Then, as with the IRAC data, we interpolate across masked regions using the median of pixels at the same galactocentric radius. A handful of cases lacked IRAC images and are so heavily covered by foreground stars that WISE struggles to recover the galaxy. In these cases, we used only the median profile.

\subsubsection{Mass-to-Light Ratio Estimates}

\begin{figure*}[ht!]
\begin{center}
\includegraphics[width=0.45\textwidth]{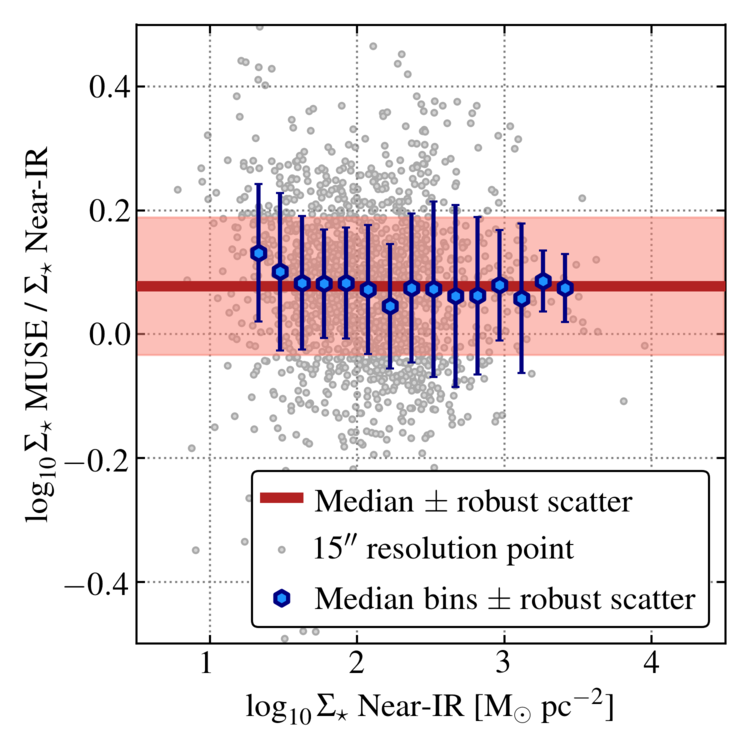}
\includegraphics[width=0.45\textwidth]{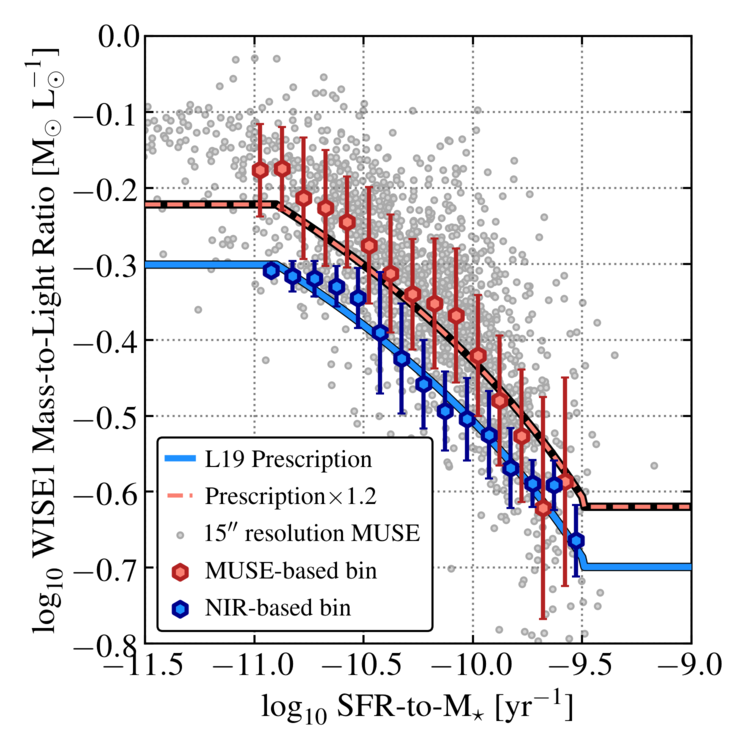}
\end{center}
\vspace*{-10px}
\caption{\textbf{Stellar mass estimates and near-IR mass-to-light ratios compared to results from PHANGS--MUSE.} Comparison between our near infrared-based stellar mass estimates and stellar masses based on full spectral fitting of PHANGS--MUSE data in the wavelength range $4850{-}7000\AA$ for 19 targets (E.~Emsellem et al.\ in preparation). The \textit{left} panel shows the ratio between MUSE-based $\Sigma_\star$ and near IR-based $\Sigma_\star$ at $15\arcsec$ resolution. Gray points show individual lines of sight. Blue points show median and robustly-estimated scatter for data binned by $\Sigma_\star$. The red line and shaded region show the overall median ratio of ${\sim}0.08$~dex, i.e., a factor of $1.2$, with about $\pm 0.1$~dex $\approx \pm 30\%$ scatter. The \textit{right} panel shows the near-IR mass-to-light ratio, \mtolwise , as a function of local specific star formation rate, SFR/$M_\star$, estimated at $15\arcsec$ resolution. Individual gray points show the $\mtolwise$ implied by the MUSE fitting. Red and blue points with error bars show the binned $\mtolwise$ in the region with MUSE coverage for our near-IR data (blue) and implied by MUSE (red, binning the gray points). Lines show the adopted \mtolwise\ prescription from \citet{LEROY19} and a version scaled by $1.2\times$, which describes the binned MUSE data reasonably well. Our fiducial stellar masses follow the blue line showing the \citet{LEROY19} prescription in order to retain the link to the SDSS via \citet{SALIM16, SALIM18}.
\label{fig:muse_check}}
\end{figure*}

We convert from IRAC $3.6$~\micron\ or WISE $3.4$~\micron\ intensity to stellar mass surface density, $\Sigma_\star$, by multiplying the measured intensity with a local estimate of the mass-to-light ratio, \mtolwise , so that
\begin{equation}
\label{eq:sigmastar_irac}
\frac{\Sigma_{\star}}{1~{\rm M_\odot~pc}^{-2} } \approx 350 \left( \frac{\mtolwise}{0.5} \right) \left(\frac{I_{\rm 3.6\mu m}}{1~{\rm MJy~sr^{-1}}} \right)~\cos i
\end{equation}
\noindent for IRAC1 data or
\begin{equation}
\label{eq:sigmastar_wise}
\frac{\Sigma_{\star}}{1~{\rm M_\odot~pc}^{-2} } \approx 330 \left( \frac{\mtolwise}{0.5} \right) \left(\frac{I_{\rm 3.4\mu m}}{1~{\rm MJy~sr^{-1}}} \right)~\cos i
\end{equation}
\noindent for WISE data. The $\cos i$ factor accounts for the inclination of the galaxy and \mtolwise\ is the near-infrared mass-to-light ratio in units of \mtolunits . The slightly different pre\-factors between Equations \ref{eq:sigmastar_irac} and \ref{eq:sigmastar_wise} reflect differences in the bandpass of the two instruments. We assume that modulo the different pre-factors, the same mass-to-light ratio \mtolwise\ describes both $3.4$~\micron\ to $3.6$~\micron. We predict \mtolwise\ from an empirical fit relating \mtolwise\ to the local SFR-to-WISE1 color from \citet{LEROY19}, 
\begin{equation}
\label{eq:pred_mtol}
\mtolwise~\left[ \mtolunits \right] =
\begin{cases}
    0.5 & \text{if}~ Q < a \\
    0.5 + b \, \left( Q - a\right) & \text{if}~a < Q <c \\
    0.2 & \text{if}~Q>c
\end{cases}
\end{equation}
\noindent where $Q = \log_{10} {\rm SFR}/( \nu_\textrm{ W1} L_{\nu ,\textrm{ W1}})$ and $\nu_\textrm{W1}$ refers to the frequency corresponding to the $3.4\mu$m central wavelength of the WISE1 band. Here $a=-11.0$, $b=-0.375$, and $c=-10.2$.

\medskip

\textbf{Placing PHANGS--ALMA in a larger context:} \citet{LEROY19} derived Equation~\eqref{eq:pred_mtol} from an empirical fit relating the measurements and fitting results in the GSWLC. \citet{SALIM16} fit multi-band NUV, optical, and IR photometry using the CIGALE population synthesis code \citep[][]{BOQUIEN19} and the \citet{BRUZUAL03} models to estimate SFR and $M_\star$ for $\gtrsim 650{,}000$ SDSS galaxies. They showed excellent agreement with the stellar masses in the previous standard JHU-MPA Value Added Catalog \citep[see][]{KAUFFMANN03,TREMONTI04}\footnote{\url{https://www.sdss.org/dr14/spectro/galaxy_mpajhu/}}. Using a similar framework, \citet{LEROY19} also estimated $M_\star$ and SFR for $\sim 15{,}000$ galaxies likely to lie within $50$~Mpc. Thus, adopting this approach to estimate stellar mass (and SFR below) maximizes our ability to place PHANGS--ALMA objects in a larger context. In addition to matching the SDSS main galaxy sample via the GSWLC, this specifically places PHANGS--ALMA on a nearly-matched mass scale to the GASS and xCOLD GASS surveys \citep{SAINTONGE17,CATINELLA18}. These also built on SDSS stellar mass estimates from the JHU-MPA Value Added Catalog and leveraged combinations of GALEX and WISE for SFR estimates. Because we also construct these estimates for all local galaxies with a CO map, we are in a position to compare these galaxies, PHANGS--ALMA, and a larger set of literature galaxy property measurements.

\medskip

\textbf{Using SFR/\boldmath$M_\star$ to predict the mass-to-light ratio:} Equation~\eqref{eq:pred_mtol} specifically uses the ratio of SFR-to-WISE1 luminosity, i.e., a ``specific star formation rate-like'' quantity, to predict \mtolwise . In the GSWLC as well as other theoretical and empirical studies \citep[e.g.,][]{BELL03,KANNAPPAN13,TELFORD20} some cognate of the specific star formation rate is a strong predictor of the age of the stellar population and thus variations in the near infrared mass-to-light ratio. Because calculating the specific star formation rate requires knowing $M_\star$, this creates a degenerate situation and we predict \mtolwise\ from the ratio of SFR-to-WISE1 luminosity. Based on comparison with the GSWLC, this predicts \mtolwise\ with ${\sim}0.1$~dex scatter. See appendix~A of \citet{LEROY19} for more details.  

\medskip

\textbf{Radial Profile-based \boldmath\mtolwise\ ---} We use Equation~\eqref{eq:pred_mtol} to predict a radial profile of \mtolwise . We first match the resolution of these all data sets at $15\arcsec$. Then, we calculate radial profiles of WISE1 emission and SFR, either from FUV+WISE4, NUV+WISE4, or only WISE4 emission as available.  Then we apply Equation~\eqref{eq:pred_mtol} to all rings. We smooth the resulting profile with a 2~kpc kernel to avoid noise in the estimate. In the outer parts of galaxies where we do not recover SFR at sufficient signal to noise, we adopt the median \mtolwise\ calculated outside $0.5~r_{25}$.

This radial profile-based approach reflects the reality that stellar populations change as a function of galactocentric radius \citep[e.g.,][among many other examples]{WATKINS16,DALE20}. Because of the use of low resolution azimuthal averages, it will be less sensitive to mass-to-light ratio variations associated with young stars or dust, e.g., in spiral arms or bars \citep[see][]{QUEREJETA15}. Overall, we found that the azimuthal averages offered a good compromise between isolating distinct regions of a galaxy and achieving good signal to noise. This refines the approach in \citet{LEROY19}. They used global colors that trace SFR/$M_\star$ to estimate \mtolwise . The largest difference arises for targets with powerful nuclear sources, either starburst or AGN, which can affect the global colors. Any such bright, contaminating source can bias the global colors to suggest high SFR/$M_\star$ and effectively obscure an underlying older stellar population.

\medskip

\textbf{Validation against PHANGS--MUSE:} PHANGS--MUSE (E.~Emsellem et al.\ in preparation; P.I. E.~Schinnerer) mapped $19$ PHANGS--ALMA targets with the MUSE integral field unit on the Very Large Telescope. PHANGS--MUSE delivers estimates of the stellar mass based on fitting the $4850{-}7000\AA$ spectrum of individual regions using stellar population synthesis models. Specifically, the fitting uses the E-MILES libraries \citep{VAZDEKIS16} and adopts the same \citet{CHABRIER03} initial mass function (IMF) as the \citet{SALIM16,SALIM18} fits, so that the IMF is matched across both treatments. These fits offer stellar mass estimates independent of the near-IR based approaches above. The PHANGS--MUSE data cover only a limited number of targets and do not span the entire stellar disk of their targets. Therefore, we make use of these to explore the uncertainty in our $\mtolwise$ and $\Sigma_\star$ values by comparing to a an independent estimate with a distinct set of systematic uncertainties. After convolving the PHANGS--MUSE internal release~v2 to $15\arcsec$ resolution, we match the astrometric grid of our IRAC- and WISE-based measurements to compare $\Sigma_\star$ from MUSE to our estimate. We also calculate the \mtolwise\ implied by MUSE by dividing the MUSE-based stellar mass by the WISE1 intensity.

We compare the MUSE results to our fiducial measurements in Figure~\ref{fig:muse_check}. The left panel shows overall excellent agreement between our near-IR based $\Sigma_\star$ and the MUSE-based values. Over two orders of magnitude in $\Sigma_\star$, the ratio between the two methods remains roughly fixed, with only a modest offset and moderate scatter. Specifically, we find a median offset of ${\sim}0.08$~dex or a factor of ${\sim}1.2$, and ${\sim}0.11$~dex or $30\%$ point-to-point scatter.

The right panel of Figure~\ref{fig:muse_check} shows that the MUSE fitting results in a similar trend of \mtolwise\ versus SFR/$M_\star$ as what we adopt for the near-IR data. The PHANGS--MUSE fitting implies a steady decline in \mtolwise\ as a function of increasing SFR/$M_\star$. The slope of the decline resembles that applied to the near-IR maps (blue). The higher $\Sigma_\star$ found in the right panel means that the actual \mtolwise\ values implied are shifted to ${\sim}20\%$ higher values than those we predict based on \citet{LEROY19}. Mostly, the offset appears to be a simple multiplicative shift. The ``reddest,'' lowest SFR/$M_\star$ regions in the MUSE data also show higher \mtolwise , approaching $1~\mtolunits$, than implied by our adopted formula. 

Overall, Figure~\ref{fig:muse_check} suggests excellent agreement between two independent ways to estimate the mass. The PHANGS--MUSE results provide a key validation of our adopted \mtolwise\ treatment. To transfer from the ``PHANGS near-IR'' approach used to compute sample properties to the PHANGS--MUSE system one would need to scale the near-IR based stellar mass or $\Sigma_\star$ up by a factor of ${\sim}1.2$. Alternatively, one could scale the PHANGS--MUSE mass down by a similar factor.

\subsubsection{Checks on Stellar Mass Estimates}

\begin{deluxetable}{lcc}
\tabletypesize{\footnotesize}
\tablecaption{Comparison Among Stellar Mass Estimates \label{tab:mstar_check}}
\tablewidth{0pt}
\tablehead{
\colhead{Other Estimate} &
\colhead{$\log_{10}$ Median\tablenotemark{a}} &
\colhead{$\log_{10}$ Scatter} \\[-5px]
\colhead{} &
\colhead{[dex]} &
\colhead{[dex]}
}
\startdata
\hline
Integrated photometry & $-0.07$ & $0.06$ \\
WISE1 vs. IRAC1 & $0.00$ & $0.03$ \\
WISE1 + fixed \mtolwise & $0.06$ & $0.13$ \\
S$^4$G ICA + fixed \mtolwise & $0.00$ & $0.11$ \\
\hline
PHANGS--MUSE\tablenotemark{b} & $0.08$ & $0.11$ \\
\hline
\enddata
\tablenotetext{a}{Reported as $\log_{10}$ of the median of the other estimate over our estimate for galaxies where both are available.}
\tablenotetext{b}{For PHANGS--MUSE only the statistics refer to a line-of-sight by line-of-sight comparison at $15\arcsec$ resolution. For all other cases we refer to comparison of integrated $M_\star$.}
\tablecomments{``Integrated photometry'': Our fiducial estimate divided by those from \citet{LEROY19} based only on integrated photometry. ``WISE1 vs. IRAC1'': comparing results using WISE1 vs. IRAC1 and otherwise treating the data identically. ``WISE1 + fixed \mtolwise '': Using WISE1 but with only a fixed mass-to-light ratio of $0.35$~\mtolunits , not the radially varying or galaxy-dependent prescription used in our fiducial estimates. ``S$^4$G ICA + fixed \mtolwise '': `old stellar emission' maps from \citet{QUEREJETA15} using a fixed \mtolwise\ of $0.5$~\mtolunits . The table reports comparisons for a sample of local galaxies that have CO mapping including all PHANGS--ALMA targets. The results restricting to just PHANGS--ALMA are almost identical. See also Figure~\ref{fig:mstar_check}.}
\end{deluxetable}

\begin{figure}[ht!]
\begin{center}
\includegraphics[width=0.495\textwidth]{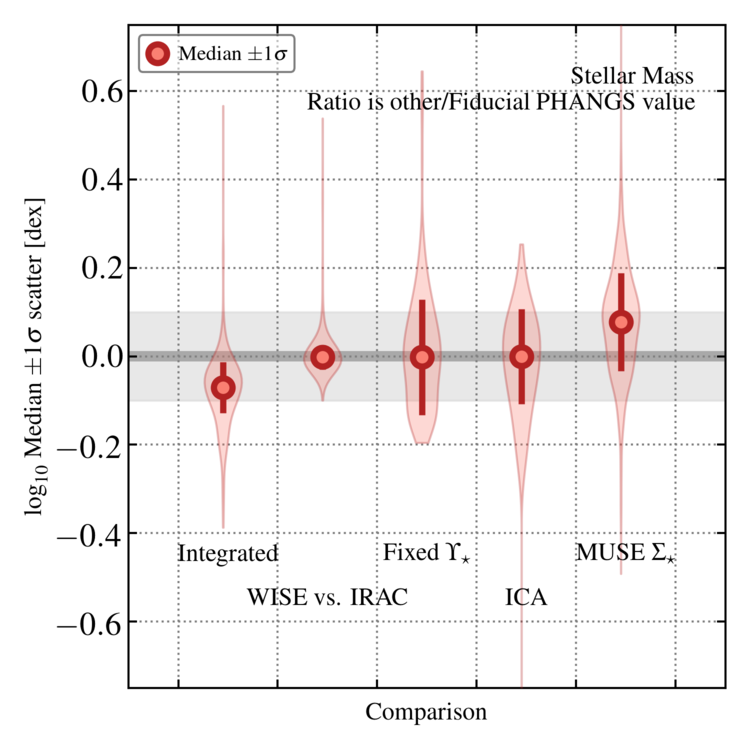}
\end{center}
\vspace*{-10px}
\caption{\textbf{Comparisons among stellar mass estimates.} Median (point), $\pm 1\sigma$ scatter (error bar), and distribution (shaded violin plot) for the ratios of stellar mass estimated in various ways. In each case, we consider the ratio of our adopted estimate, which uses IRAC- or WISE-based near-IR data and a radially varying mass-to-light ratio estimated following Equation \ref{eq:pred_mtol}, to some other estimate. From left to right we show ``Integrated'' as comparison between our estimates and those from \citet{LEROY19} using only integrated photometry; ``WISE vs.\ IRAC'' comparing results using IRAC vs.\ WISE and treating the data identically. The results are highly consistent except for a few outliers. ``Fixed \mtolwise :'' Showing results using \textit{only} WISE1 intensity with no varying mass-to-light ratio. ``ICA:'' Using independent component analysis-based `old stellar emission' maps from \citet{QUEREJETA15} with a fixed mass-to-light ratio. ``MUSE $\Sigma_\star$:'' $15''$ resolution comparison between our mass surface density estimates and PHANGS--MUSE spectral fitting results, available for a subset of targets. The gray line and shaded region show perfect agreement and $\pm 0.1$~dex scatter. See also Table~\ref{tab:mstar_check}.
\label{fig:mstar_check}}
\end{figure}

Figure~\ref{fig:mstar_check} and Table~\ref{tab:mstar_check} provide further checks on our $M_\star$ estimates. Our fiducial masses use IRAC or WISE near-IR maps with the radially varying estimate of the mass-to-light ratio described above. In Figure~\ref{fig:mstar_check} we adjust this approach and compare alternative estimates to our values. From left to right, we compare to (1) masses calculated only from integrated colors and luminosities by \citet{LEROY19}, (2) masses calculated using only IRAC or WISE data, i.e., checking whether adopting one data set or the other makes a difference; (3) masses calculated without any variable mass-to-light ratio, i.e., with a single fixed \mtolwise\ of $0.35$~\mtolunits\ applied to all data; and (4) masses calculated applying a fixed mass-to-light ratio of $0.5$~\mtolunits\ to the contaminant-corrected ``old stellar''\footnote{We refer to the first component of their independent component analysis (ICA) as `old stellar emission' following the interpretation in \citet{MEIDT12} and \citet{QUEREJETA15}. For brevity, we label this ``ICA'' in the plots and table.} maps of \citet{QUEREJETA15}. The final column, (5), compares the surface density, $\Sigma_\star$, estimated in our adopted approach to that derived from PHANGS--MUSE spectral modeling. Unlike the other cases, which consider whole galaxies, this comparison treats each $15''$ line of sight independently.

Table~\ref{tab:mstar_check} summarizes the numerical results. Qualitatively, we see that shifting from the estimates based on integrated photometry from \citet{LEROY19} to resolved estimates of \mtolwise\ here increases the overall stellar mass by only ${\sim}20\%$ on average but can matter much more in a few cases. These tend to be cases with strong nuclear concentrations, but also a few cases where our more careful treatment of the data here has improved the masking. Beside a few edge cases, it makes a negligible difference to the overall $M_\star$ whether we use IRAC or WISE as long as the treatment of the data remains the same. The adoption of variable \mtolwise\ does matter. Without the variable \mtolwise , masses scatter by $\pm 0.13$~dex or $35\%$ relative to our derived stellar masses. Similarly, if we integrate the contaminant-corrected `old stellar emission' component maps from \citet{QUEREJETA15} and apply a fixed \mtolwise , we find $\pm 0.11$~dex or $30\%$ scatter relative to our adopted values.

In Figure~\ref{fig:mstar_check} and Table~\ref{tab:mstar_check} we have already chosen our typical mass-to-light ratios to remove systematic offsets between our adopted approach and the comparison data. Thus two key points from the comparison are:

\begin{enumerate}
\item We find $\mtolwise = 0.35$~\mtolunits\ on average.
\item When using the ``old stellar emission'' maps derived by \citet{QUEREJETA15} from combining IRAC 3.6$\mu$m and $4.5\mu$m emission, we suggest to apply $\mtolwise = 0.5$~\mtolunits\ for consistency with our adopted system. The difference between this value and the average $\mtolwise = 0.35$~\mtolunits\ reflects the removal of contaminants from the \citet{QUEREJETA15} maps.
\end{enumerate}

Although we set the median to agree, the scatter between our adopted system and either the \citet{QUEREJETA15} or fixed \mtolwise\  is still significant. In general, our checks show that any substantial shift in methodology induces ${\sim}0.1$~dex, or $\approx 25\%$, scatter in the measurements. This uncertainty dramatically exceeds any statistical uncertainty associated with the photometry and we suggest adopting $\pm 0.1$~dex as a realistic systematic uncertainty on stellar masses of individual galaxies \citep[in good agreement with the level of uncertainty due to age variations found in the IRAC-based ICA maps by][]{MEIDT14}. It also seems reasonable to adopt comparable uncertainty to describe the overall calibration of our stellar mass scale.

Despite these uncertainties, we emphasize again that (1) our calibration scheme anchors our mass scale to the \citet{SALIM16,SALIM18} and \citet{LEROY19} estimates, and through these to work on the SDSS main galaxy sample, and (2) that our adopted \mtolwise\ treatment seems to qualitatively agree with results from PHANGS--MUSE, while showing a mild $0.08$~dex systematic offset.

\subsection{Stellar Disk Size}
\label{sec:size}

The size of the stellar disk plays a key role in many analyses. Using the same $\Sigma_\star$ values used to calculate $M_\star$, we derive two size estimates for each target. First, we measure the effective radius, $R_e$, that contains half of the stellar mass of the galaxy. The effective radius is a standard measure of galaxy size but can be sensitive to stellar bulges, nuclear mass concentrations, and other details of inner galaxy structure. Therefore to complement $R_e$, we measure the scale length of the stellar disk, $l_\star$, outside the galaxy center. This disk scale length is related to the metallicity gradient \citep[e.g.,][]{SANCHEZ14} and stellar disk scale height \citep[e.g.,][]{KREGEL02,SALO15,SUN20}, among other applications. In Table \ref{tab:sample_phys} we also report the $25^{\rm th}$ magnitude isophotal $B$-band radius, $r_{25}$, from RC3 \citep{DEVAUCOULEURS91} via HyperLEDA.

\subsubsection{Measurement of \texorpdfstring{$R_e$}{Re} and \texorpdfstring{$l_\star$}{lstar}}

We measure $R_e$ and $l_\star$ from the $7.5\arcsec$ resolution estimates of $\Sigma_\star$ described in \S\ref{sec:stellarmass}. Using these data, we construct a radial profile using the orientations in Table \ref{tab:sample_orient}. In order to lower our sensitivity to the adopted inclination and resolution, we consider only points within $\pm 75^\circ$ of the major axis, except for the central $30\arcsec$. Within the central $30\arcsec$, we include all position angles\footnote{We consider this necessary to ensure that we remain sensitive to any massive nuclear features, which can strongly affect the estimated $R_e$. Based on visual inspection of all profiles we do not see strong imprints of this change in procedure in the derived profiles, and as shown in \S \ref{sec:sizecheck} our estimates match previous profile-based size estimates very well.} Then, we measure the azimuthally averaged intensity using the mean within $30\arcsec$ and the median intensity in each ring at larger radii. The use of the median suppresses any unmasked foreground stars at the expense of also losing some sensitivity to galactic structure. We use the mean in the central aperture to retain full sensitivity to any bright nuclear emission when calculating $R_e$. Next, we subtract a background value from the radial profile, which is just the median of all radial bins outside $2~r_{25}$. Finally, we integrate these profiles assuming azimuthal symmetry and use this integral to identify the half-mass radius, $R_e$, the radius encompassing $90\%$ of the mass, $R_{90}$, and the stellar scale length, $l_\star$. The integral for calculating $R_e$ extends out to $2~r_{25}$, which we verified by eye in each case to represent an apparent convergence of the radial profile. We report the estimated $R_e$ and $l_\star$ in Table \ref{tab:sample_orient}.

To fit $l_\star$, we identify a fitting range over which the radial profile appears exponential. By default, this range spans from $0.4$ to $1.0~r_{25}$, i.e., excluding roughly the central two scale lengths. We examined each profile by eye and adjusted this range to reflect where the disk appears exponential over a large range of radii outside the galaxy center. This approach does a good job of matching sophisticated disk/bulge decompositions \citep[e.g.,][]{SALO15}. This also closely resembles the approach adopted by \citet{SANCHEZ14}, a study that we use to estimate metallicity gradients.

\subsubsection{Translation Between Different Size Estimates}

We measure effective radii and stellar scale lengths. The $25^{\rm th}$ magnitude $B$-band isophotal radius also remains widely used, particularly for studies of local galaxies. In Table \ref{tab:size_check} and Figure \ref{fig:size_check} we report translations between different size measurements for our galaxies: half-mass radius, $R_e$, profile-based scale length, $l_\star$, and isophotal radius, $r_{25}$ \citep[we adopt from RC3 via HyperLEDA]{DEVAUCOULEURS91,MAKAROV14}. For reference, for a pure exponential disk, $l_\star$, should relate to the effective radius via

\begin{equation}
\label{eq:relstar_theory}
R_e^{\rm disk} = 1.68~l_\star \quad {\rm or} \quad l_\star = 0.60~R_e^{\rm disk}~.
\end{equation}

\noindent In practice, we find that our measured $R_e$ and $l_\star$ deviate from this ideal case. We show this in Figure~\ref{fig:lstar_re} and report the typical ratio and scatter for our measurements in Table~\ref{tab:size_check}. We find that in practice, the relationship in our sample can be better described via

\begin{equation}
\label{eq:relstar_real}
R_e^{\rm disk} = 1.41~l_\star \quad {\rm or} \quad l_\star = 0.71~R_e^{\rm disk}~.
\end{equation}

\noindent The difference between these two formulae simply reflects that our targets are not ideal exponential disks, and that many galaxies harbor compact inner structures. These drive $R_e$ to smaller values but do not affect~$l_\star$. As Figure~\ref{fig:lstar_re} shows, even Equation~\eqref{eq:relstar_real} still misses a significant minority of outliers that have high $l_\star$ compared to $R_e$, reflecting strong central mass concentrations relative to an exponential disk.

The $25^{\rm th}$ magnitude $B$-band isophotal radius, $r_{25}$, remains in common use when studying nearby galaxies. Table~\ref{tab:size_check} and Figure~\ref{fig:size_check} show that $r_{25}$ exhibits a well-defined scaling with our two size measures: 
\begin{equation}
\label{eq:sizes}
R_e \approx r_{25}/3.08 \quad {\rm and} \quad l_\star \approx r_{25}/4.3
\end{equation}
\noindent on average, both with ${\sim}0.14$~dex scatter.

\subsubsection{Checks on Stellar Disk Sizes}
\label{sec:sizecheck}
\begin{figure*}[ht!]
\begin{center}
\includegraphics[width=0.475\textwidth]{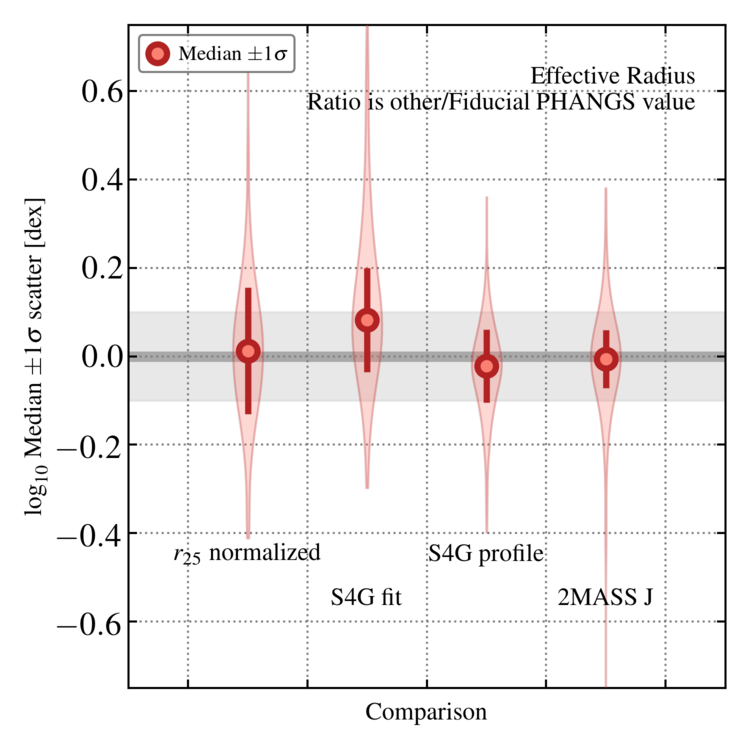}
\includegraphics[width=0.475\textwidth]{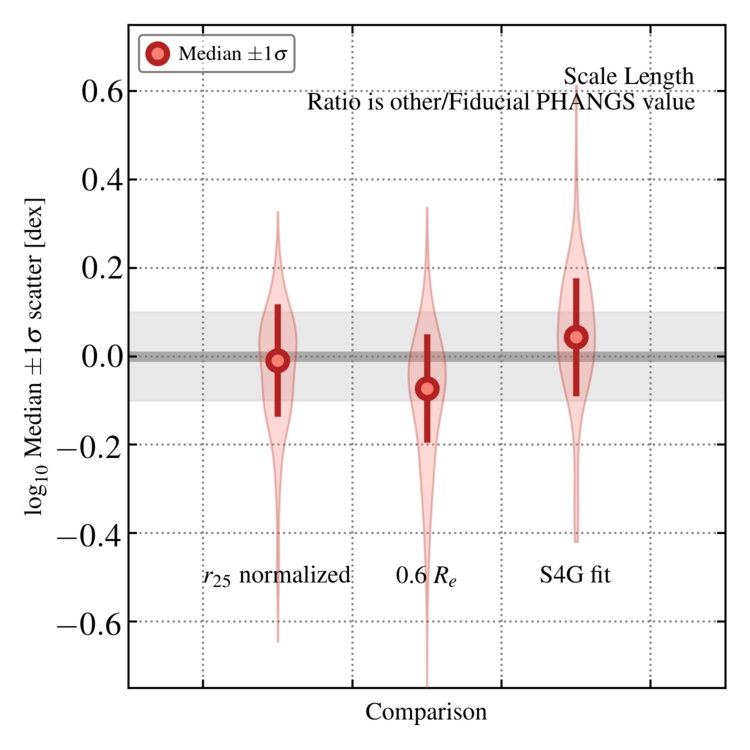}
\end{center}
\vspace*{-10px}
\caption{\textbf{Stellar disk sizes in different systems.} Comparisons between our estimated $R_e$ and $l_\star$ and literature measurements; see also Table~\ref{tab:size_check}. The \textit{left} panel compares our measured $R_e$ to literature size measurements: the isophotal radius $r_{25}$ scaled according to Equation \ref{eq:sizes} to predict $R_e$, estimates from \citet[][``S$^4$G fit'']{SALO15} and \citet[][``S$^4$G profile'']{MUNOZMATEOS15}, and estimates from 2MASS. The S$^4$G profile-based sizes and 2MASS $J$-band sizes show good agreement with our estimates. The \textit{right} panel shows results for our scale length $l_\star$ fits. Here we have only the \citet{SALO15} literature values to compare to. We do find overall agreement, but see more scatter than for $R_e$.
\label{fig:size_check}}
\end{figure*}

\begin{figure}[ht!]
\fig{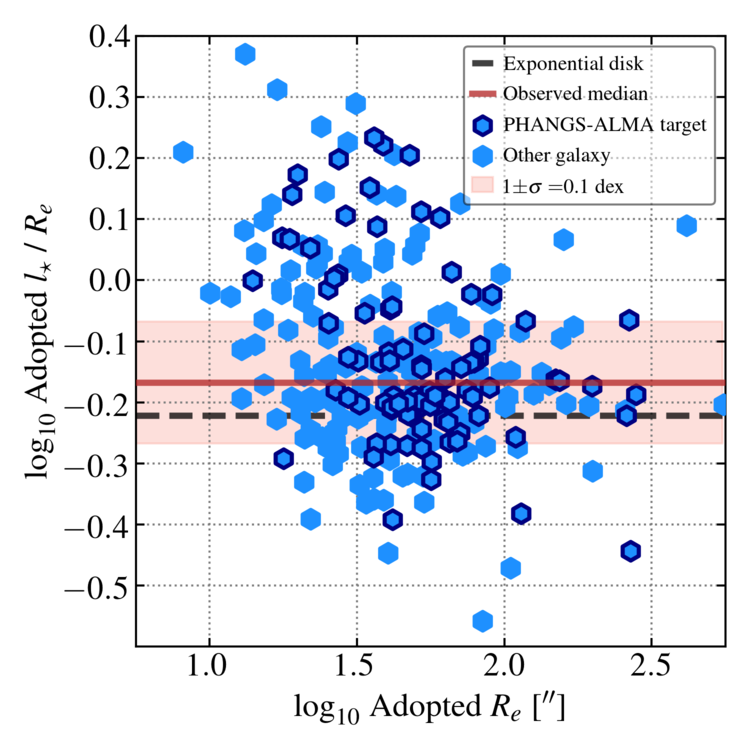}{0.475\textwidth}{}
\vspace*{-10px}
\caption{\textbf{Scale length and effective radius.} The fit exponential scale length, $l_\star$, divided by the effective radius, $R_e$, measured from the same profile. The black dashed line shows expectations for a pure exponential disk (Eq.~\ref{eq:relstar_theory}). Points above this line show some central concentration compared to a pure exponential disk, and this central concentration drives $R_e$ to smaller values. The $l_\star$ fits should better capture the exponential portion of the profile. The axes are correlated, and the detailed trend is less important than the qualitative point: that many galaxies with small $R_e$ have bright nuclear concentrations, and that the fit $l_\star$ indicates a much larger disk in some of these cases. That larger $l_\star$ will be the relevant quantity for disk structure calculations or metallicity gradient estimation.
\label{fig:lstar_re}}
\end{figure}

We compare our estimated sizes to literature size estimates for the same galaxies and show the results in Table~\ref{tab:size_check} and Figure~\ref{fig:size_check}. We compare our measured sizes to:

\begin{enumerate}
\item The $25^{\rm th}$ magnitude $B$-band isophotal radius from RC3 via HyperLEDA \citep{DEVAUCOULEURS91,MAKAROV14}. This remains the most widely used measure of galaxy size for studies of local galaxies, and the results of this comparison have already been used to derive Equation \ref{eq:sizes}.

\item Effective radii estimated by the 2MASS Large Galaxy Atlas \citep[LGA;][]{JARRETT03}. They measured $R_e$ at $J$, $H$, and $K$ band for many of our galaxies and a large sample of additional galaxies. These are profile-based sizes using near-IR data and we expect our measurements to match them well.

\item Effective radii estimated from the S$^4$G near-IR data by \citet{MUNOZMATEOS15}. These are also profile-based measurements, in many cases using the same data with slightly distinct processing, assumptions, and orientations. We expect to match these values well.

\item The results of single- and multi-component fits to the S$^4$G data by \citet{SALO15}. They present one-component S\'ersic profiles, which yield an effective radius. They also fit exponential disk components to most targets. When they fit an \texttt{expdisk} component, their $h_r$ captures the exponential scale length and represent the most direct comparison for our $l_\star$.

\end{enumerate}

Overall, Figure~\ref{fig:size_check} and Table~\ref{tab:size_check} show that our measured sizes match measurements from other sources well. As expected, we match previous profile-based estimates of the size using near-IR data well. In 2MASS, we find best agreement with the $J$-band size, with a median ratio of 2MASS-to-PHANGS $R_e$ of $0.99$. We find slightly smaller sizes in 2MASS $H$ and $K$ band, likely reflecting the poorer sensitivity in those bands, but the agreement is also excellent for those bands. Our match to the \citet{MUNOZMATEOS15} sizes is also good, we find an $0.95$ ratio on average.

On average, the effective radii from the S\'ersic model fits by \citet{SALO15} 
are ${\sim}21\%$ larger than our $R_e$. This likely reflects the fact that fitting a single S\'ersic profile to the entire galaxy de-emphasizes the contribution of any distinct, massive central component. By-eye checks of galaxies with strongly discrepant size estimates between us and \citet{SALO15} reveal that such galaxies often have central mass concentrations. These same discrepancies arise comparing the \citet{SALO15} values to \citet{MUNOZMATEOS15} ones. Those two measurements use the same data, but the single-component fitting approach of \citet{SALO15} tends to yield larger $R_e$ than the profile fits of \citet{MUNOZMATEOS15}, a $26\%$ higher $R_e$ for matched galaxies, on average, with ${\sim}0.1$~dex scatter.

Despite the good match on average, we find just under ${\sim}0.1$~dex or about $25\%$ scatter between our measurements and those of \citet{MUNOZMATEOS15} or \citet{JARRETT13}, and we adopt $\pm0.1$~dex as a realistic $1\sigma$ uncertainty on $R_e$. The scatter between estimates reflects differences in the adopted orientations and sensitivity to many aspects of the exact method used. The method of defining the total flux, choice to focus on the major axis or whole galaxy, choice to apply a radially varying mass-to-light ratio, and a host of other minor decisions all affect the derived $R_e$. To some extent, this reflects that galaxies are not axisymmetric, thin, tilted disks with a single orientation and a simple radial profile. It also highlights how $R_e$, which depends on an estimate of the total flux and focuses on the inner, often complex, part of the galaxy is a less stable size measure than an isophotal radius for local galaxies.

Figure~\ref{fig:size_check} and Table~\ref{tab:size_check} show that our stellar scale lengths agree reasonably well with those from \citet{SALO15}. Their sophisticated two dimensional decomposition tends to yield ${\sim}11\%$ larger scale lengths than our fits to the radial profiles. Our approach focuses on fitting the part of the profile roughly in the range $0.4{-}1.0~r_{25}$ that looks approximately exponential. This might lead us to emphasize more cleanly declining parts of the disk and avoid regions with flatter profiles. Regardless, the overall agreement with \citet{SALO15} is also good, and $l_\star$ is likely somewhat more uncertain than $R_e$.

\begin{deluxetable}{lcc}[th!]
\tabletypesize{\footnotesize}
\tablecaption{Comparison Among Size Estimates \label{tab:size_check}}
\tablewidth{0pt}
\tablehead{
\colhead{Other Estimate} &
\colhead{Median\tablenotemark{a}} &
\colhead{$\log_{10}$ Scatter} \\[-5px]
\colhead{} &
\colhead{[dex]} &
\colhead{[dex]}
}
\startdata
\hline
\multicolumn{3}{c}{Effective radius $R_e$\tablenotemark{b}} \\
\hline
$r_{25}$ (255\tablenotemark{c}) & $3.08$ & $0.14$ \\
2MASS $J$ (158) & $0.99$ & $0.07$ \\
S$^4$G profile fit (203) & $0.95$ & $0.08$ \\
S$^4$G S\'ersic fit (199) & $1.21$ & $0.12$ \\
\hline
\multicolumn{3}{c}{Scale length $l_\star$\tablenotemark{d}} \\
\hline
$r_{25}$ (253) & $4.3$ & $0.13$ \\
S$^4$G $h_\star$ fit (69) & $1.11$ & $0.14$ \\
Our $R_e$ (253) & $1.41$ & $0.12$ \\
\enddata
\tablenotetext{a}{Reported as median of the other estimate over our estimate for galaxies where both are available.}
\tablenotetext{b}{Effective radius calculated from stellar mass radial profile considering all data.}
\tablenotetext{c}{Number of galaxies available for that comparison.}
\tablenotetext{d}{Stellar disk scale length considering only the exponential part of the profile.}
\tablecomments{The table reports comparisons for a sample of local galaxies that have CO mapping including all PHANGS--ALMA targets. The results restricting to just PHANGS--ALMA are almost identical (Figure~\ref{fig:size_check}).}
\end{deluxetable}

\subsection{Star Formation Rate}
\label{sec:sfr}

\begin{figure*}[ht!]
\begin{center}
\includegraphics[width=0.95\textwidth]{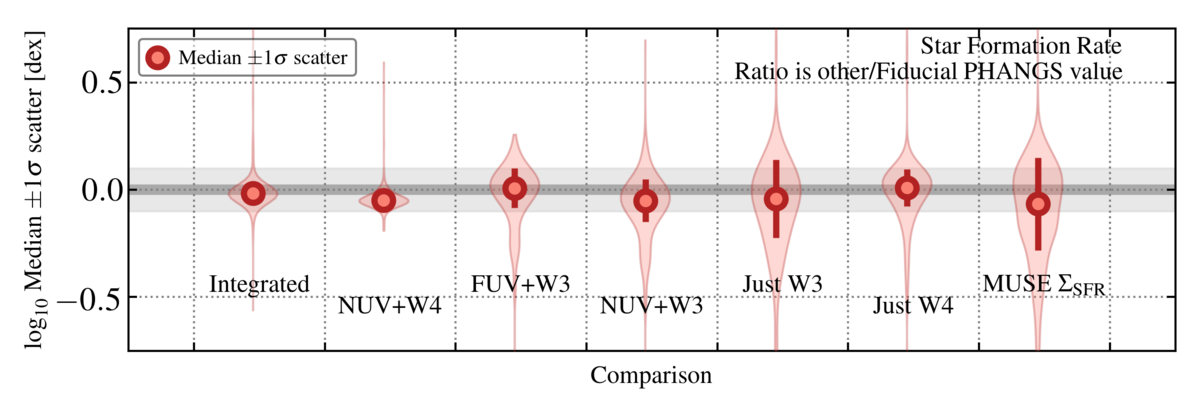}
\end{center}
\vspace*{-10px}
\caption{\textbf{Comparisons among star formation rate estimates.} Median (point), $\pm 1\sigma$ scatter (error bar), and distribution (shaded violin plot) for the ratios among star formation rates estimated in various ways. In each case, we consider the ratio of some other estimate to our adopted estimate. From left to right, we compare our fiducial values (usually FUV+WISE4, see text) to ``Integrated'': results using integrated photometry from \citet{LEROY19}, ``NUV+W4'': results preferring the GALEX NUV band instead of FUV and hybridizing with WISE4; ``FUV+W3'': results using WISE3 instead of WISE4; ``NUV+W3''" same, but now using NUV and WISE3; ``Just W3'': estimates based on only WISE3 emission; ``Just W4'': estimates based on only WISE4 emission; ``MUSE $\Sigma_{\textrm{SFR}}$'': estimates based on Balmer-decrement corrected H$\alpha$ emission from PHANGS--MUSE. Unlike the other estimates, which are all compared for whole galaxies, the MUSE $\Sigma_{\rm SFR}$ comparison considers each $15''$ line of sight. See also Table~\ref{tab:sfr_check}.
\label{fig:sfr_check}}
\end{figure*}

\begin{deluxetable}{lccc}
\tabletypesize{\scriptsize}
\tablecaption{Adopted SFR Calibrations and Uncertainties \label{tab:sfr_calibs}}
\tablewidth{0pt}
\tablehead{
\colhead{Band} &
\colhead{$\log_{10} C$\tablenotemark{a}} &
\colhead{RMS Scatter\tablenotemark{b}} &
\colhead{RMS Bias\tablenotemark{c}} \\
\colhead{} &
\colhead{$\left(\dfrac{{\rm M}_\odot~{\rm yr}^{-1}}{{\rm erg~s}^{-1}}\right)$} &
\colhead{(dex)} &
\colhead{(dex)}
}
\startdata
FUV\tablenotemark{d} & $-43.42$ & $0.1$ & $0.09$\tablenotemark{d} \\
NUV\tablenotemark{d} & $-43.24$ & $0.1$ & $\ldots$\tablenotemark{d} \\
\hline
WISE4 with FUV & $-42.73$ & $0.17$ & $0.06$ \\
WISE4 with NUV & $-42.79$ & $0.18$ & $0.08$ \\
WISE4 alone & $-42.63$ & $0.17$ & $0.15$ \\
\hline
WISE3 with FUV & $-42.79$ & $0.21$ & $0.44$ \\
WISE3 with NUV & $-42.86$ & $0.22$ & $0.47$ \\
WISE3 alone & $-42.70$ & $0.2$ & $0.37$ \\
\enddata
\tablenotetext{a}{$C$ is the factor to convert $\nu L_\nu$ to SFR \citep[notation follows][]{KENNICUTT12}.}
\tablenotetext{b}{Galaxy-to-galaxy scatter of $C$, in dex, from comparison to \citet{SALIM18}. See \citet{LEROY19} for more details.}
\tablenotetext{c}{``Bias'' here refers to the systematic scatter, in dex, among the empirical calibration coefficients to be applied to the WISE data across the SFR-$M_\star$ plane. This is calculated by binning the \citet{SALIM18} data into cells of SFR and M$_\star$, calculating $C$ for each cell, and then measuring the rms scatter across the binned data. This indicates the amount of systematic uncertainty in the empirical calibration of the tracer. See \citet{LEROY19} for more details.}
\tablenotetext{d}{The coefficients applied to FUV and NUV do not depend on which IR band is used.}
\tablecomments{From \citet{LEROY19} based on the GSWLC \citep{SALIM16,SALIM18}. These coefficients represent a best effort to anchor our SFR estimates to both a large population of local galaxies \citep{LEROY19} and the very large SDSS-based catalog of \citet{SALIM16,SALIM18}.}
\tablenotemark{d}{The bias in the calibration coefficient to convert extinction-corrected FUV emission from the \citet{SALIM18} fits to SFR for cells in SFR-$M_\star$ space with at least $50$ galaxies. The NUV calibration factor is extrapolated from FUV, so no independent bias is calculated.}
\end{deluxetable}

We estimate the star formation rate (SFR) using GALEX FUV, GALEX NUV maps, WISE3 ($12$~\micron), and WISE4 ($22$~\micron) maps. We adopt the prescription suggested by \citet{LEROY19} and calibrated to match results from the population synthesis modeling in the GSWLC \citep{SALIM16,SALIM18}. In brief:

\begin{enumerate}
\item We adopt a linear combination of FUV and WISE4 ($22$~\micron) light whenever both bands are available. If FUV is not available but NUV is, we use NUV emission. If no GALEX data are available, we use only WISE4. In this paper we use WISE3 emission only to construct alternative estimates used to check our data (see below). Table \ref{tab:sample_phys} list the combination of bands used to estimate SFR for each galaxy.

\item For each band, we convert from luminosity, $\nu L_\nu$, to SFR using a conversion factor, $C$. 

\item The conversion factor for NUV and FUV are the same no matter which IR band is used (see \citealt{SALIM07} and \citealt{LEROY19}). In practice this results from fits to the \citet{SALIM16} results. Physically, this reflects an assumption that the recent star formation history does not vary too much across our sample. The conversion factor for WISE3 or WISE4 depends on the band with which it is combined.

\item We add together the UV (if present) and mid-IR terms to obtain the total SFR. The choice to use a linear combination sacrifices some accuracy in exchange for a more stable estimator. 
\end{enumerate}

\subsubsection{Checks on Star Formation Rates}

\begin{deluxetable}{lcc}[ht!]
\tabletypesize{\footnotesize}
\tablecaption{Comparison Among Star Formation Rate Estimates \label{tab:sfr_check}}
\tablewidth{0pt}
\tablehead{
\colhead{Other Estimate} &
\colhead{$\log_{10}$ Median\tablenotemark{a}} &
\colhead{$\log_{10}$ Scatter} \\[-5px]
\colhead{} &
\colhead{[dex]} &
\colhead{[dex]}
}
\startdata
\hline
Integrated photometry & $-0.02$ & $0.02$ \\
NUV+WISE4 & $-0.05$ & $0.01$ \\
FUV+WISE3 & $0.01$ & $0.09$ \\
NUV+WISE3 & $-0.05$ & $0.10$ \\
WISE3 only & $-0.04$ & $0.18$ \\
WISE4 only & $0.0$ & $0.09$ \\
\hline
\enddata
\tablenotetext{a}{Reported as median of the other estimate over our estimate for galaxies where both are available.}
\tablecomments{All columns report $\log_{10}$ median and scatter among galaxy-integrated SFR estimates. Each row compares a different indicator to our adopted SFR. ``Integrated photometry'': our fiducial estimated divided by those from \citet{LEROY19} based only on integrated photometry. The other entries refer to changes in the adopted band combination. The table reports comparisons for a sample of local galaxies that have CO mapping including all PHANGS--ALMA targets. The results restricting to just PHANGS--ALMA are almost identical. See also Figure~\ref{fig:sfr_check}.}
\end{deluxetable}

Tables \ref{tab:sfr_calibs} and~\ref{tab:sfr_check} and Figure~\ref{fig:sfr_check} illustrate the uncertainty in our estimated SFR. Table~\ref{tab:sfr_calibs} reports the scatter and bias implied by adopting a single linear calibration and attempting to match the full \citet{SALIM18} data set. Here ``bias'' refers to the rms scatter in the implied calibration factor, $C$, as a function of $M_\star$ and SFR. To calculate this, \citet{LEROY19} binned the \citet{SALIM18} data into cells of SFR and $M_\star$, solved for $C$, in each cell, and then calculated the rms scatter across all populated cells in the SFR-$M_\star$ plane. Thus, this number captures the degree of systematic uncertainty implied by applying a single value of $C$ to the whole galaxy population. This is discussed in detail by \citet{LEROY19}, but here we emphasize two key points. First, using WISE4 only shows notably higher bias than WISE4 combined with a UV band. This reflects the fact that low mass galaxies, in particular, often show less obscuration compared to high mass galaxies. The importance of a hybrid approach to star formation rate tracers has been reviewed by \citet{KENNICUTT12} and \citet{CALZETTI13}.

The other important point from Table~\ref{tab:sfr_calibs} relates to the use of WISE3. WISE3 data have better sensitivity and resolution than the WISE4 data, and below we will use them to aperture correct our CO data. However, as Table~\ref{tab:sfr_calibs} shows, the appropriate calibration to translate WISE3 to SFR shows strong trends as a function of stellar mass and specific star formation rate. This is consistent with the significant contribution of PAHs to the WISE3 band and the known dependence of PAH emission on metallicity \citep[e.g.,][]{ENGELBRACHT06} and radiation field \citep[e.g.,][]{CHASTENET19}. This is a strong effect, greater than a factor of~$2$. We do note that this will mainly affect extremes in the galaxy population, i.e., low mass, low metallicity galaxies and very low and high SFR/$M_\star$ galaxies.

For some PHANGS--ALMA science applications, the higher resolution and improved signal to noise of the WISE3 data will be important. Indeed, we used WISE3 to define the field of view for exactly these reasons. When we need such a ratio to be quantitative, we will generally estimate a WISE3-to-WISE4 ratio for that galaxy or location and use this as an empirical correction to apply the better-calibrated WISE4-based SFR prescriptions.

Figure~\ref{fig:sfr_check} and Table~\ref{tab:sfr_check} mostly bear out these trends within our own data. In this figure we vary the adopted band combination and compare to our fiducial SFR estimate. We also compare the value from \citet{LEROY19} based only on integrated photometry. Both the table and the figure show that all band combinations have consistent calibrations on average. However, if we do not hybridize WISE with some UV band, then we find a higher galaxy-to-galaxy scatter, ${\sim}0.1$~dex using only WISE4. If we substitute WISE3 for WISE4, we also see ${\sim}0.1$~dex scatter relative to our fiducial estimates. Finally, while using WISE3 alone yields a matched result on average, though, we find almost $0.2$~dex galaxy-to-galaxy scatter relative to our fiducial estimates.

\subsubsection{UV+IR and PHANGS--MUSE Balmer-Decrement}

\begin{figure*}[ht!]
\begin{center}
\includegraphics[width=0.45\textwidth]{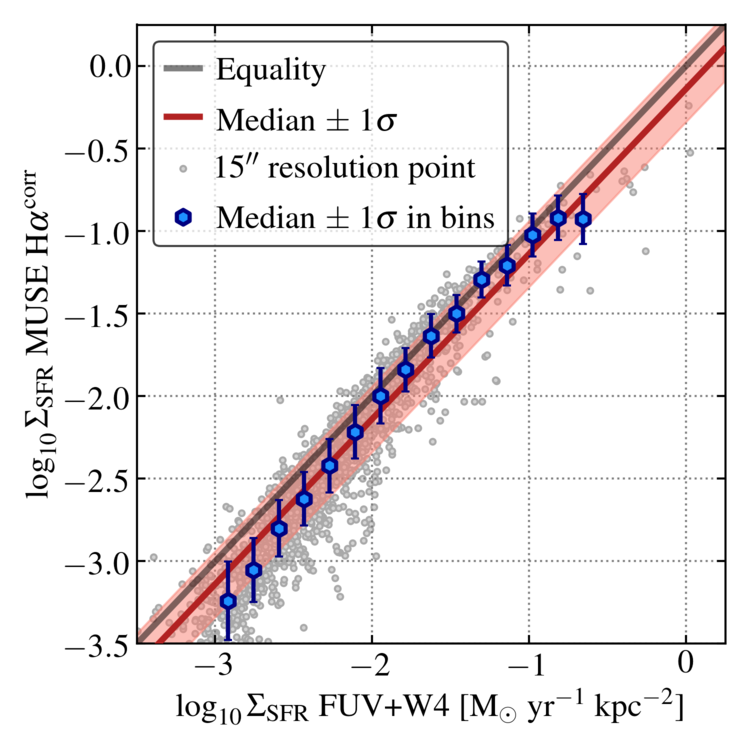}
\includegraphics[width=0.45\textwidth]{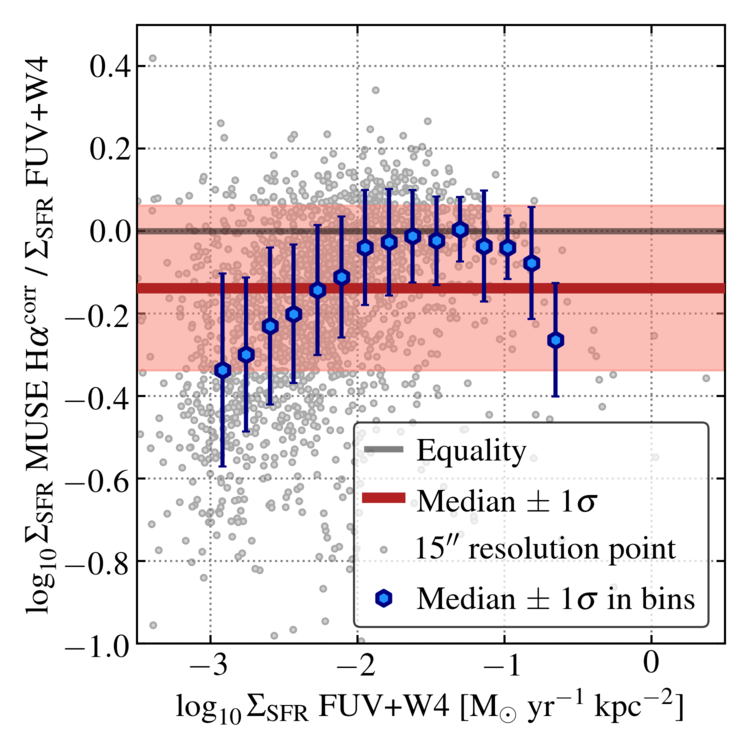}
\end{center}
\vspace*{-10px}
\caption{\textbf{Comparisons between $\Sigma_{\rm SFR}$ estimated from UV+IR and from PHANGS--MUSE Balmer decrement-corrected H$\alpha$.} For each $15''$ sightline in the field of view of PHANGS--MUSE, we compare $\Sigma_{\rm SFR}$ estimated from H$\alpha$ extinction-corrected using the Balmer decrement, H$\alpha^{\rm corr}$ ($y$-axis) to $\Sigma_{\rm SFR}$ estimated from combining FUV and mid-IR $22\mu$m emission ($x$-axis). The H$\alpha$+H$\beta$ measurements are only available over a limited field-of-view in $19$ targets, but represent a high quality, independent reference. The \textit{left} panel shows the two estimates plotted against one another. The \textit{right} panel shows the ratio of the two estimates. The two $\Sigma_{\rm SFR}$ estimates correlate well, but show a $\approx 20{-}30\%$ systematic offset. We also observe a systematic trend in the ratio as a function of $\Sigma_{\rm SFR}$, with the FUV+WISE4 estimate likely somewhat overestimating $\Sigma_{\rm SFR}$ at low values. For a more detailed exploration of these results see F. Belfiore et al. (in preparation).
\label{fig:muse_sfr_check}}
\end{figure*}

For $19$ of our targets, PHANGS--MUSE (E. Emsellem et al. in preparation) offers an independent, high-quality estimate of $\Sigma_{\rm SFR}$ from H$\alpha$ recombination line emission corrected for extinction using the Balmer decrement. An in-depth exploration of the differences between these two methods is presented by F. Belfiore et al. (in prepation). Here we present a brief comparison of $\Sigma_{\rm SFR}$ estimated from the PHANGS--MUSE H$\alpha$ and H$\beta$ maps at $15''$ resolution and the UV+IR $\Sigma_{\rm SFR}$ estimates available for the whole sample.

For this comparison, we compare our best $\Sigma_{\rm SFR}$ estimates, which are from FUV+WISE4 for all $19$ MUSE targets, to $\Sigma_{\rm SFR}$ estimated from H$\alpha$ and H$\beta$ measured from PHANGS--MUSE at $15''$ resolution. We use the two recombination lines to estimate $A_{\rm H\alpha}$, the extinction affecting the H$\alpha$ line, by assuming a screen geometry, a \citet{CARDELLI89} extinction curve, and Case B recombination (see F. Belfiore et al. (in preparation) for more details). Then we place the two $15''$ resolution $\Sigma_{\rm SFR}$ maps on the same astrometric grids and compare results.

Figure \ref{fig:muse_sfr_check} shows the results of this comparison. As the left panel shows, the two estimates correlate extremely well, showing a Spearman rank correlation coefficient of $\approx 0.95$ despite using completely different approaches to estimate $\Sigma_{\rm SFR}$. However, the estimators do yield numerically different estimates. On average the FUV+WISE4 estimate is $20{-}30\%$ higher than the H$\alpha$-based estimate. Specifically, treating all points equally yields a median H$\alpha$-to-FUV+WISE4 ratio of $\approx 0.73$, or $-0.14$~dex, with $\approx 0.2$~dex, or $60\%$ rms scatter about the median. Considering the binned measurements, shown in blue, we find a median ratio of $0.84$ or $-0.07$~dex with $\approx 0.1$ rm scatter in the median value from bin-to-bin. Given the independent approaches to estimate $\Sigma_{\rm SFR}$, this offset still represents reasonable agreement, and any average offset could be accounted for by adjusting the empirically-derived coefficient on the WISE4 term in the FUV+WISE4 indicator. 

The Figure shows that the offset and scatter between the two tracers does reflect an average offset and a systematic trend, such that the two tracers agree well at $\Sigma_{\rm SFR} \gtrsim 10^{-2}$~M$_\odot$~yr$^{-1}$~kpc$^2$ but show increasing divergence below this value. Specifically, at low $\Sigma_{\rm SFR}$, the FUV+WISE4 SFR indicator predicts a higher $\Sigma_{\rm SFR}$ than the H$\alpha$-based estimate. Figure \ref{fig:muse_sfr_check} illustrates a systematic trend with $\Sigma_{\rm SFR}$, and F. Belfiore et al. (in preparation) show that a similar trend holds as a function of the ratio $\Sigma_{\rm SFR}/\Sigma_\star$, a quantity that traces the local specific star formation rate. 

The sense of the trend agrees with the expectation that the $22\mu$m emission captured in the WISE4 band may contain significant emission not directly associated with star formation, and that this contamination will be stronger in regions with low star formation activity \citep[e.g., see][]{GROVES12,LEROY12,BOQUIEN16,SIMONIAN17}. Given the multiwavelength coverage in PHANGS, it may be possible to correct these low-intensity sight-lines using physical or empirical prescriptions for the IR cirrus \citep[e.g.,][]{LEROY12,DAVIS14}. Alternatively, high resolution mid-IR observations from the \textit{James Webb Space Telescope} offer a path forward to morphologically and structurally isolate emission associated with star forming regions \citep[e.g.,][]{CALZETTI05,BOQUIEN16}.

We refer the reader to F. Belfiore et al. (in preparation) for more details but emphasize three main points from the comparison. First, the overall scaling between the two tracers appears very good, and the median offset of $20{-}30\%$ represents good agreement given the independent approaches. Second, the clear systematic trends suggest that the FUV+WISE4 maps be treated with caution at low $\Sigma_{\rm SFR} \sim 10^{-3}$ M$_\odot$~yr$^{-1}$~kpc$^2$. A cirrus correction to the WISE4 data might improve the situation. However, third, the coefficients on the FUV+WISE4 estimator have been set to ensure average consistency with the GSWLC and the larger SDSS sample \citet{SALIM16,SALIM18}. For this work, we use the coefficients ``as is'' and note the comparison to PHANGS--MUSE as an important indicator of the systematic uncertainty.

\subsection{CO(2--1) Luminosity}
\label{sec:colum}

We use the PHANGS--ALMA data to estimate the integrated CO luminosity of each target. To do this, we integrate a $17\arcsec$ resolution version of the ALMA \cotwo\ cube. Working with a low resolution version of the data makes signal identification easier among. We apply a wide mask to the data before integrating\footnote{This is not exactly the same as the ``broad mask'' described in Sections \ref{sec:processing} and \ref{sec:products}. The mask that we use here is designed to also apply to the HERA data used for checking the fluxes and to compare to other literature CO measurements. We do not expect this detail to have any important effect on the derived CO luminosities.}. We find that the results derived using this mask agree well with those applying no masking at all. Using no mask only adds more noise to the measurement. We report the integrated \cotwo\ luminosities for each galaxy with in Tables \ref{tab:sample_phys}.

\subsubsection{Aperture Correction}
\label{sec:apcorr}

\begin{figure*}[ht!]
\begin{center}
\includegraphics[width=0.45\textwidth]{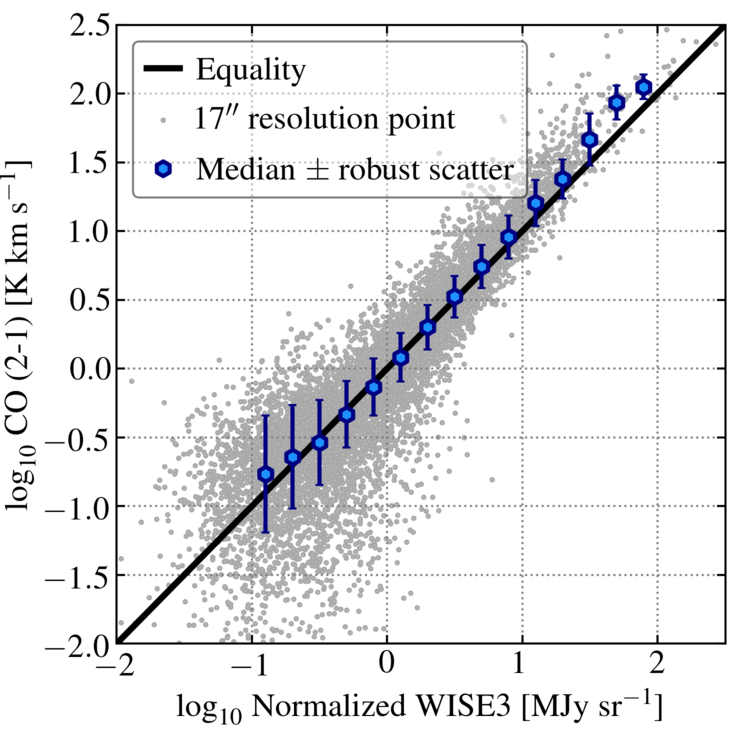}
\includegraphics[width=0.45\textwidth]{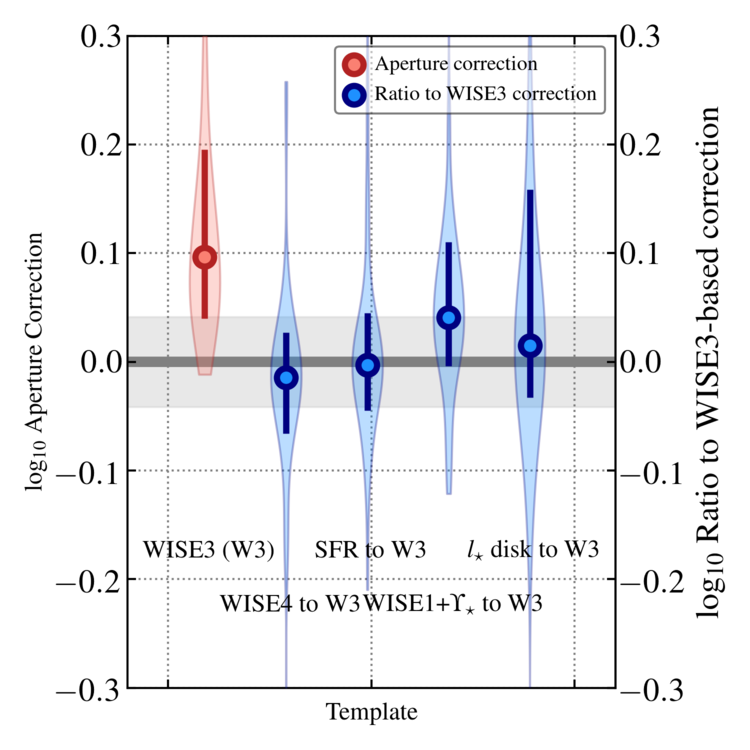}
\end{center}
\vspace*{-10px}
\caption{\textbf{Calculations of aperture corrections to infer total CO luminosity.} PHANGS--ALMA covers most of the area of active star formation but not the entire disk of each target. We derive an aperture correction to correct from our measured $L_{\rm CO}$ to the global $L_{\rm CO}$ for each target. To do this, we use WISE3 intensity as a template for CO intensity. The \textit{left} panel shows the motivation for this choice, plotting \cotwo\ intensity as a function of WISE3 intensity at $17''$ resolution within the PHANGS--ALMA coverage. A single normalization factor has been derived and applied to correct each galaxy for differences in the median WISE3-to-\cotwo\ ratio. After this normalization, the two intensities show a linear correlation coefficient of $0.88$, exhibiting a stunningly tight correlation across three decades in intensity. The \textit{right} panel shows the derived aperture corrections for WISE3 (red) and the ratio between aperture corrections derived using other templates and that from WISE3. WISE3 represents the clear best option, but if we adopted another reasonable choice like WISE4, stellar mass, or star formation rate then this would imply ${\sim}10\%$ scatter (gray band) on average and a bias of ${\lesssim}10\%$.
\label{fig:apcorr_check}}
\end{figure*}

\begin{deluxetable}{lcc}[ht!]
\tabletypesize{\footnotesize}
\tablecaption{PHANGS--ALMA \cotwo\ Aperture Corrections \label{tab:apcorr_check}}
\tablewidth{0pt}
\tablehead{
\colhead{Template} &
\colhead{Correction\tablenotemark{a}} &
\colhead{Correlation\tablenotemark{b}}
}
\startdata
\hline
WISE3 (adopted) & $1.25~(1.10{-}1.57)$ & $0.82$ \\
WISE4 & $1.16~(1.04{-}1.59)$ & $0.77$ \\
SFR (FUV+W4)\tablenotemark{c} & $1.19~(1.07{-}1.84)$ & $0.79$ \\
WISE1+$\mtolwise$ & $1.41~(1.19{-}1.86)$ & $0.53$ \\
Exponential disk ($0.2~r_{25}$) & $1.16~(1.07{-}1.55)$ & $0.43$ \\
Exponential disk ($l_\star$) & $1.36~(1.10{-}1.90)$  & $0.41$ \\
Exponential disk ($0.6~R_e$) & $1.18~(1.05{-}1.47)$ & $0.18$ \\
\hline
\enddata
\tablenotetext{a}{Median and $16^{\rm th}$ to $84^{\rm th}$ percentile range for the aperture correction to be applied to the PHANGS--ALMA data.}
\tablenotetext{b}{Linear correlation coefficient relating the template band to CO intensity in the region with coverage. Each galaxy is first normalized by the median template-to-CO ratio.}
\tablenotetext{c}{Best SFR estimate. This uses FUV+WISE4 when available. Otherwise we adopt NUV+WISE4 if available and WISE4 otherwise.}
\tablecomments{This table reports results only for PHANGS--ALMA. The correction gives the implied aperture corrections. Correlation indicates how well each band linearly predicts \cotwo\ emission for each individual galaxy inside the coverage.}
\end{deluxetable}

Our maps only cover the area of active star formation, defined by an IR intensity contour (\S \ref{sec:observations}). Though we expect most CO emission to lie within our target area, faint CO emission can continue well into the outer disks of galaxies \citep[e.g.,][]{YOUNG95,BRAINE07,SCHRUBA11}. To account for this, we derive an aperture correction that can be applied to our measured integrated \cotwo\ to yield the global $L_{\rm CO}$ for the galaxy. 

After considering several options, we chose to calculate this aperture correction by using the WISE3 map as a template for CO emission. We calculate the sum of WISE3 emission over the whole galaxy and the sum of WISE3 emission over just the region with CO coverage. In Table \ref{tab:sample_phys} we report this WISE3-based aperture correction, but note that we leave it to the user to actually apply the correction. The $L_{\rm CO}$ that we report reflects only the direct integration of our maps.

Table~\ref{tab:apcorr_check} and Figure~\ref{fig:apcorr_check} show the motivation for choosing WISE3. Inside the PHANGS--ALMA footprint, we construct several possible ``templates'' that could be used for the aperture correction. Here a template refers to a quantity that might linearly trace CO intensity. We consider WISE3 intensity, WISE4 intensity, $\Sigma_{\rm SFR}$ from FUV+WISE4, and $\Sigma_\star$ estimated from WISE1+\mtolwise . We also consider three plausible versions of an exponential disk model: one adopting a scale length of $0.2\,r_{25}$ \citep{YOUNG95,LEROY08,BOLATTO17}, one adopting our fit exponential scale length, and one adopting $0.6\,R_e$ (see Equation~\ref{eq:relstar_theory}). These size-based corrections resemble those used, e.g., by \citet{LISENFELD11} and \citet{SAINTONGE11}.

Table~\ref{tab:apcorr_check} shows that WISE3 exhibits the strongest linear correlation coefficient with CO intensity of any considered quantity. After removing a single median WISE3-to-CO ratio from each galaxy, the WISE3 emission exhibits a linear correlation coefficient of $0.82$ with \cotwo\ intensity across the PHANGS sample. This stunningly strong correlation has also been seen and investigated in detail by \citet{CHOWN21} in the EDGE--CALIFA \coone\ data. We confirm their finding that $I_{\rm CO}$ shows a strong, linear correlation with WISE3 within galaxies. After removing a galaxy-to-galaxy offset, the correlation between WISE3 emission and \cotwo\ intensity is even stronger than that relating CO to $\Sigma_{\rm SFR}$ estimated from FUV and WISE4 emission. Figure~\ref{fig:apcorr_check} shows this correlation. After galaxy-by-galaxy normalization, WISE3 tracks \cotwo\ emission almost linearly with $<0.1$~dex scatter across three decades in intensity. We do caution that as discussed above, the ratio of WISE3 emission to SFR shows strong galaxy-to-galaxy variations. These variations correlate with metallicity and specific star formation rate \citep{LEROY19}. Our application avoids much of this issue by removing an overall galaxy-to-galaxy scaling. The PHANGS--ALMA data offer the possibility to explore the empirical and physical nature of the WISE3--CO correlation in much more detail in future works.

WISE3 offers an outstanding option to derive aperture corrections and we derive a median correction of ${\sim}1.25$ for PHANGS--ALMA, with most targets yielding corrections between $1.1$ and $1.5$. In other words, the WISE3 emission implies that our maps usually cover $\sim 70{-}90\%$ of the total CO emission. Table~\ref{tab:apcorr_check} and the right panel of Figure~\ref{fig:apcorr_check} show that we would have arrived at similar aperture corrections using only WISE4 emission or our FUV+WISE4-based $\Sigma_{\rm SFR}$ estimates. Our stellar mass maps imply slightly higher corrections, on average, but $\Sigma_\star$ also correlates notably less well with $I_{\rm CO}$ than the SFR and mid-IR tracers. The right panel in Figure~\ref{fig:apcorr_check} shows that had we adopted one of these other plausible templates, we might expect the aperture correction to scatter by $\pm 10\%$ (the gray band) relative to our preferred WISE3 values. We consider that the true uncertainty on the aperture correction is slightly less than this because of the strength of the WISE3--CO correlation.

Aperture corrections that assume an exponential disk structure have the large advantage that they can be applied with almost no multi-wavelength data. However, simple exponential disk models show a notably worse correlation with $I_{\rm CO}$ than any of the other templates. Among these, the template using our fitted $l_\star$ and the one using a scale length of $0.2~r_{25}$ offer approximately equally good corrections. On average, these disk models yield about the same magnitude of correction factor as our other templates. However, Table~\ref{tab:apcorr_check} implies that they offer less precision, and Figure~\ref{fig:apcorr_check} shows that the derived correction using an $l_\star$ scale length exponential disk scatters by $10{-}20\%$ relative to the WISE3 correction.

\subsubsection{Checks on CO(2--1) Luminosity}
\label{sec:comphera}

\begin{figure*}[ht!]
\begin{center}
\includegraphics[width=0.495\textwidth]{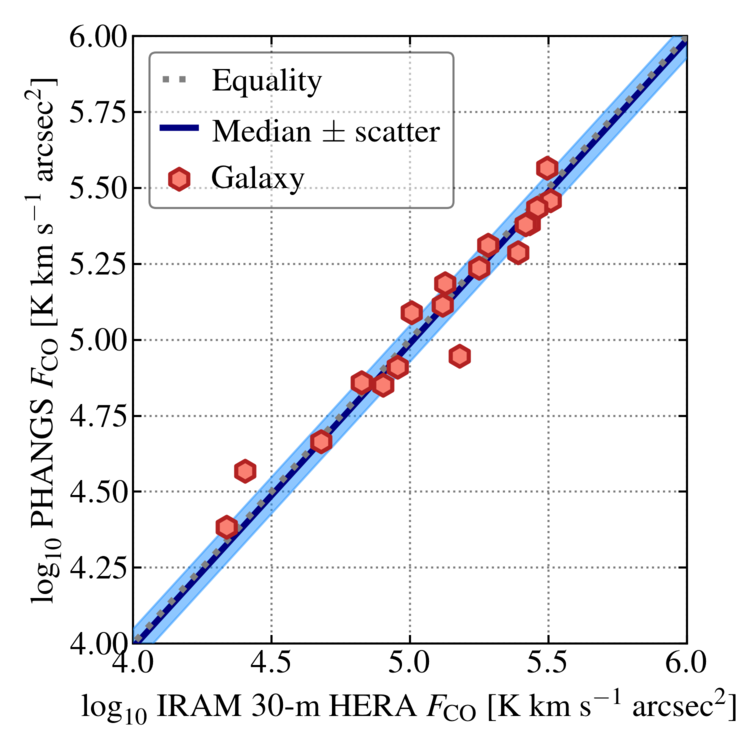}
\includegraphics[width=0.495\textwidth]{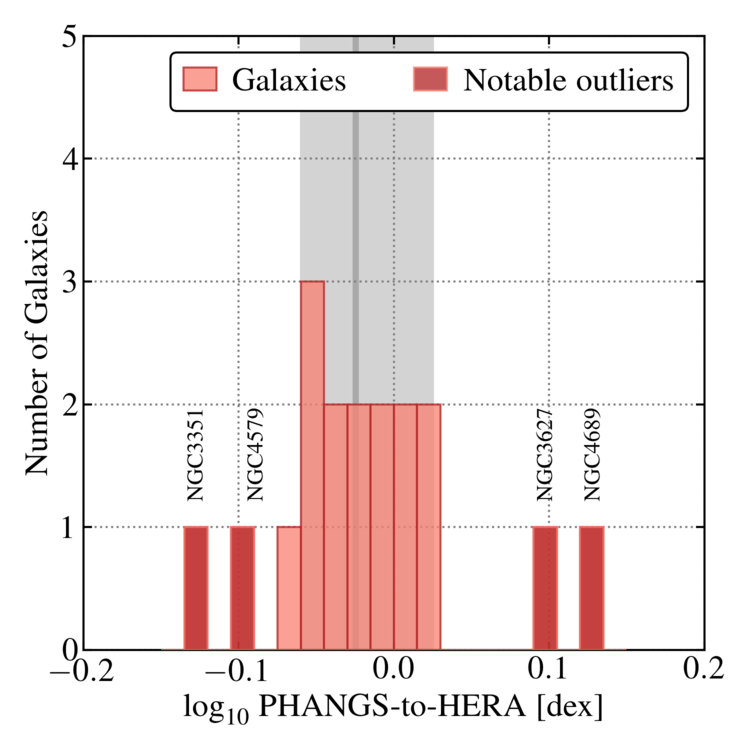}
\includegraphics[width=0.495\textwidth]{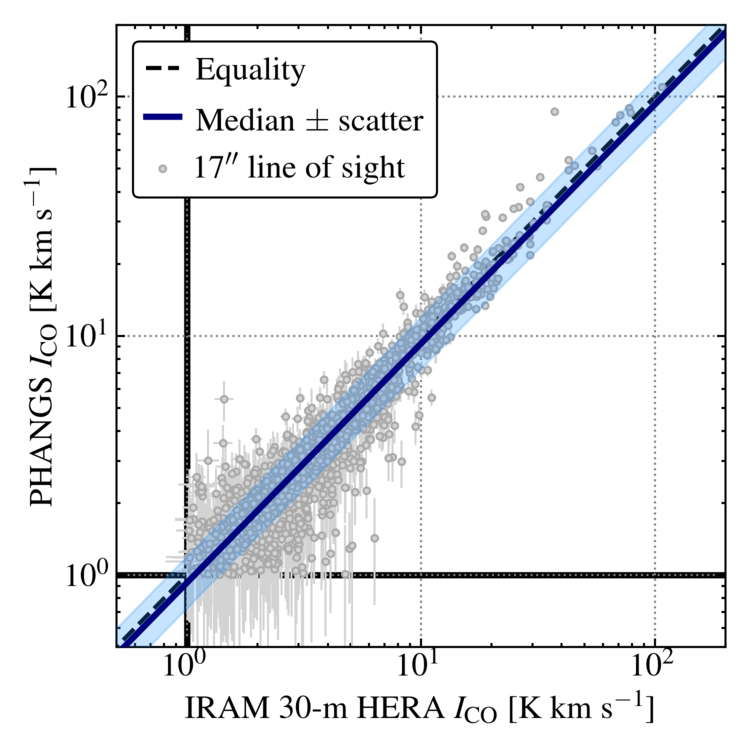}
\includegraphics[width=0.495\textwidth]{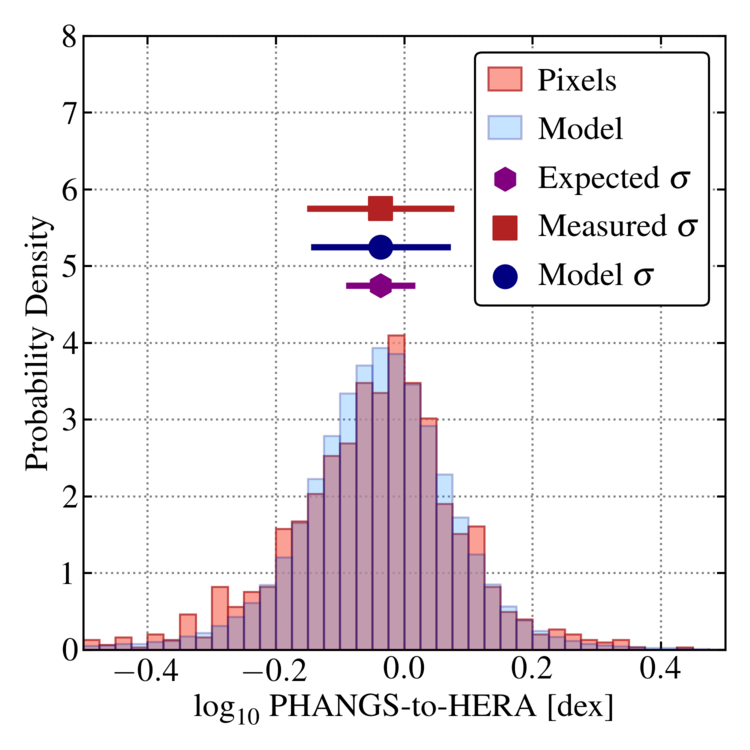}
\end{center}
\vspace*{-10px}
\caption{\textbf{Comparisons between CO(2--1) measurements from PHANGS--ALMA and previous IRAM \mbox{30-m} CO(2--1) mapping.} All four panels compare measurements from our \cotwo\ data to \cotwo\ observations targeting the same galaxies using the HERA array receiver on the IRAM \mbox{30-m} telescope \citep[from HERACLES and follow-up work;][and A.~Schruba et al.\ in preparation]{LEROY09,LEROY13}. The \textit{top left} panel compares integrated fluxes; in this figure we have applied our WISE3-based aperture correction to our PHANGS measurements. The \textit{bottom left} panel compares line-integrated intensity between the two data sets at $17\arcsec$ resolution. The right hand panels show ratios between the two data sets, the \textit{top right} one for galaxy-integrated ratios and the \textit{bottom right} one for pixel-by-pixel ratios. Overall the agreement is good. The median galaxy-integrated ratio is $0.95$ ($-0.02$~dex) with $0.057$~dex, or about $14\%$, galaxy-to-galaxy scatter. We mark four strong outliers, listed in the text. The median pixel-wise ratio is $0.92$ ($0.037$~dex) with $0.11$~dex, or about $29\%$, scatter. The model shown in the bottom panel includes the measured statistical uncertainty and adopts a $20\%$ gain uncertainty. We attribute most of the non-statistical scatter to gain uncertainties associated with the IRAM \mbox{30-m} observations, though formally these represent only relative calibration and image construction uncertainties.
\label{fig:co21_check}}
\end{figure*}

Figure~\ref{fig:co21_check} compares our derived CO fluxes and intensities to previous \cotwo\ mapping using the IRAM \mbox{30-m} telescope\footnote{This part of the work is based on observations carried out with the IRAM 30m telescope. IRAM is supported by INSU/CNRS (France), MPG (Germany) and IGN (Spain).}. In total, PHANGS--ALMA covers $18$ galaxies that have also been mapped in the \cotwo\ line using the HERA array receiver on the IRAM \mbox{30-m} telescope \citep{SCHUSTER07}. These maps are part of the HERACLES survey \citep{LEROY09,LEROY13} and a follow-up survey (A.~Schruba et al.\ in preparation). They cover wider area than the ALMA maps, but with higher noise levels. The calibration of the IRAM \mbox{30-m} maps depends on ``chopper wheel'' observations and a main beam efficiency correction, while both the ALMA interferometric and total power data are pinned to monitoring of ALMA calibrators with monitored fluxes.

The IRAM \mbox{30-m} maps have native resolution of $13.3\arcsec$, so we must smooth the PHANGS--ALMA data make a direct comparison. For this exercise, we used a $17\arcsec$ version of both cubes. We aligned both cubes for each galaxy onto the same astrometric and spectroscopic grid. We then apply an identical mask to each galaxy in each survey to create matched resolution, matched velocity-interval integrated intensity maps, along with associated uncertainty maps.

The top left panel in Figure~\ref{fig:co21_check} compares the integrated fluxes between the IRAM and ALMA maps. For the calculation in the top left panel, we apply the WISE3-based aperture corrections discussed above. We find a median PHANGS-to-HERA ratio of $0.97$, or $-0.01$~dex, with a robustly-estimated galaxy-to-galaxy scatter of $0.06$~dex, or about $15\%$. The right panel shows galaxy integrated ratios. We calculate these only integrating over the area where both maps have coverage. These also show good agreement. The statistical uncertainty in these ratios is low, $< 1\%$ in almost all cases, so the observed scatter reflects mostly calibration or other systematic uncertainties. The 16$^{\rm th}$ to 84$^{\rm th}$ percentile range, illustrated in gray, spans from $0.87$ ($-0.06$~dex) to $1.06$ ($0.025$~dex) with the median $0.97$ ($-0.01$~dex). These suggest an overall $\lesssim 5\%$ difference in the calibration scale with about $\pm 10{-}15\%$ calibration uncertainty in any given galaxy. We mark four galaxies with strongly discrepant fluxes in Figure~\ref{fig:co21_check}: NGC~3351 (PHANGS-to-HERA ratio of $0.74$), NGC~3627 ($1.24$), NGC~4579 ($0.80$), and NGC~4689 ($1.32$). The discrepancy in NGC~3627, which was observed by the IRAM \mbox{30-m} under non-ideal conditions shortly after HERA commissioning, has been noted and explored by \citet{DENBROK21}. They also noted a similar discrepancy between new and old IRAM \mbox{30-m} mapping projects for NGC~5194 (M51) which is not in our sample. \citeauthor{DENBROK21} suggest that issues with the HERA data most likely account for these strong outliers.

The bottom rows in Figure~\ref{fig:co21_check} explore the pixel-to-pixel correspondence between the PHANGS and HERA data at $17''$ resolution. Here we only consider where both maps have coverage and $I_{\rm CO} > 1$ K~km~s$^{-1}$\footnote{This corresponds to S/N $\gtrsim 5$ in the HERACLES maps, which represent the limiting signal to noise.} in both maps, a total of $1{,}161$ independent $17\arcsec$ lines of sight. Again, we find good overall agreement, here spanning two decades in CO intensity. The bottom right panel shows the distribution of ratios for individual pixels. Here we find a median ratio of $0.93$ ($-0.03$~dex) and a robustly estimated scatter of $\pm 0.11$~dex or about $\pm 28\%$. Thus, the pixel data indicate a slight offset in overall flux scale, now with PHANGS ${\sim}7\%$ fainter than HERA.

Based only on the statistical errors associated with individual data, we expect a scatter of about $\pm 0.05$~dex or about $\pm 12\%$, a range that we indicate by a blue bar in Figure~\ref{fig:co21_check}. This is significantly smaller than the measured scatter, again indicating the presence of systematic uncertainties. The blue histogram shows a model that treats these systematic uncertainties as a $\pm 20\%$ point-to-point multiplicative uncertainty and also incorporates the measured statistical errors. Overall this matches the measured scatter well. 

This $\pm 20\%$, or $\pm 0.11$~dex, point-to-point scatter is somewhat higher than the galaxy-to-galaxy scatter seen in the top row. Indeed, if we subtract the galaxy-to-galaxy scatter and the statistical scatter from the measured scatter in quadrature we are left with about $\pm 0.06{-}0.07$~dex (about $\pm 15\%$) point-to-point unaccounted scatter. Our best estimate is that this reflects uncertainties in the image reconstruction in both data sets. The HERA array receiver data suffer from variable pixel gains \citep[see][]{LEROY09,DENBROK21} while the ALMA data have been partially reconstructed from interferometric imaging. Based on a detailed analysis and visual inspection of the $5$~galaxies in \citet{DENBROK21}, the HERA pixel gain uncertainties appear to be responsible for the most serious issues, and we primarily attribute this unaccounted scatter to the HERA data. This will appear, e.g., as mapping artifacts like striping or variations in the gain across the map. Such issues are common in single dish maps, especially those made with array receivers, and the identified magnitude of uncertainty, $\pm 15\%$, agrees with that inferred from jackknife tests carried out on the HERACLES data \citep[][]{LEROY09}.

Overall, this comparison paints the following picture: 

\begin{enumerate}
    \item The two data sets agree very well to first order.
    \item There appears to be a ${\sim}3{-}7\%$ systematic offset in the flux scale between PHANGS and the HERA data, with the PHANGS data showing slightly lower flux. Our best estimate is that the ALMA data are better calibrated.
    \item We observe ${\sim}15\%$ galaxy-to-galaxy scatter in the ratio of ALMA-to-HERA fluxes, which reflects the combined calibration uncertainty in the ALMA and IRAM \mbox{30-m} data. This is largely consistent with the $10{-}15\%$ calibration uncertainty expected for the HERA data \citep{LEROY09,DENBROK21} and the $5{-}10\%$ uncertainty associated with ALMA band~6 observations.
    \item At $17\arcsec$ resolution, the point-to-point scatter between the two data sets is ${\sim}0.11$~dex. This is higher than expected from only the statistical noise or the galaxy-to-galaxy flux calibration uncertainties. We consider that this $0.11$~dex scatter reflects roughly equal contributions from statistical uncertainty, flux calibration uncertainty, and point-to-point gain variations in the single dish map. The latter arises from pixel-to-pixel gain variations in the array receiver \citep[see][]{DENBROK21} and uncertainties in interferometric image reconstruction.
\end{enumerate}

\subsection{Literature \texorpdfstring{\hi}{HI} Masses}
\label{sec:himass}

We draw integrated \hi\ masses from the literature, specifically the homogenized compilation by HyperLEDA. We derive the \hi\ mass from HyperLEDA's $21$~cm flux, applying no correction for opacity effects. Over the next year, we expect new and archival $21$~cm line imaging using the VLA (P.I.\ D.~Utomo) to improve our estimates of the \hi\ mass for our targets.

\subsection{PHANGS--ALMA Sample Properties}
\label{sec:propsummary}

\begin{figure*}
\begin{center}

\includegraphics[width=0.45\textwidth]{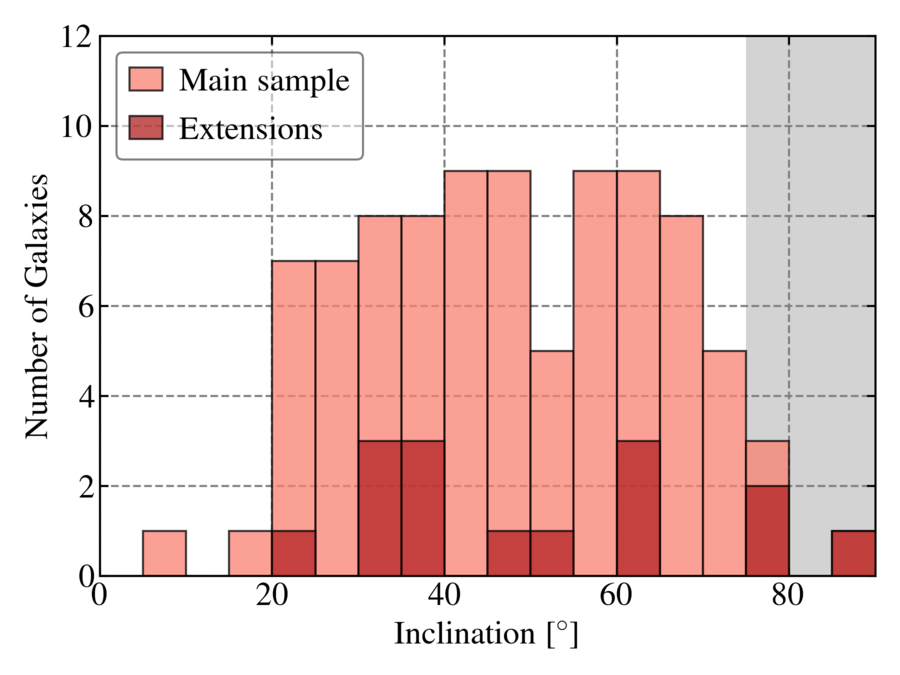}
\includegraphics[width=0.45\textwidth]{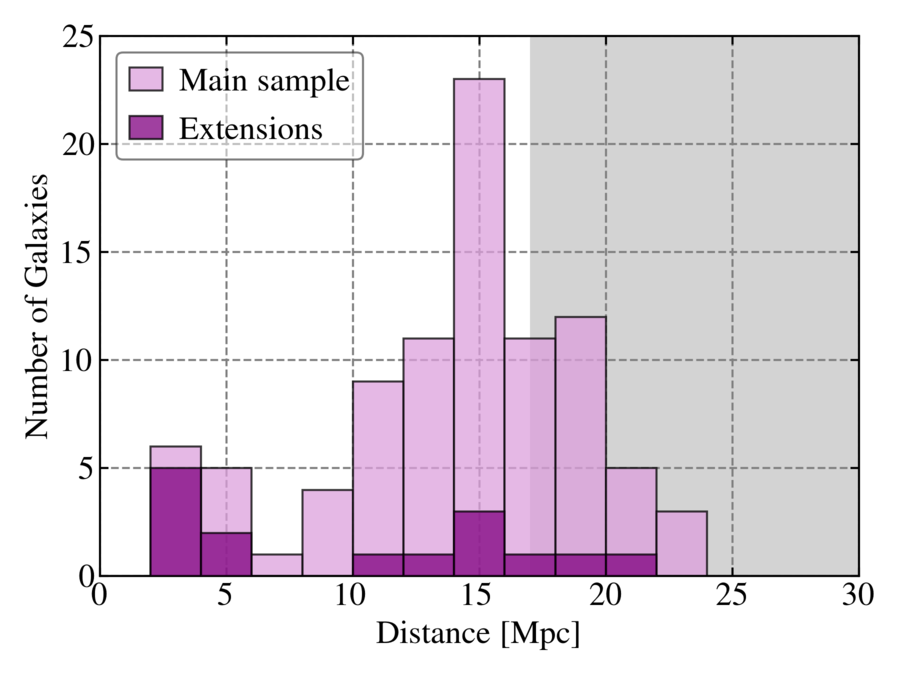}

\includegraphics[width=0.45\textwidth]{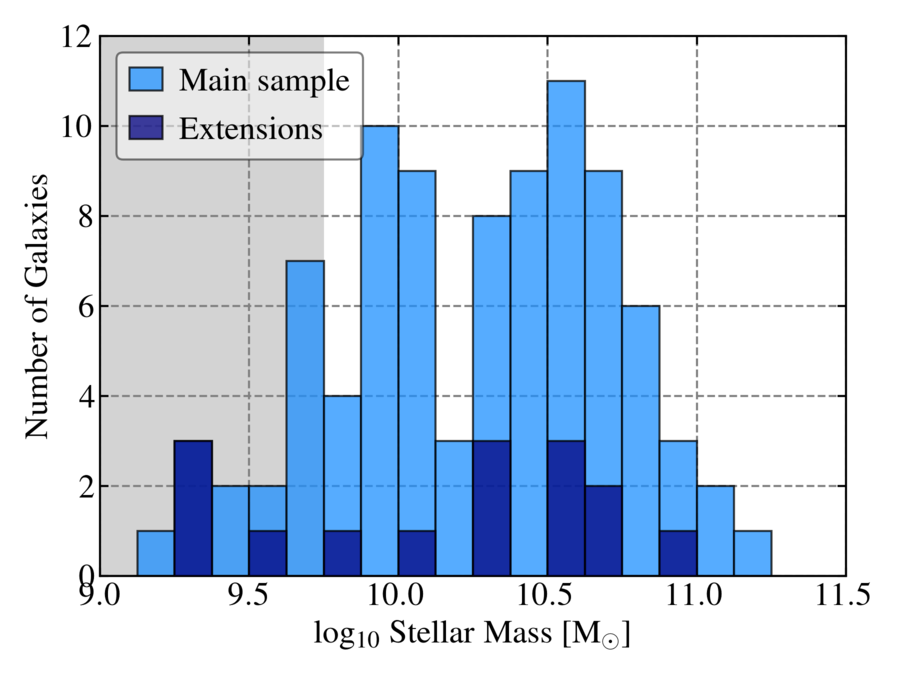}
\includegraphics[width=0.45\textwidth]{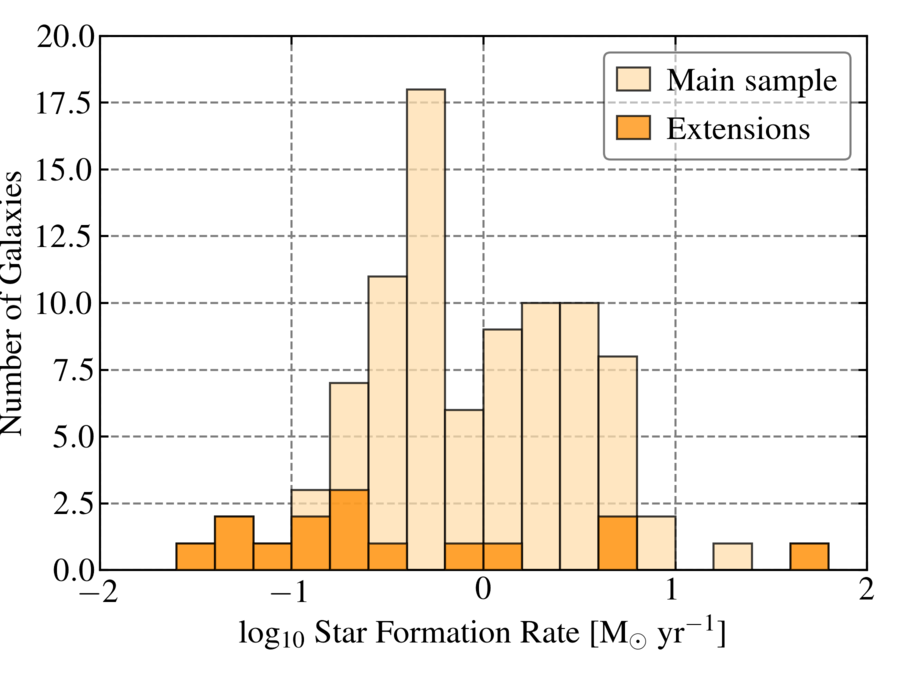}

\includegraphics[width=0.45\textwidth]{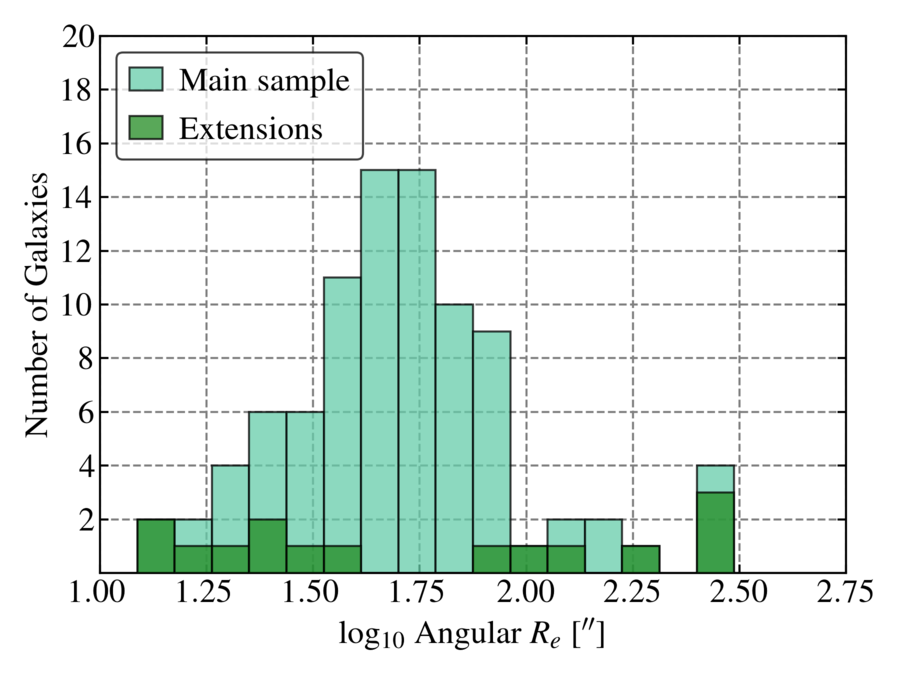}
\includegraphics[width=0.45\textwidth]{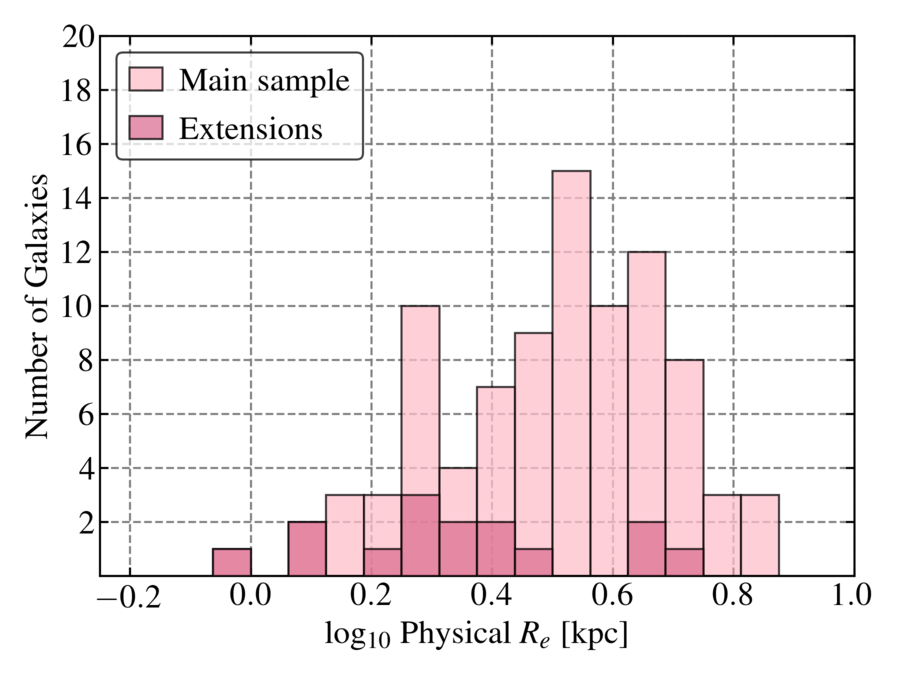}

\end{center}
\caption{\textbf{Histograms showing PHANGS--ALMA sample properties.} The \textit{top row} shows inclination and distance. Inclinations come mostly from \citet{LANG20} and we adopt distances from \citet{ANAND21}. The \textit{second row} shows stellar mass (\S \ref{sec:stellarmass}) and star formation rate (\S \ref{sec:sfr}). The \textit{third row} shows the stellar half mass radii (\S \ref{sec:size}) in both angular and physical units. In panels relevant to sample selection, gray shading marks the region excluded by the main sample selection (\S \ref{sec:sample}). Note that our best estimates of each quantity have improved since selection, so that we selected some galaxies that we now believe miss the selection criteria. The darker histogram shows properties of galaxies in survey extensions that focus on early-type and very nearby galaxies.
\label{fig:hists_1}}
\end{figure*}

\begin{figure*}
\begin{center}

\includegraphics[width=0.45\textwidth]{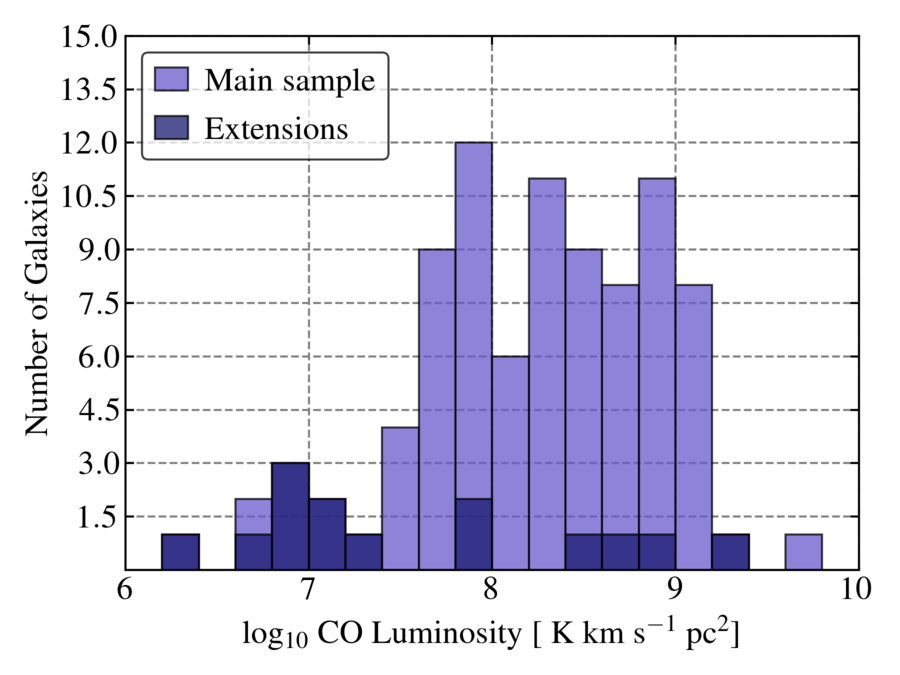}
\includegraphics[width=0.45\textwidth]{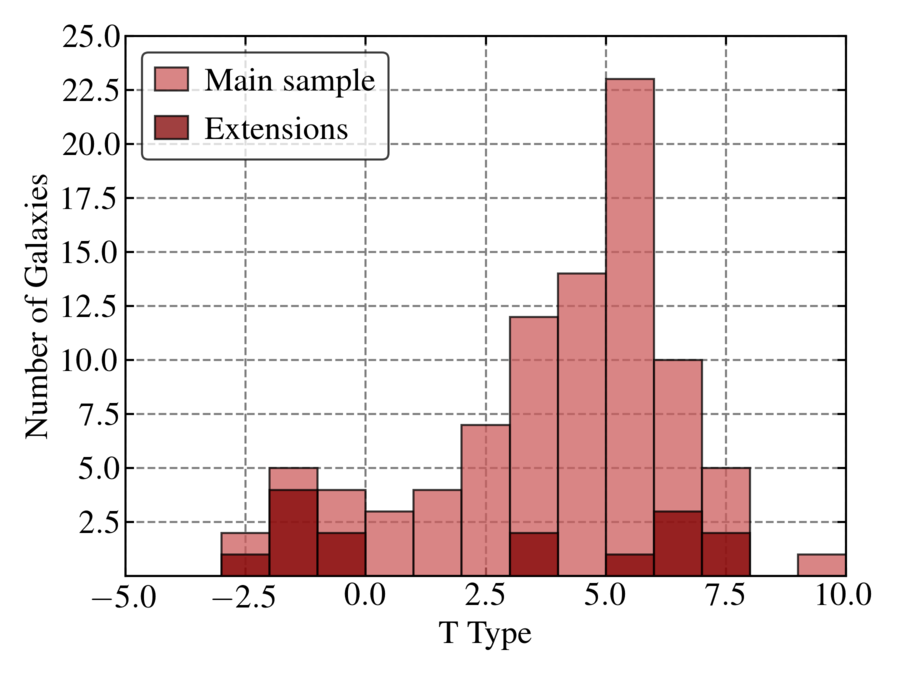}

\includegraphics[width=0.45\textwidth]{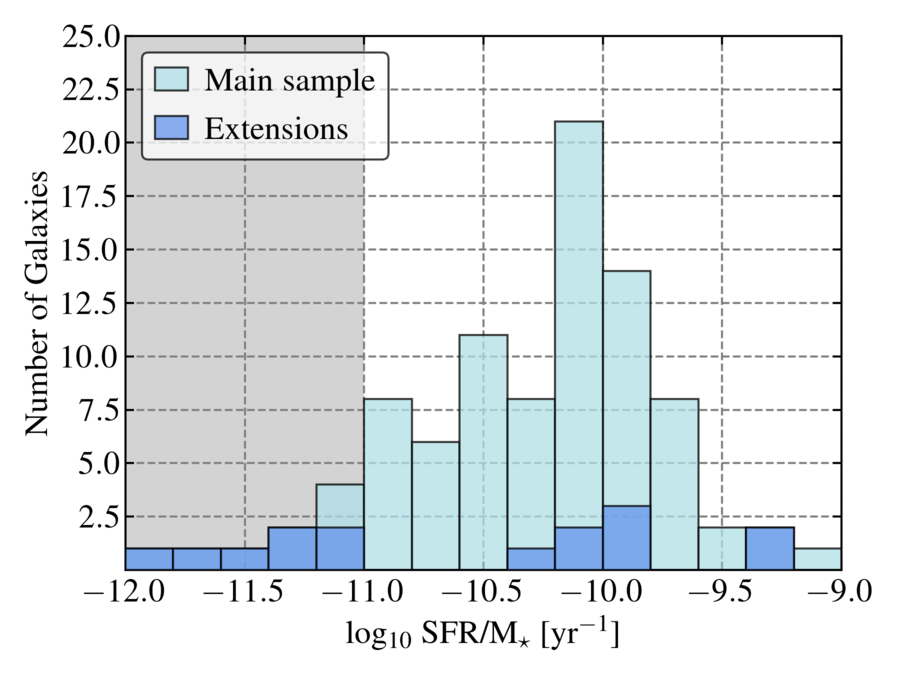}
\includegraphics[width=0.45\textwidth]{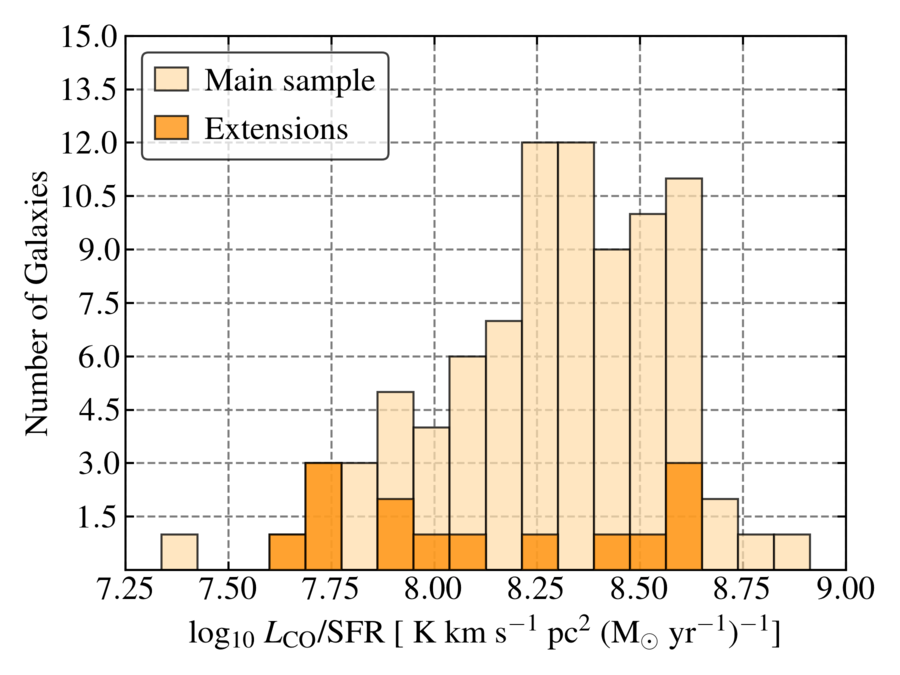}

\includegraphics[width=0.45\textwidth]{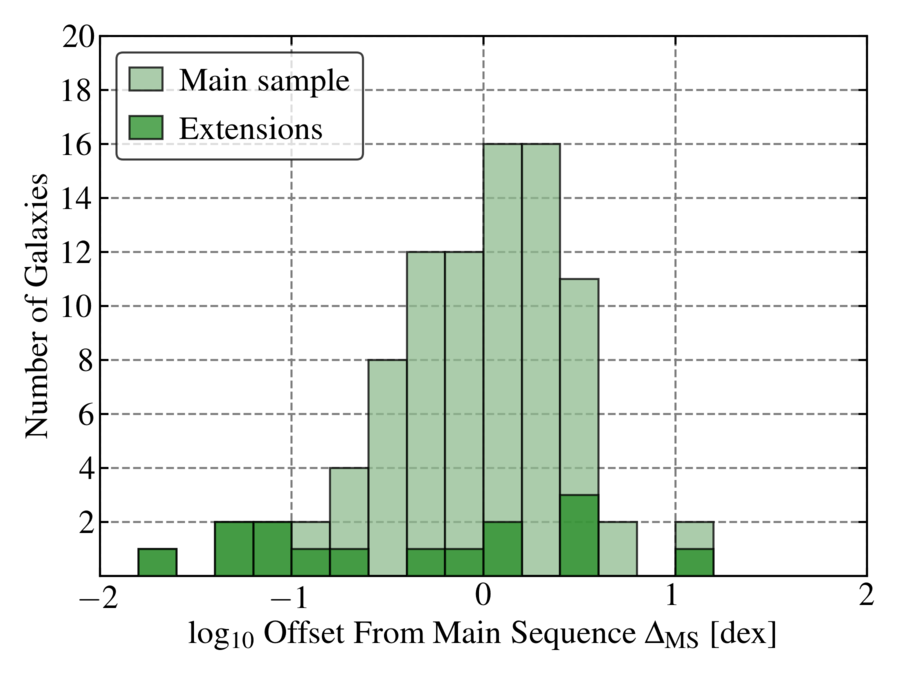}
\includegraphics[width=0.45\textwidth]{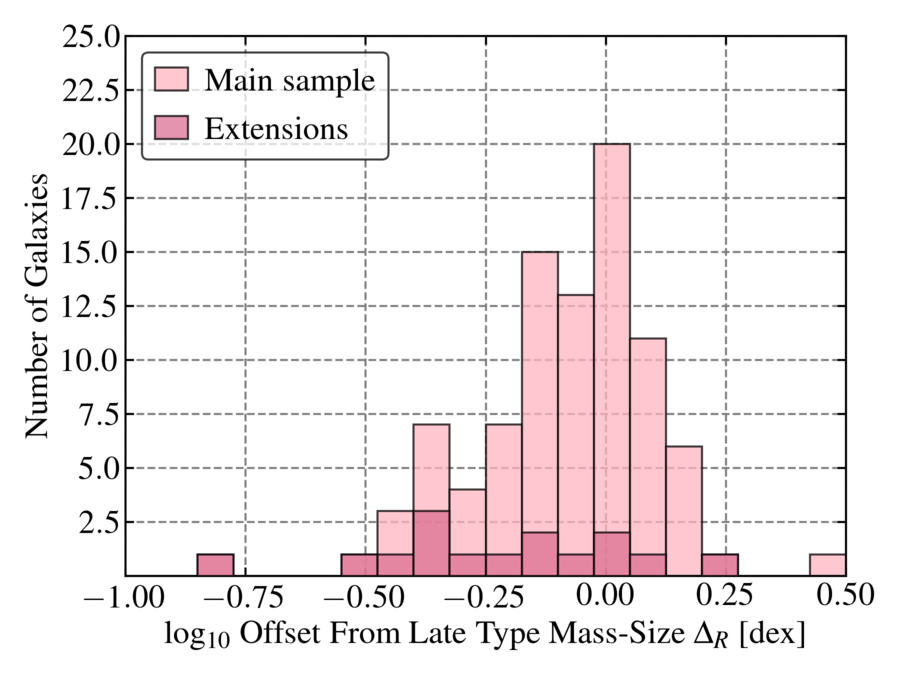}

\end{center}
\caption{\textbf{Histograms showing PHANGS--ALMA sample properties.}
The \textit{top row} shows CO luminosity (\S \ref{sec:colum}) and morphological T-type code from HyperLEDA \citep{PATUREL03,MAKAROV14}. The \textit{middle row} shows the ratio of SFR-to-$M_\star$, i.e., the specific star formation rate, and the ratio of $L_{\rm CO}$-to-SFR, which can be scaled by $\alpha_{\rm CO}$ to calculate the molecular gas depletion time. The \textit{bottom row} shows offsets from two scaling relations: the SFR-$M_\star$ ``star-forming main sequence'' and the mass--size relation. Aperture corrections have been applied to the $L_{\rm CO}$ estimates used in the plots (\S \ref{sec:apcorr}). In panels relevant to sample selection, gray shading marks the region excluded by the main sample selection (\S \ref{sec:sample}). Note that our best estimates of each quantity have improved since selection, so that we selected some galaxies that we now believe miss the selection criteria. The darker histogram shows properties of galaxies in survey extensions that focus on early-type and very nearby galaxies.
\label{fig:hists_2}}
\end{figure*}

\begin{figure*}
\begin{center}
\includegraphics[width=0.75\textwidth]{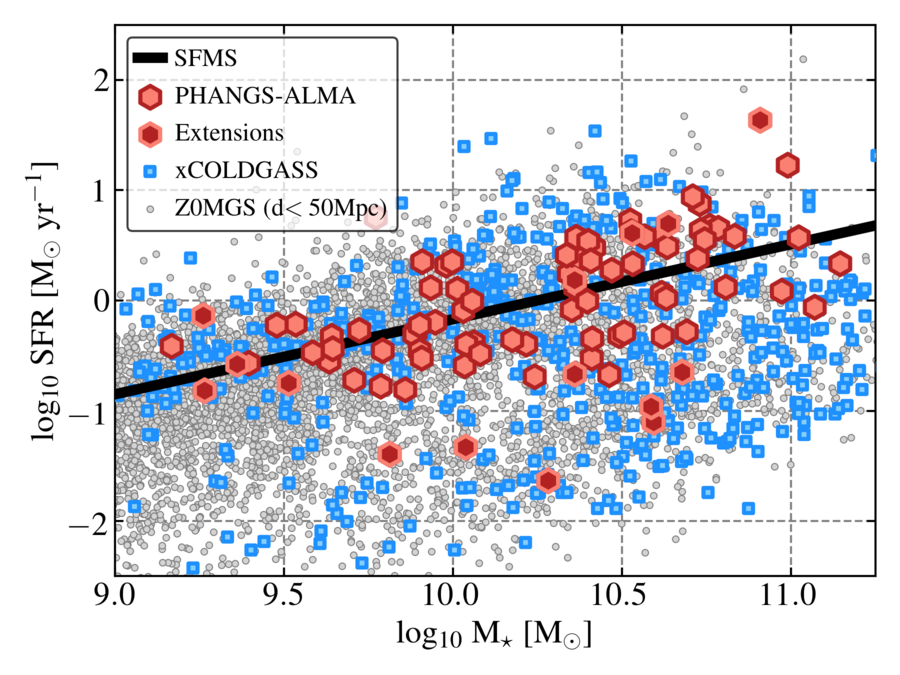}
\includegraphics[width=0.75\textwidth]{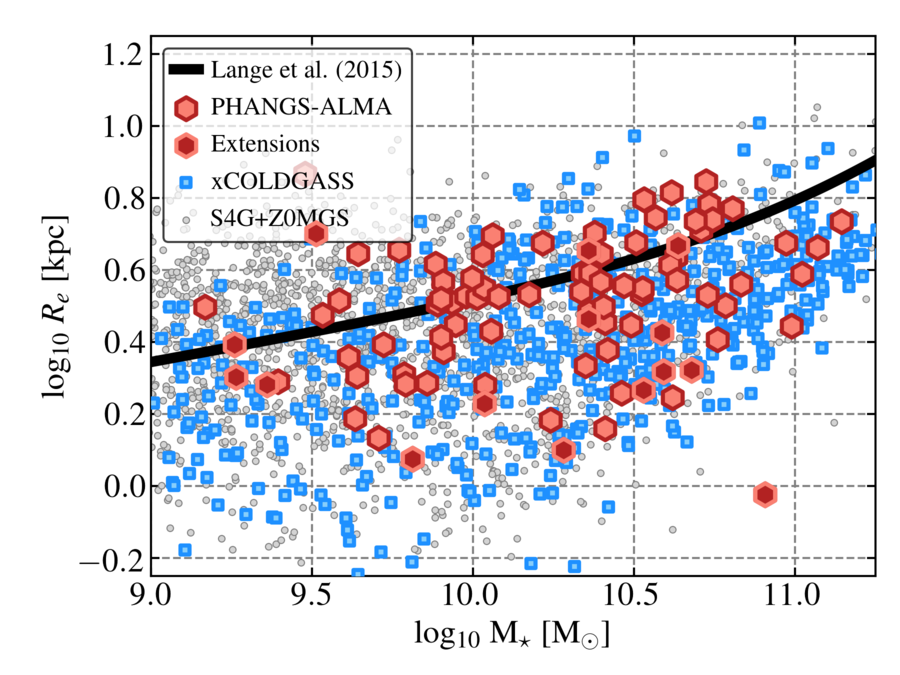}
\end{center}
\vspace*{-10px}
\caption{\textbf{PHANGS--ALMA on the star-forming main sequence and size--mass relation.} PHANGS--ALMA targets in SFR${-}M_\star$ and $R_e{-}M_\star$ space.
PHANGS--ALMA targets appear as red hexagons. For comparison, galaxies observed as part of the xCOLD~GASS survey are represented by blue squares \citep{SAINTONGE17}. We show a large sample of local galaxies as gray dots. In the SFR${-}M_\star$ space these are all galaxies with $d<50$~Mpc and measurement from \citet{LEROY19}. In the lower panel, we combined masses from \citet{LEROY19} with sizes from \citet{MUNOZMATEOS15}, with the latter being the limiting quantity. Black lines show scaling relations relevant to late-type galaxies. In the top panel we show the star-forming main sequence as a solid black line, with the specific formula from \cite{LEROY19}. In the bottom panel we show the $r$-band result for late-type galaxies from \citet{LANGE15}. We show the distribution of offsets from both relations in Figure~\ref{fig:hists_2}.
\label{fig:sample_sfms_re}}
\end{figure*}

\begin{deluxetable}{lccccc}[ht!]
\tabletypesize{\footnotesize}
\tablecaption{PHANGS--ALMA Observed Property Summary \label{tab:prop_summary}}
\tablewidth{0pt}
\tablehead{
\colhead{Property} &
\multicolumn{5}{c}{Percentile} \\
\colhead{} &
\colhead{5} &
\colhead{16} &
\colhead{50} &
\colhead{84} &
\colhead{95}
}
\startdata
$i$ [$^\circ$] & 23 & 30 & 47 & 65 & 72 \\
$d$ [Mpc] & $3.7$ & $9.9$ & $15.1$ & $19.3$ & $20.7$ \\
$\log_{10}$ $M_\star$ [M$_\odot$] & $9.45$ & $9.74$ & $10.36$ & $10.72$ & $10.94$ \\
$\log_{10}$ SFR [M$_\odot$ yr$^{-1}$] & $-0.89$ & $-0.57$ & $-0.09$ & $0.58$ & $0.74$ \\
$\log_{10}$ $R_e$ [kpc] & $0.17$ & $0.29$ & $0.53$ & $0.68$ & $0.78$ \\
$\log_{10} L_{\rm CO}$\tablenotemark{a} & $6.97$ & $7.66$ & $8.29$ & $8.95$ & $9.11$ \\
Morphological T-type & $-1.7$ & $1.2$ & $4.0$ & $6.0$ & $7.0$ \\
$\log_{10}$ SFR/$M_\star$ [yr$^{-1}$] & $-11.27$ & $-10.82$ & $-10.18$ & $-9.86$ & $-9.62$ \\
$\log_{10} L_{\rm CO}$/SFR\tablenotemark{a} & $7.71$ & $7.96$ & $8.31$ & $8.56$ & $8.63$ \\
$\Delta_{\rm MS}$ [dex] & $-1.02$ & $-0.48$ & $0.01$ & $0.41$ & $0.57$ \\
$\Delta_R$ [dex] & $-0.40$ & $-0.26$ & $-0.07$ & $0.09$ & $0.16$ \\
\enddata
\tablenotetext{a}{$L_{\rm CO}$/SFR has units of K km s$^{-1}$ pc$^2$ (M$_\odot$~yr$^{-1}$)$^{-1}$ and $L_{\rm CO}$ has units of K km s$^{-1}$ pc$^{2}$. Add $\log_{10} \alpha_{\rm CO}^{2-1}$ to the quoted value to convert to $M_{\rm mol}$ or molecular gas depletion time, $\tau_{\rm dep}^{\rm mol}$. A fiducial Milky Way $\alpha_{\rm CO}^{2-1} = 6.7$ \acounits\ \citep[][with $R_{21} = 0.65$]{BOLATTO13B} implies $\log_{10} \alpha_{\rm CO}^{2-1} = 0.83$.}
\tablecomments{See Figures~\ref{fig:hists_1} and~\ref{fig:hists_2}. Values report percentiles in the full set of targets, including extensions and archival data.}
\end{deluxetable}

Figures \ref{fig:hists_1} and~\ref{fig:hists_2} show distributions of the derived sample properties reported in Tables \ref{tab:sample_orient} and~\ref{tab:sample_phys}. Figure~\ref{fig:sample_sfms_re} shows the PHANGS--ALMA sample in two commonly-considered parameter spaces for galaxy evolution: ${\rm SFR}{-}M_\star$ and $R_{e}{-}M_\star$ space. In both plots, we distinguish the survey extension targets from the original sample. In addition to the quantities discussed already in this section, we show the ratios of $L_{\rm CO}$-to-SFR and SFR-to-$M_\star$. We apply the aperture correction to $L_{\rm CO}$ before constructing these ratios. We also show the distribution of morphological T-type code \citep[from HyperLEDA][]{MAKAROV14}. Table~\ref{tab:prop_summary} summarizes the distributions of properties for the PHANGS--ALMA sample.

\medskip

\textbf{PHANGS--ALMA on Scaling Relations:} In Figure~\ref{fig:sample_sfms_re} we plot PHANGS--ALMA in red hexagons along with galaxies from the xCOLD GASS survey \citep[blue points, from][]{SAINTONGE17} and a large sample of galaxies with $d < 50$~Mpc in ${\rm SFR}{-}M_\star$ (top) and $R_e{-}M_\star$ space (bottom). In both plots we indicate fiducial scaling relations for late-type galaxies. In the top panel, this is the result of a linear fit to ${\sim} 4{,}000$ local galaxies with $M_\star$ and SFR estimated from WISE and GALEX data that closely resemble our calculations \citep{LEROY19}. We use this to define the star-forming main sequence as
\begin{equation}
\label{eq:sfms}
\log_{10} \frac{{\rm SFR_{\rm MS}}}{{\rm 1~M_\odot~yr}^{-1}} = -0.32 \left( \log_{10} \frac{M_\star}{{\rm 1~M_\odot}} - 10.0 \right) - 10.17~.
\end{equation}
\noindent This is almost identical to the main sequence fit by \citet{CATINELLA18} for the GASS sample, which has slope $-0.344$ and a value of $-10.17$ at $\log_{10} M_\star\,{\rm [M_\odot]} = 10$.

In the bottom panel, we show the mass--size relation found by \citet[][]{LANGE15} for late-type galaxies as a black line. We plot their fiducial $r$-band relation, which uses the functional form defined by \citet{SHEN03}. Note that the relation applies specifically to late-types but we do not restrict the galaxies plotted by morphology or S\'ersic index. The large population of xCOLD GASS outliers at high $M_\star$ simply reflects the inclusion of early-type galaxies in that survey.

We calculate offsets from both scaling relations for each PHANGS--ALMA galaxy and show the distributions of $\Delta_{\rm MS}$ and $\Delta_R$ in Figure~\ref{fig:hists_2}. These are calculated as
\begin{equation}
\label{eq:deltams}
\Delta_{\rm MS} \left[ {\rm dex} \right] = \log_{10} {\rm SFR} - \log_{10} {\rm SFR_{\rm MS}} \left( M_\star \right)~,
\end{equation}
\noindent and
\begin{equation}
\label{eq:deltar}
\Delta_R \left[ {\rm dex} \right] = \log_{10} R_e - \log_{10} R_e^{\rm L15} \left( M_\star \right)~,
\end{equation}
\noindent where $R_{\rm e}^{\rm L15} \left( M_\star \right)$ refers to the effective radius predicted from the \citet{LANGE15} relation.

We also refer the reader back to Figure~\ref{fig:intscaling} where we place PHANGS--ALMA on several other scaling relations: molecular gas depletion time versus stellar mass, molecular gas mass fraction versus stellar mass, and molecular gas surface density versus star formation rate surface density and stellar mass surface density. Those figures also show xCOLD GASS and the full sample of local CO mapping targets considered in this section.

\medskip

\textbf{PHANGS--ALMA Sample Properties:} PHANGS--ALMA aimed to select all relatively massive, nearby galaxies close to the ``main sequence'' of star-forming galaxies (\S\ref{sec:sample}). Despite the small volume available within $d < 17$~Mpc and $-75^\circ \delta < +25^\circ$, Figures~\ref{fig:hists_1} through~\ref{fig:sample_sfms_re} and Table~\ref{tab:prop_summary} show that the PHANGS--ALMA sample does achieve that goal. We achieve good sampling of more than a decade in stellar mass $9.5 \lesssim \log_{10} M^\star \lesssim 10.75$. We also achieve good sampling of SFR. Our sample property estimates show good agreement with previous results, and our sample scatters around well-established scaling relations for disk galaxies. Our galaxies are mostly spirals, including both early- and late-type spirals, but not many irregular galaxies. We also cover a decade in specific star formation rate and offset from the star-forming main sequence, but a somewhat narrower range, a factor of roughly three, in $L_{\textrm{CO}}/\textrm{SFR}$. Despite uncertainties in distance and evolving distance estimates, most of our targets do lie within $19$~Mpc, with ${\sim}16\%$ scattered out to ${\sim}21$~Mpc. And by construction our targets are overwhelmingly at moderate inclination, allowing resolved multiwavelength studies.

The figures, including the contrast with xCOLD GASS, also highlight the shortcomings of the sample for studying galaxy evolution. We do not include many extreme starbursts due to the simple fact that these tend to be rare. The classic local starburst galaxies, the members of the IRAS bright galaxy sample \citep[e.g.,][]{SANDERS03}, or GOALS sample \citep{ARMUS09} mostly lie at much greater distances. We do include many galaxies that host starburst nuclei, including some classic examples like NGC~0253 and NGC~4945. But those nuclear bursts are not strong enough to displace their parent galaxy far from the star-forming main sequence. We also include only a limited number of early-type galaxies. Massive early-type galaxies are relatively rare in the local volume and in any case these were excluded by our sample selection, though our extensions have begun to build in this direction. Finally, we do not push far into the regime of dwarf galaxies, with good coverage in mass down to only $\log_{10} M_\star\,[{\rm M}_\odot] \approx 9.5$. At lower masses, conversion factor effects become large and the required integration times rise, which would require adapting our observing strategy.

All of these directions are interesting and important. Starburst galaxies host high density, high pressure gas \citep[e.g.,][]{DOWNES98}. Dwarf galaxies host sparse, faint, and isolated molecular clouds \citep[e.g.,][]{FUKUI10}, while early-type galaxies show evidence for stabilized molecular gas disks with little or no star formation \citep{DAVIS14,DAVIS17}. However, we emphasize that most stars form in ``main sequence'' galaxies within a decade of $\log_{10} M^\star\,[{\rm M}_\odot] \approx 10.6$ \citep[e.g.,][]{KARIM11,LESLIE20}. PHANGS--ALMA does a very good job of providing the first cloud-scale survey of a representative sample of such star-forming main sequence galaxies.

\section{Observations}
\label{sec:observations}

To meet the Science Goals described in \S\ref{sec:motivation}, PHANGS--ALMA mapped the distribution and kinematics of molecular gas with high sensitivity, high physical resolution, and good fidelity across \ntarget\ nearby galaxies. To do this, we observed the $J=2\rightarrow1$ transition of CO arising from the galactic disk in each target. We used the ALMA main array of \mbox{12-m} dishes in a compact configuration, which yields ${\sim}1\arcsec$ resolution at Band~6 ($\nu \approx 230$~GHz), corresponding to ${\sim}100$~pc linear resolution, or about the size of a massive GMC at the typical distance to our targets. We covered the area of active star formation in each target using large mosaics that consisted of ${\sim}100{-}450$ \mbox{12-m} pointings per galaxy. Despite short integration times for each pointing, we still achieve good $u{-}v$ coverage and sensitivities with the main array. Because nearby galaxies are extended and complex structured, we supplement the main array observations using both parts of the Morita Atacama Compact Array (ACA), the \mbox{7-m} array and the single dish, or ``total power'' (TP), antennas.

This section describes and motivates our adopted strategy (\S\ref{sec:obsstrat}). We also report details related to the actual execution of the observations (\S\ref{sec:obsexec}).

\subsection{Observing Strategy}
\label{sec:obsstrat}

\subsubsection{Spectral Setup and Choice to Target CO(2--1)}

\begin{figure*}[ht!]
\begin{center}
\includegraphics[width=0.75\textwidth]{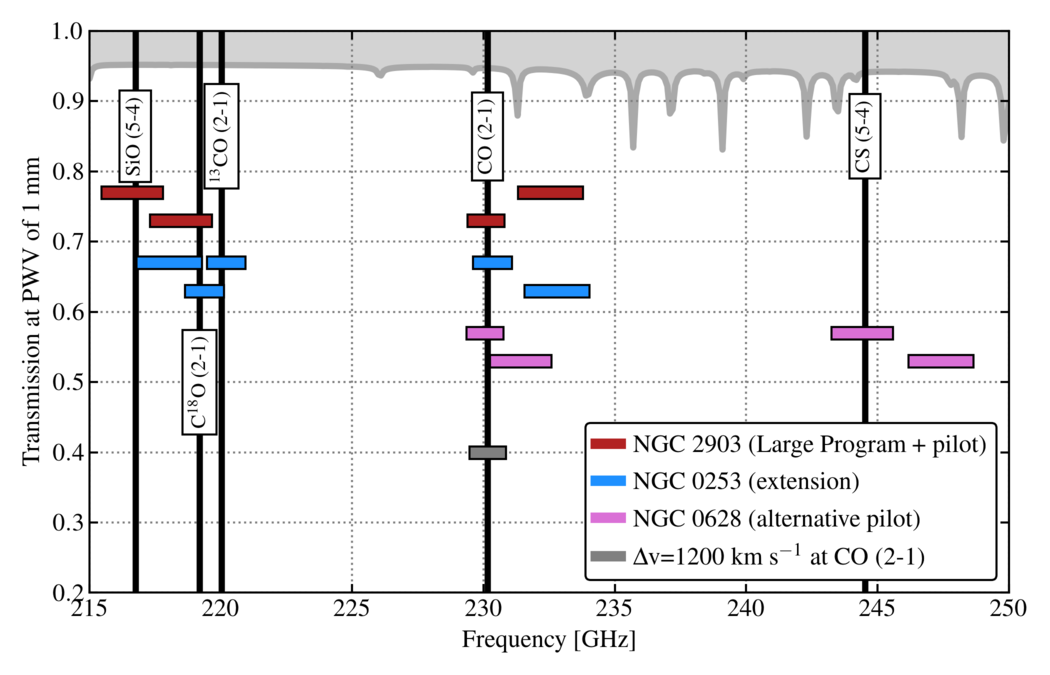}
\end{center}
\caption{\textbf{PHANGS--ALMA spectral setup.} Colored horizontal bars show the spectral windows observed using ALMA's Band~6 receiver. The top, red bars show the setup for Large Program, the follow-up completion programs, and the two main pilot surveys (see Table~\ref{tab:projects} for details). This covers \cotwo, \mbox{C$^{18}$O(2--1)}, and allocates the remaining bandwidth to continuum; though we mark \mbox{SiO(5--4)}, it is generally too faint to be detected by our observations. Added flexibility for programs started from Cycle~6 onwards allowed us to also observe $^{13}$\cotwo\ during some extensions to the project, as shown in the middle, blue bars. Pink bars show an alternative setup used for early pilot observations of NGC~0628. In this case we observe the high critical density line \mbox{CS(5--4)} instead of the rare isotopologue \mbox{C$^{18}$O(2--1)}. The gray shaded region at the top of the figure indicates the transmission of the atmosphere at a typical precipitable water vapor (PWV) level of 1~mm. Solid black lines indicate the Doppler-shifted frequency of observed spectral lines at a typical recessional velocity of $v_{\rm LSR} = 500$~km~s$^{-1}$.
\label{fig:spectralsetup}}
\end{figure*}

We target the $J=2\rightarrow1$ rotational transition of CO, \cotwo. This line is readily excited at temperatures and densities that characterize molecular clouds, and its spatial distribution in nearby galaxies correlates well with the fundamental \coone\ transition \citep[e.g., see][]{BRAINE92,LEROY09, LEROY13,KODA20,DENBROK21}. This choice represents a key practical element of our observing strategy. Given typical observed \cotwo-to-\coone\ line ratios of $R_{21} \approx 0.5{-}1$ \citep{DENBROK21,LEROY13} and considering the array configurations and system temperatures at ALMA, mapping \cotwo\ is a factor of ${\sim}2{-}4$ times faster than mapping \coone\ to the same mass surface density sensitivity and angular resolution. Moreover, the $\nu = 230.538$~GHz rest frequency of $^{12}$\cotwo\ lies in a favorable part of the atmosphere for ALMA, and can be observed effectively almost any time that ALMA is on sky.

We visualize the spectral setup of our observations in Figure~\ref{fig:spectralsetup}. We observed \cotwo\ with a bandwidth of $937.5$~MHz and a channel width of $\Delta \nu = 244$~kHz. After taking into account the online Hanning smoothing, this corresponds to a native channel width of ${\sim}0.6$ km~s$^{-1}$ and a total velocity coverage of ${\sim}1200$~km~s$^{-1}$. This channel width easily resolves emission lines from individual molecular clouds, which have typical full line width of ${\sim}10$~km~s$^{-1}$. The velocity coverage easily encompasses the full emission from each of our targets. In practice, we image the cubes at a velocity resolution of ${\sim}2.5$~km~s$^{-1}$ to reduce data volume and improve the signal to noise and quality of the deconvolution.

We allocated the other correlator resources in several different ways over the course of the project. During the Large Program and the two main pilot projects (see Table~\ref{tab:projects}), we configured the spectral setup to observe the mm continuum with the other three windows. We configured each to have the maximum possible $1.875$~GHz bandwidth. We placed one of these windows to cover the \mbox{C$^{18}$O(2--1)} line at a spectral resolution of 2~MHz, corresponding to ${\sim}5{-}6$~km~s$^{-1}$ after taking the Hanning smoothing into account. The C$^{18}$O isotopologue is several hundred times less abundant than $^{12}$C$^{16}$O \citep[e.g.,][]{WILSON94} and the \mbox{C$^{18}$O(1--0)} line is often $30{-}100$ times fainter than \coone\ in local galaxies \citep[e.g.,][]{JIMENEZDONAIRE19}. Therefore the C$^{18}$O line represents a target of opportunity that may be recovered in a few bright regions or via stacking. Though we imaged these data, this line is not included in the first PHANGS--ALMA data release.

In one early set of observations targeting NGC~0628, also illustrated in Figure~\ref{fig:spectralsetup}, we targeted \mbox{CS(5--4)} instead of \mbox{C$^{18}$O(2--1)}. The main practical difference is that this places the main \cotwo\ line in the lower sideband instead of the upper sideband, which affects the spectral variation of the noise \citep[see][]{LEROY21a}.

ALMA began allowing more flexible Band~6 spectral setups in Cycle~6 after the execution of most PHANGS--ALMA observations. Therefore, in the cases of some of the PHANGS--ALMA extensions using the ACA \mbox{7-m} array, we also targeted the $^{13}$\cotwo\ line. The $^{13}$CO isotopologue is only ${\sim}30{-}120$ times less abundant than $^{12}$CO \citep[e.g.,][]{WILSONROOD94,MILAM05}, and the $^{13}$CO line is only ${\sim}10$ times fainter than $^{12}$CO \citep[e.g.,][]{ROMANDUVAL16,CORMIER18}. This makes the $^{12}$CO/$^{13}$CO ratio a useful probe of optical depth and excitation \citep[e.g.,][]{CORMIER18}. As shown in Figure~\ref{fig:spectralsetup}, observing $^{13}$CO and C$^{18}$O together requires using two somewhat narrower $937.5$~MHz windows.

Finally, we note that ALMA employs ``Doppler setting'' rather than online Doppler tracking and records data in the topocentric frame. Our observed bandwidth is large enough that there is never any problem keeping the line emission from the galaxy in the full spectral window. However, this does imply that we must employ spectral regridding to construct the final velocity cubes.

\subsubsection{Target Angular Resolution and Inclusion of Short Spacing Observations}

We target a FWHM resolution of $1\arcsec$, which corresponds to ${\sim}100$~pc resolution at $20$~Mpc or ${\sim}80$~pc at the outer edge of our selection function (see \S\ref{sec:sample}). We picked this resolution so that an individual beam matches the size of an individual massive GMC or association of massive molecular clouds. During Cycle~5, when the PHANGS--ALMA Large Program was approved, this resolution at Band~6 and $\nu \approx 230$~GHz corresponded to the \mbox{C43-2} array configuration. In practice, because ALMA executes observations within some tolerance, many of our observations also occurred in configurations similar to \mbox{C43-1} or \mbox{C43-3}. Several factors related to data processing mean that our final data products have typical angular and physical resolutions slightly coarser than $1\arcsec$ (see \S\ref{sec:obsexec}).

Our galaxies typically have full optical sizes of several arcminutes (\S\ref{sec:sampleprops}). CO emission often shows extended, complex structure across the full area of a galaxy. The \mbox{C43-2} main configuration recovers emission only from scales of $\lesssim 10\arcsec$. Thus short- and zero-spacing information is important to accurately reconstruct the intensity distribution \citep[e.g., see the case of M51;][and see details for PHANGS--ALMA data processing in \citealt{LEROY21a}]{KODA09, PETY13}. Therefore, we observed all targets with both components of the Morita Atacama Compact Array (ACA), the compact array of \mbox{7-m} dishes and the \mbox{12-m} single-dish telescopes used for ``total power.''

The ratio of observing time among the main array, \mbox{7-m} array, and total power antennas followed the standard observatory recommendations. For our fiducial configuration \mbox{C43-2}, the ALMA Cycle~5 technical handbook recommended an observing time ratio of ${1\!:\!5\!:\!8.5}$ among the \mbox{12-m} main array, the \mbox{7-m} array, and the total power antennas.

In extension projects, we observed a subset of very nearby targets using only the ACA \mbox{7-m} and total power arrays\footnote{Projects 2018.1.01321.S, 2018.A.00062.S, and 2019.1.01235.S in Table~\ref{tab:projects}.}. These projects targeted galaxies within $5$~Mpc, and typically at ${\sim}3{-}4$~Mpc, so that the ${\sim}7\arcsec$ resolution of the \mbox{7-m} array at $\nu = 230$~GHz already corresponds to $100{-}140$~pc, reasonably matched to the main array resolutions for more distant targets.

\subsubsection{Target Sensitivity}

PHANGS--ALMA integrates for $20{-}30$~s per pointing, quickly covering each field in each large mosaic with a modest integration time. To arrive at this strategy, we targeted a fiducial rms noise level of $7.5$~mJy beam$^{-1}$ or better at the frequency of the redshifted \cotwo\ line and a $5$~km~s$^{-1}$ channel width. At our nominal $1\arcsec$ target resolution, this target sensitivity corresponds to an rms brightness temperature sensitivity of ${\sim}0.17$~K. 

We can also express our target sensitivity as a mass surface density sensitivity via:

\begin{equation}
\label{eq:sd}
\Sigma_{\rm mol} = \alpha_{\rm CO}^{1-0}~R_{21}^{-1}~I_{\rm CO (2-1)}~\cos i~.
\end{equation}

\noindent Here $\alpha_{\rm CO}^{1-0}$ is the \coone\ conversion factor, $R_{21}$ is the \cotwo-to-\coone\ line ratio in Kelvin units, $i$ is the inclination of the galaxy, and $I_{\rm CO (2-1)}$ is the line-integrated \cotwo\ intensity in K~km~s$^{-1}$. For a standard Milky Way $\alpha_{\rm CO}^{1-0} = 4.35$~\acounits\ \citep{BOLATTO13A}, a \cotwo-to-\coone\ ratio of $R_{21} = 0.65$ \citep{DENBROK21,LEROY13}, and a galaxy with inclination $i$, 
\begin{equation}
\label{eq:mwsd}
\Sigma_{\rm mol}~\text{[M$_\odot$~pc$^{-2}$]} = 6.7~I_{\rm CO (2-1)}~\text{[K~km~s$^{-1}$]}~\cos i~.
\end{equation}

\noindent Then, for our target per-channel sensitivity of $0.17$~K and if we assume a total line width of $\Delta v = 5$~km~s$^{-1}$ for the spectral line then the $1\sigma$ noise in $\Sigma_{\rm mol}$ is $\Sigma_{\rm mol} \approx 5.7$~M$_\odot$~pc$^{-2}$ before any inclination correction.

For comparison, typical GMC surface densities range from a few times $10$~M$_\odot$~pc$^{-2}$ to a few times $100$~M$_\odot$~pc$^{-2}$. Therefore, this sensitivity level allows us to detect every location where a GMC fills a reasonable fraction of the synthesized beam at good significance.

Under the same assumptions for $\alpha_{\rm CO}^{1-0} = 4.35$ \acounits, $R_{21} = 0.65$, and fiducial line width, our target sensitivity corresponds to $1\sigma$ point mass sensitivity of about $4 \times 10^3$~M$_\odot$, $1.5 \times 10^4$~M$_\odot$, and $4.4 \times 10^4$~M$_\odot$ at distances of $5$, $10$, and $17$~Mpc, respectively. Our chosen target sensitivity is motivated by the canonical value for the characteristic mass of GMCs in the inner Milky Way, $M_{\rm mol} \gtrsim 10^5$~M$_\odot$ \citep[e.g.,][]{BLITZ93}, and thus, even in our most distant targets, we expect to detect individual moderately massive GMCs ($M_{\rm mol} \gtrsim 10^5$~M$_\odot$) at $2{-}3\sigma$ significance. Note that this $\Delta v = 5$~km~s$^{-1}$ is the fiducial value used for sensitivity calculations. As described above, we observed at finer spectral resolution than this and our nominal data products are constructed using ${\sim}2.5$~km~s$^{-1}$ channels.

The achieved surface brightness sensitivity depends sensitively on the final beam size of the image, and our final resolutions tend to be slightly coarser than $1\arcsec$. Therefore, in practice our images tend to have somewhat better sensitivity than this $0.17$~K. We discuss the achieved properties of the PHANGS--ALMA data more below.

This target sensitivity drove our main array integration times to low values, $\approx 12{-}24$~s per pointing. Integrating on a single pointing under typical Band~6 conditions, achieving $7.5$~mJy~beam$^{-1}$ per $5$~km~s$^{-1}$ channel takes ALMA ${\sim}30$~sec using 43 antennas. After accounting for the effect of overlapping mosaic pointings, the actual nominal integration time per field dropped to ${\sim}12$~sec. Quantization of ALMA integration times and scheduling block creation tended to increase this per field integration time to $\approx 18{-}24$~s, which set our final sensitivity (see \S \ref{sec:sensitivity}).

\subsubsection{Target Area}

\begin{figure*}[ht!]
\begin{center}
\includegraphics[width=0.475\textwidth]{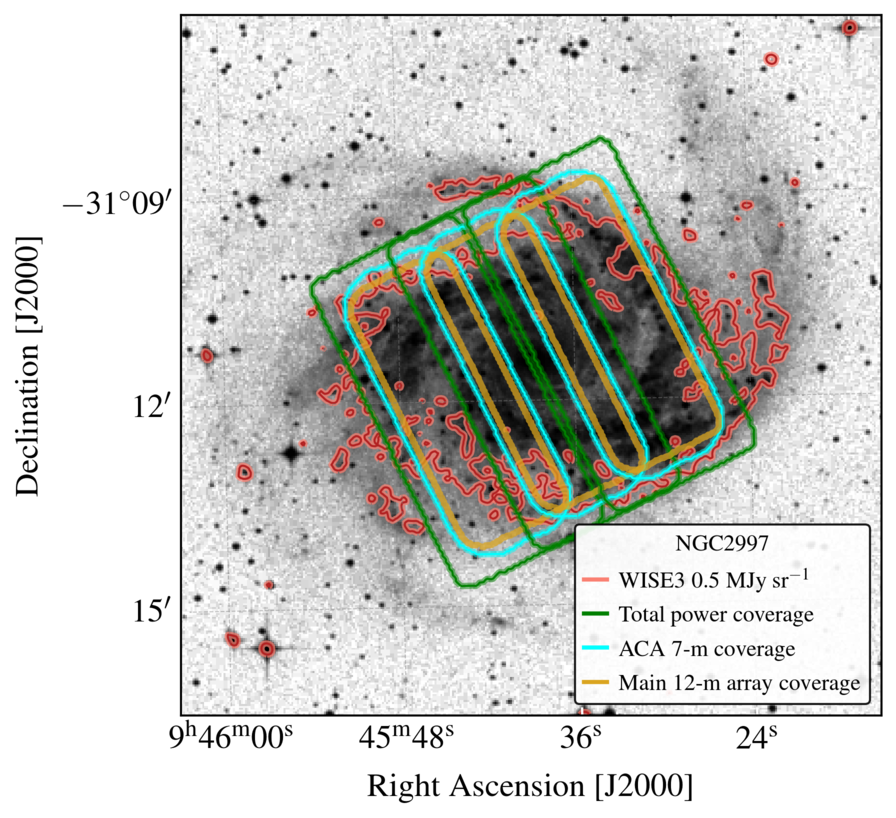}
\includegraphics[width=0.475\textwidth]{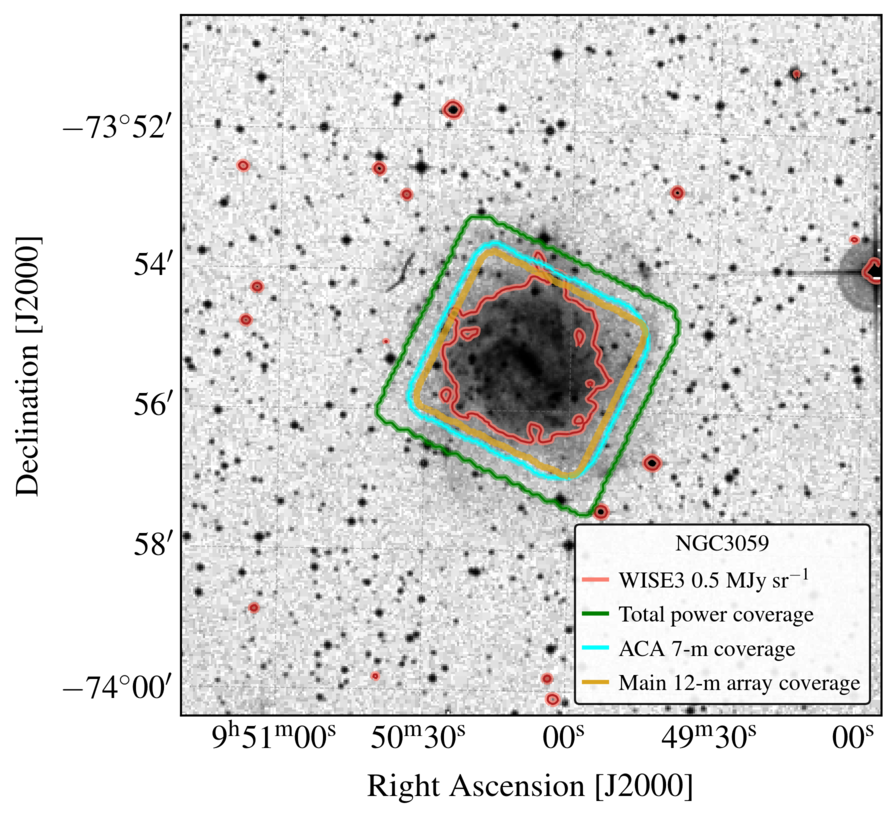}
\end{center}
\caption{\textbf{PHANGS--ALMA coverage illustrated.} Digitized Sky Survey images of two of our targets: (\textit{left}) NGC~2997, which we observe in three ``parts,'' each covered by separate ALMA Science Goals, and (\textit{right}) NGC~3059, which we covered with a single Science Goal. Red contours show the WISE Band~3 ($\lambda = 12~\mu$m) $0.5$~MJy~sr$^{-1}$ contour, which we use to define our target area. Contours indicate the area covered by the total power, \mbox{7-m}, and \mbox{12-m} array mosaics.
\label{fig:coverage}}
\end{figure*}

Our targets all have large extent compared to the $30\arcsec$ primary beam of the \mbox{12-m} ALMA antennas at $\nu = 230$~GHz. We cover them using mosaics consisting of many individual pointings, typically $100{-}300$ per galaxy. To select the exact target area, we use mid-IR emission, which we take to indicate the presence of recent star formation and thus the likely presence of molecular gas (see \S\ref{sec:apcorr} for more discussion on this topic).

Specifically, when designing the field of view for the Large Program, we aimed to cover the full area of each galaxy that shows WISE3 intensity above $0.5$~MJy~sr$^{-1}$ at $7.5\arcsec$ resolution. The pilot programs targeted the area with WISE4 intensity above $1$~MJy~sr$^{-1}$ at $15\arcsec$ resolution, which yielded qualitatively similar results. Follow the tight scaling between mid-IR intensity and CO intensity discussed in \S \ref{sec:apcorr}, this design implies that we will cover all parts of the galaxy where $I_{\rm CO}^{2-1} \gtrsim 1$~K~km~s$^{-1}$ at low resolution. Roughly, this means we cover all region where we expect molecular gas to dominate the cold gas budget in the galaxy \citep[e.g., see][]{LEROY08} and also all regions where the average intensity over a large area exceeds our $1\sigma$ sensitivity at high resolution. 

As illustrated in Figure~\ref{fig:coverage}, we defined the actual coverage region for each galaxy as a rotated rectangle, with the position angle and extent matched to the WISE contour by eye. In a few cases, a rectangle was not a good match to the exact WISE contour, e.g., because a small amount of mid-IR emission would force us to dramatically expand the mosaic and thus the integration time. In these cases, we made a judgment call to adjust the exact field of view to a more practical value.

In practice, as \S\ref{sec:apcorr}, Table~\ref{tab:apcorr_check}, and Figure~\ref{fig:coverage} show, our adopted field of view typically encloses $80\%$ of the WISE3 emission and almost $90\%$ of the WISE4 emission from each target. The aperture corrections in Tables~\ref{tab:sample_phys} give a more detailed estimate of how well our field of view captures the area of interest for each target. These values should indicate the fraction of total CO emission captured by our field of view. Qualitatively, our maps are more compact than those made by array receivers on single-dish telescopes \citep[e.g., HERACLES;][]{LEROY09}, but they do cover almost all of the active star formation and molecular gas in each target.

Within each rectangle, the ALMA observing tool places hexagonally-packed fields with pointings spaced by $\lambda/\sqrt{3} D$ or ${\sim}13\arcsec$. During each of our observing cycles, ALMA allowed no more than $150$ pointings per ``Science Goal.'' This restriction placed a maximum size on individual mosaics, and also defined the field of view for the \mbox{7-m} and TP observations. Many of our targets required two or even three of these maximum-sized 150 pointing mosaics to cover the mid-IR defined regions. For these targets, we observed the galaxy in several ``parts.'' All parts shared the same spectral setup, but we defined each part as a separate Science Goal with a distinct field of view. The fields of view of different Science Goals targeting the same galaxy overlapped one another by ${\sim}10\arcsec$ along the shared edge to ensure uniform sensitivity. The separate Science Goals were scheduled, observed, and mostly reduced separately. The left panel in Figure~\ref{fig:coverage} illustrates a three-part case, NGC~2997. In this galaxy, ALMA observed each of the three rectangular fields separately. The figure demonstrates the overlap between fields that we use to achieve a good final image of the whole galaxy. The right panel shows NGC~3059, where a single $<150$~pointing mosaic covers the whole galaxy. Both panels show the rotated rectangular fields that we use and the final coverage of the different arrays, with the total power data slightly extended relative to the ACA \mbox{7-m} data and both covering the footprint of the \mbox{12-m} data.

As we discuss below, observing the galaxies in separate parts led to uneven resolution in some cases. Our processing matches the angular resolution of the parts at the coarsest common beam, which led to a modest overall loss in resolution in our data products. In practice, a modest difference in surface brightness sensitivity between the individual parts of a mosaic is likely to be the most noticeable consequence of these multi-part observations.

Following standard ALMA practice, the total power observations used an ``on-the-fly mapping'' approach to cover the area of interest. Individual observations used multiple antennas to cover the same area. The total power observations were descended from the same Science Goals as the interferometric observations. As a result, when a galaxy was observed in multiple parts spread across several Science Goals, the total power observations were similarly divided to cover individual parts of the galaxy.

\subsubsection{Notes on Archival Observations}

We periodically reviewed the ALMA archive to find all observations that match our PHANGS--ALMA setup. We searched for observations of nearby, $d \lesssim 20$~Mpc, galaxies targeting \cotwo\ and covering all or most of the galaxy's star-forming area at ${\sim}100$~pc resolution. As of our latest review, in autumn 2020, we identified four galaxies that closely matched the PHANGS--ALMA setup and included these in our sample. The targets are NGC~1365, NGC~5128 (Centaurus~A), NGC~5236 (M83), and NGC~7793; and we list the project codes and PIs in Table~\ref{tab:projects}. We include NGC~1365 ``as~is'' from the archive. For Centaurus~A, the \cotwo\ part of the total power data in the archive was not usable due to a problem in the observations. We obtained new ACA mapping in ALMA Cycle~7 (P.I.\ C.~Faesi, see Table~\ref{tab:projects}). NGC~5236 was observed over two cycles using 9 separate Science Goals. The data have large extent, but otherwise match the PHANGS--ALMA setup exactly. The archival NGC~7793 observations obtained only \mbox{12-m} main array data. We obtained \mbox{7-m} and total power data in ALMA Cycle~7 (P.I.\ C.~Faesi, see Table~\ref{tab:projects}).

For this version of a PHANGS--ALMA data release, we considered only the \cotwo\ line and only processed wide-area maps of star-forming galaxies. By including the \coone\ and \cothree\ lines and relaxing the areal coverage requirements, many more archival observations of molecular line emission in nearby galaxies could be included. The PHANGS--ALMA data processing pipeline \citep{LEROY21a} can straightforwardly handle these data, and we intend to return to this in future work.

\subsection{Execution of Observations}
\label{sec:obsexec}

 \begin{deluxetable*}{lcccccccc} 
 \tablecaption{PHANGS--ALMA Total Power Observation Log \label{tab:obslog_tp}} 
 \tablewidth{0pt} 
 \tabletypesize{\footnotesize} 
 \tablehead{ 
 \colhead{Target} & 
 \colhead{Project} & 
 \colhead{Start} & 
 \colhead{End} & 
 \colhead{Min. El.} & 
 \colhead{Max. El.} & 
 \colhead{$N_{\rm ant}$} & 
 \colhead{PWV} & 
 \colhead{Gain}  
 \\ 
 \colhead{} & 
 \colhead{} & 
 \colhead{(MJD)} & 
 \colhead{(MJD)} & 
 \colhead{($^\circ$)} & 
 \colhead{($^\circ$)} & 
 \colhead{} & 
 \colhead{(mm)} &  
 \colhead{(Jy/K)} 
 }  
\startdata 
ESO097-013 & 2018.1.01321.S & 58551.230 & 58551.281 &  43.0 &  47.0 & 3 & 1.80 &  41.2\\ 
ESO097-013 & 2018.1.01321.S & 58551.282 & 58551.333 &  47.0 &  48.0 & 3 & 1.80 &  41.1\\ 
IC1954 & 2017.1.00886.L & 58124.013 & 58124.062 &  60.0 &  61.0 & 3 & 1.10 &  44.4\\ 
IC1954 & 2017.1.00886.L & 58126.046 & 58126.095 &  57.0 &  61.0 & 3 & 2.00 &  44.3\\ 
IC1954 & 2017.1.00886.L & 58132.986 & 58133.036 &  60.0 &  61.0 & 3 & 2.30 &  44.5\\ 
IC1954 & 2017.1.00886.L & 58133.036 & 58133.085 &  56.0 &  61.0 & 3 & 2.20 &  44.6\\ 
IC1954 & 2017.1.00886.L & 58133.951 & 58134.000 &  56.0 &  61.0 & 3 & 3.10 &  43.9\\ 
IC1954 & 2017.1.00886.L & 58134.001 & 58134.051 &  59.0 &  61.0 & 3 & 3.00 &  44.2\\ 
IC1954 & 2017.1.00886.L & 58136.028 & 58136.077 &  56.0 &  61.0 & 3 & 1.80 &  44.1\\ 
IC5273 & 2017.1.00886.L & 58115.863 & 58115.920 &  70.0 &  75.0 & 3 & 2.80 &  43.5\\ 
\enddata 
\tablecomments{This table is a stub. The full version of the table appears as a machine-readable table 
in the online version of the paper. Columns give: Target---the target of the observations;  
Project---the ALMA project code; Start and End---the beginning  
and end of the observations reported as Modified Julian Day (00:00 on Jan~1, 2018 is MJD 58119);   
Min.\ and Max.\ El.---the minimum and maximum elevation  
during the observation; $N_{\rm ant}$---the number of antennas participating in the observations; PWV---the precipitable  
water vapor during the observation; Gain---the Jansky per Kelvin calibration for the observation.} 
\end{deluxetable*}

 \begin{deluxetable*}{lcccccccccc} 
 \tablecaption{PHANGS--ALMA 7{-}m Observation Log \label{tab:obslog_7m}} 
 \tablewidth{0pt} 
 \tabletypesize{\footnotesize} 
 \tablehead{ 
 \colhead{Target} & 
 \colhead{Project} & 
 \colhead{Start} & 
 \colhead{End} & 
 \colhead{Min. El.} & 
 \colhead{Max. El.} & 
 \colhead{$N_{\rm ant}$} & 
 \colhead{PWV} & 
 \colhead{Min. $u{-}v$} &  
 \colhead{Med. $u{-}v$} &  
 \colhead{Max. $u{-}v$}  
 \\ 
 \colhead{} & 
 \colhead{} & 
 \colhead{(MJD)} & 
 \colhead{(MJD)} & 
 \colhead{($^\circ$)} & 
 \colhead{($^\circ$)} & 
 \colhead{} & 
 \colhead{(mm)} &  
 \colhead{(m)} &  
 \colhead{(m)} &  
 \colhead{(m)} 
 }  
\startdata 
CIRCINUS\_1 & 2018.1.01321 & 58501.398 & 58501.428 &  43.5 &  60.7 & 12 & 2.67 &   7.9 &  20.3 &  37.4\\ 
CIRCINUS\_2 & 2018.1.01321 & 58501.429 & 58501.459 &  44.6 &  70.5 & 12 & 2.45 &   8.2 &  20.9 &  36.8\\ 
IC1954 & 2017.1.00886 & 58055.221 & 58055.281 &  52.3 &  63.3 & 11 & 1.14 &   8.2 &  21.1 &  43.8\\ 
IC1954 & 2017.1.00886 & 58056.211 & 58056.270 &  53.5 &  61.2 & 11 & 1.15 &   8.3 &  21.0 &  43.7\\ 
IC1954 & 2017.1.00886 & 58055.281 & 58055.338 &  49.0 &  75.8 & 11 & 1.08 &   7.3 &  20.9 &  43.9\\ 
IC5273 & 2017.1.00886 & 58032.085 & 58032.136 &  63.1 &  80.4 & 11 & 0.51 &   8.6 &  24.0 &  47.7\\ 
IC5273 & 2017.1.00886 & 58037.970 & 58038.020 &  55.5 &  77.1 & 10 & 1.40 &   7.5 &  20.8 &  46.0\\ 
IC5273 & 2017.1.00886 & 58038.045 & 58038.095 &  51.8 &  85.0 & 10 & 1.38 &   8.8 &  21.9 &  47.5\\ 
IC5332 & 2015.1.00925 & 57559.388 & 57559.445 &  65.5 &  77.0 & 8 & 0.84 &   8.7 &  21.2 &  42.7\\ 
IC5332 & 2015.1.00925 & 57559.445 & 57559.501 &  54.7 &  73.2 & 8 & 0.75 &   8.0 &  21.2 &  43.1\\ 
\enddata 
\tablecomments{This table is a stub. The full version of the table appears as a machine readable table 
in the online version of the paper. Columns give: Target---the target of the observations;  
Project---the ALMA project code; Start and End---the beginning  
and end of the observations reported as Modified Julian Day (00:00 on Jan~1, 2018 is MJD 58119);   
Min.\ and Max.\ El.---the minimum and maximum elevation  
during the observation; $N_{\rm ant}$---the number of antennas participating in the observations; PWV---the precipitable  
water vapor during the observation; Min., Med., and Max.\ $u{-}v$---minimum, median, and maximum baseline.} 
\end{deluxetable*}

 \begin{deluxetable*}{lcccccccccc} 
 \tablecaption{PHANGS--ALMA 12{-}m Observation Log \label{tab:obslog_12m}} 
 \tablewidth{0pt} 
 \tabletypesize{\footnotesize} 
 \tablehead{ 
 \colhead{Target} & 
 \colhead{Project} & 
 \colhead{Start} & 
 \colhead{End} & 
 \colhead{Min. El.} & 
 \colhead{Max. El.} & 
 \colhead{$N_{\rm ant}$} & 
 \colhead{PWV} & 
 \colhead{Min. $u{-}v$} &  
 \colhead{Med. $u{-}v$} &  
 \colhead{Max. $u{-}v$}  
 \\ 
 \colhead{} & 
 \colhead{} & 
 \colhead{(MJD)} & 
 \colhead{(MJD)} & 
 \colhead{($^\circ$)} & 
 \colhead{($^\circ$)} & 
 \colhead{} & 
 \colhead{(mm)} &  
 \colhead{(m)} &  
 \colhead{(m)} &  
 \colhead{(m)} 
 }  
\startdata 
IC1954 & 2017.1.00886 & 58299.543 & 58299.575 &  55.2 &  64.1 & 44 & 0.67 &  13.1 &  87.9 & 303.4\\ 
IC5273 & 2017.1.00886 & 58256.379 & 58256.409 &  54.6 &  61.9 & 45 & 1.24 &  12.4 &  82.5 & 284.4\\ 
IC5273 & 2017.1.00886 & 58264.398 & 58264.428 &  59.2 &  70.5 & 44 & 1.24 &  13.8 &  86.6 & 297.8\\ 
IC5332 & 2015.1.00925 & 57558.316 & 57558.360 &  51.1 &  65.6 & 43 & 0.71 &  13.1 & 201.0 & 641.0\\ 
NGC0628 & 2012.1.00650 & 56316.907 & 56316.949 &  36.3 &  83.1 & 29 & 3.74 &  12.4 & 118.2 & 367.9\\ 
NGC0628 & 2012.1.00650 & 56781.615 & 56781.672 &  46.1 &  59.8 & 30 & 2.34 &  13.5 & 171.1 & 509.5\\ 
NGC0628 & 2012.1.00650 & 56781.687 & 56781.743 &  36.7 &  60.1 & 30 & 2.56 &  12.0 & 154.7 & 532.4\\ 
NGC0628 & 2012.1.00650 & 56782.487 & 56782.546 &  28.7 &  75.6 & 32 & 1.75 &  16.9 & 139.6 & 455.7\\ 
NGC0628 & 2012.1.00650 & 56794.721 & 56794.769 &  19.8 &  68.1 & 32 & 0.44 &  12.7 & 161.8 & 638.9\\ 
NGC0628 & 2012.1.00650 & 56994.130 & 56994.189 &  29.4 &  55.9 & 33 & 0.62 &  11.6 &  75.3 & 279.4\\ 
\enddata 
\tablecomments{This table is a stub. The full version of the table appears as a machine readable table 
in the online version of the paper. Columns give: Target---the target of the observations;  
Project---the ALMA project code; Start and End---the beginning  
and end of the observations reported as Modified Julian Day (00:00 on Jan~1, 2018 is MJD 58119);   
Min.\ and Max.\ El.---the minimum and maximum elevation  
during the observation; $N_{\rm ant}$---the number of antennas participating in the observations; PWV---the precipitable  
water vapor during the observation; Min., Med., and Max.\ $u{-}v$---minimum, median, and maximum baseline.} 
\end{deluxetable*}

Including all pilots and extensions, observations for PHANGS--ALMA spanned six years, from early 2013 until the end of 2019. Tables~\ref{tab:obslog_tp},~\ref{tab:obslog_7m}, and~\ref{tab:obslog_12m} report details of the individual interferometric and total power observations that passed observatory quality assurance and were included in our data processing. In the tables, we report the target name, the relevant project code, the date of the observation, the duration of the observation, the number of antennas used, the elevation range, and the precipitable water vapor (PWV) associated with the observation. For the interferometric observations, we also report the minimum, maximum, and median baseline length associated with the observation. For the total power observations, we report the Jy/K gain factor supplied by the observatory.

\begin{figure*}[ht!]
\begin{center}
\includegraphics[width=0.95\textwidth]{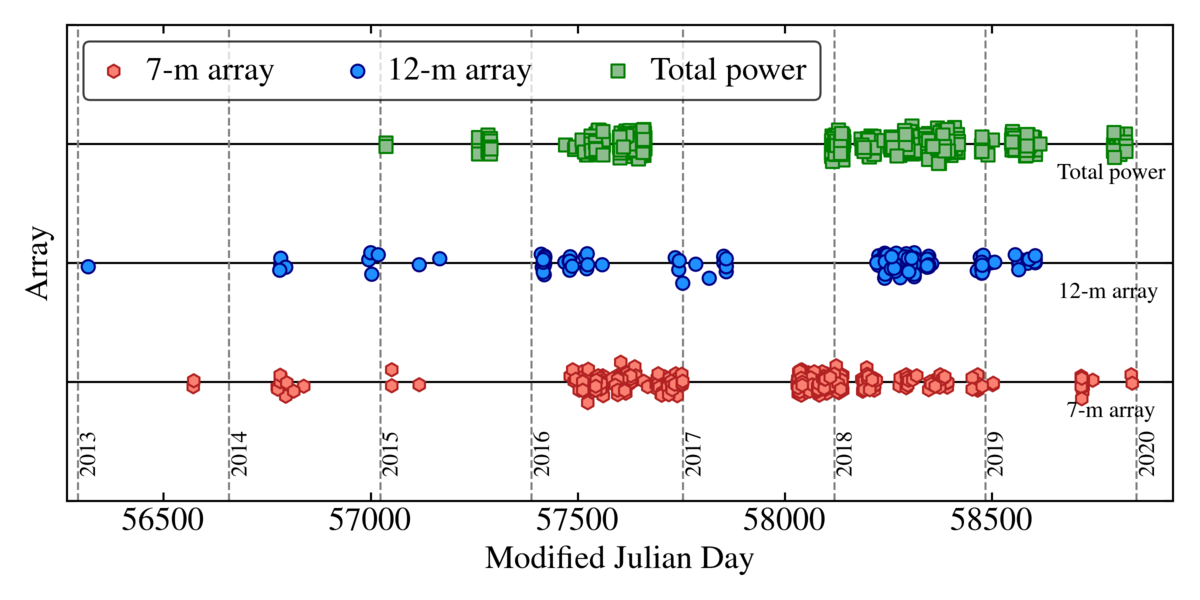}
\end{center}
\caption{\textbf{PHANGS--ALMA observing dates (including archival observations for the 12{-}m and 7{-}m arrays but not the total power data).} Date of observations of PHANGS-ALMA observations with the \mbox{12-m}, \mbox{7-m}, and total power arrays. We include all pilot and archival observations processed as part of the survey described in this paper.
\label{fig:obsdate}}
\end{figure*}

We visualize the dates and the properties of individual observations in Figures~\ref{fig:obsdate},~\ref{fig:obscond}, and~\ref{fig:basedist}. Table~\ref{tab:obscond} summarizes the observations, giving the typical duration, observing conditions, and $u{-}v$ coverage with each array. We also report the total number of observations, $184$ for the \mbox{12-m} array, $479$ for the \mbox{7-m} array, and $823$ for the total power antennas. Taking into the account the duration of each observation, the total array time for the survey, including pilots, archival observations\footnote{Currently the archival total power observations are not included in these plots and logs.}, extensions, and the Large Program, is $177$~hours in the main array, $652$ hours in the \mbox{7-m} array, and $855$ hours of total power observations (typically using $2{-}3$ antennas at a time). As Figure~\ref{fig:obsdate} illustrates, the bulk of these observations occurred in 2016--2019, coinciding with the execution of the two main pilot programs and the Large Program.

\subsubsection{Calibration}

For the interferometric data, calibration followed the standard ALMA procedures with no special requirements. The observatory selected the primary and secondary calibrators that were used to calibrate the bandpass, flux scale, and phase response of each antenna. The cycle time also followed standard observatory procedures, and the observations employed the standard ALMA calibrations to measure and correct for atmospheric water vapor and to measure the system temperature.

For the total power data, we provided the observatory with reference ``OFF'' positions. We chose these to be well-separated from the galaxy, typically $5\arcmin{-}10\arcmin$ away and always well outside the optical radius of the galaxy. We verified from mid-IR WISE and optical imaging that the OFF position did not coincide with any emission from the galaxy or a neighboring galaxy. For \cotwo\ the recessional velocity of our galaxies means that they are well-separated from any foreground Milky Way emission.

In a subset of targets the \cotwo\ total power observations showed significant contamination by an atmospheric ozone feature at $229.575$~GHz. In cases where the ozone feature overlapped the emission from the galaxy, this sometimes contaminated the total power data to the point where they could not be used. We present a discussion of this effect in the appendix of \citet{LEROY21a} and in \citet{HERRERA20}, and a more detailed summary is presented in two memos by A.~Usero\footnote{Available at \url{https://sites.google.com/view/phangs/publications}~.}. This telluric ozone contamination reflects imperfect sky subtraction. To improve the situation, we worked with the observatory to re-observe the most strongly affected cases using a reference fixed in elevation, rather than in the equatorial frame. These observations were also conducted with more stringent requirements on the weather, ensuring less atmospheric contamination. We mark the affected targets in Table~\ref{tab:sample_obs}. See \citet{LEROY21a} for more details on the effect and processing.

\subsubsection{Observing Time and Conditions}

\begin{deluxetable}{lccc}[ht!]
\tabletypesize{\footnotesize}
\tablecaption{PHANGS--ALMA \cotwo\ Observing Conditions \label{tab:obscond}}
\tablewidth{0pt}
\tablehead{
\colhead{Quantity} &
\multicolumn{3}{c}{Array} \\
\colhead{} &
\colhead{\mbox{12-m}} &
\colhead{\mbox{7-m}} &
\colhead{TP} 
}
\startdata
\hline
Number of observations & 184 & 479 & 823 \\
Total time [h] & 177 & 652 & 855 \\
\hline
\multicolumn{4}{l}{Median conditions during observations\tablenotemark{a} ... } \\
... duration [h] & 
$1.0^{+0.2}_{-0.2}$ & 
$1.4^{+0.1}_{-0.2}$ & 
$1.2^{+0.1}_{-0.4}$ \\
... PWV\tablenotemark{b} [mm] & 
$1.4^{+0.8}_{-0.7}$ & 
$1.0^{+0.8}_{-0.5}$ & 
$0.9^{+1.0}_{-0.4}$ \\
... mean elevation [$^\circ$] & 
$60^{+9}_{-7}$ & 
$58^{+10}_{-7}$ & 
$54^{+16}_{-9}$ \\
... min. baseline [m] & 
$13^{+2}_{-1}$ &
$8^{+0.7}_{-0.8}$ &
\nodata \\
... median baseline [m] & 
$97^{+34}_{-14}$ &
$21^{+2}_{-1}$ &
\nodata \\
... max. baseline [m] &
$388^{+99}_{-91}$ &
$44^{+4}_{-5}$ & 
\nodata \\
\hline
\enddata
\tablenotetext{a}{Values quote median and error bars give $16\%{-}84\%$ range.}
\tablenotetext{b}{Precipitable water vapor.}
\end{deluxetable}

\begin{figure*}[ht!]
\begin{center}
\includegraphics[width=0.45\textwidth]{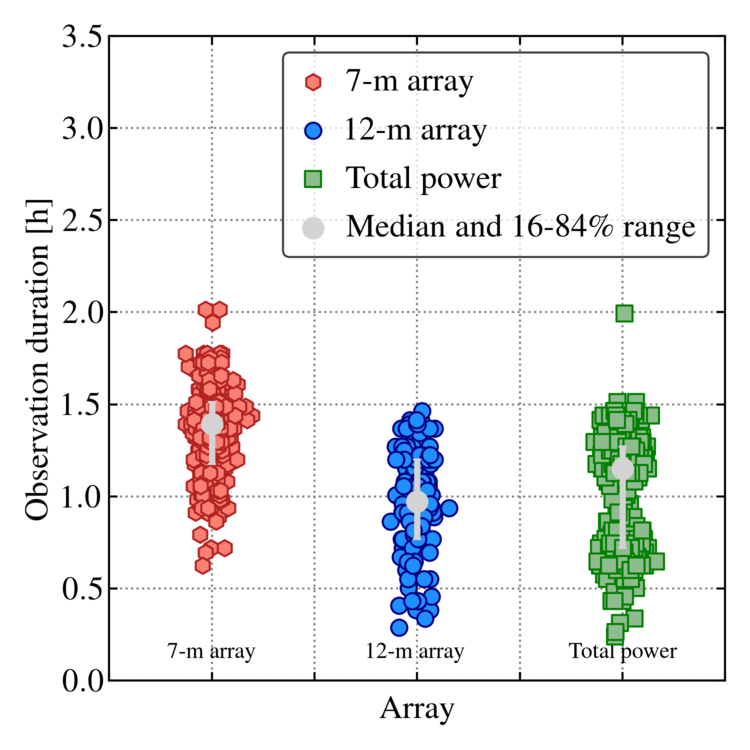}
\includegraphics[width=0.45\textwidth]{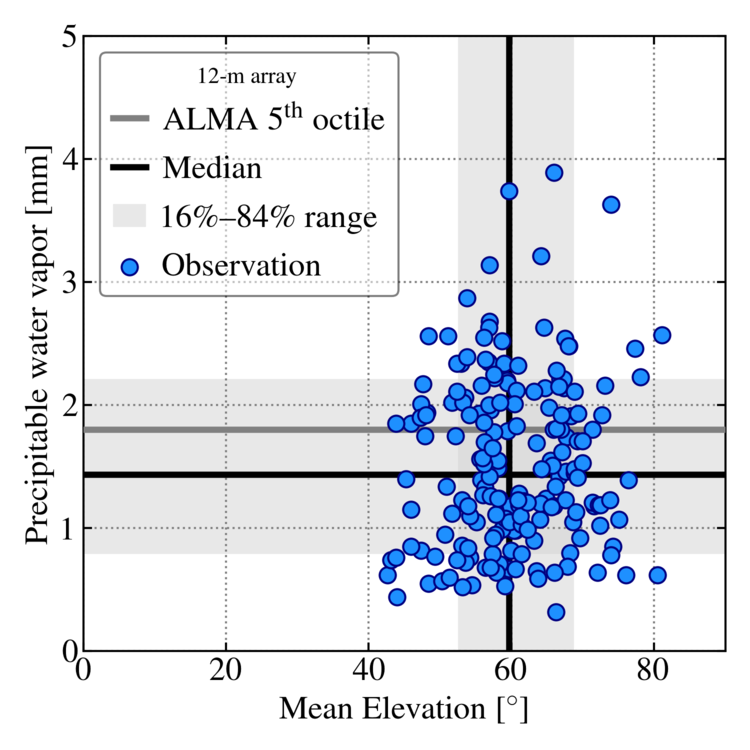}
\includegraphics[width=0.45\textwidth]{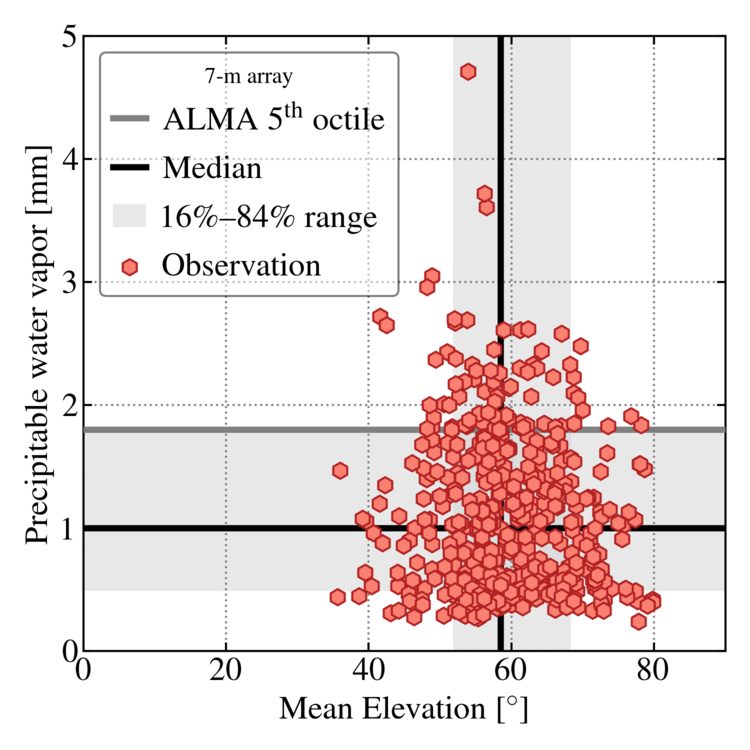}
\includegraphics[width=0.45\textwidth]{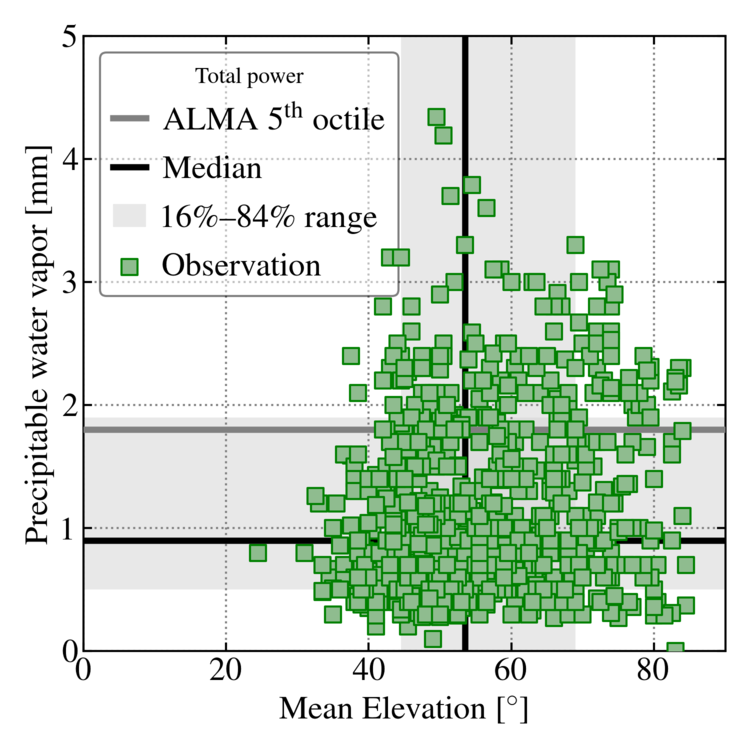}
\end{center}
\caption{\textbf{PHANGS--ALMA observing conditions for individual observations.} \textit{Top left:} Duration, from start to finish, of individual observing blocks observed with the \mbox{12-m} array, \mbox{7-m} array, and total power telescopes. Gray points and error bars show the median and $16\%{-}84\%$ range of the duration of individual observations, which are typically $1{-}1.5$~h. The remaining three panels show the mean elevation and precipitable water vapor (PWV) measured in the atmosphere for each individual block. Gray regions show the $16\%{-}84\%$ range of each quantity and black lines show the median. A~thick gray line shows ALMA's 5$^{\rm th}$ octile of PWV, ${\sim}1.8$~mm,  the nominal threshold for Band~6 ($230$~GHz) observations.
\label{fig:obscond}}
\end{figure*}

Figure~\ref{fig:obscond} visualizes the conditions of our individual observations, which are also summarized in Table~\ref{tab:obscond} and reported in detail in  Tables~\ref{tab:obslog_tp},~\ref{tab:obslog_7m}, and~\ref{tab:obslog_12m}. In brief, individual observations typically lasted $1{-}1.4$~h across all arrays, and most observations took place at mean elevation of $50{-}70^\circ$ with precipitable water vapor (PWV) $\sim 0.5{-}2$~mm. The tables and figures report more exact numbers, but overall this reflects almost ideal scheduling on behalf of the array. The high elevation imply limited shadowing, and minimal atmospheric contamination, and the PWV of the observations is almost always less than the ${\sim}1.8$~mm associated with ALMA's 5$^{\rm th}$ octile of weather conditions, the nominal threshold for 230~GHz (Band~6) observations.

\subsubsection{Achieved $u{-}v$ coverage and synthesized beam}

\begin{figure}[ht!]
\begin{center}
\includegraphics[width=0.45\textwidth]{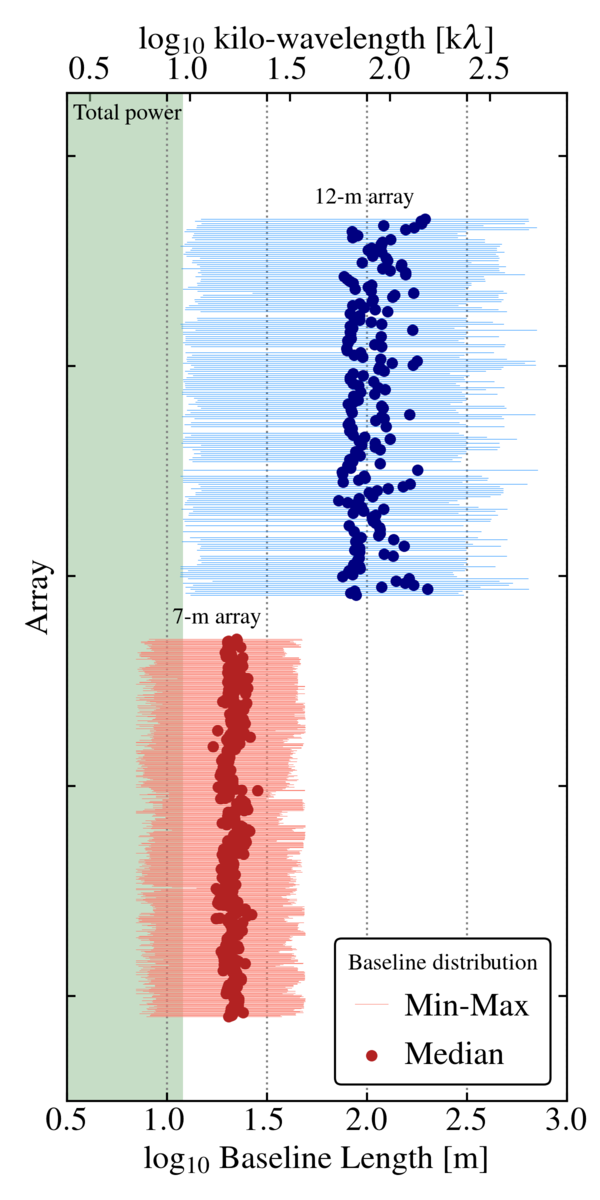}
\end{center}
\caption{\textbf{PHANGS--ALMA baseline lengths.}
\label{fig:basedist} Illustration of $u{-}v$ baseline lengths in the \mbox{12-m} (blue, upper) and \mbox{7-m} (red, lower) arrays. Each line shows one observation, with the circle showing the median value and the lines showing the full minimum to maximum baseline length in the data. A~green shaded region indicates the \mbox{12-m} diameter of the total power antennas. For a more detailed look at the distribution of collecting area vs.\ baseline length in one case, see Figure~\ref{fig:uvresponse}.}
\end{figure}

\begin{figure*}[ht!]
\begin{center}
\includegraphics[width=0.475\textwidth]{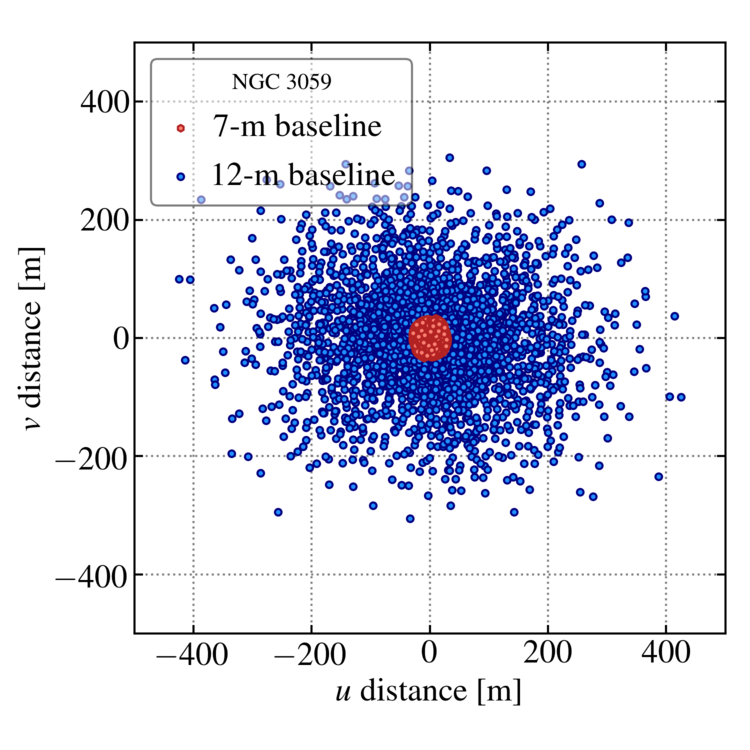}
\includegraphics[width=0.475\textwidth]{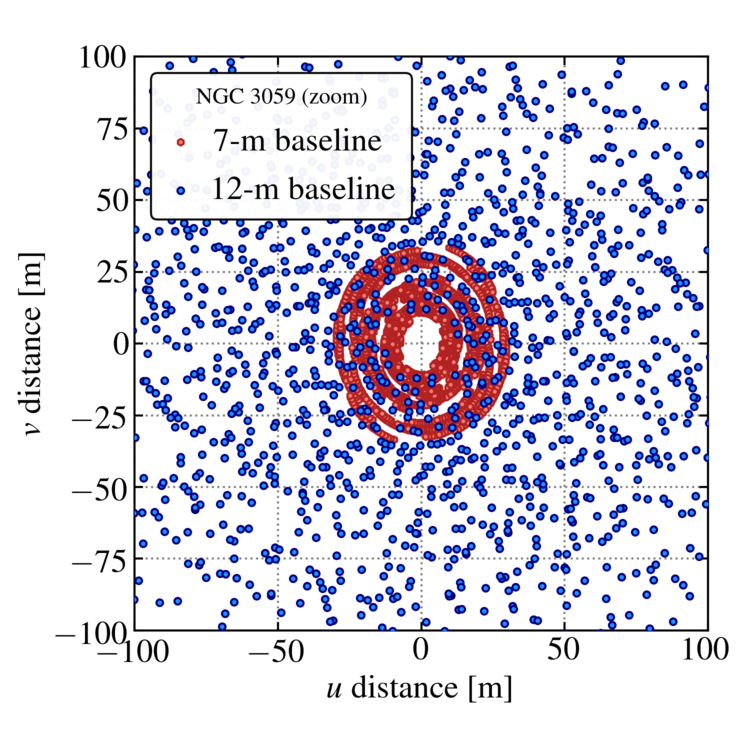}
\end{center}
\caption{\textbf{PHANGS--ALMA \boldmath{$u{-}v$} coverage example.} Example plots of $u{-}v$ plane coverage for two pointings in NGC~3059, one covered by the ACA \mbox{7-m} array (red points) and one covered by the main \mbox{12-m} array (blue points). Both panels show the same data, with the \textit{right} hand panel showing a zoom-in to the inner part of the $u{-}v$ plane. Figure~\ref{fig:uvresponse} shows the collecting area vs.\ baseline length for these data and Figure~\ref{fig:psfs} show the synthesized beams for NGC~3059.
\label{fig:uvcoverage}}
\end{figure*}

\begin{figure}[ht!]
\begin{center}
\includegraphics[width=0.475\textwidth]{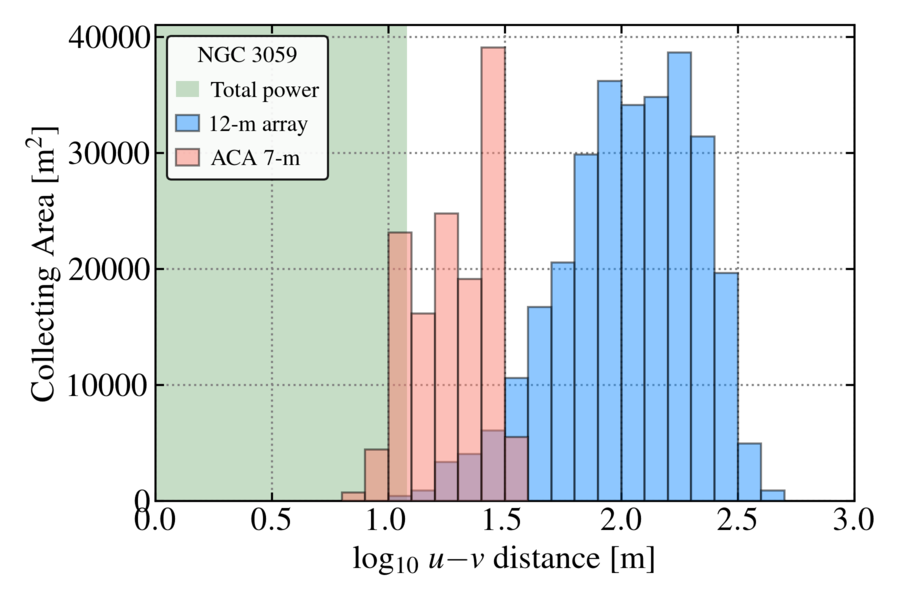}
\end{center}
\caption{\textbf{PHANGS--ALMA distribution of collecting area vs. \boldmath{$u{-}v$} distance for individual integrations.} For the same pointings in Figure~\ref{fig:uvcoverage}, we show the distributions of collecting area vs.\ $\log_{10}$ $u{-}v$ baseline length for individual integrations from the \mbox{12-m} array (blue) and \mbox{7-m} array observations. The figure illustrates good overlap between the three components of ALMA and good sensitivity to a range of spatial scales. Note that the shaded total power region only indicates the $u{-}v$ range of the total power antennas, not the amount of collecting area.
\label{fig:uvresponse}}
\end{figure}

\begin{figure*}[ht!]
\begin{center}
\includegraphics[width=0.475\textwidth]{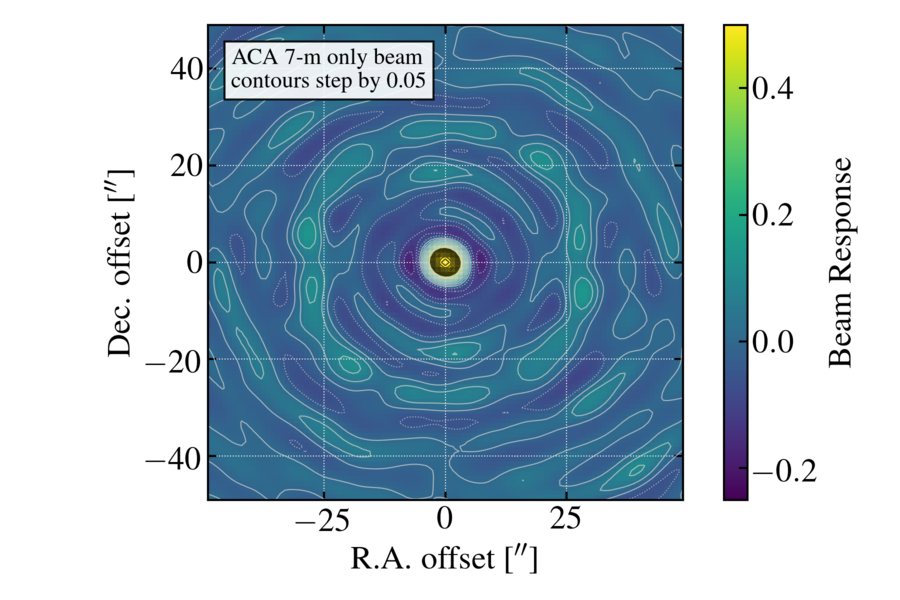}
\includegraphics[width=0.475\textwidth]{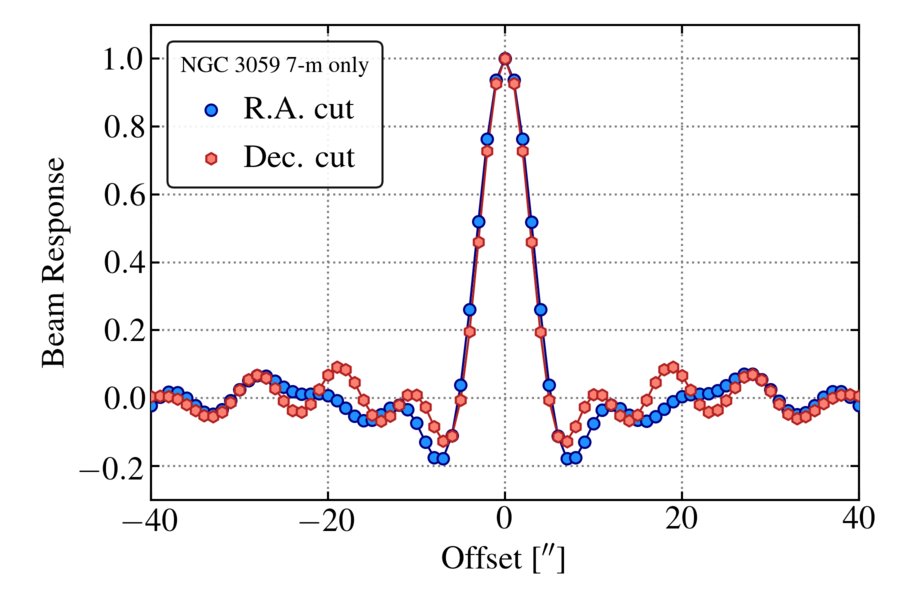}
\includegraphics[width=0.475\textwidth]{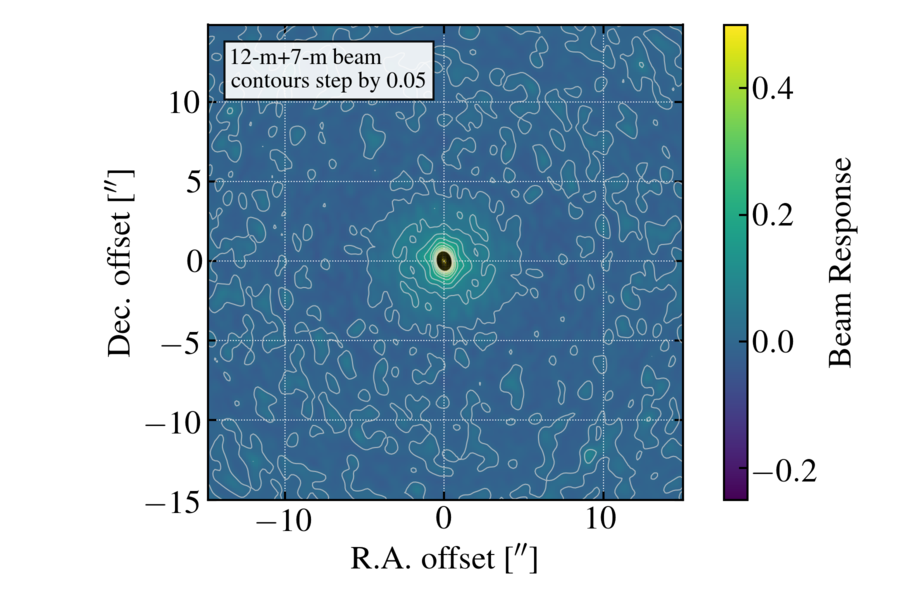}
\includegraphics[width=0.475\textwidth]{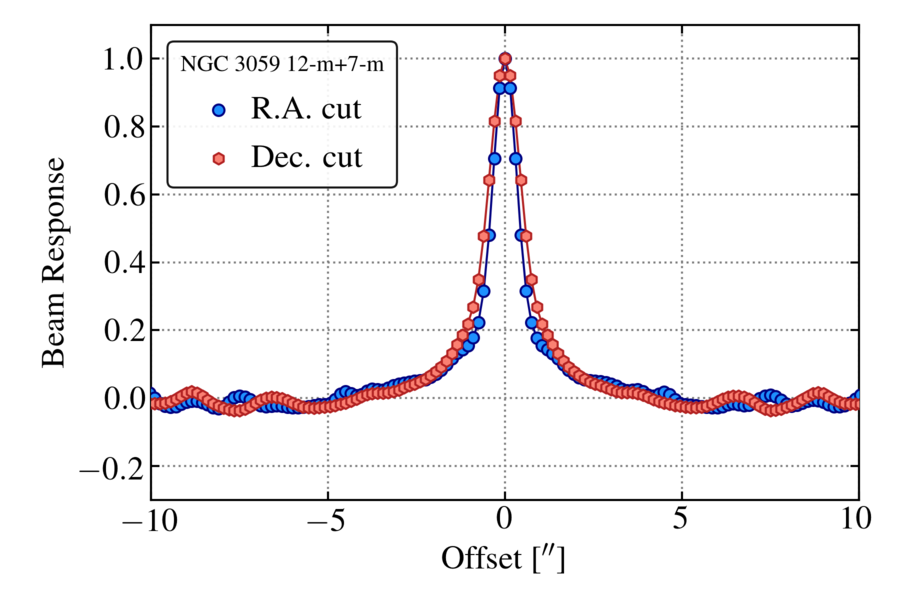}
\end{center}
\caption{\textbf{PHANGS--ALMA synthesized beam examples.} Synthesized beams for one channel in our NGC~3059 \cotwo\ cube produced by our \texttt{CASA} imaging for the (\textit{top row}) ACA \mbox{7-m}-only and (\textit{bottom row}) \mbox{12-m}+\mbox{7-m} combined arrays. The \textit{left} column shows images of the synthesized beam response for each case, with contours spaced by by $0.05$. The \textit{right} column shows response cuts along right ascension and declination through the synthesized beam.
\label{fig:psfs}}
\end{figure*}

Tables~\ref{tab:obslog_12m} through~\ref{tab:obscond} report the minimum, maximum, and median $u{-}v$ baseline length of each observation. We visualize this in Figure~\ref{fig:basedist}, which shows these values for each observation. On average, the \mbox{12-m} baselines span $13{-}388$~m with median baseline length $97$~m. In terms of wavelength, this translates to a range of ${\sim}10{-}3000$~k$\lambda$, corresponding to ${\sim}20\arcsec{-}0.7\arcsec$, with a mean of ${\sim}75$~k$\lambda$, corresponding to ${\sim}2.8\arcsec$. The \mbox{7-m} data typically span baselines $8{-}44$~m with a mean of $21$~m. This maps to $6{-}34$~k$\lambda$, corresponding to $33\arcsec{-}6\arcsec$, with a mean of $16$~k$\lambda$, corresponding to ${\sim}13\arcsec$. Figure~\ref{fig:basedist} shows that there is some scatter in the median \mbox{12-m} baseline length across our data, and significant, about $\pm 25\%$, scatter in the maximum baseline for the \mbox{12-m} data. However, the \mbox{7-m} coverage remains stable, consistent with the fixed positions of the ACA, and in all cases there is good overlap between the \mbox{12-m} baseline range, the \mbox{7-m} baseline range, and the diameter of the total power antennas.

The baseline range only gives the approximate response of the array. Figures~\ref{fig:uvcoverage}, \ref{fig:uvresponse}, and~\ref{fig:psfs} show the $u{-}v$ coverage, distribution of collecting area vs.\ baseline distance, and achieved synthesized beam in more detail. All of these figures focus on one example galaxy, NGC~3059, which also appears in Figure~\ref{fig:coverage}.

Figures~\ref{fig:uvcoverage} and~\ref{fig:uvresponse} show the $u{-}v$ plane coverage for one \mbox{7-m} pointing (red) and one \mbox{12-m} pointing (blue). In Figure~\ref{fig:uvcoverage}, we see that the \mbox{7-m} data, which combine several distinct observations, achieve good rotation synthesis. The \mbox{12-m} data do not, but the figure shows the excellent instantaneous $u{-}v$ coverage of ALMA. The \mbox{12-m} baselines do an outstanding job of filling out the $u{-}v$ plane even without long enough integrations to achieve rotation synthesis. In Figure~\ref{fig:uvresponse} we plot the same data as in Figure~\ref{fig:uvcoverage}, but now showing the distribution of collecting area as a function of $u{-}v$ baseline length for individual integrations. We use collecting area as a proxy for sensitivity, assuming fixed atmospheric conditions and integration length. The figure shows a relatively even distribution of sensitivity as a function of $\log_{10} u{-}v$ distance, and so spatial scale, out to ${\sim}300$~m. Similar to Figure~\ref{fig:basedist}, but now in sharper detail, we see good sensitivity to a range of spatial scales and overlap between the three types of telescopes (note that we only indicate the $u{-}v$ range of the total power data, not the collecting area).

Figure~\ref{fig:psfs} shows the sampling of the $u{-}v$ coverage. Here we show the synthesized beam for a single channel in NGC~3059. This is the same target used for Figures~\ref{fig:uvcoverage} and~\ref{fig:uvresponse}, but the beams here represent the average result across the whole mosaic because \texttt{CASA} does not track spatial variations of the beam. We show the beam for both the \mbox{7-m} array alone and the combined \mbox{12-m}+\mbox{7-m} array.

Both beams show a reasonable Gaussian core, but the \mbox{7-m}-only beam shows significantly worse positive and negative sidelobes. This is consistent with the significantly worse performance of \mbox{7-m}-only imaging compared to \mbox{12-m}+\mbox{7-m} imaging discussed by \citet{LEROY21a}. Overall, the \mbox{12-m}+\mbox{7-m} synthesized beam, which corresponds to our key data product, looks symmetric without clearly visible pathologies in the sidelobes.

\subsection{Mosaic sensitivity}

\begin{figure*}[ht!]
\begin{center}
\includegraphics[width=0.475\textwidth]{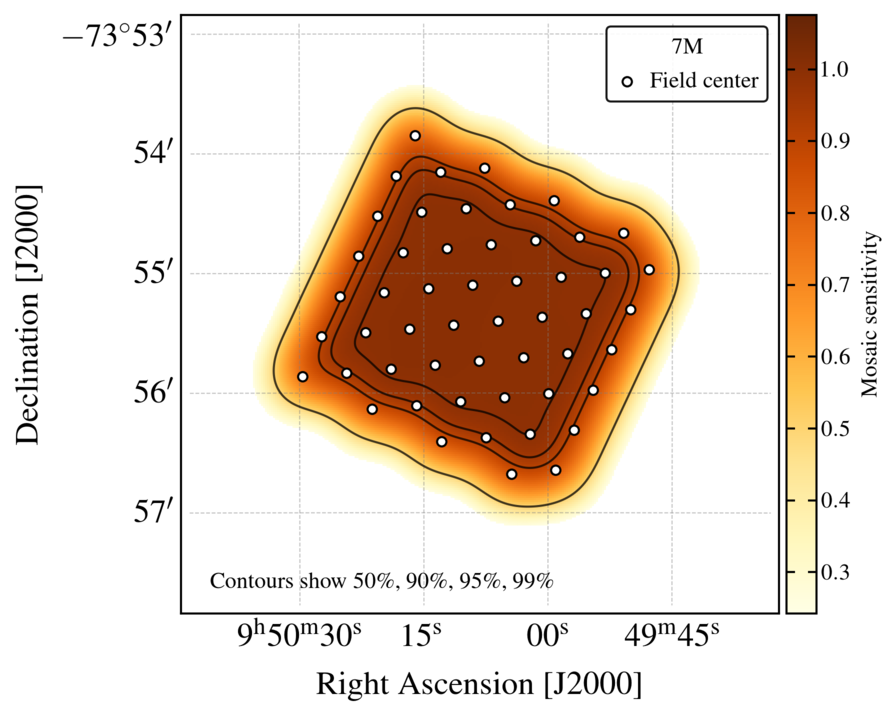}
\includegraphics[width=0.475\textwidth]{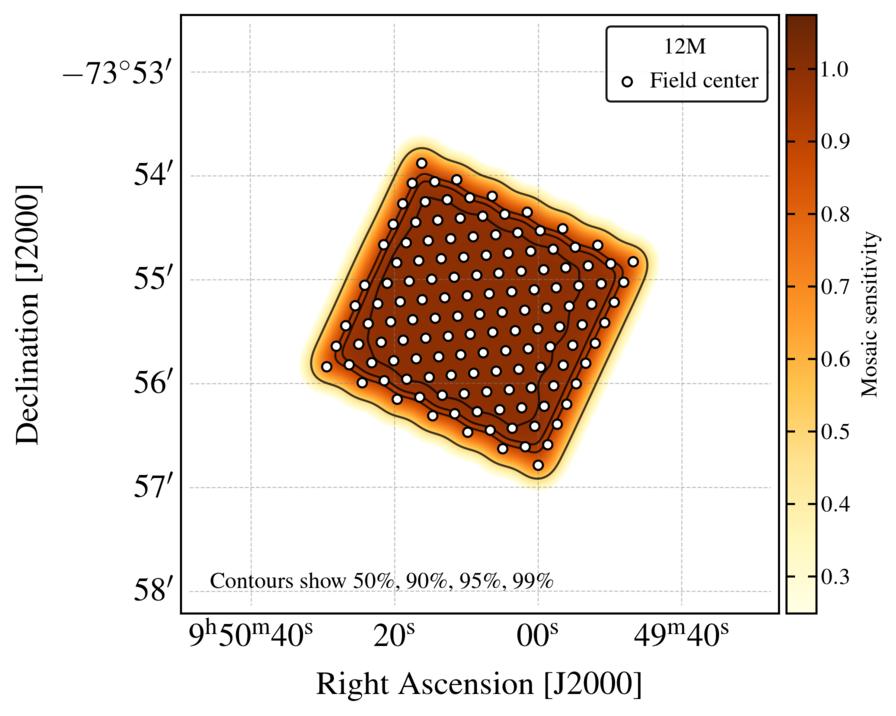}
\end{center}
\caption{\textbf{Example of PHANGS--ALMA mosaic sensitivity.} Examples of mosaic sensitivity, meaning combined primary beam response of all observations, for the \mbox{7-m} (\textit{left}) and \mbox{12-m}+\mbox{7-m} (\textit{right}) observations of one channel in one PHANGS-ALMA target, NGC~3059. The image shows the primary beam response, which is proportional to the inverse of the noise in the final image, i.e., a response of $0.5$ translates to $2\times$ higher noise. The black-and-white markers show the locations of individual pointings and contours show the $50\%$, $90\%$, and $95\%$, and $99\%$ response. The figure illustrates the mosaic pointings and that the combined ALMA response is even across the field.
\label{fig:response}}
\end{figure*}

Figure~\ref{fig:response} illustrates the combined primary beam response, i.e., the mosaic sensitivity, of the \mbox{7-m} data and \mbox{12-m}+\mbox{7-m} data for one channel in one ALMA data cube. The figure shows that ALMA achieves an even sensitivity across the area of the galaxy and demonstrates the Nyquist-sampled pointing spacing used by ALMA. The outer $0.5$ response contour indicates the typical extent of our imaging, though in a few cases we use a more stringent primary beam cutoff. This even coverage, at the few percent level, is typical of our targets, with only a few cases showing more uneven coverage due to flagging or issues with the execution of observations.

\section{Data Processing}
\label{sec:processing}

\begin{figure*}[ht!]
\begin{center}
\includegraphics[width=0.8\textwidth]{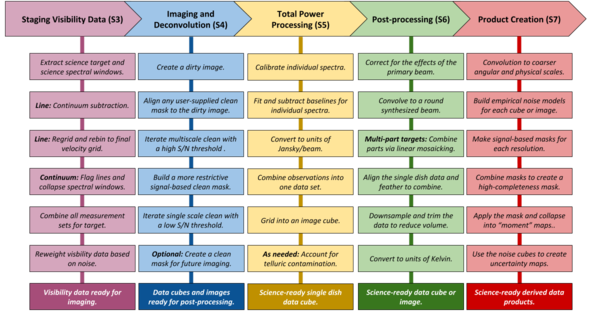}
\end{center}
\caption{\textbf{PHANGS--ALMA Processing Pipeline.} \label{fig:flowchart} Reproduced \textit{exactly} from \citet{LEROY21a}. The figure illustrates the processing steps used to process the $u{-}v$ and total power data delivered by ALMA into final science-ready data products. \citet{LEROY21a} present the processing pipeline and discuss the motivation, implementation, and limitations of each step in detail.}
\end{figure*}

\subsection{Summary}

\citet{LEROY21a} describe the reduction and data processing for PHANGS--ALMA. That paper explains our approaches to calibration, $u{-}v$ data processing, imaging, post-processing of the imaged data, short-spacing correction, data product creation, and quality assurance. It includes an overview, illustrations of each step, and detailed discussions of our motivation for each step. The accompanying \texttt{CASA}+\texttt{python} data pipeline is also publicly available\footnote{URL provided as soon as that paper is published and sent to the arxiv.}. \citet{HERRERA20} describe our calibration and imaging procedure for the total power data. Given these detailed presentations elsewhere, we only summarize the procedure here.

Figure~\ref{fig:flowchart}, reproduced exactly from \citet{LEROY21a}, summarizes the overall workflow of the PHANGS--ALMA processing pipeline. For the interferometric data, we download each observation and apply the ALMA observatory pipeline calibration. Based on inspection of the $u{-}v$ data, we found that, in general, additional by-hand flagging does not appreciably improve the quality of the final data products. We also performed several checks related to the overall calibration. We found that the internal stability of the total power appears excellent, with rms scatter of ${\sim}3\%$ from observation to observation, and that the fluxes of the secondary calibrators derived from the 7{-}m and 12{-}m array observations agree well with expectations from the ALMA calibrator database. As discussed above, we did identify a minor, $2{-}5\%$ net magnitude, issue regarding the overall calibration of a subset of the total power data. We expect this to be fixed in future releases of PHANGS--ALMA but at the moment this represents a ``known issue.''

Next, we staged the calibrated $u{-}v$ data for imaging. This step involves continuum subtraction, spectral regridding and rebinning, and an empirical reweighting of the data. For the current release of PHANGS--ALMA, we carried out these steps in \texttt{CASA} version \mbox{5.6.1}, though our tests showed little influence of the adopted \texttt{CASA} version on our final results.

We imaged the staged $u{-}v$ data using the \texttt{CASA} task \texttt{tclean}. We use a two stage deconvolution process. In the first stage, we used the ``multiscale clean'' \citep{CORNWELL08} deconvolution algorithm with no clean mask or a very extended clean mask, and we imposed a residual threshold of 4~times the rms noise in the cube. In other words, we carry out a multiscale clean and clean the image until the maximum residual, i.e., not deconvolved emission, has signal to noise of~4. Then, we construct a more restrictive clean mask based on the location of emission in the current version of the deconvolved image. Within this more restrictive clean mask, we continue deconvolving using a classic ``Hogbom clean'' \citep{HOGBOM74}. We run this second stage of deconvolution, stopping when the residual, i.e., not deconvolved emission, within the clean mask has a peak $\leq 1\sigma$ or the amount of deconvolved emission changes by a negligible amount over successive calls to clean. In practice, the signal-to-noise criterion is essentially always reached.

After imaging, we correct the image cube for the response of the primary beam. The image products often have elliptical synthesized beams at arbitrary position angles. We convolve these so that the final images have a round, Gaussian synthesized beam shape. For galaxies observed in multiple separate parts by several different Science Goals, we image the data separately and then combine the different parts via a noise-weighted linear mosaicking operation at this stage. Before doing this, we smooth all of the parts of a galaxy to share a common synthesized beam. At this stage, each galaxy has a single deconvolved, primary beam-corrected interferometric image cube with a round synthesized beam.

Next, we combine the total power and interferometric data using the \text{CASA} implementation of the ``feather'' algorithm \citep[e.g., see][]{COTTON17}. Feathering combines the total power and interferometer data in the Fourier domain, ensuring that in the final image the low-spatial-frequency information, that is crucial to determine the overall flux of the image, is set by the total power data. Before this stage, we calibrate, baseline subtract, and image the total power data following the procedures described in \citet{HERRERA20} and \citet{LEROY21a}. For a subset of galaxies, this processing included additional steps to deal with the terrestrial ozone contamination described above.

At this stage, we have our final data cubes, which we convert from units of Jy~beam$^{-1}$ to units of Kelvin. We use these to construct a series of data products for use in scientific analysis. First, we convolve the cubes to a series of fixed physical and angular resolutions, which allows the whole sample to be subject to rigorous comparative analysis. The exact suite of resolutions has varied over time. For the current release we create cubes with FWHM synthesize beams of $2\arcsec$, $7.5\arcsec$, $11\arcsec$, and $15\arcsec$, as well as $60$~pc, $90$~pc, $120$~pc, $150$~pc, $500$~pc, $750$~pc, and $1$~kpc. When convolving to fixed physical resolution, we take into account the distance to the galaxy as compiled by \citet{ANAND21}, discussed in Section~\ref{sec:distance}, and reported in Table~\ref{tab:sample_orient}.

For each cube at each resolution, we construct a three dimensional noise model, bootstrapped from regions of the cube that contain little or no signal. Then, we combine the noise models and the data cubes to construct a series of masks that identify the likely location of real CO emission inside each cube. We construct two sets of masks, high confidence ``strict'' masks and high completeness ``broad'' masks. The strict masks are constructed for each cube at each resolution. Following \citet{ROSOLOWSKY06} and \citet{LEROY21a}, these strict masks initially consider only locations in the cube where emission exceeds signal-to-noise of $4$ over two consecutive velocity channels. Then, we extend these masks to include any adjacent regions where the emission exceeds a signal-to-noise of $2$ over two consecutive velocity channels. We refer to these strict masks as ``high confidence'' because this construction procedure renders them very likely to include mostly real emission. The broad masks are created from the union of strict masks across all resolutions. These masks aim for ``high completeness'' in the sense that they tend to include almost all emission in the cube, as determined by comparison to the integrated flux from the total power data.

Finally, we apply the masks to the cubes and noise models and ``collapse'' the cubes to produce a series of two-dimensional maps and associated uncertainty maps. We discuss the specific products more below, but briefly these ``moment maps'' include images of integrated intensity, peak intensity along the line of sight, intensity-weighted mean velocity, and the spectral width of the CO line along the line of sight. We produce an associated uncertainty map for each moment map.

For more details on each step of this process see \citet{LEROY21a} and the associated, public PHANGS--ALMA post-processing pipeline.

\subsection{Validation, known issues, and limitations}

We engaged in a mixture of automated and manual quality assurance of the data and validation of the PHANGS--ALMA pipeline. A broad cross-section of the team participated in several rounds of manual inspection, looking at the cubes, associated imaging products, and derived data products. For each version of the data after early 2019, we ran automated regression tests to quantify changes from version to version and to ensure that we understood the reason for all changes. Finally, to validate our pipeline, we carried out end-to-end tests using the ALMA simulator in \texttt{CASA}. These tests allowed us to verify that the pipeline output represents a close match to the known input to the simulator.

These tests, and other checks, showed that our data processing has yielded stable results over time, that the results pass manual inspections by experts, and that the output from the pipeline replicates known input to good precision. However, there are still several known limitations of our approach and issues related to data processing that we note here.

\medskip

\begin{enumerate}
\item \textbf{Both the spectral and angular resolution of the data could be improved at the cost of signal-to-noise.} The convolution to a round beam and matching beams between parts of multi-part mosaics both somewhat inflate the synthesized beam size. We also use a $u{-}v$ weighting scheme with intermediate ``robustness parameter.'' All of these choices improve the surface brightness sensitivity, but they degrade the resolution by a modest amount. Our cubes also use a ${\sim}2.5$~km~s$^{-1}$ channel width, which is coarser than the native spectral resolution. By altering these choices, future processing of the data could improve the spatial and spectral resolution at the cost of signal-to-noise. This may be of particular interest for bright regions like the centers of galaxies.

\item \textbf{Deconvolution of \mbox{7-m}-only imaging depends sensitively on the signal-to-noise and structure of the data.} Both our real data and end-to-end tests based on simulations show that the image reconstruction performs poorly using low or moderate signal-to-noise data from the \mbox{7-m} array alone. The effect is explored in detail in \citet{LEROY21a}, and it is largely alleviated by short-spacing correction. However, the \mbox{7-m}-only data for many galaxies appear unreliable. We also view the combined \mbox{7-m} and total power data as less reliable than the full combined \mbox{12-m}, \mbox{7-m}, and total power data.

\item \textbf{There is a small known bias in the total power flux calibration.} As mentioned above and discussed in \citet{LEROY21a}, there was a small lag incorporating upgrades to the surface accuracy of the total power antennas into the observatory-provided calibration. As a result, a subset of our total power data sets apply an observatory-provided calibration known to be high by ${\sim}7\%$. The net effect is that a subset of cubes has overall fluxes biased high by $2{-}5\%$. This will be fixed in future releases.

\item \textbf{The data processing introduces some mild dependence of noise on frequency.} As discussed in \citet{LEROY21a}, the regridding and rebinning parts of the $u{-}v$ staging procedure introduce a mild spectral variation into the noise of the final data cubes. The overall magnitude is ${\sim}10\%$ across the full bandwidth of the cubes. The effect seems unavoidable in the current version of \texttt{CASA}, but we expect this to be addressed in the future.

\end{enumerate}

Several other issues arose during processing that we have largely addressed, including the telluric ozone contamination of the total power data. To our knowledge the issues above represent the main outstanding issues and limitations of the data.

\section{Data properties and key products}
\label{sec:products}

\begin{deluxetable}{lc}[ht!]
\tabletypesize{\footnotesize}
\tablecaption{PHANGS--ALMA \cotwo\ Cube Summary \label{tab:cubeprops}}
\tablewidth{0pt}
\tablehead{
\colhead{Quantity} &
\colhead{Value}
}
\startdata
\hline
\multicolumn{2}{l}{Targets} \\
... \mbox{12-m}+\mbox{7-m}+TP & 81 \\
... \mbox{12-m}+\mbox{7-m}+TP, $\theta < 165$~pc\tablenotemark{a} & 78 \\
... \mbox{7-m}+TP or \mbox{7-m} & 9 \\
... \mbox{7-m}+TP or \mbox{7-m}, $\theta < 165$~pc\tablenotemark{a} & 7 
\\
\hline
Properties of \mbox{12-m}+\mbox{7-m}+TP data... & median$^{+1\sigma}_{-1\sigma}$ (min--max) \\
\hline
\multicolumn{2}{l}{Resolution...} \\
... angular [\arcsec] & 
$1.3^{+0.4}_{-0.2}$ ($0.6{-}2.1$) \\
... physical [pc] & 
$98^{+31}_{-35}$ ($26{-}183$) 
\\
\multicolumn{2}{l}{$1\sigma$ noise in the cube ...} \\
... mJy~beam$^{-1}$ native res. & 
$6.2_{-1.8}^{+2.4}$ ($4.4{-}8.6$)
\\
... mK native res. & 
$85_{-41}^{+36}$ ($4.4{-}8.6$)
\\
... mK at $150$~pc res. &
$50_{-20}^{+26}$ ($16{-}114$)
\\
... mass\tablenotemark{b,c} native res. [$10^3$~M$_\odot$] & 
$21^{+13}_{-10}$ ($1.5{-}76$)
\\
... mass\tablenotemark{b,c} at $150$~pc res. [$10^3$~M$_\odot$] & 
$31^{+16}_{-12}$ ($10{-}69$)
\\
\multicolumn{2}{l}{Implied $1\sigma$ surface brightness noise\tablenotemark{c} ...} \\
... K~km~s$^{-1}$ native res. & 
$0.30^{+0.13}_{-0.14}$ ($0.12{-}1.3$)
\\
... K~km~s$^{-1}$ at 150~pc res. & 
$0.18^{+0.09}_{-0.07}$ ($0.06{-}0.41$)
\\
... $\Sigma_{\rm mol}$ \tablenotemark{b,c} native res. [M$_\odot$~pc$^{-2}$] &
$2.0^{+0.8}_{-1.0}$ ($0.8{-}8.7$)
\\
... mass\tablenotemark{b,c} at 150~pc res. [M$_\odot$~pc$^{-2}$] &
$1.2^{+0.6}_{-0.5}$ ($0.4{-}2.7$)
\\
\multicolumn{2}{l}{Completeness\tablenotemark{d} ...} \\
... at native resolution & $61^{+20}_{-24}\%$ ($0{-}94$\%)
\\
... at 150~pc resolution & 
$67_{-26}^{+18}\%$ ($0{-}100$\%)
\\
\hline
Area mapped & \\
\hline
Physical area & \\
... total for survey [kpc$^2$] & 10,650 \\
... mean \mbox{12-m}+\mbox{7-m}+TP [kpc$^2$] & 124 \\
... mean \mbox{7-m}+TP [kpc$^2$] & 71
\\
Angular area & \\
... total for survey [arcmin$^2$] & 1,050 \\
... mean \mbox{12-m}+\mbox{7-m}+TP [arcmin$^2$] & 7.0 \\
... mean \mbox{7-m}+TP [arcmin$^2$] & 54 \\
\enddata
\tablenotetext{a}{We adopt a $10\%$ tolerance when convolving to a fixed physical resolution, so that all maps with $\theta < 165$~pc have ``150 pc'' data products.}
\tablenotetext{b}{When calculating mass, we adopt a Milky Way $\alpha_{\rm CO}^{1-0}$ and $R_{21} = 0.65$. See text and Equations~\ref{eq:sd} and~\ref{eq:mass}.
}
\tablenotetext{c}{When calculating line-integrated quantities, including mass and surface brightness, we adopt a full line width of $\Delta v = 5$~km~s$^{-1}$. This implies an improvement of $\sqrt{2}$ in the surface brightness sensitivity.}
\tablenotetext{d}{``Completeness'' is defined here as the ratio of the sum of flux in the strictly masked moment~0 map to the direct sum of the whole cube or the broad map. Completeness is not random but correlates with the overall brightness of the galaxy. See \S\ref{sec:sensitivity}.}
\end{deluxetable}

\subsection{Properties of the final cubes}

\startlongtable
 \begin{deluxetable*}{lccccccccccl} 
 \tablecaption{PHANGS-ALMA Cube Properties  \label{tab:sample_obs}} 
 \tablewidth{0pt} 
 \tabletypesize{\footnotesize} 
 \tablehead{ 
 \colhead{Galaxy} & 
 \colhead{Arrays} & 
 \multicolumn{2}{c}{Resolution} & 
 \multicolumn{2}{c}{Area Mapped} & 
 \multicolumn{3}{c}{Noise} & 
 \multicolumn{2}{c}{Completeness} & 
 \colhead{Notes}  
 \\ 
\cmidrule(lr){3-4}\cmidrule(lr){5-6}\cmidrule(lr){7-9}\cmidrule(lr){10-11} \colhead{} & 
 \colhead{} & 
 \colhead{Angular} & 
 \colhead{Physical} & 
 \colhead{Angular} & 
 \colhead{Physical} & 
 \colhead{Natve} & 
 \colhead{150~pc} & 
 \colhead{Natve} & 
 \colhead{Native} & 
 \colhead{150~pc} & 
 \colhead{}   
 \\ 
 \colhead{} & 
 \colhead{} & 
 \colhead{($''$)} & 
 \colhead{(pc)} & 
 \colhead{(arcmin$^2$)} & 
 \colhead{(kpc$^2$)} & 
 \colhead{(mK)} & 
 \colhead{(mK)} & 
 \colhead{$\left(\frac{\rm mJy}{{\rm beam}^{-1}}\right)$} & 
 \colhead{(\%)} &  
 \colhead{(\%)} &  
 \colhead{}  
 }  
\startdata 
NGC 0247  & 7m+TP & $8.51$  & $153.1$  & $ 80.7$  & $ 93.9$  & $  23$  & $  23$  & $73.9$  & $ 7.2$  & $ 7.2$  & \\ 
NGC 0253  & 7m+TP & $8.37$  & $150.2$  & $ 98.7$  & $114.4$  & $  36$  & $  36$  & $108.2$  & $85.8$  & $85.8$  & \\ 
NGC 0300  & 7m+TP & $8.18$  & $ 82.8$  & $ 57.0$  & $ 21.1$  & $  35$  & $  13$  & $100.4$  & $36.7$  & $51.3$  & \\ 
NGC 0628  & 12m+7m+TP & $1.12$  & $ 53.5$  & $ 14.7$  & $120.7$  & $ 115$  & $  41$  & $ 6.3$  & $44.7$  & $64.5$  & \\ 
NGC 0685  & 12m+7m+TP & $1.69$  & $163.0$  & $  4.8$  & $162.4$  & $  40$  & $  40$  & $ 4.8$  & $36.5$  & $36.5$  & \\ 
NGC 1068  & 7m+TP & $8.69$  & $588.6$  & $  6.5$  & $106.6$  & $  18$  & \nodata  & $58.9$  & $96.6$  & \nodata  & \\ 
NGC 1097  & 12m+7m+TP & $1.70$  & $111.7$  & $ 13.4$  & $208.7$  & $  52$  & $  37$  & $ 6.4$  & $79.9$  & $82.7$  & \\ 
NGC 1087  & 12m+7m+TP & $1.60$  & $123.1$  & $  6.8$  & $145.4$  & $  66$  & $  52$  & $ 7.3$  & $64.7$  & $67.5$  & \\ 
NGC 1313  & 7m & $7.93$  & $166.0$  & $ 55.7$  & $ 88.0$  & $  28$  & $  28$  & $75.0$  & $11.6$  & $11.6$  & \\ 
NGC 1300  & 12m+7m+TP & $1.23$  & $113.1$  & $ 13.0$  & $396.8$  & $ 102$  & $  76$  & $ 6.6$  & $48.5$  & $52.4$  & \\ 
NGC 1317  & 12m+7m+TP & $1.59$  & $147.1$  & $  1.7$  & $ 52.1$  & $  46$  & $  46$  & $ 4.9$  & $89.2$  & $90.0$  & \\ 
IC 1954   & 12m+7m+TP & $1.56$  & $ 97.1$  & $  4.0$  & $ 55.6$  & $  45$  & $  27$  & $ 4.7$  & $64.1$  & $70.7$  & \\ 
NGC 1365  & 12m+7m+TP & $1.38$  & $130.8$  & $  7.3$  & $237.2$  & $ 112$  & $  95$  & $ 9.1$  & $75.3$  & $76.4$  & \\ 
NGC 1385  & 12m+7m+TP & $1.27$  & $105.9$  & $  6.8$  & $171.3$  & $  82$  & $  55$  & $ 5.7$  & $60.9$  & $65.5$  & \\ 
NGC 1433  & 12m+7m+TP & $1.10$  & $ 99.1$  & $ 11.8$  & $346.0$  & $ 119$  & $  76$  & $ 6.2$  & $51.9$  & $57.3$  & \\ 
NGC 1511  & 12m+7m+TP & $1.45$  & $107.1$  & $  3.6$  & $ 71.2$  & $  67$  & $  45$  & $ 6.0$  & $75.6$  & $81.0$  & \\ 
NGC 1512  & 12m+7m+TP & $1.03$  & $ 94.5$  & $  7.9$  & $235.5$  & $ 120$  & $  70$  & $ 5.5$  & $46.6$  & $53.1$  & \\ 
NGC 1546  & 12m+7m+TP & $1.28$  & $109.6$  & $  2.8$  & $ 73.9$  & $  62$  & $  41$  & $ 4.4$  & $88.4$  & $91.6$  & \\ 
NGC 1559  & 12m+7m+TP & $1.25$  & $117.5$  & $  6.8$  & $217.6$  & $  88$  & $  66$  & $ 5.9$  & $60.8$  & $64.5$  & \\ 
NGC 1566  & 12m+7m+TP & $1.25$  & $107.6$  & $ 11.9$  & $314.3$  & $  90$  & $  64$  & $ 6.1$  & $81.1$  & $92.3$  & \\ 
NGC 1637  & 12m+7m+TP & $1.39$  & $ 78.9$  & $  6.2$  & $ 71.7$  & $  36$  & $  17$  & $ 3.0$  & $77.0$  & $83.5$  & \\ 
NGC 1672  & 12m+7m+TP & $1.93$  & $181.7$  & $  8.1$  & $258.2$  & $  85$  & \nodata  & $13.7$  & $72.8$  & \nodata  & \\ 
NGC 1809  & 12m+7m+TP & $1.41$  & $136.0$  & $  2.4$  & $ 79.4$  & $  96$  & $  86$  & $ 8.2$  & $32.3$\tablenotemark{F}  & $33.2$\tablenotemark{F}  & \\ 
NGC 1792  & 12m+7m+TP & $1.92$  & $150.9$  & $  9.1$  & $203.0$  & $  46$  & $  46$  & $ 7.3$  & $88.5$  & $88.5$  & \\ 
NGC 2090  & 12m+7m+TP & $1.30$  & $ 73.8$  & $  3.0$  & $ 35.6$  & $  99$  & $  52$  & $ 7.2$  & $60.4$  & $66.7$  & \\ 
NGC 2283  & 12m+7m+TP & $1.31$  & $ 87.0$  & $  5.4$  & $ 85.4$  & $  84$  & $  51$  & $ 6.2$  & $34.5$  & $36.5$  & \\ 
NGC 2566  & 12m+7m+TP & $1.28$  & $145.3$  & $  6.8$  & $317.1$  & $  88$  & $  88$  & $ 6.2$  & $74.7$  & $74.7$  & \\ 
NGC 2775  & 12m+7m+TP & $1.09$  & $122.6$  & $  5.5$  & $248.0$  & $ 134$  & $ 108$  & $ 6.9$  & $27.7$  & $33.0$  & \\ 
NGC 2835  & 12m+7m+TP & $0.84$  & $ 50.0$  & $  7.9$  & $ 99.4$  & $ 239$  & $  74$  & $ 7.4$  & $19.2$  & $25.8$  & \\ 
NGC 2903  & 12m+7m+TP & $1.45$  & $ 70.5$  & $ 15.5$  & $131.4$  & $  71$  & $  37$  & $ 6.6$  & $78.1$  & $83.7$  & \\ 
NGC 2997  & 12m+7m+TP & $1.77$  & $120.5$  & $ 19.0$  & $318.1$  & $  42$  & $  32$  & $ 5.7$  & $78.3$  & $80.9$  & \\ 
NGC 3059  & 12m+7m+TP & $1.22$  & $119.9$  & $  7.5$  & $259.3$  & $  95$  & $  75$  & $ 6.1$  & $58.6$  & $61.9$  & \\ 
NGC 3137  & 12m+7m+TP & $1.51$  & $120.0$  & $  3.4$  & $ 76.2$  & $  62$  & $  45$  & $ 6.1$  & $54.0$  & $59.4$  & \\ 
NGC 3239  & 12m+7m+TP & $1.28$  & $ 67.5$  & $  2.9$  & $ 28.5$  & $ 132$  & $  73$  & $ 9.4$  & $ 1.7$\tablenotemark{F}  & $ 0.9$\tablenotemark{F}  & \\ 
NGC 3351  & 12m+7m+TP & $1.46$  & $ 70.7$  & $  7.8$  & $ 65.2$  & $ 108$  & $  46$  & $10.0$  & $60.2$  & $71.7$  & \\ 
NGC 3489  & 12m+7m+TP & $0.75$  & $ 42.9$  & $  0.7$  & $  8.1$  & $ 100$  & $  27$  & $ 2.4$  & $50.1$\tablenotemark{F}  & $58.4$\tablenotemark{F}  & \\ 
NGC 3511  & 12m+7m+TP & $1.80$  & $121.5$  & $  6.4$  & $105.0$  & $  44$  & $  34$  & $ 6.1$  & $75.4$  & $77.5$  & \\ 
NGC 3507  & 12m+7m+TP & $1.36$  & $155.2$  & $  5.1$  & $239.9$  & $  79$  & $  79$  & $ 6.3$  & $40.0$  & $40.0$  & \\ 
NGC 3521  & 12m+7m+TP & $1.33$  & $ 85.5$  & $ 12.6$  & $186.4$  & $  61$  & $  35$  & $ 4.7$  & $85.2$  & $89.2$  & \\ 
NGC 3596  & 12m+7m+TP & $1.22$  & $ 66.9$  & $  4.6$  & $ 49.3$  & $ 128$  & $  67$  & $ 8.3$  & $47.4$  & $59.9$  & \\ 
NGC 3599  & 12m+7m+TP & $0.66$  & $ 63.2$  & $  0.7$  & $ 23.0$  & $ 120$  & $  50$  & $ 2.2$  & $59.4$\tablenotemark{F}  & $54.8$\tablenotemark{F}  & \\ 
NGC 3621  & 12m+7m+TP & $1.82$  & $ 62.4$  & $ 12.2$  & $ 51.4$  & $  39$  & $  16$  & $ 5.6$  & $83.4$  & $87.8$  & \\ 
NGC 3626  & 12m+7m+TP & $1.17$  & $114.1$  & $  1.9$  & $ 66.2$  & $ 156$  & $ 114$  & $ 9.3$  & $33.9$\tablenotemark{F}  & $65.4$\tablenotemark{F}  & \\ 
NGC 3627  & 12m+7m+TP & $1.63$  & $ 89.2$  & $ 11.8$  & $128.4$  & $  80$  & $  45$  & $ 9.1$  & $80.8$  & $85.3$  & \\ 
NGC 4207  & 12m+7m+TP & $1.22$  & $ 93.2$  & $  1.6$  & $ 34.7$  & $ 132$  & $  85$  & $ 8.5$  & $68.9$  & $73.7$  & \\ 
NGC 4254  & 12m+7m+TP & $1.78$  & $113.1$  & $ 12.7$  & $184.1$  & $  64$  & $  45$  & $ 8.7$  & $79.4$  & $82.4$  & \\ 
NGC 4293  & 12m+7m+TP & $1.16$  & $ 89.0$  & $  3.5$  & $ 73.5$  & $ 130$  & $  80$  & $ 7.6$  & $66.7$  & $70.5$  & \\ 
NGC 4298  & 12m+7m+TP & $1.59$  & $114.7$  & $  4.2$  & $ 79.8$  & $  38$  & $  25$  & $ 4.1$  & $82.2$  & $85.7$  & \\ 
NGC 4303  & 12m+7m+TP & $1.81$  & $149.3$  & $  7.8$  & $190.5$  & $  82$  & $  82$  & $11.5$  & $73.1$  & $73.1$  & \\ 
NGC 4321  & 12m+7m+TP & $1.67$  & $122.9$  & $ 13.7$  & $269.0$  & $  89$  & $  69$  & $10.7$  & $69.2$  & $72.2$  & \\ 
NGC 4424  & 12m+7m+TP & $1.14$  & $ 89.4$  & $  1.4$  & $ 30.6$  & $ 151$  & $  89$  & $ 8.5$  & $61.4$  & $69.4$  & \\ 
NGC 4457  & 12m+7m+TP & $1.11$  & $ 81.3$  & $  2.2$  & $ 42.7$  & $ 101$  & $  52$  & $ 5.4$  & $73.6$  & $80.7$  & \\ 
NGC 4459  & 12m+7m+TP & $0.64$  & $ 48.9$  & $  0.7$  & $ 14.5$  & $ 113$  & $  34$  & $ 2.0$  & $84.7$\tablenotemark{F}  & $79.8$\tablenotemark{F}  & \\ 
NGC 4476  & 12m+7m+TP & $0.87$  & $ 73.9$  & $  0.7$  & $ 18.2$  & $  70$  & $  32$  & $ 2.3$  & $85.1$\tablenotemark{F}  & $97.9$\tablenotemark{F}  & \\ 
NGC 4477  & 12m+7m+TP & $0.62$  & $ 47.2$  & $  0.7$  & $ 14.6$  & $ 125$  & $  41$  & $ 2.0$  & $52.2$\tablenotemark{F}  & $46.3$\tablenotemark{F}  & \\ 
NGC 4496A & 12m+7m+TP & $1.25$  & $ 90.3$  & $  4.2$  & $ 77.8$  & $ 104$  & $  68$  & $ 7.0$  & $29.6$  & $29.2$  & \\ 
NGC 4535  & 12m+7m+TP & $1.56$  & $119.1$  & $  7.9$  & $167.2$  & $  83$  & $  60$  & $ 8.6$  & $60.6$  & $66.9$  & \\ 
NGC 4536  & 12m+7m+TP & $1.48$  & $116.3$  & $ 10.2$  & $228.5$  & $  44$  & $  33$  & $ 4.2$  & $80.8$  & $84.7$  & \\ 
NGC 4540  & 12m+7m+TP & $1.37$  & $104.8$  & $  1.9$  & $ 39.4$  & $ 111$  & $  76$  & $ 9.0$  & $45.5$  & $52.6$  & \\ 
NGC 4548  & 12m+7m+TP & $1.69$  & $132.7$  & $  6.4$  & $143.1$  & $  49$  & $  41$  & $ 6.1$  & $40.6$  & $42.1$  & \\ 
NGC 4569  & 12m+7m+TP & $1.69$  & $128.9$  & $  6.3$  & $132.8$  & $  47$  & $  38$  & $ 5.9$  & $81.6$  & $83.2$  & \\ 
NGC 4571  & 12m+7m+TP & $1.18$  & $ 85.0$  & $  3.9$  & $ 73.7$  & $ 118$  & $  70$  & $ 7.1$  & $18.8$  & $31.7$  & \\ 
NGC 4579  & 12m+7m+TP & $1.79$  & $182.7$  & $  8.2$  & $304.5$  & $  46$  & \nodata  & $ 6.4$  & $60.3$  & \nodata  & \\ 
NGC 4596  & 12m+7m+TP & $0.65$  & $ 49.6$  & $  0.7$  & $ 14.3$  & $  99$  & $  30$  & $ 1.8$  & $93.8$\tablenotemark{F}  & $179.9$\tablenotemark{F}  & \\ 
NGC 4654  & 12m+7m+TP & $1.72$  & $182.8$  & $  8.1$  & $330.1$  & $  52$  & \nodata  & $ 6.6$  & $82.7$  & \nodata  & \\ 
NGC 4689  & 12m+7m+TP & $1.18$  & $ 86.0$  & $  5.8$  & $111.0$  & $ 111$  & $  63$  & $ 6.7$  & $51.5$  & $62.7$  & \\ 
NGC 4694  & 12m+7m+TP & $1.17$  & $ 89.3$  & $  1.1$  & $ 23.9$  & $ 143$  & $  82$  & $ 8.4$  & $24.8$\tablenotemark{F}  & $28.9$\tablenotemark{F}  & \\ 
NGC 4731  & 12m+7m+TP & $1.53$  & $ 98.4$  & $  3.9$  & $ 58.7$  & $  39$  & $  23$  & $ 3.9$  & $42.6$\tablenotemark{F}  & $47.2$\tablenotemark{F}  & \\ 
NGC 4781  & 12m+7m+TP & $1.31$  & $ 71.7$  & $  5.4$  & $ 58.0$  & $  46$  & $  20$  & $ 3.4$  & $65.4$  & $75.6$  & \\ 
NGC 4826  & 12m+7m+TP & $1.26$  & $ 26.9$  & $  6.4$  & $ 10.6$  & $  77$  & $  18$  & $ 5.3$  & $87.6$  & $93.6$  & \\ 
NGC 4941  & 12m+7m+TP & $1.59$  & $115.3$  & $  3.5$  & $ 66.2$  & $  40$  & $  28$  & $ 4.4$  & $57.0$  & $63.3$  & \\ 
NGC 4951  & 12m+7m+TP & $1.25$  & $ 91.2$  & $  2.8$  & $ 53.2$  & $  91$  & $  54$  & $ 6.2$  & $47.7$  & $52.4$  & \\ 
NGC 4945  & 7m+TP & $7.90$  & $132.9$  & $ 61.5$  & $ 62.7$  & $  40$  & $  32$  & $107.3$  & $85.1$  & $87.1$  & \\ 
NGC 5042  & 12m+7m+TP & $1.33$  & $107.8$  & $  3.5$  & $ 83.4$  & $  79$  & $  54$  & $ 6.0$  & $28.1$  & $31.2$  & \\ 
NGC 5068  & 12m+7m+TP & $1.04$  & $ 26.2$  & $ 15.3$  & $ 35.0$  & $ 185$  & $  38$  & $ 8.7$  & $33.4$  & $46.5$  & \\ 
NGC 5134  & 12m+7m+TP & $1.23$  & $119.0$  & $  2.8$  & $ 94.7$  & $  87$  & $  66$  & $ 5.7$  & $41.3$  & $45.6$  & \\ 
NGC 5128  & 7m+TP & $8.13$  & $145.5$  & $ 60.0$  & $ 69.2$  & $  35$  & $  35$  & $100.9$  & $76.8$  & $76.8$  & \\ 
NGC 5236  & 12m+7m+TP & $2.14$  & $ 50.7$  & $ 56.7$  & $114.7$  & $  43$  & $  16$  & $ 8.6$  & $87.8$  & $92.1$  & \\ 
NGC 5248  & 12m+7m+TP & $1.29$  & $ 93.1$  & $  8.5$  & $159.4$  & $  92$  & $  62$  & $ 6.6$  & $71.8$  & $76.2$  & \\ 
CIRCINUS & 7m+TP & $7.32$  & $149.0$  & $ 23.4$  & $ 34.9$  & $  55$  & $  55$  & $128.1$  & $77.1$  & $77.7$  & \\ 
NGC 5530  & 12m+7m+TP & $1.13$  & $ 66.9$  & $  5.7$  & $ 72.0$  & $ 114$  & $  56$  & $ 6.2$  & $47.5$  & $57.3$  & \\ 
NGC 5643  & 12m+7m+TP & $1.30$  & $ 79.9$  & $ 10.6$  & $143.7$  & $  77$  & $  43$  & $ 5.6$  & $69.7$  & $76.0$  & \\ 
NGC 6300  & 12m+7m+TP & $1.08$  & $ 60.4$  & $  7.8$  & $ 88.2$  & $ 116$  & $  54$  & $ 5.8$  & $64.7$  & $73.0$  & \\ 
NGC 6744  & 12m+7m+TP & $1.13$  & $ 51.6$  & $ 15.8$  & $118.2$  & $ 187$  & $  69$  & $10.4$  & $35.8$  & $60.3$  & \\ 
IC 5273   & 12m+7m+TP & $1.76$  & $120.7$  & $  3.5$  & $ 60.2$  & $  33$  & $  25$  & $ 4.4$  & $53.0$  & $56.1$  & \\ 
NGC 7456  & 12m+7m+TP & $1.68$  & $127.8$  & $  2.2$  & $ 45.7$  & $  40$  & $  32$  & $ 4.9$  & $31.6$\tablenotemark{F}  & $33.9$\tablenotemark{F}  & \\ 
NGC 7496  & 12m+7m+TP & $1.68$  & $152.0$  & $  4.4$  & $129.7$  & $  36$  & $  36$  & $ 4.3$  & $73.3$  & $73.3$  & \\ 
IC 5332   & 12m+7m+TP & $0.74$  & $ 32.1$  & $  7.8$  & $ 53.4$  & $ 366$  & $ 102$  & $ 8.6$  & $ 2.5$\tablenotemark{F}  & $ 7.1$\tablenotemark{F}  & \\ 
NGC 7743  & 12m+7m+TP & $0.64$  & $ 63.2$  & $  0.7$  & $ 23.6$  & $  76$  & $  26$  & $ 1.3$  & $58.1$\tablenotemark{F}  & $60.8$\tablenotemark{F}  & \\ 
NGC 7793  & 7m+TP & $7.99$  & $140.2$  & $ 41.2$  & $ 45.7$  & $  34$  & $  34$  & $94.9$  & $32.3$  & $32.3$  & \\ 
\enddata 
\tablenotetext{F}{Faint galaxy, completeness is unreliable. See text.}
 \tablecomments{ 
Properties of the PHANGS--ALMA data cubes. Columns --- name of the galaxy; best available array combination;
physical and angular resolution of the native resolution cube;
noise in surface brightness (mK) units at the native resolution and $150$~pc resolution, if available;
noise in mJy~beam$^{-1}$ units in the native-resoltion cube;
completeness, defined as the ratio of flux inside the strict mask to the sum of the cube or the broad mask. See text.
at native resolution and 150~pc resolution.
 } 
\end{deluxetable*}

\begin{figure}[ht!]
\begin{center}
\includegraphics[width=0.475\textwidth]{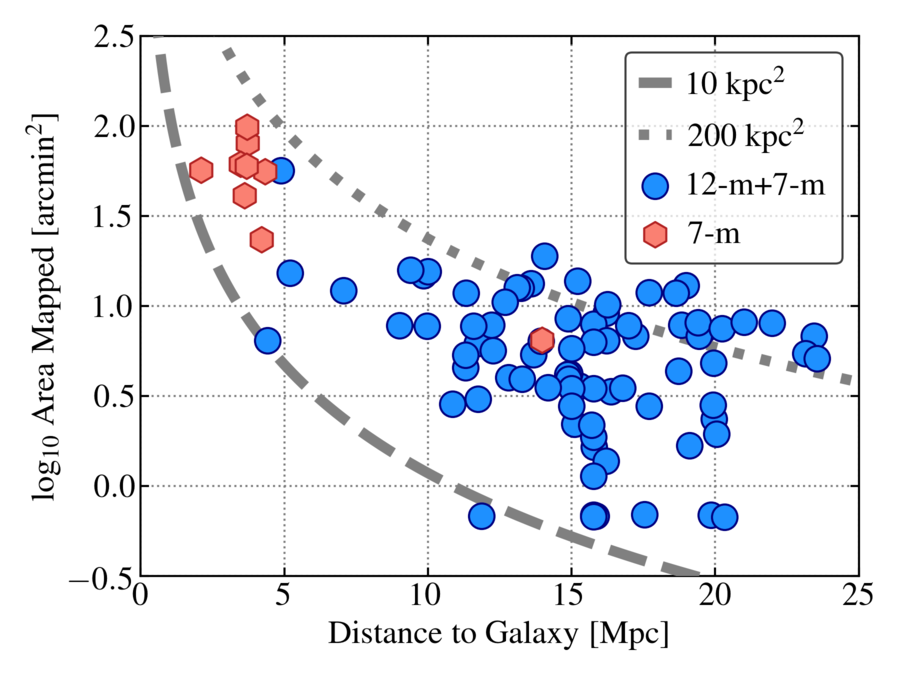}
\end{center}
\caption{\textbf{Area Mapped.} Angular area mapped for each PHANGS--ALMA target as a function of distance to the target. Galaxies mapped using the \mbox{12-m}+\mbox{7-m} array appear in blue, and those mapped with the \mbox{7-m}-only appear in red. The dotted and dashed lines show physical areas of 200~kpc$^2$ and 10~kpc$^2$.
\label{fig:areamapped}}
\end{figure}

Table \ref{tab:cubeprops} summarizes the properties of our final cubes, which we report in detail in Table~\ref{tab:sample_obs}. For each target, we note the best available array combination, angular and physical resolution, noise, and completeness. When translating angular to physical beam size and coverage area, we use the distances reported in Table~\ref{tab:sample_orient} and compiled by \citet{ANAND21}. The reported noise represents the median value in the three dimensional noise cube, and should be characteristic for the cube. Here we define completeness, $f_{\rm flux}$, as 

\begin{equation}
\label{eq:comp}
f_{\rm flux} = \frac{\textstyle \sum \text{strict mom0}}{\textstyle \text{max} \bigl(\sum \text{broad~mom0},~\sum \text{cube} \bigr)}~,
\end{equation}

\noindent that is, the ratio of flux inside the ``strict,'' high confidence, high signal-to-noise based mask to the total flux in the cube. The ``max'' in the denominator uses the larger of the sum inside the ``broad'' mask or the cube; we adopt this to help account for some instability in the direct sum of cubes with faint CO emission (see below). Thus, the completeness, $f_{\rm flux}$, represents the fraction of the total flux recovered at good signal-to-noise.

Figure~\ref{fig:areamapped} and Table~\ref{tab:cubeprops} summarize the area mapped. In total, PHANGS--ALMA surveyed ${\sim}1050$~arcmin$^2$, slightly larger than the angular size of the Moon. At the distances to our targets, this translates to ${\sim}10{,}650$~kpc$^2$. On average, the individual maps cover $124$~kpc$^2$ for the \mbox{12-m}+\mbox{7-m}+TP and $71$~kpc$^2$ for \mbox{7-m}+TP data. This corresponds to linear map sizes of ${\sim}11$~kpc or typical coverage out to $r_{\rm gal} \approx 5{-}6$~kpc.  Because the \mbox{7-m}-only targets tend to lie much closer, they cover on average 54~arcmin$^2$, compared to $7$~arcmin$^2$, on average for the \mbox{12-m}+\mbox{7-m} data.

\subsubsection{Resolution}

\begin{figure*}[ht!]
\begin{center}
\includegraphics[width=0.475\textwidth]{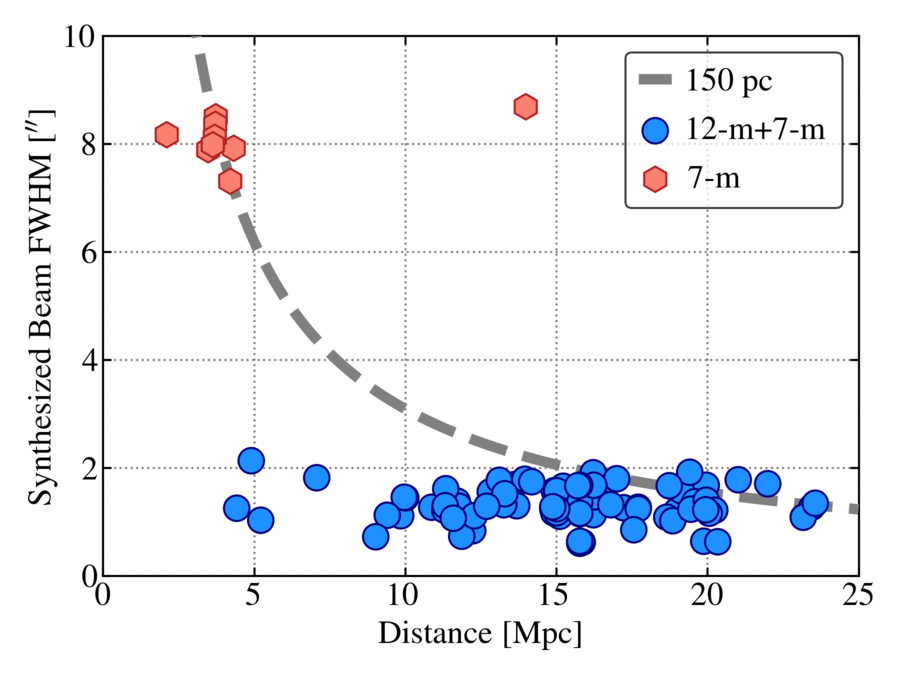}
\includegraphics[width=0.475\textwidth]{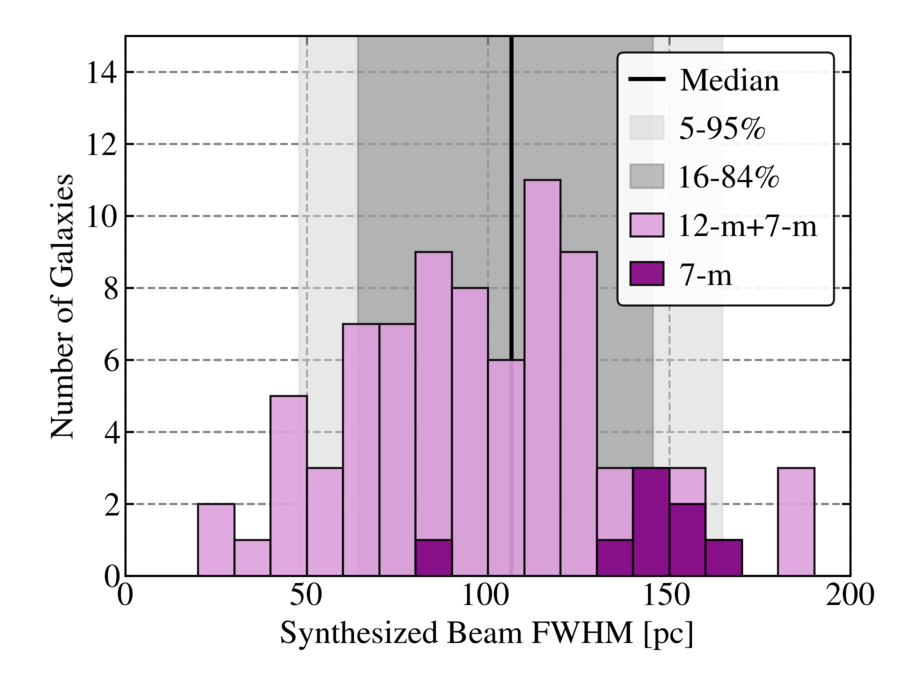}
\includegraphics[width=0.475\textwidth]{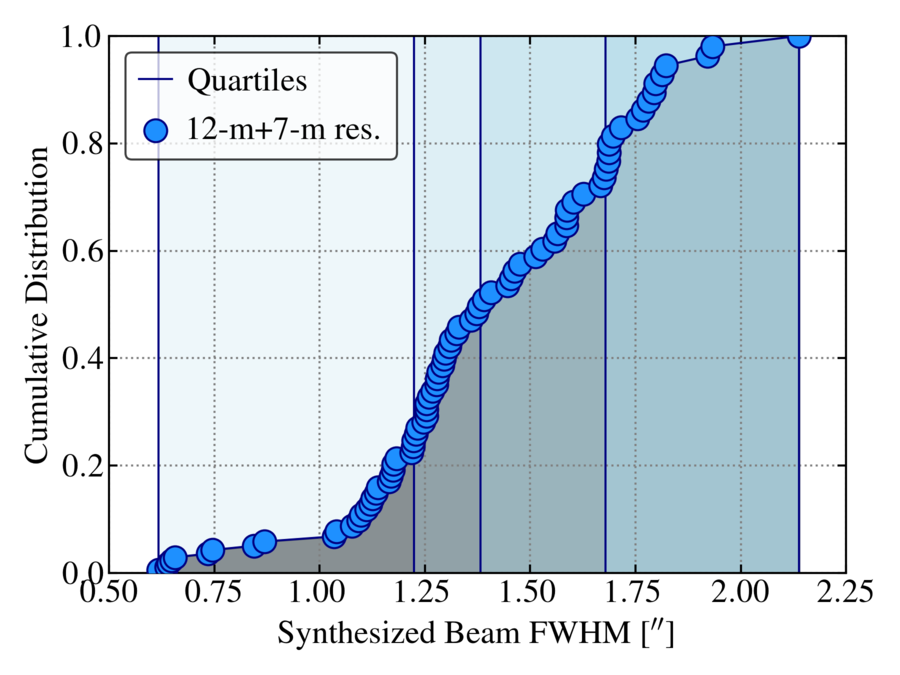}
\includegraphics[width=0.475\textwidth]{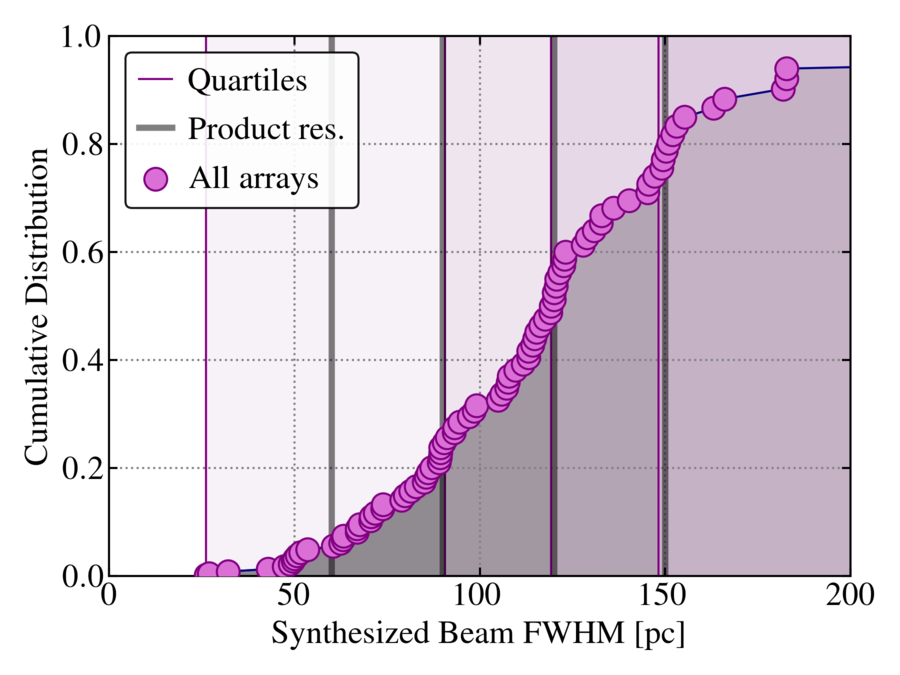}
\end{center}
\caption{\textbf{Angular and Physical Resolution of PHANGS--ALMA.} The \textit{top left} panel shows the FWHM angular resolution of each final data cube as a function of the distance to that galaxy. We mark \mbox{7-m} data and \mbox{12-m}+\mbox{7-m} data with separate symbols, and a dashed line shows $150$~pc. Most of the \mbox{7-m}+TP data target the nearest $d < 5$~Mpc galaxies, and so achieve high, ${\sim}150$~pc resolution. The single large distance \mbox{7-m}-only data set is NGC~1068. The \textit{top right} panel shows the distribution of FWHM physical synthesized beam size. The \textit{bottom row} shows the cumulative distributions of angular (\textit{bottom left}) and physical (\textit{bottom right}) beam sizes. Lines and shaded regions show the quartiles in each distribution. In the bottom right panel, solid vertical lines show the physical resolutions used to construct data products.
\label{fig:resolution}}
\end{figure*}

Figure~\ref{fig:resolution} and Table~\ref{tab:cubeprops} summarize the resolutions reported in Table~\ref{tab:sample_obs}. For our $81$ \mbox{12-m}+\mbox{7-m}+TP data sets, the median angular resolution is $1.3\arcsec$ with $70\%$ of the data between $1.1{-}1.7\arcsec$. At the distance to our targets, this translates to a median physical resolution of $98$~pc with $70\%$ of the data between $60$~pc and $120$~pc. The top left panel in Figure~\ref{fig:resolution} shows that most of the \mbox{7-m}-only data sets target very nearby galaxies, $d < 5$~Mpc. As a result, they achieve similar physical resolutions to the \mbox{12-m}+\mbox{7-m}+TP observations of our more distant sample members, with a typical resolution of ${\sim}150$~pc for these very nearby galaxies. The lone large-distance \mbox{7-m}-only data set is NGC~1068, which has not yet been mapped in \cotwo\ using the \mbox{12-m} array \citep[but has been mapped in \cothree\ using ALMA by][]{GARCIABURILLO14}.

We construct data products at a series of fixed physical resolutions, in addition to the native angular resolution and several coarser angular resolutions. For the \mbox{12-m}+\mbox{7-m} data we adopt FWHM beam size $\theta = 60$, $90$, $120$, and $150$~pc at the distance to the galaxy. Reflecting uncertainties in the distance determination, we allow a $\pm 10\%$ tolerance when labeling a galaxy as having that resolution; that is, we label a map with $\theta = 91$~pc or $110$~pc as a ``100~pc'' resolution map. The bottom right panel of Figure~\ref{fig:resolution} and Table~\ref{tab:cubeprops} show the motivation for these choices: $78$ of our $81$ \mbox{12-m}+\mbox{7-m} galaxies and $7$ of $9$ \mbox{7-m}-only galaxies have $\theta < 165$~pc, allowing us to construct a ``$150$~pc resolution'' map. Meanwhile, roughly half of our targets have physical resolution better than $\theta = 120$~pc and $25\%$ have physical resolution better than $\theta = 90$~pc.

\subsubsection{Sensitivity}
\label{sec:sensitivity}

\begin{figure*}[ht!]
\begin{center}
\includegraphics[width=0.475\textwidth]{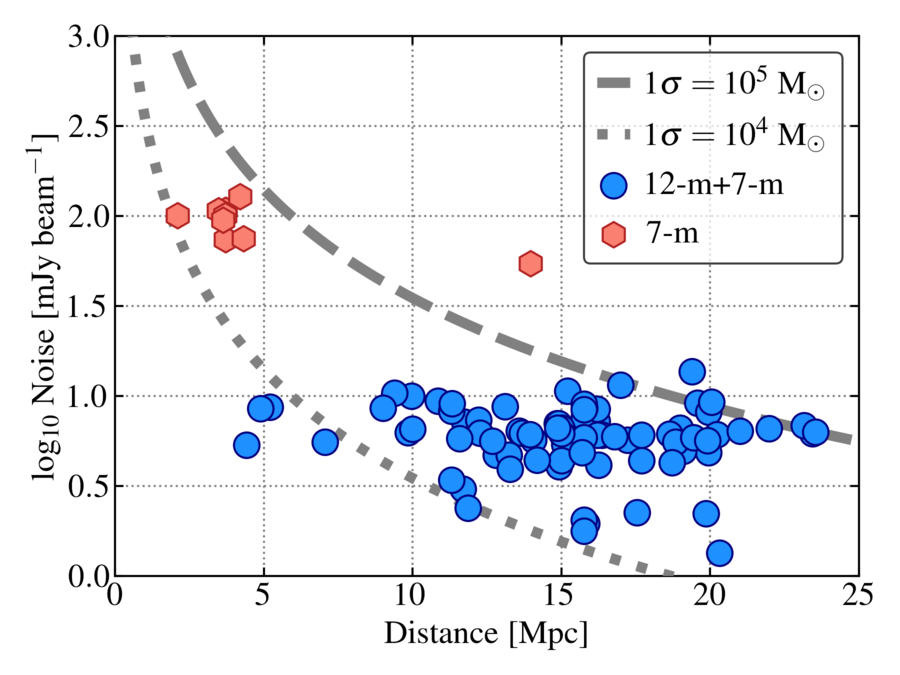}
\includegraphics[width=0.475\textwidth]{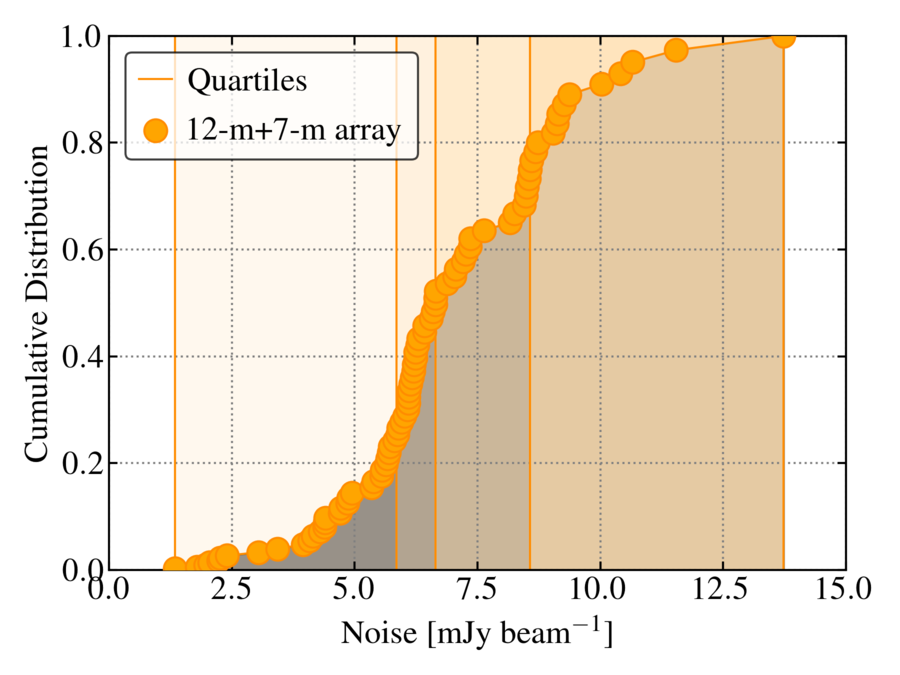}
\includegraphics[width=0.475\textwidth]{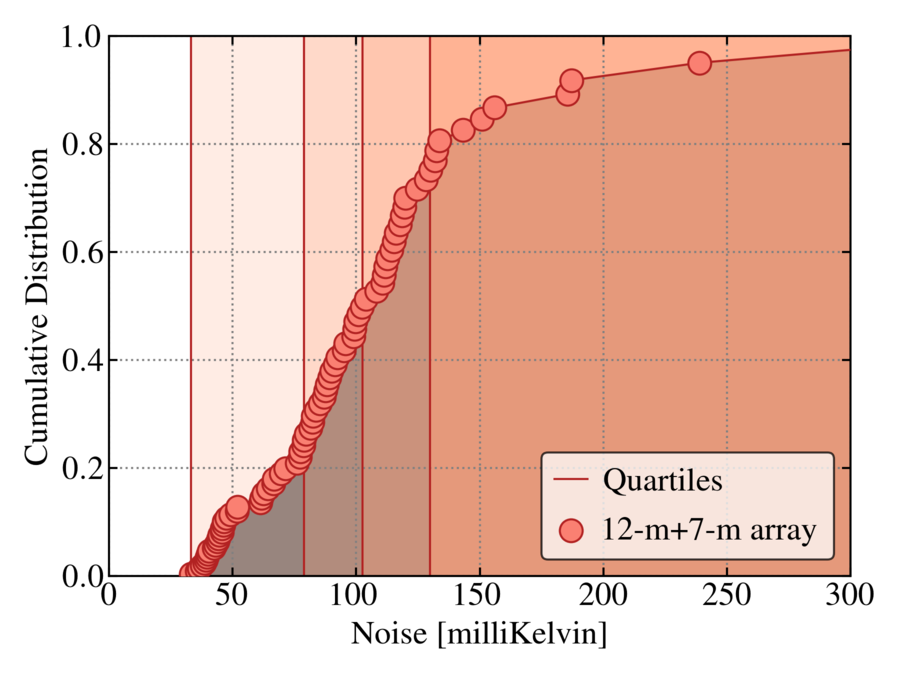}
\includegraphics[width=0.475\textwidth]{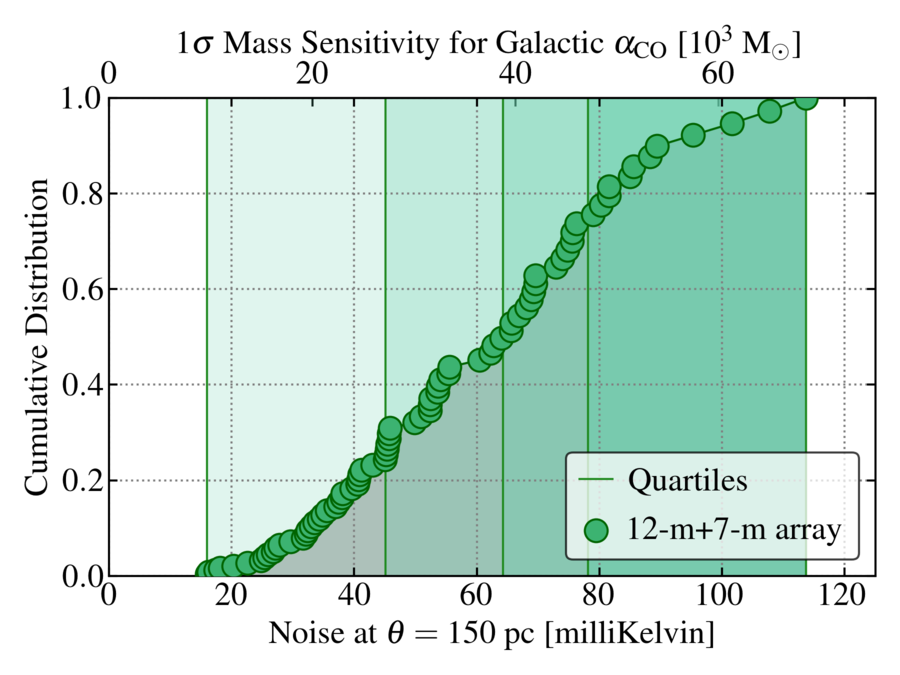}
\end{center}
\caption{\textbf{Sensitivity of PHANGS--ALMA.} The \textit{top left} panel shows the median rms noise in each $\delta v \approx 2.5$~km~s$^{-1}$ channel of each final data cube (at their native resolution) as a function of the distance to that galaxy. We mark \mbox{7-m} data and \mbox{12-m}+\mbox{7-m} data with separate symbols. Dashed lines show the cases where the noise equates to a $1\sigma$ mass sensitivity of $10^4$~M$_\odot$ or $10^5$~M$_\odot$, assuming $\alpha_{\rm CO} = 4.35$~\acounits , $R_{21} = 0.65$, and $\Delta v = 5$~km~s$^{-1}$. Most of our cubes have $1\sigma$ sensitivities between $10^4$~M$_\odot$ and $10^5$~M$_\odot$ per beam. The other panels show the cumulative distributions of noise in the \mbox{12-m}+\mbox{7-m} data in (\textit{top right}) Jy~beam$^{-1}$ units, (\textit{bottom left}) surface brightness units, i.e., milliKelvin, at the native resolution, and (\textit{bottom right}) surface brightness units at a common 150~pc resolution. The upper axis in the last panel also shows the corresponding $1\sigma$ mass sensitivity assuming the same $\alpha_{\rm CO} = 4.35$~\acounits , $R_{21} = 0.65$, and $\Delta v = 5$~km~s$^{-1}$.
\label{fig:sensitivity}}
\end{figure*}

\begin{figure}[ht!]
\begin{center}
\includegraphics[width=0.475\textwidth]{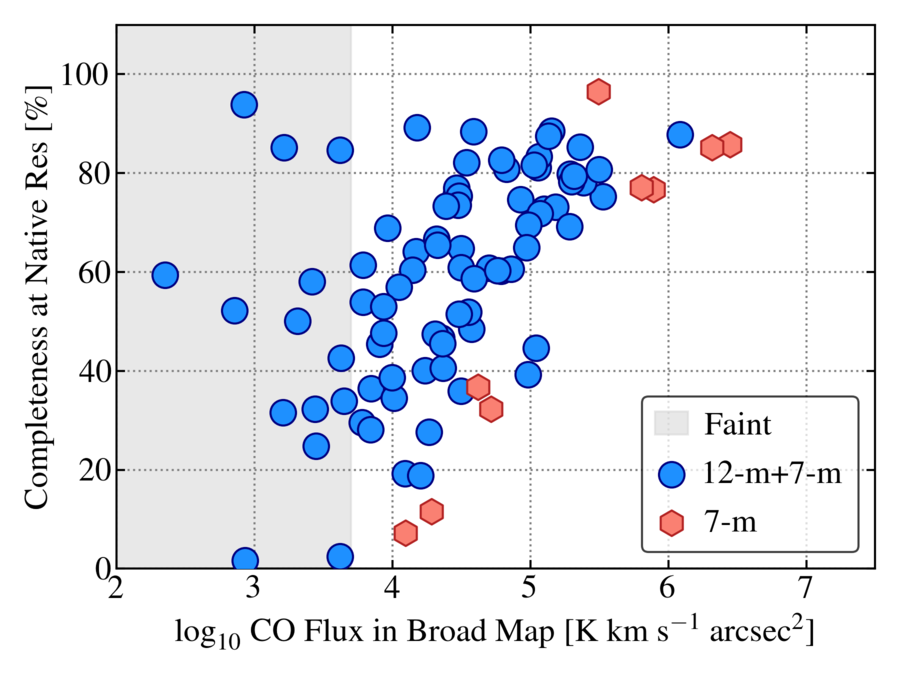}
\end{center}
\caption{\textbf{Completeness of the PHANGS--ALMA high signal-to-noise maps at the native resolution.} We defined completeness as the ratio of flux within the high confidence ``strict'' mask to the integrated flux in the cube or the high completeness ``broad'' mask. Thus the number reflects the fraction of the total flux captured at moderate- to high signal-to-noise in each cube. The completeness correlates with total flux, to first order tracking the overall brightness of CO in the target.
\label{fig:completeness}}
\end{figure}

Figure~\ref{fig:sensitivity} and Table~\ref{tab:cubeprops} summarize the sensitivity of our data. The top left panel in Figure~\ref{fig:sensitivity} shows the point source sensitivity in Jy~beam$^{-1}$ units as a function of the distance to the source. In that panel, we plot two lines showing the cases where the noise and distance equate to $1\sigma$ mass sensitivity of $10^4$ and $10^5$~M$_\odot$ for a line width of $\Delta v = 5$~km~s$^{-1}$. As above, we translate from line flux to molecular mass via:
\begin{equation}
\label{eq:mass}
M_{\rm mol}~[ {\rm M}_\odot ] = 2.63 \times 10^3~\frac{\alpha_{\rm CO}^{1-0}}{\alpha_{\rm CO}^{\rm MW}}~\frac{1}{R_{21}}~D_{\rm Mpc}^2~F_{\rm CO}^{2-1}
\end{equation}
\noindent where $F_{\rm CO}^{2-1}$ is the line-integrated \cotwo\ flux in Jy~km~s$^{-1}$ of a source at $D_{\rm Mpc}$ distance. Again $\alpha_{\rm CO}^{1-0}$ is the CO-to-H$_2$ conversion factor and $R_{21}$ is the \cotwo-to-\coone\ line ratio in Kelvins, so that $R_{21}=1$ refers to the thermal case. For reference, if the flux is instead in units of K~km~s$^{-1}$~arcsec$^2$, the mass can be calculated via

\begin{equation}
\label{eq:masskelvin}
M_{\rm mol}~[ {\rm M}_\odot ] = 1.02 \times 10^2~\frac{\alpha_{\rm CO}^{1-0}}{\alpha_{\rm CO}^{\rm MW}}~\frac{1}{R_{21}}~D_{\rm Mpc}^2~F_{\rm CO}^{\prime~2-1}
\end{equation}

\noindent where now $F_{\rm CO}^{\prime~2-1}$ is the line-integrated \cotwo\ flux in units of K~km~s$^{-1}$~arcsec$^2$.

Note that when converting our sensitivity into mass units, e.g., in Figure~\ref{fig:sensitivity} and Table~\ref{tab:cubeprops}, we assume a fiducial line width of $5$~km~s$^{-1}$ for the line but we measure the initial noise in $2.54$~km~s$^{-1}$ channels.  Noise in Jy~beam$^{-1}$ improves by $\sqrt{2}$ due to averaging before calculating $F_{\rm CO} = \sigma \times \Delta v$ with $\sigma$ the noise in Jy~beam$^{-1}$ averaged across the line.

The table and figure show that our typical rms point source sensitivity is $6.2$~mJy~beam$^{-1}$ per $2.54$~km~s$^{-1}$ channel, about $\sqrt{2}$ times better than our minimum sensitivity of $7.5$~mJy~beam$^{-1}$ per $5$~km~s$^{-1}$ channel (see \S\ref{sec:obsstrat}). This is partially due to the convolutions during our post-processing, but mostly reflecting that our observations typically achieve ${\sim}24$~s per pointing, which translates to ${\sim}1$~min of effective integration time after accounting for overlap of mosaic fields. There is about $\pm 30\%$ rms scatter in the point source sensitivity across our sample.

Our achieved sensitivity translates to a median point mass sensitivity of $2 \times 10^4$~M$_\odot$ for a Galactic $\alpha_{\rm CO}$ and $R_{21} = 0.65$. As discussed above, this means that, on average, we will detect individual GMCs, which are often defined to have mass $>10^5$~M$_\odot$, at signal-to-noise of $5$ or higher.

Our median surface brightness sensitivity is $85$~mK per $2.54$~km~s$^{-1}$ channel, again better than the nominal target of $170$~mK per $5$~km~s$^{-1}$ channel. This surface brightness sensitivity depends sharply on the achieved angular resolution, with rms noise $\propto \theta^{-2}$ even for fixed point source sensitivity. This number thus partially also reflects that our achieved median resolution is ${\sim}1.3\arcsec$ while the target was $1\arcsec$. Because the surface brightness sensitivity depends on both point source sensitivity and achieved angular resolution, our data show a $\pm 50\%$ scatter about the median value.

Again assuming a full line width $\Delta v = 5$~km~s$^{-1}$, this $85$~mK per $2.54$~km~s$^{-1}$ channel translates to $1\sigma$ surface brightness sensitivity of ${\sim}0.30$~K~km~s$^{-1}$. Following Equation~\ref{eq:mwsd}, this translates to median $1\sigma$ mass surface density sensitivity of $\Sigma_{\rm mol} = 2.0$~M$_\odot$~pc$^{-2}$. As discussed above, typical resolved GMCs have surface densities of a few times $10{-}100$~M$_\odot$~pc$^{-2}$. Thus our maps should recover CO emission everywhere where GMCs fill an appreciable fraction of the beam. 

As discussed above, $150$~pc is a common physical scale for most of the data. When we convolve to this scale, the typical surface brightness sensitivity is $1\sigma = 50$~mK, again with $\pm 50\%$ scatter. The corresponding $1\sigma$ point mass sensitivity is $3.1 \times 10^4$~M$_\odot$, and the $1\sigma$ mass surface density sensitivity is $\Sigma_{\rm mol} = 1.2$~M$_\odot$~pc$^{-2}$.

Together these sensitivity numbers can all be roughly summarized as: the PHANGS--ALMA CO maps are sensitive to individual massive giant molecular clouds at the $3{-}10\sigma$ level. This has been key to the already published studies of \citet{SUN18,SUN20B} and \citet{ROSOLOWSKY21}, which have used PHANGS--ALMA to study populations of individual clouds across our sample.

\subsubsection{Completeness}

Table~\ref{tab:cubeprops} and Figure~\ref{fig:completeness} illustrate the completeness, $c$, of the data. Following Equation~\ref{eq:comp}, we define ``completeness'' as the ratio of the flux detected within the ``strict'' mask to the total flux in the cube, estimated from either a direct sum or using the ``broad'' mask. This completeness thus measures how much of the flux in the cube is detected at moderate signal to noise at a given sensitivity and resolution. Figure~\ref{fig:completeness} shows completeness for the native resolution and Tables~\ref{tab:cubeprops} and \ref{tab:sample_obs} also report values at $150$~pc resolution.

At our native resolution, the \mbox{12-m}+\mbox{7-m}+TP data have median completeness of ${\sim}61\%$, but with a wide range. This improves to ${\sim}67\%$ at 150~pc resolution, but still shows a wide range. Figure~\ref{fig:completeness} shows that the completeness variations track the overall flux of the target, with brighter galaxies yielding higher recovery. A few of the galaxies are so faint that a simple sum of the cube no longer yields a flux consistent with integrating over the broad, high completeness map. This typically occurs when $F_{\rm CO} \lesssim 10^{3.5}$~K~km~s$^{-1}$~arcsec$^2$. We mark this regime in Figure~\ref{fig:completeness} and note galaxies that have flux lower than this in Table~\ref{tab:sample_obs}. To avoid instability when calculating $c$, we use the maximum flux of either the direct sum of the cube or the broad map in Equation~\ref{eq:comp}.

Excluding these ``faint'' galaxies increases the median completeness in the sample by ${\sim}5\%$, but Figure~\ref{fig:completeness} shows that there is still a trend with brightness at higher flux: we detect ${\sim}60{-}90\%$ of the flux at native resolution in high-flux galaxies. Meanwhile for galaxies with intermediate integrated fluxes, the completeness varies across almost the whole range of possible values, but is lower on average.

Overall these levels of flux recovery are very good, and in line with previous measurements on Local Group and related data by \citet{LEROY16} and previous measurements in PHANGS--ALMA by \citet{SUN18,SUN20B}. The completeness increases dramatically as we degrade the resolution of the data, with almost all area in PHANGS--ALMA detected at ${\sim}15\arcsec$ or ${\sim}1$~kpc resolution (recall that this is expected because PHANGS--ALMA targets regions with detectable mid-IR dust emission).

Note that \citet{SUN20B} give equations for the selection function associated with our strict masking technique. These can be useful for modeling the properties of the emission not included in the strict mask.

\subsection{Description of high level data products}

\citet{LEROY21a} describe the creation of the PHANGS--ALMA data products, and we refer the reader to that paper for details. Here we provide a high level summary of the products that make up the PHANGS--ALMA data release:

\begin{enumerate}
\item \textbf{Cubes at multiple resolutions.} We convolve the native resolution, post-processed data cubes to a series of common physical and angular resolutions. In addition to the native resolution, we convolve all data to angular resolutions of $\theta = 2\arcsec$, $7.5\arcsec$, $11\arcsec$, and $15\arcsec$ (FWHM). Almost all of the \mbox{12-m}+\mbox{7-m} data can reach each of these angular resolutions, while the \mbox{7-m}-only data sets can typically only reach $11\arcsec$ and $15\arcsec$.

We also convolve each cube to a FWHM resolution corresponding to $60$, $90$, $120$, $150$, $500$, $750$, and $1000$~pc using the distances compiled by \citet{ANAND21} and reported in Table~\ref{tab:sample_orient}. As discussed above the first four values roughly correspond to the quartiles of the distribution of physical resolutions for our native resolution \mbox{12-m}+\mbox{7-m} data set.

During these convolutions we impose a $10\%$ tolerance. That is, if the FWHM beam of a cube is already within $\pm 10\%$ of the target resolution, we do not convolve but label the cube as belonging to that fixed-resolution subset. We do not alter the cube metadata in these cases, so that, e.g., the FITS header \texttt{BMAJ} and \texttt{BMIN} keywords still reflect the true values.

Because most scientific applications involve measuring surface brightness or integrating over an aperture larger than the beam, we present the PHANGS--ALMA cubes in units of Kelvin. Conversions to Jy~beam$^{-1}$ are provided as part of the header metadata.

\item \textbf{Noise models for each cube.} For each data cube at each resolution, we construct a three dimensional noise model. This model gives our best estimate of the rms noise at each position--position--velocity pixel. We construct the noise model by treating the spectral and spatial variations of the noise as separable and then using iterative outlier rejection to isolate the parts of the cube most likely to be signal free. When discussing the noise statistics above, we quote the median value across the whole noise model, but see \citet{LEROY21a} for illustrations of the spatial and spectral variations of the noise.

\item \textbf{``Strict'' and ``broad'' masks for each cube.} Combining the cubes and the noise models, we construct two sets of three dimensional masks that identify the location of likely signal. Our ``strict'' masks are \textit{high confidence}, meaning that they include only pixels with at least moderate signal to noise at the relevant resolution. The strict masks should be used in calculations that require secure detections. These are the basis of our completeness calculations above.

We also construct ``broad'' masks that include all regions of the cube likely to contain signal. These are \textit{high completeness} masks, in the sense that we expect them to include essentially all emission from the galaxy. To build the broad masks, we use the union of the strict masks at all resolutions, from the native resolution up to $15\arcsec$ and $1$~kpc. The coarse resolution masks play a crucial role here because we detect emission along almost every line of sight in those cubes. Including the high resolution masks in the union tends to matter most for high line width, compact features, which mostly occur in galaxy centers. Within the cubes, the broad masks typically look like an extended, ``puffed~up,'' version of the galaxy rotation curve in position-position velocity space.

Because the broad mask represents a union of all strict masks, there is only a single broad mask, which we apply across all resolutions. We reproject it onto different three dimensional astrometric grids as necessary.

\item \textbf{Integrated intensity maps.} We apply the masks to the cubes and integrate along the spectral axis to produced two-dimensional maps of line-integrated intensity (``moment~0''), in units of K~km~s$^{-1}$. We create versions of these maps for each resolution and for both the strict and the broad masks. The ``strict'' integrated intensity maps should be used as our best map of securely detected CO emission from each galaxy. The ``broad'' integrated intensity maps represent our most complete map of CO emission from each galaxy. The broad maps will include some negative pixels, reflecting that they include regions dominated by noise but likely to contain faint emission. Integrating across these broad maps should yield the total flux in the cube to good approximation.

The integrated intensity, $I_{\rm CO}^{2-1}$ in units of K~km~s$^{-1}$, can be translated into an estimate of molecular gas mass surface density, $\Sigma_{\rm mol}$ in units of M$_\odot$~pc$^{-2}$, following Equation~\ref{eq:sd} or Equation~\ref{eq:mwsd} for standard assumptions. At the resolution of PHANGS--ALMA it remains ambiguous whether or not to apply an inclination correction, $\cos i$, when estimating $\Sigma_{\rm mol}$. There is a long history of assuming individual molecular clouds to have an isotropic, spherically symmetric geometry \citep[e.g.,][]{SOLOMON87,ROSOLOWSKY06}. However, our resolution is of the same order as the ${\sim}100$~pc disk scale height \citep{HEYER15,SUN20}. For now, the user will need to make their own best judgment on this topic.

For each integrated intensity map, we also propagate the noise model into maps of the associated statistical uncertainty in $I_{\rm CO}^{2-1}$ \citep[see details in][]{LEROY21a}.

\item \textbf{Peak intensity maps.} We also calculate maps of the peak intensity, $I_{\rm CO}^{2-1}$ along each line of sight. These maps of peak intensity, also called ``peak temperature'' or ``moment~8,'' are useful to highlight the detailed structure of emission in the cube, though they have a less straightforward physical interpretation than the line-integrated intensity maps. 

To calculate the peak intensity, we first calculate the full velocity range covered by the relevant mask for each sightline. Then for each line of sight, we find the maximum intensity at any velocity in this range and record this as the peak intensity.

We also create a version of the peak intensity map after convolving the spectral axis of the cube with a 12.5~km~s$^{-1}$ boxcar kernel. Because $12.5$~km~s$^{-1}$ corresponds to a typical full line width for a bright molecular cloud, this spectral smoothing essentially represents running a matched filter across the spectral axis.

We present the peak intensity maps in units of Kelvin.

\item \textbf{Velocity fields.} We calculate the intensity-weighted mean velocity of emission, frequently referred to as ``moment~1,'' along each line of sight. 

We present two versions of the intensity-weighted mean velocity field. The first uses only emission within the strict mask. These strictly masked velocity maps present the mean velocity associated with all securely detected emission.

The second set of velocity maps consider more emission and so cover a larger area. Because noise spikes can contribute spurious outliers to the velocity field, we apply several prior expectations when constructing this velocity field. These ``moment~1 with prior'' maps begin with the broad mask. We calculate the intensity-weighted mean velocity using the broad masks. This typically includes many spurious velocities associated with low signal-to-noise or even signal-free regions. Therefore, we blank all values from the map that either (1) have signal to noise less than two in the integrated intensity map or (2) deviate from a ``prior expectation'' velocity field by more than some tolerance, $\pm 30$~km~s$^{-1}$ by default. For this release, we use the velocity field calculated using the strict mask at lower resolution.

We present all velocity fields in units of km~s$^{-1}$ and also calculate maps of associated statistical uncertainty by propagating errors from the noise model cubes.

\item \textbf{Line width maps.} We calculate two sets of line width maps, both using only the strict masks. First, we calculate the rms velocity dispersion about the intensity-weighted mean velocity, often referred to as ``moment~2.'' This calculation diverges in the presence of noise, so we only present the calculation for the strictly masked case.

The second moment can be sensitive to clipping and unstable in the presence of noise. We also calculate and report the ``effective width'' following the definition of \citet{HEYER01} and discussed in \citet{LEROY16} and \citet{SUN18,SUN20B}. In this measurement, the line width $\sigma_\mathrm{ew} \equiv \int I_{\rm CO}\, \mathrm{d}v / (\sqrt{2 \pi} I_\mathrm{CO}^\mathrm{peak})$. In other words, the line width is defined as the ratio of the line integrated intensity to the peak intensity. In the limit of a well-resolved, high signal-to-noise Gaussian line profile the two line width measurements match. The effective width is more robust to the presence of noise but also sensitive to the requirement to resolve the line. The ratio between the two line width measurements can also indicate deviations from a single Gaussian spectral shape.

For both line width estimates, we present the result in units of km~s$^{-1}$ and also calculate and deliver associated maps of statistical uncertainty. For this delivery, we do not correct the measurements for the line spread function. This can be an important effect for narrow lines and should be addressed by the user \citep[e.g., following][]{LEROY16,KOCH18B,SUN20,ROSOLOWSKY21}.
\end{enumerate}

This already represents a rich set of data products that characterize the CO spectral line position-by-position across each target. We also provide several supporting data products, including maps indicating the coverage of the ALMA observations. There are also several clear next steps. In the inner regions of galaxies and in regions of colliding flows \citep[e.g., bar ends, see][]{BESLIC21}, multi-component decomposition like that presented in \citet{HENSHAW20} represents a better treatment of the spectrum than single component methods. In the future, we also intend to use the rotation curves fit in \citet{LANG20} to refine the ``priors'' use to construct the velocity fields \citep[e.g., see][]{COLOMBO14B}.
Our next round of products will also likely include ``shuffled'' cubes appropriate for spectral stacking \citep[e.g., see][]{SCHRUBA11}.

\section{Atlas}
\label{sec:atlas}

\begin{figure*}[ht!]
\begin{center}
\includegraphics[width=0.75\textwidth]{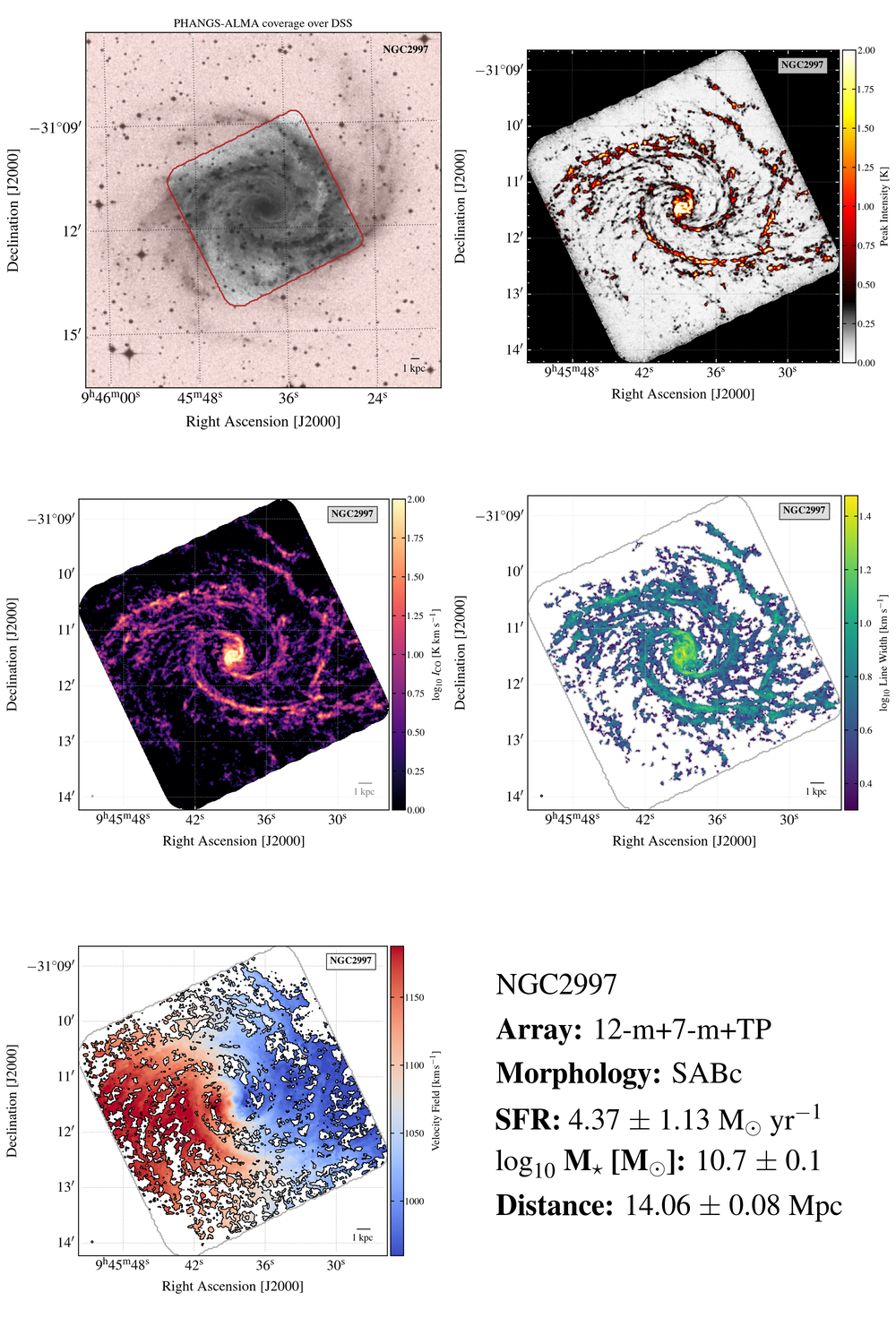}
\end{center}
\caption{\textbf{PHANGS--ALMA Image Atlas.} PHANGS--ALMA data products for one galaxy, NGC~2997. \textit{Top left:} ALMA mapping coverage, indicated by the unshaded area, over the DSS optical image. \textit{Top right:} Peak intensity image. \textit{Middle left:} Integrated intensity (``moment 0'') image on a logarithmic stretch, calculated summing over the ``broad,'' high completeness, mask (\S \ref{sec:products}). \textit{Middle right:} Line width, here $1\sigma$ estimated from the effective width. \textit{Bottom left:} intensity weighted mean velocity field constructed using a low resolution prior. The bottom right panel provides key information for the galaxy. Similar figures for all $90$ PHANGS--ALMA targets appear as part of an online figure set. \label{fig:atlas}}
\end{figure*}

\figsetstart
\figsetnum{A}
\figsettitle{Atlas of PHANGS--ALMA maps.}
\figsetgrpstart
\figsetgrpnum{A.34}
\figsetgrptitle{PHANGS--ALMA atlas view of NGC2997}
\figsetplot{ngc2997_combined.png}
\figsetgrpnote{}
\figsetgrpend
\figsetgrpstart
\figsetgrpnum{A.90}
\figsetgrptitle{PHANGS--ALMA atlas view of NGC7793}
\figsetplot{ngc7793_combined.png}
\figsetgrpnote{}
\label{figset:atlas}
\figsetgrpend
\figsetend

A figure set in the online version of this paper presents an atlas of data products for the PHANGS--ALMA targets, which we illustrate with an example in Figure \ref{fig:atlas}. For each galaxy, we illustrate the area mapped by ALMA over the DSS optical image of the galaxy (\textit{top left}). Then we show four ALMA data products:

\begin{enumerate}
\item The peak intensity map calculated over a 12.5~km~s$^{-1}$ window (\textit{top right}) highlights detailed structure of each galaxy. Arms, bars, filamentary structure, and individual cloud complexes are all highly visible in these maps. Because the maps show the maximum over the full velocity extent of the mask, these offer a direct view of the contents of the data cubes.

\item The integrated intensity map (\textit{middle left}) on a logarithmic stretch shows the distribution of CO~(2-1) emission in the galaxy. Here we show the integrated intensity map constructed using the ``broad'', i.e., high completeness, mask. Therefore these maps show almost all emission from the galaxy, though note that we do set the lower end of the color stretch to 1~K~km~s$^{-1}$, cloud to our nominal $3\sigma$ limit (see Table \ref{tab:cubeprops}).

\item A map of line width (\textit{middle right}) on a logarithmic stretch. In the atlas, we show $1\sigma$ line widths calculated using the effective width. Because the line width is challenging to calculate in the presence of noise, these maps use the ``strict'', i.e., high confidence, masks. 

\item The velocity field appears in the \textit{bottom left} panel. We show the ``moment~1 with prior'', i.e., the intensity-weighted mean velocity calculated within the broad, high completeness mask and then pruned using a low resolution prior estimate of the velocity. This represents our velocity field with the highest covering fraction and illustrates both large and small-scale systematic motions in the galaxy. 
\end{enumerate}

Along with these images, each figure notes a few key properties of each galaxy along with the best available array combination.

\section{Summary}
\label{sec:summary}

This paper presents PHANGS--ALMA, the first large $\theta \approx 1\arcsec \approx 100$~pc resolution \cotwo\ survey of a representative nearby galaxy population. The paper describes:

\begin{enumerate}
\item \textbf{The scientific motivation for PHANGS--ALMA (\S\ref{sec:motivation}).} We summarize how PHANGS--ALMA builds on previous CO mapping efforts in the local universe, highlight that many key physics related to both star formation and galaxy evolution occur at or near the scale of individual giant molecular clouds (GMCs), and outline the high-level science goals of PHANGS--ALMA:

$\bullet$ Measure the  demographics  of  molecular  clouds, and measure how GMC populations depend on host galaxy and location in a galaxy.

$\bullet$ Measure the star formation efficiency per free fall time, $\epsilon_{\rm ff}$, at cloud scales. Measure how $\epsilon_{\rm ff}$ depends on the density, dynamical state, and turbulence in molecular clouds.

$\bullet$ Quantify  the  ``violent  cycling''  between  phases  of the  star formation  process. Use  this  to  constrain the life cycle of clouds and feedback.

$\bullet$ Measure  how  the  self-regulated, large scale structure of galaxy disks emerges from a medium made of individual clouds and star-forming regions.

$\bullet$ Measure  the  motions,  flows,  and  organization  of cold gas in galaxies at $100{−}1{,}000$~pc scales.

\item \textbf{The sample selection (\S\ref{sec:sample}).} PHANGS--ALMA attempts to target all relatively massive, actively star-forming galaxies within $17$~Mpc that are not heavily inclined. This means that the sample selection is unbiased, easy to understand, and simple to attempt to reproduce. It also means that a secondary goal of PHANGS--ALMA is that ALMA provide a high quality CO map of all local galaxies. Our estimates of key galaxy properties, including distance and stellar mass, have evolved some since the original sample selection. We show that while uncertainties in galaxy properties do affect our exact sample, we include approximately the expected number of targets. We also show that the targets do a good job of spanning the $z=0$ main sequence of star-forming galaxies and the mass radius relation for late-type galaxies.

\item \textbf{Calculation and presentation of the properties of the sample galaxies (\S\ref{sec:sampleprops}).} We present estimates the physical properties of the PHANGS--ALMA targets: size, mass, CO luminosity, and star formation rate (SFR). We draw distances from the recent work by \citet{ANAND21}. We compare estimates of these quantities using different methods and provide quantitative translations between different methods of estimating size, mass, and SFR. This includes new work on the size and stellar mass estimation.

We attempt to present numbers on a system that is self-consistent with work on the SDSS main galaxy sample by \citet{SALIM16} and \citet{SALIM18}, which is itself broadly consistent with earlier work on the SDSS main galaxy sample. To verify this overall consistency, we show a good agreement between the properties of the PHANGS--ALMA targets and previous measurements of the main sequence of star-forming galaxies and the mass-radius relation.

PHANGS--ALMA typically covers only ${\sim}70\%$ of the star formation activity, and so presumably missed ${\sim}30\%$ of the molecular gas. To account for this, we provide aperture corrections that can be used to correct the measured CO luminosity to the full CO luminosity for comparison to other global properties or unresolved CO surveys. We consider several aperture correction templates, including an exponential disk model and several bands. We find that WISE3 $12~\mu$m emission provides the best template for such an aperture correction, in good agreement with recent work by \citet{CHOWN21}.

\item \textbf{A summary of the PHANGS--ALMA observations (\S\ref{sec:observations}).} We observe the area of active star formation, as gauged from mid-infrared (WISE $12~\mu$m) emission. To cover this area, we use large mosaics. In many cases where the galaxy exceeds the area covered by a single $150$~pointing mosaic we break a galaxy into parts and observe each part separately, combining the data into a single cube in post-processing. The observations spanned $6$~years and the typical elevation and atmospheric conditions were excellent, $> 50^\circ$ and $1{-}1.4$~mm of precipitable water vapor. Even though the total observing time on any given field using the \mbox{12-m} array was short, the achieve $u{-}v$ coverage for the combined \mbox{12-m}+\mbox{7-m} data appears excellent and yields a synthesized beam without dramatic sidelobe features.

\item \textbf{A short summary of the data processing (\S\ref{sec:processing}).} We provide a short overview of the PHANGS--ALMA post-processing and product creation pipeline. The full pipeline is presented in \citet{LEROY21a} and will be publicly available along with the PHANGS--ALMA data products.

\item \textbf{A description of the properties of the PHANGS--ALMA cubes and data products (\S\ref{sec:products}).} We cover $81$ targets with the combined \mbox{12-m}, \mbox{7-m}, and total power arrays and $9$ targets with the \mbox{7-m} array or combined \mbox{7-m} plus total power arrays. A typical \mbox{12-m}+\mbox{7-m}+TP map covers ${\sim}7$~arcmin$^2$ or $124$~kpc$^2$. In total, the survey maps ${\sim}1050$~arcmin$^2$ or $10{,}650$~kpc$^2$.

For the \mbox{12-m}+\mbox{7-m}+TP data, the median angular resolution is $1.3\arcsec$, the channel width is $2.54$~km~s$^{-1}$, and the typical $1\sigma$ sensitivity is ${\sim}6.2$~mJy~beam$^{-1}$ or 85~mK. These values translate to a median physical resolution of $100$~pc and $1\sigma$ sensitivity to mass of ${\sim}2 \times 10^4$~M$_\odot$ for typical Milky Way conditions. Across almost the whole survey, $\theta = 150$~pc represents a common physical resolution achievable for most data. At the native resolution, our cubes recover a median ${\sim}60\%$ of the total flux in the noise at good signal to noise.

\item \textbf{An atlas showing the data for each target (\S\ref{sec:atlas}).} A figure set shows the areal coverage, peak and integrated intensity, velocity field, and line width maps for each galaxy.

\end{enumerate}

\acknowledgments

We thank the anonymous referee and the editors for a rapid, constructive review during a difficult time. 

This work was carried out as part of the PHANGS collaboration.

The work of A.K.L., J.S., and D.U. was partially supported by the National Science Foundation (NSF) under Grants No.1615105, 1615109, and 1653300, as well as by the National Aeronautics and Space Administration (NASA) under ADAP grants NNX16AF48G and NNX17AF39G. C.M.F. acknowledges support from the NSF under Award No. 1903946. A.S. is supported by an NSF Astronomy and Astrophysics Postdoctoral Fellowship under award AST-1903834.

ER acknowledges the support of the Natural Sciences and Engineering Research Council of Canada (NSERC), funding reference number RGPIN-2017-03987, and computational support from Compute Canada.

D.L., T.S., E.S., C.M.F., K.S., and T.G.W. acknowledge funding from the European Research Council (ERC) under the European Union’s Horizon 2020 research and innovation programme (grant agreement No. 694343).

CH, AH,and JP acknowledge support by the Programme National “Physique et Chimie du Milieu Interstellaire” (PCMI) of CNRS/INSU with INC/INP co-funded by CEA and CNES.  
AH acknowledges support by the Programme National Cosmology et Galaxies (PNCG) of CNRS/INSU with INP and IN2P3, co-funded by CEA and CNES. 

AU and AG-R acknowledge support from the Spanish funding grants AYA2016-79006-P (MINECO/FEDER) and PID2019-108765GB-I00 (MICINN). 
AU acknowledges support from the Spanish funding grant PGC2018-094671-B-I00 (MCIU/AEI/FEDER).
MQ acknowledges support from the research project  PID2019-106027GA-C44 from the Spanish Ministerio de Ciencia e Innovaci\'on.

M.C. and J.M.D.K. gratefully acknowledge funding from the German Research Foundation (DFG) through an Emmy Noether Research Group (grant number KR4801/1-1). M.C., J.M.D.K., and J.J.K. gratefully acknowledge funding from the DFG Sachbeihilfe (grant number KR4801/2-1). J.M.D.K. gratefully acknowledges funding from the European Research Council (ERC) under the European Union's Horizon 2020 research and innovation programme via the ERC Starting Grant MUSTANG (grant agreement number 714907).

FB, ATB, IB, JdB, JP acknowledge funding from the European Union’s Horizon 2020 research and innovation programme (grant agreement No 726384/EMPIRE).

CE acknowledges funding from the Deutsche Forschungsgemeinschaft (DFG) Sachbeihilfe, grant number BI1546/3-1.

R.S.K.\ , S.C.O.G.\ and M.C.S.\ acknowledge financial support from the DFG via the collaborative research center (SFB 881, Project-ID 138713538) ``The Milky Way System” (subprojects A1, B1, B2, and B8). They also acknowledge subsidies from the Heidelberg Cluster of Excellence {\em STRUCTURES} in the framework of Germany’s Excellence Strategy (grant EXC-2181/1 - 390900948) and funding from the ERC via the ERC Synergy Grant {\em ECOGAL} (grant 855130).

KK and FS gratefully acknowledge funding from the DFG in the form of an Emmy Noether Research Group (grant number KR4598/2-1). 

EW acknowledges support from the Deutsche Forschungsgemeinschaft (DFG, German Research Foundation) -- Project-ID 138713538 -- SFB 881 (``The Milky Way System'', subproject P2). 

PSB acknowledges  support  from  the  research project PID2019-107427GB-C31  from the Spanish Ministerio de Ciencia  Innovaci\'on.

The work of E.C.O. was supported by the NSF under grant 
No. AST-1713949 and  NASA under grant No. NNX17AG26G.

This paper makes use of the following ALMA data, which have been processed as part of the PHANGS--ALMA \cotwo\ survey: \\
\noindent ADS/JAO.ALMA\#2012.1.00650.S, \linebreak % (N628/M74)
ADS/JAO.ALMA\#2013.1.00803.S, \linebreak % (N5128/CenA)
ADS/JAO.ALMA\#2013.1.01161.S, \linebreak % (N1365 + N5236/M83)
ADS/JAO.ALMA\#2015.1.00121.S, \linebreak % (N5236/M83)
ADS/JAO.ALMA\#2015.1.00782.S, \linebreak % (N1313 + N7793)
ADS/JAO.ALMA\#2015.1.00925.S, \linebreak % (pilot low mass)
ADS/JAO.ALMA\#2015.1.00956.S, \linebreak % (pilot high mass)
ADS/JAO.ALMA\#2016.1.00386.S, \linebreak % (N5236/M83)
ADS/JAO.ALMA\#2017.1.00392.S, \linebreak % (low mass follow-up)
ADS/JAO.ALMA\#2017.1.00766.S, \linebreak % (early-type)
ADS/JAO.ALMA\#2017.1.00886.L, \linebreak % (large program)
ADS/JAO.ALMA\#2018.1.00484.S, \linebreak % (early-type)
ADS/JAO.ALMA\#2018.1.01321.S, \linebreak % (N253, N300, Circinus)
ADS/JAO.ALMA\#2018.1.01651.S, \linebreak % (main sample follow-up)
ADS/JAO.ALMA\#2018.A.00062.S, \linebreak % (ACA-only nearby)
ADS/JAO.ALMA\#2019.1.01235.S, \linebreak % (local sample follow up)
ADS/JAO.ALMA\#2019.2.00129.S, \linebreak % (NGC 1068)
ALMA is a partnership of ESO (representing its member states), NSF (USA), and NINS (Japan), together with NRC (Canada), NSC and ASIAA (Taiwan), and KASI (Republic of Korea), in cooperation with the Republic of Chile. The Joint ALMA Observatory is operated by ESO, AUI/NRAO, and NAOJ. The National Radio Astronomy Observatory is a facility of the National Science Foundation operated under cooperative agreement by Associated Universities, Inc.

\software{
\texttt{aplpy},
ALMA Calibration Pipeline (L.~Davis et al.\ in preparation),
\texttt{CASA} \citep{MCMULLIN07},
\texttt{numpy} \citep{NUMPY2006}, 
\texttt{scipy} \citep{SCIPY2020}, 
\texttt{astropy} \citep{ASTROPY1,ASTROPY2},
IDL Astronomy User's Library \citep{IDLASTRO}, 
\texttt{cprops} \citep{ROSOLOWSKY06},
GILDAS \citet{PETY2005},
PHANGS--ALMA Pipeline \citep{LEROY21a},
PHANGS--ALMA Total Power Pipeline \citep{HERRERA20},
\texttt{R} \citep{RMANUAL},
\texttt{spectral-cube} \citep{SPECTRALCUBE2020},
\texttt{radio-beam} (\url{https://github.com/radio-astro-tools/radio-beam})
}

%\bibliography{akl}

\begin{thebibliography}{}
\expandafter\ifx\csname natexlab\endcsname\relax\def\natexlab#1{#1}\fi

\bibitem[{{Agertz} {et~al.}(2013){Agertz}, {Kravtsov}, {Leitner}, \&
  {Gnedin}}]{AGERTZ13}
{Agertz}, O., {Kravtsov}, A.~V., {Leitner}, S.~N., \& {Gnedin}, N.~Y. 2013,
  \apj, 770, 25

\bibitem[{{Alatalo} {et~al.}(2013){Alatalo}, {Davis}, {Bureau}, {Young},
  {Blitz}, {Crocker}, {Bayet}, {Bois}, {Bournaud}, {Cappellari}, {Davies}, {de
  Zeeuw}, {Duc}, {Emsellem}, {Khochfar}, {Krajnovi{\'c}}, {Kuntschner},
  {Lablanche}, {Morganti}, {McDermid}, {Naab}, {Oosterloo}, {Sarzi}, {Scott},
  {Serra}, \& {Weijmans}}]{ALATALO13}
{Alatalo}, K., {Davis}, T.~A., {Bureau}, M., {et~al.} 2013, \mnras, 432, 1796

\bibitem[{{Anand} {et~al.}(2021){Anand}, {Lee}, {Van Dyk}, {Leroy},
  {Rosolowsky}, {Schinnerer}, {Larson}, {Kourkchi}, {Kreckel}, {Scheuermann},
  {Rizzi}, {Thilker}, {Tully}, {Bigiel}, {Blanc}, {Boquien}, {Chandar}, {Dale},
  {Emsellem}, {Deger}, {Glover}, {Grasha}, {Groves}, {Klessen}, {Kruijssen},
  {Querejeta}, {S{\'a}nchez-Bl{\'a}zquez}, {Schruba}, {Turner}, {Ubeda},
  {Williams}, \& {Whitmore}}]{ANAND21}
{Anand}, G.~S., {Lee}, J.~C., {Van Dyk}, S.~D., {et~al.} 2021, \mnras, 501,
  3621

\bibitem[{{Armus} {et~al.}(2009){Armus}, {Mazzarella}, {Evans}, {Surace},
  {Sanders}, {Iwasawa}, {Frayer}, {Howell}, {Chan}, {Petric}, {Vavilkin},
  {Kim}, {Haan}, {Inami}, {Murphy}, {Appleton}, {Barnes}, {Bothun}, {Bridge},
  {Charmandaris}, {Jensen}, {Kewley}, {Lord}, {Madore}, {Marshall},
  {Melbourne}, {Rich}, {Satyapal}, {Schulz}, {Spoon}, {Sturm}, {U}, {Veilleux},
  \& {Xu}}]{ARMUS09}
{Armus}, L., {Mazzarella}, J.~M., {Evans}, A.~S., {et~al.} 2009, \pasp, 121,
  559

\bibitem[{{Astropy Collaboration} {et~al.}(2013){Astropy Collaboration},
  {Robitaille}, {Tollerud}, {Greenfield}, {Droettboom}, {Bray}, {Aldcroft},
  {Davis}, {Ginsburg}, {Price-Whelan}, {Kerzendorf}, {Conley}, {Crighton},
  {Barbary}, {Muna}, {Ferguson}, {Grollier}, {Parikh}, {Nair}, {Unther},
  {Deil}, {Woillez}, {Conseil}, {Kramer}, {Turner}, {Singer}, {Fox}, {Weaver},
  {Zabalza}, {Edwards}, {Azalee Bostroem}, {Burke}, {Casey}, {Crawford},
  {Dencheva}, {Ely}, {Jenness}, {Labrie}, {Lim}, {Pierfederici}, {Pontzen},
  {Ptak}, {Refsdal}, {Servillat}, \& {Streicher}}]{ASTROPY1}
{Astropy Collaboration}, {Robitaille}, T.~P., {Tollerud}, E.~J., {et~al.} 2013,
  \aap, 558, A33

\bibitem[{{Astropy Collaboration} {et~al.}(2018){Astropy Collaboration},
  {Price-Whelan}, {Sip{\H{o}}cz}, {G{\"u}nther}, {Lim}, {Crawford}, {Conseil},
  {Shupe}, {Craig}, {Dencheva}, {Ginsburg}, {Vand erPlas}, {Bradley},
  {P{\'e}rez-Su{\'a}rez}, {de Val-Borro}, {Aldcroft}, {Cruz}, {Robitaille},
  {Tollerud}, {Ardelean}, {Babej}, {Bach}, {Bachetti}, {Bakanov}, {Bamford},
  {Barentsen}, {Barmby}, {Baumbach}, {Berry}, {Biscani}, {Boquien}, {Bostroem},
  {Bouma}, {Brammer}, {Bray}, {Breytenbach}, {Buddelmeijer}, {Burke},
  {Calderone}, {Cano Rodr{\'\i}guez}, {Cara}, {Cardoso}, {Cheedella}, {Copin},
  {Corrales}, {Crichton}, {D'Avella}, {Deil}, {Depagne}, {Dietrich}, {Donath},
  {Droettboom}, {Earl}, {Erben}, {Fabbro}, {Ferreira}, {Finethy}, {Fox},
  {Garrison}, {Gibbons}, {Goldstein}, {Gommers}, {Greco}, {Greenfield},
  {Groener}, {Grollier}, {Hagen}, {Hirst}, {Homeier}, {Horton}, {Hosseinzadeh},
  {Hu}, {Hunkeler}, {Ivezi{\'c}}, {Jain}, {Jenness}, {Kanarek}, {Kendrew},
  {Kern}, {Kerzendorf}, {Khvalko}, {King}, {Kirkby}, {Kulkarni}, {Kumar},
  {Lee}, {Lenz}, {Littlefair}, {Ma}, {Macleod}, {Mastropietro}, {McCully},
  {Montagnac}, {Morris}, {Mueller}, {Mumford}, {Muna}, {Murphy}, {Nelson},
  {Nguyen}, {Ninan}, {N{\"o}the}, {Ogaz}, {Oh}, {Parejko}, {Parley}, {Pascual},
  {Patil}, {Patil}, {Plunkett}, {Prochaska}, {Rastogi}, {Reddy Janga},
  {Sabater}, {Sakurikar}, {Seifert}, {Sherbert}, {Sherwood-Taylor}, {Shih},
  {Sick}, {Silbiger}, {Singanamalla}, {Singer}, {Sladen}, {Sooley},
  {Sornarajah}, {Streicher}, {Teuben}, {Thomas}, {Tremblay}, {Turner},
  {Terr{\'o}n}, {van Kerkwijk}, {de la Vega}, {Watkins}, {Weaver}, {Whitmore},
  {Woillez}, {Zabalza}, \& {Astropy Contributors}}]{ASTROPY2}
{Astropy Collaboration}, {Price-Whelan}, A.~M., {Sip{\H{o}}cz}, B.~M., {et~al.}
  2018, \aj, 156, 123

\bibitem[{{Baldry} {et~al.}(2008){Baldry}, {Glazebrook}, \&
  {Driver}}]{BALDRY08}
{Baldry}, I.~K., {Glazebrook}, K., \& {Driver}, S.~P. 2008, \mnras, 388, 945

\bibitem[{{Barbarino} {et~al.}(2015){Barbarino}, {Dall'Ora}, {Botticella},
  {Della Valle}, {Zampieri}, {Maund}, {Pumo}, {Jerkstrand}, {Benetti},
  {Elias-Rosa}, {Fraser}, {Gal-Yam}, {Hamuy}, {Inserra}, {Knapic}, {LaCluyze},
  {Molinaro}, {Ochner}, {Pastorello}, {Pignata}, {Reichart}, {Ries},
  {Riffeser}, {Schmidt}, {Schmidt}, {Smareglia}, {Smartt}, {Smith},
  {Sollerman}, {Sullivan}, {Tomasella}, {Turatto}, {Valenti}, {Yaron}, \&
  {Young}}]{BARBARINO15}
{Barbarino}, C., {Dall'Ora}, M., {Botticella}, M.~T., {et~al.} 2015, \mnras,
  448, 2312

\bibitem[{{Bell} \& {de Jong}(2001)}]{BELL01}
{Bell}, E.~F., \& {de Jong}, R.~S. 2001, \apj, 550, 212

\bibitem[{{Bell} {et~al.}(2003){Bell}, {McIntosh}, {Katz}, \&
  {Weinberg}}]{BELL03}
{Bell}, E.~F., {McIntosh}, D.~H., {Katz}, N., \& {Weinberg}, M.~D. 2003, \apjs,
  149, 289

\bibitem[{{Be\v{s}li\'c} {et~al.}(2021){Be\v{s}li\'c}, {Barnes}, {Bigiel},
  {Puschnig}, {Pety}, {Herrera Contreras}, {Leroy}, {Usero}, {Schinnerer},
  {Meidt}, {Emsellem}, {Hughes}, {Faesi}, {Kreckel}, {Belfiore}, {Chevance},
  {den Brok}, {Eibensteiner}, {Glover}, {Grasha}, {Jimenez-Donaire}, {Klessen},
  {Kruijssen}, {Liu}, {Pessa}, {Querejeta}, {Rosolowsky}, {Saito}, {Santoro},
  {Schruba}, {Sormani}, \& {Williams}}]{BESLIC21}
{Be\v{s}li\'c}, I., {Barnes}, A.~T., {Bigiel}, F., {et~al.} 2021, \mnras\
  submitted

\bibitem[{{Bigiel} {et~al.}(2008){Bigiel}, {Leroy}, {Walter}, {Brinks}, {de
  Blok}, {Madore}, \& {Thornley}}]{BIGIEL08}
{Bigiel}, F., {Leroy}, A., {Walter}, F., {et~al.} 2008, \aj, 136, 2846

\bibitem[{{Blanton} \& {Moustakas}(2009)}]{BLANTON09}
{Blanton}, M.~R., \& {Moustakas}, J. 2009, \araa, 47, 159

\bibitem[{{Blitz}(1993)}]{BLITZ93}
{Blitz}, L. 1993, in Protostars and Planets III, ed. E.~H. {Levy} \& J.~I.
  {Lunine}, 125--161

\bibitem[{{Blitz} {et~al.}(2007){Blitz}, {Fukui}, {Kawamura}, {Leroy},
  {Mizuno}, \& {Rosolowsky}}]{BLITZ07}
{Blitz}, L., {Fukui}, Y., {Kawamura}, A., {et~al.} 2007, in Protostars and
  Planets V, ed. B.~{Reipurth}, D.~{Jewitt}, \& K.~{Keil}, 81--96

\bibitem[{{Blitz} \& {Rosolowsky}(2006)}]{BLITZ06}
{Blitz}, L., \& {Rosolowsky}, E. 2006, \apj, 650, 933

\bibitem[{{Boissier} {et~al.}(2003){Boissier}, {Prantzos}, {Boselli}, \&
  {Gavazzi}}]{BOISSIER03}
{Boissier}, S., {Prantzos}, N., {Boselli}, A., \& {Gavazzi}, G. 2003, \mnras,
  346, 1215

\bibitem[{{Bolatto} {et~al.}(2008){Bolatto}, {Leroy}, {Rosolowsky}, {Walter},
  \& {Blitz}}]{BOLATTO08}
{Bolatto}, A.~D., {Leroy}, A.~K., {Rosolowsky}, E., {Walter}, F., \& {Blitz},
  L. 2008, \apj, 686, 948

\bibitem[{{Bolatto} {et~al.}(2013{\natexlab{a}}){Bolatto}, {Wolfire}, \&
  {Leroy}}]{BOLATTO13B}
{Bolatto}, A.~D., {Wolfire}, M., \& {Leroy}, A.~K. 2013{\natexlab{a}}, \araa,
  51, 207

\bibitem[{{Bolatto} {et~al.}(2013{\natexlab{b}}){Bolatto}, {Warren}, {Leroy},
  {Walter}, {Veilleux}, {Ostriker}, {Ott}, {Zwaan}, {Fisher}, {Weiss},
  {Rosolowsky}, \& {Hodge}}]{BOLATTO13A}
{Bolatto}, A.~D., {Warren}, S.~R., {Leroy}, A.~K., {et~al.} 2013{\natexlab{b}},
  \nat, 499, 450

\bibitem[{{Bolatto} {et~al.}(2017){Bolatto}, {Wong}, {Utomo}, {Blitz}, {Vogel},
  {S{\'a}nchez}, {Barrera-Ballesteros}, {Cao}, {Colombo}, {Dannerbauer},
  {Garc{\'{\i}}a-Benito}, {Herrera-Camus}, {Husemann}, {Kalinova}, {Leroy},
  {Leung}, {Levy}, {Mast}, {Ostriker}, {Rosolowsky}, {Sandstrom}, {Teuben},
  {van de Ven}, \& {Walter}}]{BOLATTO17}
{Bolatto}, A.~D., {Wong}, T., {Utomo}, D., {et~al.} 2017, \apj, 846, 159

\bibitem[{{Boquien} {et~al.}(2019){Boquien}, {Burgarella}, {Roehlly}, {Buat},
  {Ciesla}, {Corre}, {Inoue}, \& {Salas}}]{BOQUIEN19}
{Boquien}, M., {Burgarella}, D., {Roehlly}, Y., {et~al.} 2019, \aap, 622, A103

\bibitem[{{Boquien} {et~al.}(2016){Boquien}, {Kennicutt}, {Calzetti}, {Dale},
  {Galametz}, {Sauvage}, {Croxall}, {Draine}, {Kirkpatrick}, {Kumari}, {Hunt},
  {De Looze}, {Pellegrini}, {Rela{\~n}o}, {Smith}, \& {Tabatabaei}}]{BOQUIEN16}
{Boquien}, M., {Kennicutt}, R., {Calzetti}, D., {et~al.} 2016, \aap, 591, A6

\bibitem[{{Bothwell} {et~al.}(2014){Bothwell}, {Wagg}, {Cicone}, {Maiolino},
  {M{\o}ller}, {Aravena}, {De Breuck}, {Peng}, {Espada}, {Hodge},
  {Impellizzeri}, {Mart{\'{\i}}n}, {Riechers}, \& {Walter}}]{BOTHWELL14}
{Bothwell}, M.~S., {Wagg}, J., {Cicone}, C., {et~al.} 2014, \mnras, 445, 2599

\bibitem[{{Braine} \& {Combes}(1992)}]{BRAINE92}
{Braine}, J., \& {Combes}, F. 1992, \aap, 264, 433

\bibitem[{{Braine} {et~al.}(2007){Braine}, {Ferguson}, {Bertoldi}, \&
  {Wilson}}]{BRAINE07}
{Braine}, J., {Ferguson}, A.~M.~N., {Bertoldi}, F., \& {Wilson}, C.~D. 2007,
  \apjl, 669, L73

\bibitem[{{Braine} {et~al.}(2018){Braine}, {Rosolowsky}, {Gratier}, {Corbelli},
  \& {Schuster}}]{BRAINE18}
{Braine}, J., {Rosolowsky}, E., {Gratier}, P., {Corbelli}, E., \& {Schuster},
  K.~F. 2018, \aap, 612, A51

\bibitem[{{Brownson} {et~al.}(2020){Brownson}, {Belfiore}, {Maiolino}, {Lin},
  \& {Carniani}}]{BROWNSON20}
{Brownson}, S., {Belfiore}, F., {Maiolino}, R., {Lin}, L., \& {Carniani}, S.
  2020, \mnras, 498, L66

\bibitem[{{Bruzual} \& {Charlot}(2003)}]{BRUZUAL03}
{Bruzual}, G., \& {Charlot}, S. 2003, \mnras, 344, 1000

\bibitem[{{Burkhart}(2018)}]{BURKHART18}
{Burkhart}, B. 2018, \apj, 863, 118

\bibitem[{{Cald{\'u}-Primo} \& {Schruba}(2016)}]{CALDUPRIMO16}
{Cald{\'u}-Primo}, A., \& {Schruba}, A. 2016, \aj, 151, 34

\bibitem[{{Calzetti}(2013)}]{CALZETTI13}
{Calzetti}, D. 2013, in Secular Evolution of Galaxies, ed.
  J.~{Falc{\'o}n-Barroso} \& J.~H. {Knapen}, 419

\bibitem[{{Calzetti} {et~al.}(2005){Calzetti}, {Kennicutt}, {Bianchi},
  {Thilker}, {Dale}, {Engelbracht}, {Leitherer}, {Meyer}, {Sosey}, {Mutchler},
  {Regan}, {Thornley}, {Armus}, {Bendo}, {Boissier}, {Boselli}, {Draine},
  {Gordon}, {Helou}, {Hollenbach}, {Kewley}, {Madore}, {Martin}, {Murphy},
  {Rieke}, {Rieke}, {Roussel}, {Sheth}, {Smith}, {Walter}, {White}, {Yi},
  {Scoville}, {Polletta}, \& {Lindler}}]{CALZETTI05}
{Calzetti}, D., {Kennicutt}, Jr., R.~C., {Bianchi}, L., {et~al.} 2005, \apj,
  633, 871

\bibitem[{{Cardelli} {et~al.}(1989){Cardelli}, {Clayton}, \&
  {Mathis}}]{CARDELLI89}
{Cardelli}, J.~A., {Clayton}, G.~C., \& {Mathis}, J.~S. 1989, \apj, 345, 245

\bibitem[{{Catinella} {et~al.}(2018){Catinella}, {Saintonge}, {Janowiecki},
  {Cortese}, {Dav{\'e}}, {Lemonias}, {Cooper}, {Schiminovich}, {Hummels},
  {Fabello}, {Ger{\'e}b}, {Kilborn}, \& {Wang}}]{CATINELLA18}
{Catinella}, B., {Saintonge}, A., {Janowiecki}, S., {et~al.} 2018, \mnras, 476,
  875

\bibitem[{{Chabrier}(2003)}]{CHABRIER03}
{Chabrier}, G. 2003, \pasp, 115, 763

\bibitem[{{Chastenet} {et~al.}(2019){Chastenet}, {Sandstrom}, {Chiang},
  {Leroy}, {Utomo}, {Bot}, {Gordon}, {Draine}, {Fukui}, {Onishi}, \&
  {Tsuge}}]{CHASTENET19}
{Chastenet}, J., {Sandstrom}, K., {Chiang}, I.~D., {et~al.} 2019, \apj, 876, 62

\bibitem[{{Chevance} {et~al.}(2020{\natexlab{a}}){Chevance}, {Kruijssen},
  {Krumholz}, {Groves}, {Keller}, {Hughes}, {Glover}, {Henshaw}, {Herrera},
  {Kim}, {Leroy}, {Pety}, {Razza}, {Rosolowsky}, {Schinnerer}, {Schruba},
  {Barnes}, {Bigiel}, {Blanc}, {Emsellem}, {Faesi}, {Grasha}, {Klessen},
  {Kreckel}, {Liu}, {Longmore}, {Meidt}, {Querejeta}, {Saito}, {Sun}, \&
  {Usero}}]{CHEVANCE21}
{Chevance}, M., {Kruijssen}, J.~M.~D., {Krumholz}, M.~R., {et~al.}
  2020{\natexlab{a}}, arXiv e-prints, arXiv:2010.13788

\bibitem[{{Chevance} {et~al.}(2020{\natexlab{b}}){Chevance}, {Kruijssen},
  {Hygate}, {Schruba}, {Longmore}, {Groves}, {Henshaw}, {Herrera}, {Hughes},
  {Jeffreson}, {Lang}, {Leroy}, {Meidt}, {Pety}, {Razza}, {Rosolowsky},
  {Schinnerer}, {Bigiel}, {Blanc}, {Emsellem}, {Faesi}, {Glover}, {Haydon},
  {Ho}, {Kreckel}, {Lee}, {Liu}, {Querejeta}, {Saito}, {Sun}, {Usero}, \&
  {Utomo}}]{CHEVANCE20}
{Chevance}, M., {Kruijssen}, J.~M.~D., {Hygate}, A. P.~S., {et~al.}
  2020{\natexlab{b}}, \mnras, 493, 2872

\bibitem[{{Chevance} {et~al.}(2020{\natexlab{c}}){Chevance}, {Kruijssen},
  {Vazquez-Semadeni}, {Nakamura}, {Klessen}, {Ballesteros-Paredes}, {Inutsuka},
  {Adamo}, \& {Hennebelle}}]{CHEVANCE20b}
{Chevance}, M., {Kruijssen}, J.~M.~D., {Vazquez-Semadeni}, E., {et~al.}
  2020{\natexlab{c}}, \ssr, 216, 50

\bibitem[{{Chown} {et~al.}(2021){Chown}, {Li}, {Parker}, {Wilson}, {Li}, \&
  {Gao}}]{CHOWN21}
{Chown}, R., {Li}, C., {Parker}, L., {et~al.} 2021, \mnras, 500, 1261

\bibitem[{{Colombo} {et~al.}(2014{\natexlab{a}}){Colombo}, {Hughes},
  {Schinnerer}, {Meidt}, {Leroy}, {Pety}, {Dobbs}, {Garc{\'\i}a-Burillo},
  {Dumas}, {Thompson}, {Schuster}, \& {Kramer}}]{COLOMBO14A}
{Colombo}, D., {Hughes}, A., {Schinnerer}, E., {et~al.} 2014{\natexlab{a}},
  \apj, 784, 3

\bibitem[{{Colombo} {et~al.}(2014{\natexlab{b}}){Colombo}, {Meidt},
  {Schinnerer}, {Garc{\'{\i}}a-Burillo}, {Hughes}, {Pety}, {Leroy}, {Dobbs},
  {Dumas}, {Thompson}, {Schuster}, \& {Kramer}}]{COLOMBO14B}
{Colombo}, D., {Meidt}, S.~E., {Schinnerer}, E., {et~al.} 2014{\natexlab{b}},
  \apj, 784, 4

\bibitem[{{Colombo} {et~al.}(2018){Colombo}, {Kalinova}, {Utomo}, {Rosolowsky},
  {Bolatto}, {Levy}, {Wong}, {Sanchez}, {Leroy}, {Ostriker}, {Blitz}, {Vogel},
  {Mast}, {Garc{\'\i}a-Benito}, {Husemann}, {Dannerbauer}, {Ellmeier}, \&
  {Cao}}]{COLOMBO18}
{Colombo}, D., {Kalinova}, V., {Utomo}, D., {et~al.} 2018, \mnras, 475, 1791

\bibitem[{{Corbelli} {et~al.}(2017){Corbelli}, {Braine}, {Bandiera},
  {Brouillet}, {Combes}, {Druard}, {Gratier}, {Mata}, {Schuster}, {Xilouris},
  \& {Palla}}]{CORBELLI17}
{Corbelli}, E., {Braine}, J., {Bandiera}, R., {et~al.} 2017, \aap, 601, A146

\bibitem[{{Corder} {et~al.}(2008){Corder}, {Sheth}, {Scoville}, {Koda},
  {Vogel}, \& {Ostriker}}]{CORDER08}
{Corder}, S., {Sheth}, K., {Scoville}, N.~Z., {et~al.} 2008, \apj, 689, 148

\bibitem[{{Cormier} {et~al.}(2018){Cormier}, {Bigiel}, {Jim{\'e}nez-Donaire},
  {Leroy}, {Gallagher}, {Usero}, {Sandstrom}, {Bolatto}, {Hughes}, {Kramer},
  {Krumholz}, {Meier}, {Murphy}, {Pety}, {Rosolowsky}, {Schinnerer}, {Schruba},
  {Sliwa}, \& {Walter}}]{CORMIER18}
{Cormier}, D., {Bigiel}, F., {Jim{\'e}nez-Donaire}, M.~J., {et~al.} 2018,
  \mnras, 475, 3909

\bibitem[{{Cornwell}(2008)}]{CORNWELL08}
{Cornwell}, T.~J. 2008, IEEE Journal of Selected Topics in Signal Processing,
  2, 793

\bibitem[{{Cotton}(2017)}]{COTTON17}
{Cotton}, W.~D. 2017, \pasp, 129, 094501

\bibitem[{{Courtois} \& {Tully}(2012)}]{COURTOIS12}
{Courtois}, H.~M., \& {Tully}, R.~B. 2012, \apj, 749, 174

\bibitem[{{Dale} {et~al.}(2020){Dale}, {Anderson}, {Bran}, {Cox}, {Drake},
  {Lee}, {Pilawa}, {Alexander Slane}, {Soto}, {Jensen}, {Sutter}, {Turner}, \&
  {Kobulnicky}}]{DALE20}
{Dale}, D.~A., {Anderson}, K.~R., {Bran}, L.~M., {et~al.} 2020, \aj, 159, 195

\bibitem[{{Dale}(2015)}]{DALE15}
{Dale}, J.~E. 2015, \nar, 68, 1

\bibitem[{{Davis} {et~al.}(2014){Davis}, {Jiang}, {Stone}, \&
  {Murray}}]{DAVIS14}
{Davis}, S.~W., {Jiang}, Y.-F., {Stone}, J.~M., \& {Murray}, N. 2014, \apj,
  796, 107

\bibitem[{{Davis} \& {McDermid}(2017)}]{DAVIS17}
{Davis}, T.~A., \& {McDermid}, R.~M. 2017, \mnras, 464, 453

\bibitem[{{de Vaucouleurs} {et~al.}(1991){de Vaucouleurs}, {de Vaucouleurs},
  {Corwin}, {Buta}, {Paturel}, \& {Fouque}}]{DEVAUCOULEURS91}
{de Vaucouleurs}, G., {de Vaucouleurs}, A., {Corwin}, Herold~G., J., {et~al.}
  1991, {Third Reference Catalogue of Bright Galaxies} (New York: Springer)

\bibitem[{{den Brok} {et~al.}(2021){den Brok}, {Chatzigiannakis}, {Bigiel},
  {Puschnig}, {Barnes}, {Leroy}, {Jimenez-Donaire}, {Usero}, {Schinnerer},
  {Rosolowsky}, {Faesi}, {Grasha}, {Hughes}, {Kruijssen}, {Liu}, {Neumann},
  {Pety}, {Querejeta}, T., {Schruba}, \& {Stuber}}]{DENBROK21}
{den Brok}, J.~S., {Chatzigiannakis}, D., {Bigiel}, F., {et~al.} 2021, \mnras\
  submitted

\bibitem[{{Dobbs} {et~al.}(2019){Dobbs}, {Rosolowsky}, {Pettitt}, {Braine},
  {Corbelli}, \& {Sun}}]{DOBBS19}
{Dobbs}, C.~L., {Rosolowsky}, E., {Pettitt}, A.~R., {et~al.} 2019, \mnras, 485,
  4997

\bibitem[{{Dobbs} {et~al.}(2014){Dobbs}, {Krumholz}, {Ballesteros-Paredes},
  {Bolatto}, {Fukui}, {Heyer}, {Low}, {Ostriker}, \&
  {V{\'a}zquez-Semadeni}}]{DOBBS14}
{Dobbs}, C.~L., {Krumholz}, M.~R., {Ballesteros-Paredes}, J., {et~al.} 2014, in
  Protostars and Planets VI, ed. H.~{Beuther}, R.~S. {Klessen}, C.~P.
  {Dullemond}, \& T.~{Henning}, 3

\bibitem[{{Donovan Meyer} {et~al.}(2012){Donovan Meyer}, {Koda}, {Momose},
  {Fukuhara}, {Mooney}, {Towers}, {Egusa}, {Kennicutt}, {Kuno}, {Carty},
  {Sawada}, \& {Scoville}}]{DONOVANMEYER12}
{Donovan Meyer}, J., {Koda}, J., {Momose}, R., {et~al.} 2012, \apj, 744, 42

\bibitem[{{Donovan Meyer} {et~al.}(2013){Donovan Meyer}, {Koda}, {Momose},
  {Mooney}, {Egusa}, {Carty}, {Kennicutt}, {Kuno}, {Rebolledo}, {Sawada},
  {Scoville}, \& {Wong}}]{DONOVANMEYER13}
---. 2013, \apj, 772, 107

\bibitem[{{Downes} \& {Solomon}(1998)}]{DOWNES98}
{Downes}, D., \& {Solomon}, P.~M. 1998, \apj, 507, 615

\bibitem[{{Druard} {et~al.}(2014){Druard}, {Braine}, {Schuster}, {Schneider},
  {Gratier}, {Bontemps}, {Boquien}, {Combes}, {Corbelli}, {Henkel}, {Herpin},
  {Kramer}, {van der Tak}, \& {van der Werf}}]{DRUARD14}
{Druard}, C., {Braine}, J., {Schuster}, K.~F., {et~al.} 2014, \aap, 567, A118

\bibitem[{{Egusa} {et~al.}(2017){Egusa}, {Mentuch Cooper}, {Koda}, \&
  {Baba}}]{EGUSA17}
{Egusa}, F., {Mentuch Cooper}, E., {Koda}, J., \& {Baba}, J. 2017, \mnras, 465,
  460

\bibitem[{{Elmegreen}(1989)}]{ELMEGREEN89}
{Elmegreen}, B.~G. 1989, \apj, 338, 178

\bibitem[{{Elmegreen}(2000)}]{ELMEGREEN00}
---. 2000, \apj, 530, 277

\bibitem[{{Elmegreen} {et~al.}(2018){Elmegreen}, {Elmegreen}, \&
  {Efremov}}]{ELMEGREEN18}
{Elmegreen}, B.~G., {Elmegreen}, D.~M., \& {Efremov}, Y.~N. 2018, \apj, 863, 59

\bibitem[{{Engargiola} {et~al.}(2003){Engargiola}, {Plambeck}, {Rosolowsky}, \&
  {Blitz}}]{ENGARGIOLA03}
{Engargiola}, G., {Plambeck}, R.~L., {Rosolowsky}, E., \& {Blitz}, L. 2003,
  \apjs, 149, 343

\bibitem[{{Engelbracht} {et~al.}(2006){Engelbracht}, {Kundurthy}, {Gordon},
  {Rieke}, {Kennicutt}, {Smith}, {Regan}, {Makovoz}, {Sosey}, {Draine},
  {Helou}, {Armus}, {Calzetti}, {Meyer}, {Bendo}, {Walter}, {Hollenbach},
  {Cannon}, {Murphy}, {Dale}, {Buckalew}, \& {Sheth}}]{ENGELBRACHT06}
{Engelbracht}, C.~W., {Kundurthy}, P., {Gordon}, K.~D., {et~al.} 2006, \apjl,
  642, L127

\bibitem[{{Espada} {et~al.}(2019){Espada}, {Verley}, {Miura}, {Israel},
  {Henkel}, {Matsushita}, {Vila-Vilaro}, {Ott}, {Morokuma-Matsui}, {Peck},
  {Hirota}, {Aalto}, {Quillen}, {Hogerheijde}, {Neumayer}, {Vlahakis}, {Iono},
  \& {Kohno}}]{ESPADA19}
{Espada}, D., {Verley}, S., {Miura}, R.~E., {et~al.} 2019, \apj, 887, 88

\bibitem[{{Evans} {et~al.}(2014){Evans}, {Heiderman}, \&
  {Vutisalchavakul}}]{EVANS14}
{Evans}, II, N.~J., {Heiderman}, A., \& {Vutisalchavakul}, N. 2014, \apj, 782,
  114

\bibitem[{{Faesi} {et~al.}(2018){Faesi}, {Lada}, \& {Forbrich}}]{FAESI18}
{Faesi}, C.~M., {Lada}, C.~J., \& {Forbrich}, J. 2018, \apj, 857, 19

\bibitem[{{Federrath} \& {Klessen}(2012)}]{FEDERRATH12}
{Federrath}, C., \& {Klessen}, R.~S. 2012, \apj, 761, 156

\bibitem[{{Federrath} \& {Klessen}(2013)}]{FEDERRATH13}
---. 2013, \apj, 763, 51

\bibitem[{{Federrath} {et~al.}(2010){Federrath}, {Roman-Duval}, {Klessen},
  {Schmidt}, \& {Mac Low}}]{FEDERRATH10}
{Federrath}, C., {Roman-Duval}, J., {Klessen}, R.~S., {Schmidt}, W., \& {Mac
  Low}, M.~M. 2010, \aap, 512, A81

\bibitem[{{Foyle} {et~al.}(2010){Foyle}, {Rix}, {Walter}, \& {Leroy}}]{FOYLE10}
{Foyle}, K., {Rix}, H.-W., {Walter}, F., \& {Leroy}, A.~K. 2010, \apj, 725, 534

\bibitem[{{Freedman} {et~al.}(2001){Freedman}, {Madore}, {Gibson}, {Ferrarese},
  {Kelson}, {Sakai}, {Mould}, {Kennicutt}, {Ford}, {Graham}, {Huchra},
  {Hughes}, {Illingworth}, {Macri}, \& {Stetson}}]{FREEDMAN01}
{Freedman}, W.~L., {Madore}, B.~F., {Gibson}, B.~K., {et~al.} 2001, \apj, 553,
  47

\bibitem[{{Freeman} {et~al.}(2017){Freeman}, {Rosolowsky}, {Kruijssen},
  {Bastian}, \& {Adamo}}]{FREEMAN17}
{Freeman}, P., {Rosolowsky}, E., {Kruijssen}, J.~M.~D., {Bastian}, N., \&
  {Adamo}, A. 2017, \mnras, 468, 1769

\bibitem[{{Fukui} {et~al.}(2020){Fukui}, {Habe}, {Inoue}, {Enokiya}, \&
  {Tachihara}}]{FUKUI20}
{Fukui}, Y., {Habe}, A., {Inoue}, T., {Enokiya}, R., \& {Tachihara}, K. 2020,
  \pasj, arXiv:2009.05077

\bibitem[{{Fukui} \& {Kawamura}(2010)}]{FUKUI10}
{Fukui}, Y., \& {Kawamura}, A. 2010, \araa, 48, 547

\bibitem[{{Fukui} {et~al.}(1999){Fukui}, {Mizuno}, {Yamaguchi}, {Mizuno},
  {Onishi}, {Ogawa}, {Yonekura}, {Kawamura}, {Tachihara}, {Xiao}, {Yamaguchi},
  {Hara}, {Hayakawa}, {Kato}, {Abe}, {Saito}, {Mano}, {Matsunaga}, {Mine},
  {Moriguchi}, {Aoyama}, {Asayama}, {Yoshikawa}, \& {Rubio}}]{FUKUI99}
{Fukui}, Y., {Mizuno}, N., {Yamaguchi}, R., {et~al.} 1999, \pasj, 51, 745

\bibitem[{{Fukui} {et~al.}(2008){Fukui}, {Kawamura}, {Minamidani}, {Mizuno},
  {Kanai}, {Mizuno}, {Onishi}, {Yonekura}, {Mizuno}, {Ogawa}, \&
  {Rubio}}]{FUKUI08}
{Fukui}, Y., {Kawamura}, A., {Minamidani}, T., {et~al.} 2008, \apjs, 178, 56

\bibitem[{{Garc{\'{\i}}a-Burillo} {et~al.}(2003){Garc{\'{\i}}a-Burillo},
  {Combes}, {Hunt}, {Boone}, {Baker}, {Tacconi}, {Eckart}, {Neri}, {Leon},
  {Schinnerer}, \& {Englmaier}}]{GARCIABURILLO03}
{Garc{\'{\i}}a-Burillo}, S., {Combes}, F., {Hunt}, L.~K., {et~al.} 2003, \aap,
  407, 485

\bibitem[{{Garc{\'\i}a-Burillo} {et~al.}(2014){Garc{\'\i}a-Burillo}, {Combes},
  {Usero}, {Aalto}, {Krips}, {Viti}, {Alonso-Herrero}, {Hunt}, {Schinnerer},
  {Baker}, {Boone}, {Casasola}, {Colina}, {Costagliola}, {Eckart}, {Fuente},
  {Henkel}, {Labiano}, {Mart{\'\i}n}, {M{\'a}rquez}, {Muller}, {Planesas},
  {Ramos Almeida}, {Spaans}, {Tacconi}, \& {van der Werf}}]{GARCIABURILLO14}
{Garc{\'\i}a-Burillo}, S., {Combes}, F., {Usero}, A., {et~al.} 2014, \aap, 567,
  A125

\bibitem[{{Gatto} {et~al.}(2015){Gatto}, {Walch}, {Low}, {Naab}, {Girichidis},
  {Glover}, {W{\"u}nsch}, {Klessen}, {Clark}, {Baczynski}, {Peters},
  {Ostriker}, {Ib{\'a}{\~n}ez-Mej{\'\i}a}, \& {Haid}}]{GATTO15}
{Gatto}, A., {Walch}, S., {Low}, M. M.~M., {et~al.} 2015, \mnras, 449, 1057

\bibitem[{{Geen} {et~al.}(2021){Geen}, {Bieri}, {Rosdahl}, \& {de
  Koter}}]{GEEN21}
{Geen}, S., {Bieri}, R., {Rosdahl}, J., \& {de Koter}, A. 2021, \mnras, 501,
  1352

\bibitem[{{Geen} {et~al.}(2016){Geen}, {Hennebelle}, {Tremblin}, \&
  {Rosdahl}}]{GEEN16}
{Geen}, S., {Hennebelle}, P., {Tremblin}, P., \& {Rosdahl}, J. 2016, \mnras,
  463, 3129

\bibitem[{{Gentry} {et~al.}(2017){Gentry}, {Krumholz}, {Dekel}, \&
  {Madau}}]{GENTRY17}
{Gentry}, E.~S., {Krumholz}, M.~R., {Dekel}, A., \& {Madau}, P. 2017, \mnras,
  465, 2471

\bibitem[{{Ginsburg} {et~al.}(2019){Ginsburg}, {Koch}, {Robitaille},
  {Beaumont}, {Adamginsburg}, {Sip{\H{o}}cz}, {ZuHone}, {Patra}, {Jones},
  {Lim}, {Stern}, {Rosolowsky}, {Earl}, {De Val-Borro}, {Jrobbfed}, {Shuokong},
  {Kepley}, {Sokolov}, {Badger}, {Maret}, {Garrido}, {Booker}, \&
  {Tollerud}}]{SPECTRALCUBE2020}
{Ginsburg}, A., {Koch}, E., {Robitaille}, T., {et~al.} 2019,
  {radio-astro-tools/spectral-cube: Release v0.4.5}, doi:10.5281/zenodo.3558614

\bibitem[{{Grasha} {et~al.}(2018){Grasha}, {Calzetti}, {Bittle}, {Johnson},
  {Donovan Meyer}, {Kennicutt}, {Elmegreen}, {Adamo}, {Krumholz}, {Fumagalli},
  {Grebel}, {Gouliermis}, {Cook}, {Gallagher}, {Aloisi}, {Dale}, {Linden},
  {Sacchi}, {Thilker}, {Walterbos}, {Messa}, {Wofford}, \& {Smith}}]{GRASHA18}
{Grasha}, K., {Calzetti}, D., {Bittle}, L., {et~al.} 2018, \mnras, 481, 1016

\bibitem[{{Gratier} {et~al.}(2010){Gratier}, {Braine}, {Rodriguez-Fernandez},
  {Israel}, {Schuster}, {Brouillet}, \& {Gardan}}]{GRATIER10}
{Gratier}, P., {Braine}, J., {Rodriguez-Fernandez}, N.~J., {et~al.} 2010, \aap,
  512, A68+

\bibitem[{{Gratier} {et~al.}(2012){Gratier}, {Braine}, {Rodriguez-Fernandez},
  {Schuster}, {Kramer}, {Corbelli}, {Combes}, {Brouillet}, {van der Werf}, \&
  {R{\"o}llig}}]{GRATIER12}
---. 2012, \aap, 542, A108

\bibitem[{{Groves} {et~al.}(2012){Groves}, {Krause}, {Sandstrom}, {Schmiedeke},
  {Leroy}, {Linz}, {Kapala}, {Rix}, {Schinnerer}, {Tabatabaei}, {Walter}, \&
  {da Cunha}}]{GROVES12}
{Groves}, B., {Krause}, O., {Sandstrom}, K., {et~al.} 2012, \mnras, 426, 892

\bibitem[{{Grudi{\'c}} {et~al.}(2020){Grudi{\'c}}, {Kruijssen},
  {Faucher-Gigu{\`e}re}, {Hopkins}, {Ma}, {Quataert}, \&
  {Boylan-Kolchin}}]{GRUDIC20}
{Grudi{\'c}}, M.~Y., {Kruijssen}, J.~M.~D., {Faucher-Gigu{\`e}re}, C.-A.,
  {et~al.} 2020, arXiv e-prints, arXiv:2008.04453

\bibitem[{{Hartmann} {et~al.}(2001{\natexlab{a}}){Hartmann},
  {Ballesteros-Paredes}, \& {Bergin}}]{BALLESTEROS01}
{Hartmann}, L., {Ballesteros-Paredes}, J., \& {Bergin}, E.~A.
  2001{\natexlab{a}}, \apj, 562, 852

\bibitem[{{Hartmann} {et~al.}(2001{\natexlab{b}}){Hartmann},
  {Ballesteros-Paredes}, \& {Bergin}}]{HARTMANN01}
---. 2001{\natexlab{b}}, \apj, 562, 852

\bibitem[{{Helfer} {et~al.}(2003){Helfer}, {Thornley}, {Regan}, {Wong},
  {Sheth}, {Vogel}, {Blitz}, \& {Bock}}]{HELFER03}
{Helfer}, T.~T., {Thornley}, M.~D., {Regan}, M.~W., {et~al.} 2003, \apjs, 145,
  259

\bibitem[{{Hennebelle} \& {Chabrier}(2011)}]{HENNEBELLE11}
{Hennebelle}, P., \& {Chabrier}, G. 2011, \apjl, 743, L29

\bibitem[{{Henshaw} {et~al.}(2020){Henshaw}, {Kruijssen}, {Longmore}, {Riener},
  {Leroy}, {Rosolowsky}, {Ginsburg}, {Battersby}, {Chevance}, {Meidt},
  {Glover}, {Hughes}, {Kainulainen}, {Klessen}, {Schinnerer}, {Schruba},
  {Beuther}, {Bigiel}, {Blanc}, {Emsellem}, {Henning}, {Herrera}, {Koch},
  {Pety}, {Ragan}, \& {Sun}}]{HENSHAW20}
{Henshaw}, J.~D., {Kruijssen}, J.~M.~D., {Longmore}, S.~N., {et~al.} 2020,
  Nature Astronomy, 4, 1064

\bibitem[{{Herrera} {et~al.}(2020){Herrera}, {Pety}, {Hughes}, {Meidt},
  {Kreckel}, {Querejeta}, {Saito}, {Lang}, {Jim{\'e}nez-Donaire}, {Pessa},
  {Cormier}, {Usero}, {Sliwa}, {Faesi}, {Blanc}, {Bigiel}, {Chevance}, {Dale},
  {Grasha}, {Glover}, {Hygate}, {Kruijssen}, {Leroy}, {Rosolowsky},
  {Schinnerer}, {Schruba}, {Sun}, \& {Utomo}}]{HERRERA20}
{Herrera}, C.~N., {Pety}, J., {Hughes}, A., {et~al.} 2020, \aap, 634, A121

\bibitem[{{Heyer} \& {Dame}(2015)}]{HEYER15}
{Heyer}, M., \& {Dame}, T.~M. 2015, \araa, 53, 583

\bibitem[{{Heyer} {et~al.}(2001){Heyer}, {Carpenter}, \& {Snell}}]{HEYER01}
{Heyer}, M.~H., {Carpenter}, J.~M., \& {Snell}, R.~L. 2001, \apj, 551, 852

\bibitem[{{Hirota} {et~al.}(2018){Hirota}, {Egusa}, {Baba}, {Kuno}, {Muraoka},
  {Tosaki}, {Miura}, {Nakanishi}, \& {Kawabe}}]{HIROTA18}
{Hirota}, A., {Egusa}, F., {Baba}, J., {et~al.} 2018, \pasj, 70, 73

\bibitem[{{H{\"o}gbom}(1974)}]{HOGBOM74}
{H{\"o}gbom}, J.~A. 1974, \aaps, 15, 417

\bibitem[{{Hopkins}(2013)}]{HOPKINS13d}
{Hopkins}, P.~F. 2013, \mnras, 428, 1950

\bibitem[{{Hopkins} {et~al.}(2012){Hopkins}, {Kere{\v s}}, {Murray},
  {Quataert}, \& {Hernquist}}]{HOPKINS12}
{Hopkins}, P.~F., {Kere{\v s}}, D., {Murray}, N., {Quataert}, E., \&
  {Hernquist}, L. 2012, \mnras, 427, 968

\bibitem[{{Hopkins} {et~al.}(2013){Hopkins}, {Narayanan}, \&
  {Murray}}]{HOPKINS13c}
{Hopkins}, P.~F., {Narayanan}, D., \& {Murray}, N. 2013, \mnras, 432, 2647

\bibitem[{{Huang} {et~al.}(2020){Huang}, {Riess}, {Yuan}, {Macri}, {Zakamska},
  {Casertano}, {Whitelock}, {Hoffmann}, {Filippenko}, \& {Scolnic}}]{HUANG20}
{Huang}, C.~D., {Riess}, A.~G., {Yuan}, W., {et~al.} 2020, \apj, 889, 5

\bibitem[{{Huang} \& {Kauffmann}(2015)}]{HUANG15}
{Huang}, M.-L., \& {Kauffmann}, G. 2015, \mnras, 450, 1375

\bibitem[{{Hughes} {et~al.}(2013{\natexlab{a}}){Hughes}, {Meidt}, {Colombo},
  {Schinnerer}, {Pety}, {Leroy}, {Dobbs}, {Garc{\'{\i}}a-Burillo}, {Thompson},
  {Dumas}, {Schuster}, \& {Kramer}}]{HUGHES13A}
{Hughes}, A., {Meidt}, S.~E., {Colombo}, D., {et~al.} 2013{\natexlab{a}}, \apj,
  779, 46

\bibitem[{{Hughes} {et~al.}(2013{\natexlab{b}}){Hughes}, {Meidt}, {Schinnerer},
  {Colombo}, {Pety}, {Leroy}, {Dobbs}, {Garc{\'{\i}}a-Burillo}, {Thompson},
  {Dumas}, {Schuster}, \& {Kramer}}]{HUGHES13B}
{Hughes}, A., {Meidt}, S.~E., {Schinnerer}, E., {et~al.} 2013{\natexlab{b}},
  \apj, 779, 44

\bibitem[{{Hunt} {et~al.}(2015){Hunt}, {Garc{\'\i}a-Burillo}, {Casasola},
  {Caselli}, {Combes}, {Henkel}, {Lundgren}, {Maiolino}, {Menten}, {Testi}, \&
  {Weiss}}]{HUNT15}
{Hunt}, L.~K., {Garc{\'\i}a-Burillo}, S., {Casasola}, V., {et~al.} 2015, \aap,
  583, A114

\bibitem[{{Ib{\'a}{\~n}ez-Mej{\'\i}a}
  {et~al.}(2016){Ib{\'a}{\~n}ez-Mej{\'\i}a}, {Mac Low}, {Klessen}, \&
  {Baczynski}}]{IBANEZMEJIA16}
{Ib{\'a}{\~n}ez-Mej{\'\i}a}, J.~C., {Mac Low}, M.-M., {Klessen}, R.~S., \&
  {Baczynski}, C. 2016, \apj, 824, 41

\bibitem[{{Inoue} \& {Fukui}(2013)}]{INOUE13}
{Inoue}, T., \& {Fukui}, Y. 2013, \apjl, 774, L31

\bibitem[{{Jackson} {et~al.}(2010){Jackson}, {Finn}, {Chambers}, {Rathborne},
  \& {Simon}}]{JACKSON10}
{Jackson}, J.~M., {Finn}, S.~C., {Chambers}, E.~T., {Rathborne}, J.~M., \&
  {Simon}, R. 2010, \apjl, 719, L185

\bibitem[{{Jameson} {et~al.}(2016){Jameson}, {Bolatto}, {Leroy}, {Meixner},
  {Roman-Duval}, {Gordon}, {Hughes}, {Israel}, {Rubio}, {Indebetouw}, {Madden},
  {Bot}, {Hony}, {Cormier}, {Pellegrini}, {Galametz}, \&
  {Sonneborn}}]{JAMESON16}
{Jameson}, K.~E., {Bolatto}, A.~D., {Leroy}, A.~K., {et~al.} 2016, \apj, 825,
  12

\bibitem[{{Jarrett} {et~al.}(2003){Jarrett}, {Chester}, {Cutri}, {Schneider},
  \& {Huchra}}]{JARRETT03}
{Jarrett}, T.~H., {Chester}, T., {Cutri}, R., {Schneider}, S.~E., \& {Huchra},
  J.~P. 2003, \aj, 125, 525

\bibitem[{{Jarrett} {et~al.}(2013){Jarrett}, {Masci}, {Tsai}, {Petty},
  {Cluver}, {Assef}, {Benford}, {Blain}, {Bridge}, {Donoso}, {Eisenhardt},
  {Koribalski}, {Lake}, {Neill}, {Seibert}, {Sheth}, {Stanford}, \&
  {Wright}}]{JARRETT13}
{Jarrett}, T.~H., {Masci}, F., {Tsai}, C.~W., {et~al.} 2013, \aj, 145, 6

\bibitem[{{Jeffreson} \& {Kruijssen}(2018)}]{JEFFRESON18}
{Jeffreson}, S. M.~R., \& {Kruijssen}, J.~M.~D. 2018, \mnras, 476, 3688

\bibitem[{{Jeffreson} {et~al.}(2020){Jeffreson}, {Kruijssen}, {Keller},
  {Chevance}, \& {Glover}}]{JEFFRESON20}
{Jeffreson}, S. M.~R., {Kruijssen}, J.~M.~D., {Keller}, B.~W., {Chevance}, M.,
  \& {Glover}, S. C.~O. 2020, \mnras, 498, 385

\bibitem[{{Jim{\'e}nez-Donaire} {et~al.}(2019){Jim{\'e}nez-Donaire}, {Bigiel},
  {Leroy}, {Usero}, {Cormier}, {Puschnig}, {Gallagher}, {Kepley}, {Bolatto},
  {Garc{\'\i}a-Burillo}, {Hughes}, {Kramer}, {Pety}, {Schinnerer}, {Schruba},
  {Schuster}, \& {Walter}}]{JIMENEZDONAIRE19}
{Jim{\'e}nez-Donaire}, M.~J., {Bigiel}, F., {Leroy}, A.~K., {et~al.} 2019,
  \apj, 880, 127

\bibitem[{{Jogee} {et~al.}(2005){Jogee}, {Scoville}, \& {Kenney}}]{JOGEE05}
{Jogee}, S., {Scoville}, N., \& {Kenney}, J.~D.~P. 2005, \apj, 630, 837

\bibitem[{{Kannappan} {et~al.}(2013){Kannappan}, {Stark}, {Eckert}, {Moffett},
  {Wei}, {Pisano}, {Baker}, {Vogel}, {Fabricant}, {Laine}, {Norris}, {Jogee},
  {Lepore}, {Hough}, \& {Weinberg-Wolf}}]{KANNAPPAN13}
{Kannappan}, S.~J., {Stark}, D.~V., {Eckert}, K.~D., {et~al.} 2013, \apj, 777,
  42

\bibitem[{{Karachentsev} {et~al.}(2004){Karachentsev}, {Karachentseva},
  {Huchtmeier}, \& {Makarov}}]{KARACHENTSEV04}
{Karachentsev}, I.~D., {Karachentseva}, V.~E., {Huchtmeier}, W.~K., \&
  {Makarov}, D.~I. 2004, \aj, 127, 2031

\bibitem[{{Karim} {et~al.}(2011){Karim}, {Schinnerer},
  {Mart{\'\i}nez-Sansigre}, {Sargent}, {van der Wel}, {Rix}, {Ilbert},
  {Smol{\v{c}}i{\'c}}, {Carilli}, {Pannella}, {Koekemoer}, {Bell}, \&
  {Salvato}}]{KARIM11}
{Karim}, A., {Schinnerer}, E., {Mart{\'\i}nez-Sansigre}, A., {et~al.} 2011,
  \apj, 730, 61

\bibitem[{{Kauffmann} {et~al.}(2003){Kauffmann}, {Heckman}, {White}, {Charlot},
  {Tremonti}, {Brinchmann}, {Bruzual}, {Peng}, {Seibert}, {Bernardi},
  {Blanton}, {Brinkmann}, {Castander}, {Cs{\'a}bai}, {Fukugita}, {Ivezic},
  {Munn}, {Nichol}, {Padmanabhan}, {Thakar}, {Weinberg}, \&
  {York}}]{KAUFFMANN03}
{Kauffmann}, G., {Heckman}, T.~M., {White}, S. D.~M., {et~al.} 2003, \mnras,
  341, 33

\bibitem[{{Kawamura} {et~al.}(2009){Kawamura}, {Mizuno}, {Minamidani},
  {Filipovi{\'c}}, {Staveley-Smith}, {Kim}, {Mizuno}, {Onishi}, {Mizuno}, \&
  {Fukui}}]{KAWAMURA09}
{Kawamura}, A., {Mizuno}, Y., {Minamidani}, T., {et~al.} 2009, \apjs, 184, 1

\bibitem[{{Keller} \& {Kruijssen}(2020)}]{KELLER20}
{Keller}, B.~W., \& {Kruijssen}, J.~M.~D. 2020, arXiv e-prints,
  arXiv:2004.03608

\bibitem[{{Kenney} {et~al.}(1992){Kenney}, {Wilson}, {Scoville}, {Devereux}, \&
  {Young}}]{KENNEY92}
{Kenney}, J.~D.~P., {Wilson}, C.~D., {Scoville}, N.~Z., {Devereux}, N.~A., \&
  {Young}, J.~S. 1992, \apjl, 395, L79

\bibitem[{{Kennicutt} \& {Evans}(2012)}]{KENNICUTT12}
{Kennicutt}, R.~C., \& {Evans}, N.~J. 2012, \araa, 50, 531

\bibitem[{{Kennicutt}(1989)}]{KENNICUTT89}
{Kennicutt}, Jr., R.~C. 1989, \apj, 344, 685

\bibitem[{{Kennicutt} {et~al.}(2007){Kennicutt}, {Calzetti}, {Walter}, {Helou},
  {Hollenbach}, {Armus}, {Bendo}, {Dale}, {Draine}, {Engelbracht}, {Gordon},
  {Prescott}, {Regan}, {Thornley}, {Bot}, {Brinks}, {de Blok}, {de Mello},
  {Meyer}, {Moustakas}, {Murphy}, {Sheth}, \& {Smith}}]{KENNICUTT07}
{Kennicutt}, Jr., R.~C., {Calzetti}, D., {Walter}, F., {et~al.} 2007, \apj,
  671, 333

\bibitem[{{Kim} \& {Ostriker}(2017)}]{KIMOSTRIKER17}
{Kim}, C.-G., \& {Ostriker}, E.~C. 2017, \apj, 846, 133

\bibitem[{{Kim} {et~al.}(2013){Kim}, {Ostriker}, \& {Kim}}]{KIM13}
{Kim}, C.-G., {Ostriker}, E.~C., \& {Kim}, W.-T. 2013, \apj, 776, 1

\bibitem[{{Kim} {et~al.}(2020{\natexlab{a}}){Kim}, {Chevance}, {Kruijssen},
  {Schruba}, {Sandstrom}, {Hygate}, {Barnes}, {Bigiel}, {Blanc}, {Cao}, {Dale},
  {Faesi}, {Glover}, {Grasha}, {Groves}, {Herrera}, {Klessen}, {Kreckel},
  {Lee}, {Leroy}, {Pety}, {Querejeta}, {Schinnerer}, {Sun}, {Usero}, \&
  {Williams}}]{KIM21}
{Kim}, J., {Chevance}, M., {Kruijssen}, J.~M.~D., {et~al.} 2020{\natexlab{a}},
  \mnras\ submitted, arXiv:2012.00019

\bibitem[{{Kim} {et~al.}(2018){Kim}, {Kim}, \& {Ostriker}}]{JGKIM18}
{Kim}, J.-G., {Kim}, W.-T., \& {Ostriker}, E.~C. 2018, \apj, 859, 68

\bibitem[{{Kim} {et~al.}(2019){Kim}, {Kim}, \& {Ostriker}}]{JGKIM19}
---. 2019, \apj, 883, 102

\bibitem[{{Kim} {et~al.}(2020{\natexlab{b}}){Kim}, {Ostriker}, \&
  {Filippova}}]{JGKIM21}
{Kim}, J.-G., {Ostriker}, E.~C., \& {Filippova}, N. 2020{\natexlab{b}}, arXiv
  e-prints, arXiv:2011.07772

\bibitem[{{Kim} {et~al.}(2020{\natexlab{c}}){Kim}, {Kim}, \&
  {Ostriker}}]{WTKIM20}
{Kim}, W.-T., {Kim}, C.-G., \& {Ostriker}, E.~C. 2020{\natexlab{c}}, \apj, 898,
  35

\bibitem[{{Klessen} \& {Glover}(2016)}]{KLESSEN16}
{Klessen}, R.~S., \& {Glover}, S. C.~O. 2016, Saas-Fee Advanced Course, 43, 85

\bibitem[{{Koch} {et~al.}(2018){Koch}, {Rosolowsky}, \& {Leroy}}]{KOCH18B}
{Koch}, E., {Rosolowsky}, E., \& {Leroy}, A.~K. 2018, Research Notes of the
  American Astronomical Society, 2, 220

\bibitem[{{Koch} \& {Rosolowsky}(2015)}]{KOCH15}
{Koch}, E.~W., \& {Rosolowsky}, E.~W. 2015, \mnras, 452, 3435

\bibitem[{{Koda} {et~al.}(2009){Koda}, {Scoville}, {Sawada}, {La Vigne},
  {Vogel}, {Potts}, {Carpenter}, {Corder}, {Wright}, {White}, {Zauderer},
  {Patience}, {Sargent}, {Bock}, {Hawkins}, {Hodges}, {Kemball}, {Lamb},
  {Plambeck}, {Pound}, {Scott}, {Teuben}, \& {Woody}}]{KODA09}
{Koda}, J., {Scoville}, N., {Sawada}, T., {et~al.} 2009, \apjl, 700, L132

\bibitem[{{Koda} {et~al.}(2020){Koda}, {Sawada}, {Sakamoto}, {Hirota}, {Egusa},
  {Boissier}, {Calzetti}, {Meyer}, {Elmegreen}, {de Paz}, {Harada}, {Ho},
  {Kobayashi}, {Kuno}, {Mart{\'\i}n}, {Muraoka}, {Nakanishi}, {Scoville},
  {Seibert}, {Vlahakis}, \& {Watanabe}}]{KODA20}
{Koda}, J., {Sawada}, T., {Sakamoto}, K., {et~al.} 2020, \apjl, 890, L10

\bibitem[{{Kourkchi} {et~al.}(2020){Kourkchi}, {Courtois}, {Graziani},
  {Hoffman}, {Pomar{\`e}de}, {Shaya}, \& {Tully}}]{KOURKCHI20}
{Kourkchi}, E., {Courtois}, H.~M., {Graziani}, R., {et~al.} 2020, \aj, 159, 67

\bibitem[{{Kourkchi} \& {Tully}(2017)}]{KOURKCHI17}
{Kourkchi}, E., \& {Tully}, R.~B. 2017, \apj, 843, 16

\bibitem[{{Kreckel} {et~al.}(2016){Kreckel}, {Blanc}, {Schinnerer}, {Groves},
  {Adamo}, {Hughes}, \& {Meidt}}]{KRECKEL16}
{Kreckel}, K., {Blanc}, G.~A., {Schinnerer}, E., {et~al.} 2016, \apj, 827, 103

\bibitem[{{Kreckel} {et~al.}(2018){Kreckel}, {Faesi}, {Kruijssen}, {Schruba},
  {Groves}, {Leroy}, {Bigiel}, {Blanc}, {Chevance}, {Herrera}, {Hughes},
  {McElroy}, {Pety}, {Querejeta}, {Rosolowsky}, {Schinnerer}, {Sun}, {Usero},
  \& {Utomo}}]{KRECKEL18}
{Kreckel}, K., {Faesi}, C., {Kruijssen}, J.~M.~D., {et~al.} 2018, \apjl, 863,
  L21

\bibitem[{{Kregel} {et~al.}(2002){Kregel}, {van der Kruit}, \& {de
  Grijs}}]{KREGEL02}
{Kregel}, M., {van der Kruit}, P.~C., \& {de Grijs}, R. 2002, \mnras, 334, 646

\bibitem[{{Kruijssen}(2012)}]{KRUIJSSEN12}
{Kruijssen}, J.~M.~D. 2012, \mnras, 426, 3008

\bibitem[{{Kruijssen} \& {Longmore}(2014)}]{KRUIJSSEN14}
{Kruijssen}, J.~M.~D., \& {Longmore}, S.~N. 2014, \mnras, 439, 3239

\bibitem[{{Kruijssen} {et~al.}(2018){Kruijssen}, {Schruba}, {Hygate}, {Hu},
  {Haydon}, \& {Longmore}}]{KRUIJSSEN18}
{Kruijssen}, J.~M.~D., {Schruba}, A., {Hygate}, A. e. P.~S., {et~al.} 2018,
  \mnras, 479, 1866

\bibitem[{{Kruijssen} {et~al.}(2019){Kruijssen}, {Schruba}, {Chevance},
  {Longmore}, {Hygate}, {Haydon}, {McLeod}, {Dalcanton}, {Tacconi}, \& {van
  Dishoeck}}]{KRUIJSSEN19}
{Kruijssen}, J.~M.~D., {Schruba}, A., {Chevance}, M., {et~al.} 2019, \nat, 569,
  519

\bibitem[{{Krumholz}(2014)}]{KRUMHOLZ14b}
{Krumholz}, M.~R. 2014, \physrep, 539, 49

\bibitem[{{Krumholz} \& {Dekel}(2012)}]{KRUMHOLZ12}
{Krumholz}, M.~R., \& {Dekel}, A. 2012, \apj, 753, 16

\bibitem[{{Krumholz} \& {McKee}(2005)}]{KRUMHOLZ05}
{Krumholz}, M.~R., \& {McKee}, C.~F. 2005, \apj, 630, 250

\bibitem[{{Krumholz} \& {McKee}(2020)}]{KRUMHOLZ20}
---. 2020, \mnras, 494, 624

\bibitem[{{Krumholz} {et~al.}(2019){Krumholz}, {McKee}, \&
  {Bland-Hawthorn}}]{KRUMHOLZ19}
{Krumholz}, M.~R., {McKee}, C.~F., \& {Bland-Hawthorn}, J. 2019, \araa, 57, 227

\bibitem[{{Kuno} {et~al.}(2007){Kuno}, {Sato}, {Nakanishi}, {Hirota}, {Tosaki},
  {Shioya}, {Sorai}, {Nakai}, {Nishiyama}, \& {Vila-Vilar{\'o}}}]{KUNO07}
{Kuno}, N., {Sato}, N., {Nakanishi}, H., {et~al.} 2007, \pasj, 59, 117

\bibitem[{{La Vigne} {et~al.}(2006){La Vigne}, {Vogel}, \&
  {Ostriker}}]{LAVIGNE06}
{La Vigne}, M.~A., {Vogel}, S.~N., \& {Ostriker}, E.~C. 2006, \apj, 650, 818

\bibitem[{{Landsman}(1993)}]{IDLASTRO}
{Landsman}, W.~B. 1993, in Astronomical Society of the Pacific Conference
  Series, Vol.~52, Astronomical Data Analysis Software and Systems II, ed.
  R.~J. {Hanisch}, R.~J.~V. {Brissenden}, \& J.~{Barnes}, 246

\bibitem[{{Lang}(2014)}]{LANG14}
{Lang}, D. 2014, \aj, 147, 108

\bibitem[{{Lang} {et~al.}(2020){Lang}, {Meidt}, {Rosolowsky}, {Nofech},
  {Schinnerer}, {Leroy}, {Emsellem}, {Pessa}, {Glover}, {Groves}, {Hughes},
  {Kruijssen}, {Querejeta}, {Schruba}, {Bigiel}, {Blanc}, {Chevance},
  {Colombo}, {Faesi}, {Henshaw}, {Herrera}, {Liu}, {Pety}, {Puschnig}, {Saito},
  {Sun}, \& {Usero}}]{LANG20}
{Lang}, P., {Meidt}, S.~E., {Rosolowsky}, E., {et~al.} 2020, \apj, 897, 122

\bibitem[{{Lange} {et~al.}(2015){Lange}, {Driver}, {Robotham}, {Kelvin},
  {Graham}, {Alpaslan}, {Andrews}, {Baldry}, {Bamford}, {Bland-Hawthorn},
  {Brough}, {Cluver}, {Conselice}, {Davies}, {Haeussler}, {Konstantopoulos},
  {Loveday}, {Moffett}, {Norberg}, {Phillipps}, {Taylor},
  {L{\'o}pez-S{\'a}nchez}, \& {Wilkins}}]{LANGE15}
{Lange}, R., {Driver}, S.~P., {Robotham}, A. S.~G., {et~al.} 2015, \mnras, 447,
  2603

\bibitem[{{Lee} {et~al.}(2016){Lee}, {Miville-Desch{\^e}nes}, \&
  {Murray}}]{LEE16}
{Lee}, E.~J., {Miville-Desch{\^e}nes}, M.-A., \& {Murray}, N.~W. 2016, \apj,
  833, 229

\bibitem[{{Lee} {et~al.}(2021){Lee}, {Whitmore}, {Thilker}, {Deger}, {Larson},
  {Ubeda}, {Anand}, {Boquien}, {Chandar}, {Dale}, {Emsellem}, {Leroy},
  {Rosolowsky}, {Schinnerer}, {Schmidt}, {Turner}, {Van Dyk}, {White},
  {Barnes}, {Belfiore}, {Bigiel}, {Blanc}, {Cao}, {Chevance}, {Congiu},
  {Egorov}, {Glover}, {Grasha}, {Groves}, {Henshaw}, {Hughes}, {Klessen},
  {Koch}, {Kreckel}, {Kruijssen}, {Liu}, {Lopez}, {Mayker}, {Meidt}, {Murphy},
  {Pan}, {Pety}, {Querejeta}, {Razza}, {Saito}, {Sanchez-Blazquez}, {Santoro},
  {Sardone}, {Scheuermann}, {Schruba}, {Sun}, {Usero}, {Watkins}, \&
  {Williams}}]{LEE21}
{Lee}, J.~C., {Whitmore}, B.~C., {Thilker}, D.~A., {et~al.} 2021, \apjs\
  submitted, arXiv:2101.02855

\bibitem[{{Leonard} {et~al.}(2003){Leonard}, {Kanbur}, {Ngeow}, \&
  {Tanvir}}]{LEONARD03}
{Leonard}, D.~C., {Kanbur}, S.~M., {Ngeow}, C.~C., \& {Tanvir}, N.~R. 2003,
  \apj, 594, 247

\bibitem[{{Leroy} {et~al.}(2006){Leroy}, {Bolatto}, {Walter}, \&
  {Blitz}}]{LEROY06}
{Leroy}, A., {Bolatto}, A., {Walter}, F., \& {Blitz}, L. 2006, \apj, 643, 825

\bibitem[{{Leroy} {et~al.}(2008){Leroy}, {Walter}, {Brinks}, {Bigiel}, {de
  Blok}, {Madore}, \& {Thornley}}]{LEROY08}
{Leroy}, A.~K., {Walter}, F., {Brinks}, E., {et~al.} 2008, \aj, 136, 2782

\bibitem[{{Leroy} {et~al.}(2009){Leroy}, {Bolatto}, {Bot}, {Engelbracht},
  {Gordon}, {Israel}, {Rubio}, {Sandstrom}, \& {Stanimirovi{\'c}}}]{LEROY09}
{Leroy}, A.~K., {Bolatto}, A., {Bot}, C., {et~al.} 2009, \apj, 702, 352

\bibitem[{{Leroy} {et~al.}(2012){Leroy}, {Bigiel}, {de Blok}, {Boissier},
  {Bolatto}, {Brinks}, {Madore}, {Munoz-Mateos}, {Murphy}, {Sandstrom},
  {Schruba}, \& {Walter}}]{LEROY12}
{Leroy}, A.~K., {Bigiel}, F., {de Blok}, W.~J.~G., {et~al.} 2012, ArXiv
  e-prints, arXiv:1202.2873

\bibitem[{{Leroy} {et~al.}(2013{\natexlab{a}}){Leroy}, {Lee}, {Schruba},
  {Bolatto}, {Hughes}, {Pety}, {Sandstrom}, {Schinnerer}, \&
  {Walter}}]{LEROY13B}
{Leroy}, A.~K., {Lee}, C., {Schruba}, A., {et~al.} 2013{\natexlab{a}}, \apjl,
  769, L12

\bibitem[{{Leroy} {et~al.}(2013{\natexlab{b}}){Leroy}, {Walter}, {Sandstrom},
  {Schruba}, {Munoz-Mateos}, {Bigiel}, {Bolatto}, {Brinks}, {de Blok}, {Meidt},
  {Rix}, {Rosolowsky}, {Schinnerer}, {Schuster}, \& {Usero}}]{LEROY13}
{Leroy}, A.~K., {Walter}, F., {Sandstrom}, K., {et~al.} 2013{\natexlab{b}},
  \aj, 146, 19

\bibitem[{{Leroy} {et~al.}(2016){Leroy}, {Hughes}, {Schruba}, {Rosolowsky},
  {Blanc}, {Bolatto}, {Colombo}, {Escala}, {Kramer}, {Kruijssen}, {Meidt},
  {Pety}, {Querejeta}, {Sandstrom}, {Schinnerer}, {Sliwa}, \&
  {Usero}}]{LEROY16}
{Leroy}, A.~K., {Hughes}, A., {Schruba}, A., {et~al.} 2016, \apj, 831, 16

\bibitem[{{Leroy} {et~al.}(2017){Leroy}, {Schinnerer}, {Hughes}, {Kruijssen},
  {Meidt}, {Schruba}, {Sun}, {Bigiel}, {Aniano}, {Blanc}, {Bolatto},
  {Chevance}, {Colombo}, {Gallagher}, {Garcia-Burillo}, {Kramer}, {Querejeta},
  {Pety}, {Thompson}, \& {Usero}}]{LEROY17A}
{Leroy}, A.~K., {Schinnerer}, E., {Hughes}, A., {et~al.} 2017, \apj, 846, 71

\bibitem[{{Leroy} {et~al.}(2019){Leroy}, {Sandstrom}, {Lang}, {Lewis}, {Salim},
  {Behrens}, {Chastenet}, {Chiang}, {Gallagher}, {Kessler}, \&
  {Utomo}}]{LEROY19}
{Leroy}, A.~K., {Sandstrom}, K.~M., {Lang}, D., {et~al.} 2019, \apjs, 244, 24

\bibitem[{{Leroy} {et~al.}(2021){Leroy}, {Hughes}, {Liu}, {Pety}, {Rosolowsky},
  {Saito}, {Schinnerer}, {Schruba}, {Usero}, {Faesi}, {Herrera}, {Chevance},
  {Hygate}, {Kepley}, {Koch}, {Querejeta}, {Sliwa}, {Will}, {Wilson}, {Anand},
  {Barnes}, {Belfiore}, {Be\v{s}li\'c}, {Bigiel}, {Blanc}, {Cao}, {Chandar},
  {Mok}, {Congiu}, {Dale}, {den Brok}, {Eibensteiner}, {Emsellem},
  {Garc\'{i}a-Rodr\'{i}guez}, {Glover}, {Grasha}, {Groves}, {Henshaw}, {Kim},
  {Klessen}, {Kreckel}, {Kruijssen}, {Larson}, {Lee}, {Mayker}, {Meidt}, {Pan},
  {Puschnig}, {Razza}, {S\'anchez-Bl'azquez}, {Sandstrom}, {Santoro},
  {Sardone}, {Scheuermann}, {Sun}, {Thilker}, {Ubeda}, {Utomo}, {Watkins}, \&
  {Williams}}]{LEROY21a}
{Leroy}, A.~K., {Hughes}, A., {Liu}, D., {et~al.} 2021, \apjs~submitted

\bibitem[{{Leslie} {et~al.}(2020){Leslie}, {Schinnerer}, {Liu}, {Magnelli},
  {Algera}, {Karim}, {Davidzon}, {Gozaliasl}, {Jim{\'e}nez-Andrade}, {Lang},
  {Sargent}, {Novak}, {Groves}, {Smol{\v{c}}i{\'c}}, {Zamorani}, {Vaccari},
  {Battisti}, {Vardoulaki}, {Peng}, \& {Kartaltepe}}]{LESLIE20}
{Leslie}, S.~K., {Schinnerer}, E., {Liu}, D., {et~al.} 2020, \apj, 899, 58

\bibitem[{{Lin} {et~al.}(2019){Lin}, {Pan}, {Ellison}, {Belfiore}, {Shi},
  {S{\'a}nchez}, {Hsieh}, {Rowland s}, {Ramya}, {Thorp}, {Li}, \&
  {Maiolino}}]{LIN19}
{Lin}, L., {Pan}, H.-A., {Ellison}, S.~L., {et~al.} 2019, \apjl, 884, L33

\bibitem[{{Lisenfeld} {et~al.}(2011){Lisenfeld}, {Espada}, {Verdes-Montenegro},
  {Kuno}, {Leon}, {Sabater}, {Sato}, {Sulentic}, {Verley}, \&
  {Yun}}]{LISENFELD11}
{Lisenfeld}, U., {Espada}, D., {Verdes-Montenegro}, L., {et~al.} 2011, \aap,
  534, A102

\bibitem[{{Longmore} {et~al.}(2013){Longmore}, {Bally}, {Testi}, {Purcell},
  {Walsh}, {Bressert}, {Pestalozzi}, {Molinari}, {Ott}, {Cortese}, {Battersby},
  {Murray}, {Lee}, {Kruijssen}, {Schisano}, \& {Elia}}]{LONGMORE13}
{Longmore}, S.~N., {Bally}, J., {Testi}, L., {et~al.} 2013, \mnras, 429, 987

\bibitem[{{Lopez} {et~al.}(2011){Lopez}, {Krumholz}, {Bolatto}, {Prochaska}, \&
  {Ramirez-Ruiz}}]{LOPEZ11}
{Lopez}, L.~A., {Krumholz}, M.~R., {Bolatto}, A.~D., {Prochaska}, J.~X., \&
  {Ramirez-Ruiz}, E. 2011, \apj, 731, 91

\bibitem[{{Lopez} {et~al.}(2014){Lopez}, {Krumholz}, {Bolatto}, {Prochaska},
  {Ramirez-Ruiz}, \& {Castro}}]{LOPEZ14}
{Lopez}, L.~A., {Krumholz}, M.~R., {Bolatto}, A.~D., {et~al.} 2014, \apj, 795,
  121

\bibitem[{{Lynds}(1970)}]{LYNDS70}
{Lynds}, B.~T. 1970, in IAU Symposium, Vol.~38, The Spiral Structure of our
  Galaxy, ed. W.~{Becker} \& G.~I. {Kontopoulos}, 26

\bibitem[{{Mac Low} \& {Klessen}(2004)}]{MACLOW04}
{Mac Low}, M.-M., \& {Klessen}, R.~S. 2004, Reviews of Modern Physics, 76, 125

\bibitem[{{Makarov} {et~al.}(2014){Makarov}, {Prugniel}, {Terekhova},
  {Courtois}, \& {Vauglin}}]{MAKAROV14}
{Makarov}, D., {Prugniel}, P., {Terekhova}, N., {Courtois}, H., \& {Vauglin},
  I. 2014, \aap, 570, A13

\bibitem[{{Martin} \& {Kennicutt}(2001)}]{MARTIN01}
{Martin}, C.~L., \& {Kennicutt}, Robert~C., J. 2001, \apj, 555, 301

\bibitem[{{McGaugh} \& {Schombert}(2014)}]{MCGAUGH14}
{McGaugh}, S.~S., \& {Schombert}, J.~M. 2014, \aj, 148, 77

\bibitem[{{McKee} \& {Ostriker}(2007)}]{MCKEE07}
{McKee}, C.~F., \& {Ostriker}, E.~C. 2007, \araa, 45, 565

\bibitem[{{McMullin} {et~al.}(2007){McMullin}, {Waters}, {Schiebel}, {Young},
  \& {Golap}}]{MCMULLIN07}
{McMullin}, J.~P., {Waters}, B., {Schiebel}, D., {Young}, W., \& {Golap}, K.
  2007, in Astronomical Society of the Pacific Conference Series, Vol. 376,
  Astronomical Data Analysis Software and Systems XVI, ed. R.~A. {Shaw},
  F.~{Hill}, \& D.~J. {Bell}, 127

\bibitem[{{Meidt} {et~al.}(2012){Meidt}, {Schinnerer}, {Knapen}, {Bosma},
  {Athanassoula}, {Sheth}, {Buta}, {Zaritsky}, {Laurikainen}, {Elmegreen},
  {Elmegreen}, {Gadotti}, {Salo}, {Regan}, {Ho}, {Madore}, {Hinz}, {Skibba},
  {Gil de Paz}, {Mu{\~n}oz-Mateos}, {Men{\'e}ndez-Delmestre}, {Seibert}, {Kim},
  {Mizusawa}, {Laine}, \& {Comer{\'o}n}}]{MEIDT12}
{Meidt}, S.~E., {Schinnerer}, E., {Knapen}, J.~H., {et~al.} 2012, \apj, 744, 17

\bibitem[{{Meidt} {et~al.}(2013){Meidt}, {Schinnerer}, {Garc{\'{\i}}a-Burillo},
  {Hughes}, {Colombo}, {Pety}, {Dobbs}, {Schuster}, {Kramer}, {Leroy}, {Dumas},
  \& {Thompson}}]{MEIDT13}
{Meidt}, S.~E., {Schinnerer}, E., {Garc{\'{\i}}a-Burillo}, S., {et~al.} 2013,
  \apj, 779, 45

\bibitem[{{Meidt} {et~al.}(2014){Meidt}, {Schinnerer}, {van de Ven},
  {Zaritsky}, {Peletier}, {Knapen}, {Sheth}, {Regan}, {Querejeta},
  {Mu{\~n}oz-Mateos}, {Kim}, {Hinz}, {Gil de Paz}, {Athanassoula}, {Bosma},
  {Buta}, {Cisternas}, {Ho}, {Holwerda}, {Skibba}, {Laurikainen}, {Salo},
  {Gadotti}, {Laine}, {Erroz-Ferrer}, {Comer{\'o}n}, {Men{\'e}ndez-Delmestre},
  {Seibert}, \& {Mizusawa}}]{MEIDT14}
{Meidt}, S.~E., {Schinnerer}, E., {van de Ven}, G., {et~al.} 2014, \apj, 788,
  144

\bibitem[{{Meidt} {et~al.}(2015){Meidt}, {Hughes}, {Dobbs}, {Pety}, {Thompson},
  {Garc{\'{\i}}a-Burillo}, {Leroy}, {Schinnerer}, {Colombo}, {Querejeta},
  {Kramer}, {Schuster}, \& {Dumas}}]{MEIDT15}
{Meidt}, S.~E., {Hughes}, A., {Dobbs}, C.~L., {et~al.} 2015, \apj, 806, 72

\bibitem[{{Meidt} {et~al.}(2018){Meidt}, {Leroy}, {Rosolowsky}, {Kruijssen},
  {Schinnerer}, {Schruba}, {Pety}, {Blanc}, {Bigiel}, {Chevance}, {Hughes},
  {Querejeta}, \& {Usero}}]{MEIDT18}
{Meidt}, S.~E., {Leroy}, A.~K., {Rosolowsky}, E., {et~al.} 2018, \apj, 854, 100

\bibitem[{{Meidt} {et~al.}(2020){Meidt}, {Glover}, {Kruijssen}, {Leroy},
  {Rosolowsky}, {Hughes}, {Schinnerer}, {Schruba}, {Usero}, {Bigiel}, {Blanc},
  {Chevance}, {Pety}, {Querejeta}, \& {Utomo}}]{MEIDT20}
{Meidt}, S.~E., {Glover}, S. C.~O., {Kruijssen}, J.~M.~D., {et~al.} 2020, \apj,
  892, 73

\bibitem[{{Milam} {et~al.}(2005){Milam}, {Savage}, {Brewster}, {Ziurys}, \&
  {Wyckoff}}]{MILAM05}
{Milam}, S.~N., {Savage}, C., {Brewster}, M.~A., {Ziurys}, L.~M., \& {Wyckoff},
  S. 2005, \apj, 634, 1126

\bibitem[{{Miville-Desch{\^e}nes} {et~al.}(2017){Miville-Desch{\^e}nes},
  {Murray}, \& {Lee}}]{MIVILLE17}
{Miville-Desch{\^e}nes}, M.-A., {Murray}, N., \& {Lee}, E.~J. 2017, \apj, 834,
  57

\bibitem[{{Mizuno} {et~al.}(2001){Mizuno}, {Rubio}, {Mizuno}, {Yamaguchi},
  {Onishi}, \& {Fukui}}]{MIZUNO01}
{Mizuno}, N., {Rubio}, M., {Mizuno}, A., {et~al.} 2001, \pasj, 53, L45

\bibitem[{{Momose} {et~al.}(2013){Momose}, {Koda}, {Kennicutt}, {Egusa},
  {Calzetti}, {Liu}, {Donovan Meyer}, {Okumura}, {Scoville}, {Sawada}, \&
  {Kuno}}]{MOMOSE13}
{Momose}, R., {Koda}, J., {Kennicutt}, Jr., R.~C., {et~al.} 2013, \apjl, 772,
  L13

\bibitem[{{Mu{\~n}oz-Mateos} {et~al.}(2015){Mu{\~n}oz-Mateos}, {Sheth},
  {Regan}, {Kim}, {Laine}, {Erroz-Ferrer}, {Gil de Paz}, {Comeron}, {Hinz},
  {Laurikainen}, {Salo}, {Athanassoula}, {Bosma}, {Bouquin}, {Schinnerer},
  {Ho}, {Zaritsky}, {Gadotti}, {Madore}, {Holwerda}, {Men{\'e}ndez-Delmestre},
  {Knapen}, {Meidt}, {Querejeta}, {Mizusawa}, {Seibert}, {Laine}, \&
  {Courtois}}]{MUNOZMATEOS15}
{Mu{\~n}oz-Mateos}, J.~C., {Sheth}, K., {Regan}, M., {et~al.} 2015, \apjs, 219,
  3

\bibitem[{{Murphy} {et~al.}(2011){Murphy}, {Condon}, {Schinnerer}, {Kennicutt},
  {Calzetti}, {Armus}, {Helou}, {Turner}, {Aniano}, {Beir{\~a}o}, {Bolatto},
  {Brandl}, {Croxall}, {Dale}, {Donovan Meyer}, {Draine}, {Engelbracht},
  {Hunt}, {Hao}, {Koda}, {Roussel}, {Skibba}, \& {Smith}}]{MURPHY11}
{Murphy}, E.~J., {Condon}, J.~J., {Schinnerer}, E., {et~al.} 2011, \apj, 737,
  67

\bibitem[{{Murray}(2011)}]{MURRAY11}
{Murray}, N. 2011, \apj, 729, 133

\bibitem[{{Nieten} {et~al.}(2006){Nieten}, {Neininger}, {Gu{\'e}lin},
  {Ungerechts}, {Lucas}, {Berkhuijsen}, {Beck}, \& {Wielebinski}}]{NIETEN06}
{Nieten}, C., {Neininger}, N., {Gu{\'e}lin}, M., {et~al.} 2006, \aap, 453, 459

\bibitem[{{Noeske} {et~al.}(2007){Noeske}, {Weiner}, {Faber}, {Papovich},
  {Koo}, {Somerville}, {Bundy}, {Conselice}, {Newman}, {Schiminovich}, {Le
  Floc'h}, {Coil}, {Rieke}, {Lotz}, {Primack}, {Barmby}, {Cooper}, {Davis},
  {Ellis}, {Fazio}, {Guhathakurta}, {Huang}, {Kassin}, {Martin}, {Phillips},
  {Rich}, {Small}, {Willmer}, \& {Wilson}}]{NOESKE07}
{Noeske}, K.~G., {Weiner}, B.~J., {Faber}, S.~M., {et~al.} 2007, \apjl, 660,
  L43

\bibitem[{{Nugent} {et~al.}(2006){Nugent}, {Sullivan}, {Ellis}, {Gal-Yam},
  {Leonard}, {Howell}, {Astier}, {Carlberg}, {Conley}, {Fabbro}, {Fouchez},
  {Neill}, {Pain}, {Perrett}, {Pritchet}, \& {Regnault}}]{NUGENT06}
{Nugent}, P., {Sullivan}, M., {Ellis}, R., {et~al.} 2006, \apj, 645, 841

\bibitem[{{Ochsendorf} {et~al.}(2017){Ochsendorf}, {Meixner}, {Roman-Duval},
  {Rahman}, \& {Evans}}]{OCHSENDORF17}
{Ochsendorf}, B.~B., {Meixner}, M., {Roman-Duval}, J., {Rahman}, M., \&
  {Evans}, Neal~J., I. 2017, \apj, 841, 109

\bibitem[{Oliphant(2006)}]{NUMPY2006}
Oliphant, T.~E. 2006, A guide to NumPy, Vol.~1 (Trelgol Publishing USA)

\bibitem[{{Onodera} {et~al.}(2010){Onodera}, {Kuno}, {Tosaki}, {Kohno},
  {Nakanishi}, {Sawada}, {Muraoka}, {Komugi}, {Miura}, {Kaneko}, {Hirota}, \&
  {Kawabe}}]{ONODERA10}
{Onodera}, S., {Kuno}, N., {Tosaki}, T., {et~al.} 2010, \apjl, 722, L127

\bibitem[{{Onodera} {et~al.}(2012){Onodera}, {Kuno}, {Tosaki}, {Muraoka},
  {Miura}, {Kohno}, {Nakanishi}, {Sawada}, {Komugi}, {Kaneko}, {Hirota}, \&
  {Kawabe}}]{ONODERA12}
---. 2012, \pasj, 64, 133

\bibitem[{{Orr} {et~al.}(2018){Orr}, {Hayward}, {Hopkins}, {Chan},
  {Faucher-Gigu{\`e}re}, {Feldmann}, {Kere{\v{s}}}, {Murray}, \&
  {Quataert}}]{ORR18}
{Orr}, M.~E., {Hayward}, C.~C., {Hopkins}, P.~F., {et~al.} 2018, \mnras, 478,
  3653

\bibitem[{{Ostriker} {et~al.}(2010){Ostriker}, {McKee}, \&
  {Leroy}}]{OSTRIKER10}
{Ostriker}, E.~C., {McKee}, C.~F., \& {Leroy}, A.~K. 2010, \apj, 721, 975

\bibitem[{{Padoan} {et~al.}(2012){Padoan}, {Haugb{\o}lle}, \&
  {Nordlund}}]{PADOAN12}
{Padoan}, P., {Haugb{\o}lle}, T., \& {Nordlund}, {\AA}. 2012, \apjl, 759, L27

\bibitem[{{Padoan} {et~al.}(2017){Padoan}, {Haugb{\o}lle}, {Nordlund}, \&
  {Frimann}}]{PADOAN17}
{Padoan}, P., {Haugb{\o}lle}, T., {Nordlund}, {\r{A}}., \& {Frimann}, S. 2017,
  \apj, 840, 48

\bibitem[{{Padoan} \& {Nordlund}(2002)}]{PADOAN02}
{Padoan}, P., \& {Nordlund}, {\AA}. 2002, \apj, 576, 870

\bibitem[{{Padoan} {et~al.}(2016){Padoan}, {Pan}, {Haugb{\o}lle}, \&
  {Nordlund}}]{PADOAN16}
{Padoan}, P., {Pan}, L., {Haugb{\o}lle}, T., \& {Nordlund}, {\AA}. 2016, \apj,
  822, 11

\bibitem[{{Paturel} {et~al.}(2003){Paturel}, {Petit}, {Prugniel}, {Theureau},
  {Rousseau}, {Brouty}, {Dubois}, \& {Cambr{\'e}sy}}]{PATUREL03}
{Paturel}, G., {Petit}, C., {Prugniel}, P., {et~al.} 2003, \aap, 412, 45

\bibitem[{{Pety}(2005)}]{PETY2005}
{Pety}, J. 2005, in SF2A-2005: Semaine de l'Astrophysique Francaise, ed.
  F.~{Casoli}, T.~{Contini}, J.~M. {Hameury}, \& L.~{Pagani}, 721--722

\bibitem[{{Pety} {et~al.}(2013){Pety}, {Schinnerer}, {Leroy}, {Hughes},
  {Meidt}, {Colombo}, {Dumas}, {Garc{\'{\i}}a-Burillo}, {Schuster}, {Kramer},
  {Dobbs}, \& {Thompson}}]{PETY13}
{Pety}, J., {Schinnerer}, E., {Leroy}, A.~K., {et~al.} 2013, \apj, 779, 43

\bibitem[{{Pierce} {et~al.}(1994){Pierce}, {Welch}, {McClure}, {van den Bergh},
  {Racine}, \& {Stetson}}]{PIERCE94}
{Pierce}, M.~J., {Welch}, D.~L., {McClure}, R.~D., {et~al.} 1994, \nat, 371,
  385

\bibitem[{{Querejeta} {et~al.}(2015){Querejeta}, {Meidt}, {Schinnerer},
  {Cisternas}, {Mu{\~n}oz-Mateos}, {Sheth}, {Knapen}, {van de Ven}, {Norris},
  {Peletier}, {Laurikainen}, {Salo}, {Holwerda}, {Athanassoula}, {Bosma},
  {Groves}, {Ho}, {Gadotti}, {Zaritsky}, {Regan}, {Hinz}, {Gil de Paz},
  {Menendez-Delmestre}, {Seibert}, {Mizusawa}, {Kim}, {Erroz-Ferrer}, {Laine},
  \& {Comer{\'o}n}}]{QUEREJETA15}
{Querejeta}, M., {Meidt}, S.~E., {Schinnerer}, E., {et~al.} 2015, \apjs, 219, 5

\bibitem[{{R Core Team}(2015)}]{RMANUAL}
{R Core Team}. 2015, R: A Language and Environment for Statistical Computing, R
  Foundation for Statistical Computing, Vienna, Austria

\bibitem[{{Rahman} {et~al.}(2011){Rahman}, {Bolatto}, {Wong}, {Leroy},
  {Walter}, {Rosolowsky}, {West}, {Bigiel}, {Ott}, {Xue}, {Herrera-Camus},
  {Jameson}, {Blitz}, \& {Vogel}}]{RAHMAN11}
{Rahman}, N., {Bolatto}, A.~D., {Wong}, T., {et~al.} 2011, \apj, 730, 72

\bibitem[{{Rahman} {et~al.}(2012){Rahman}, {Bolatto}, {Xue}, {Wong}, {Leroy},
  {Walter}, {Bigiel}, {Rosolowsky}, {Fisher}, {Vogel}, {Blitz}, {West}, \&
  {Ott}}]{RAHMAN12}
{Rahman}, N., {Bolatto}, A.~D., {Xue}, R., {et~al.} 2012, \apj, 745, 183

\bibitem[{{Rahner} {et~al.}(2017){Rahner}, {Pellegrini}, {Glover}, \&
  {Klessen}}]{RAHNER17}
{Rahner}, D., {Pellegrini}, E.~W., {Glover}, S. C.~O., \& {Klessen}, R.~S.
  2017, \mnras, 470, 4453

\bibitem[{{Rahner} {et~al.}(2019){Rahner}, {Pellegrini}, {Glover}, \&
  {Klessen}}]{RAHNER19}
---. 2019, \mnras, 483, 2547

\bibitem[{{Raskutti} {et~al.}(2016){Raskutti}, {Ostriker}, \&
  {Skinner}}]{RASKUTTI16}
{Raskutti}, S., {Ostriker}, E.~C., \& {Skinner}, M.~A. 2016, \apj, 829, 130

\bibitem[{{Raskutti} {et~al.}(2017){Raskutti}, {Ostriker}, \&
  {Skinner}}]{RASKUTTI17}
---. 2017, \apj, 850, 112

\bibitem[{{Regan} {et~al.}(2001){Regan}, {Thornley}, {Helfer}, {Sheth}, {Wong},
  {Vogel}, {Blitz}, \& {Bock}}]{REGAN01}
{Regan}, M.~W., {Thornley}, M.~D., {Helfer}, T.~T., {et~al.} 2001, \apj, 561,
  218

\bibitem[{{Reissl} {et~al.}(2018){Reissl}, {Klessen}, {Mac Low}, \&
  {Pellegrini}}]{REISSL18}
{Reissl}, S., {Klessen}, R.~S., {Mac Low}, M.-M., \& {Pellegrini}, E.~W. 2018,
  \aap, 611, A70

\bibitem[{{Rice} {et~al.}(2016){Rice}, {Goodman}, {Bergin}, {Beaumont}, \&
  {Dame}}]{RICE16}
{Rice}, T.~S., {Goodman}, A.~A., {Bergin}, E.~A., {Beaumont}, C., \& {Dame},
  T.~M. 2016, \apj, 822, 52

\bibitem[{{Roman-Duval} {et~al.}(2016){Roman-Duval}, {Heyer}, {Brunt}, {Clark},
  {Klessen}, \& {Shetty}}]{ROMANDUVAL16}
{Roman-Duval}, J., {Heyer}, M., {Brunt}, C.~M., {et~al.} 2016, \apj, 818, 144

\bibitem[{{Roman-Duval} {et~al.}(2010){Roman-Duval}, {Israel}, {Bolatto},
  {Hughes}, {Leroy}, {Meixner}, {Gordon}, {Madden}, {Paradis}, {Kawamura},
  {Li}, {Sauvage}, {Wong}, {Bernard}, {Engelbracht}, {Hony}, {Kim}, {Misselt},
  {Okumura}, {Ott}, {Panuzzo}, {Pineda}, {Reach}, \& {Rubio}}]{ROMANDUVAL10}
{Roman-Duval}, J., {Israel}, F.~P., {Bolatto}, A., {et~al.} 2010, \aap, 518,
  L74+

\bibitem[{{Rosolowsky}(2007)}]{ROSOLOWSKY07}
{Rosolowsky}, E. 2007, \apj, 654, 240

\bibitem[{{Rosolowsky} \& {Blitz}(2005)}]{ROSOLOWSKY05}
{Rosolowsky}, E., \& {Blitz}, L. 2005, \apj, 623, 826

\bibitem[{{Rosolowsky} {et~al.}(2003){Rosolowsky}, {Engargiola}, {Plambeck}, \&
  {Blitz}}]{ROSOLOWSKY03}
{Rosolowsky}, E., {Engargiola}, G., {Plambeck}, R., \& {Blitz}, L. 2003, \apj,
  599, 258

\bibitem[{{Rosolowsky} {et~al.}(2007){Rosolowsky}, {Keto}, {Matsushita}, \&
  {Willner}}]{ROSOLOWSKY07B}
{Rosolowsky}, E., {Keto}, E., {Matsushita}, S., \& {Willner}, S.~P. 2007, \apj,
  661, 830

\bibitem[{{Rosolowsky} \& {Leroy}(2006)}]{ROSOLOWSKY06}
{Rosolowsky}, E., \& {Leroy}, A. 2006, \pasp, 118, 590

\bibitem[{{Rosolowsky} {et~al.}(2021){Rosolowsky}, {Hughes}, {Leroy}, {Sun},
  {Querejeta}, {Schruba}, {Usero}, {Herrera}, {Liu}, {Pety}, {Saito},
  {Be{\v{s}}li{\'c}}, {Bigiel}, {Blanc}, {Chevance}, {Dale}, {Deger}, {Faesi},
  {Glover}, {Henshaw}, {Klessen}, {Kruijssen}, {Larson}, {Lee}, {Meidt}, {Mok},
  {Schinnerer}, {Thilker}, \& {Williams}}]{ROSOLOWSKY21}
{Rosolowsky}, E., {Hughes}, A., {Leroy}, A.~K., {et~al.} 2021, \mnras~accepted,
  arXiv:2101.04697

\bibitem[{{Rubio} {et~al.}(2015){Rubio}, {Elmegreen}, {Hunter}, {Brinks},
  {Cort{\'e}s}, \& {Cigan}}]{RUBIO15}
{Rubio}, M., {Elmegreen}, B.~G., {Hunter}, D.~A., {et~al.} 2015, \nat, 525, 218

\bibitem[{{Ruiz-Lapuente}(1996)}]{RUIZ96}
{Ruiz-Lapuente}, P. 1996, \apjl, 465, L83

\bibitem[{{Saintonge} {et~al.}(2011){Saintonge}, {Kauffmann}, {Wang}, {Kramer},
  {Tacconi}, {Buchbender}, {Catinella}, {Graci{\'a}-Carpio}, {Cortese},
  {Fabello}, {Fu}, {Genzel}, {Giovanelli}, {Guo}, {Haynes}, {Heckman},
  {Krumholz}, {Lemonias}, {Li}, {Moran}, {Rodriguez-Fernandez}, {Schiminovich},
  {Schuster}, \& {Sievers}}]{SAINTONGE11}
{Saintonge}, A., {Kauffmann}, G., {Wang}, J., {et~al.} 2011, \mnras, 415, 61

\bibitem[{{Saintonge} {et~al.}(2016){Saintonge}, {Catinella}, {Cortese},
  {Genzel}, {Giovanelli}, {Haynes}, {Janowiecki}, {Kramer}, {Lutz},
  {Schiminovich}, {Tacconi}, {Wuyts}, \& {Accurso}}]{SAINTONGE16}
{Saintonge}, A., {Catinella}, B., {Cortese}, L., {et~al.} 2016, \mnras, 462,
  1749

\bibitem[{{Saintonge} {et~al.}(2017){Saintonge}, {Catinella}, {Tacconi},
  {Kauffmann}, {Genzel}, {Cortese}, {Dav{\'e}}, {Fletcher},
  {Graci{\'a}-Carpio}, {Kramer}, {Heckman}, {Janowiecki}, {Lutz}, {Rosario},
  {Schiminovich}, {Schuster}, {Wang}, {Wuyts}, {Borthakur}, {Lamperti}, \&
  {Roberts-Borsani}}]{SAINTONGE17}
{Saintonge}, A., {Catinella}, B., {Tacconi}, L.~J., {et~al.} 2017, \apjs, 233,
  22

\bibitem[{{Saintonge} {et~al.}(2018){Saintonge}, {Wilson}, {Xiao}, {Lin},
  {Hwang}, {Tosaki}, {Bureau}, {Cigan}, {Clark}, {Clements}, {De Looze},
  {Dharmawardena}, {Gao}, {Gear}, {Greenslade}, {Lamperti}, {Lee}, {Li},
  {Micha{\l}owski}, {Mok}, {Pan}, {Sansom}, {Sargent}, {Smith}, {Williams},
  {Yang}, {Zhu}, {Accurso}, {Barmby}, {Brinks}, {Bourne}, {Brown}, {Chung},
  {Chung}, {Cibinel}, {Coppin}, {Davies}, {Davis}, {Eales}, {Fanciullo},
  {Fang}, {Gao}, {Glass}, {Gomez}, {Greve}, {He}, {Ho}, {Huang}, {Jeong},
  {Jiang}, {Jiao}, {Kemper}, {Kim}, {Kim}, {Kim}, {Ko}, {Kong}, {Lacaille},
  {Lacey}, {Lee}, {Lee}, {Lee}, {Masters}, {Oh}, {Papadopoulos}, {Park},
  {Park}, {Parsons}, {Rowland s}, {Scicluna}, {Scudder}, {Sethuram},
  {Serjeant}, {Shao}, {Sheen}, {Shi}, {Shim}, {Smith}, {Spekkens}, {Tsai},
  {Verma}, {Urquhart}, {Violino}, {Viti}, {Wake}, {Wang}, {Wouterloot}, {Yang},
  {Yim}, {Yuan}, \& {Zheng}}]{SAINTONGE18}
{Saintonge}, A., {Wilson}, C.~D., {Xiao}, T., {et~al.} 2018, \mnras, 481, 3497

\bibitem[{{Sakamoto} {et~al.}(1999){Sakamoto}, {Okumura}, {Ishizuki}, \&
  {Scoville}}]{SAKAMOTO99A}
{Sakamoto}, K., {Okumura}, S.~K., {Ishizuki}, S., \& {Scoville}, N.~Z. 1999,
  \apjs, 124, 403

\bibitem[{{Salim} {et~al.}(2018){Salim}, {Boquien}, \& {Lee}}]{SALIM18}
{Salim}, S., {Boquien}, M., \& {Lee}, J.~C. 2018, \apj, 859, 11

\bibitem[{{Salim} {et~al.}(2007){Salim}, {Rich}, {Charlot}, {Brinchmann},
  {Johnson}, {Schiminovich}, {Seibert}, {Mallery}, {Heckman}, {Forster},
  {Friedman}, {Martin}, {Morrissey}, {Neff}, {Small}, {Wyder}, {Bianchi},
  {Donas}, {Lee}, {Madore}, {Milliard}, {Szalay}, {Welsh}, \& {Yi}}]{SALIM07}
{Salim}, S., {Rich}, R.~M., {Charlot}, S., {et~al.} 2007, \apjs, 173, 267

\bibitem[{{Salim} {et~al.}(2016){Salim}, {Lee}, {Janowiecki}, {da Cunha},
  {Dickinson}, {Boquien}, {Burgarella}, {Salzer}, \& {Charlot}}]{SALIM16}
{Salim}, S., {Lee}, J.~C., {Janowiecki}, S., {et~al.} 2016, \apjs, 227, 2

\bibitem[{{Salo} {et~al.}(2015){Salo}, {Laurikainen}, {Laine}, {Comer{\'o}n},
  {Gadotti}, {Buta}, {Sheth}, {Zaritsky}, {Ho}, {Knapen}, {Athanassoula},
  {Bosma}, {Laine}, {Cisternas}, {Kim}, {Mu{\~n}oz-Mateos}, {Regan}, {Hinz},
  {Gil de Paz}, {Menendez-Delmestre}, {Mizusawa}, {Erroz-Ferrer}, {Meidt}, \&
  {Querejeta}}]{SALO15}
{Salo}, H., {Laurikainen}, E., {Laine}, J., {et~al.} 2015, \apjs, 219, 4

\bibitem[{{S{\'a}nchez} {et~al.}(2014){S{\'a}nchez}, {Rosales-Ortega},
  {Iglesias-P{\'a}ramo}, {Moll{\'a}}, {Barrera-Ballesteros}, {Marino},
  {P{\'e}rez}, {S{\'a}nchez-Blazquez}, {Gonz{\'a}lez Delgado}, {Cid Fernand
  es}, {de Lorenzo-C{\'a}ceres}, {Mendez-Abreu}, {Galbany}, {Falcon-Barroso},
  {Miralles-Caballero}, {Husemann}, {Garc{\'\i}a-Benito}, {Mast}, {Walcher},
  {Gil de Paz}, {Garc{\'\i}a-Lorenzo}, {Jungwiert}, {V{\'\i}lchez},
  {J{\'\i}lkov{\'a}}, {Lyubenova}, {Cortijo-Ferrero}, {D{\'\i}az}, {Wisotzki},
  {M{\'a}rquez}, {Bland-Hawthorn}, {Ellis}, {van de Ven}, {Jahnke},
  {Papaderos}, {Gomes}, {Mendoza}, \& {L{\'o}pez-S{\'a}nchez}}]{SANCHEZ14}
{S{\'a}nchez}, S.~F., {Rosales-Ortega}, F.~F., {Iglesias-P{\'a}ramo}, J.,
  {et~al.} 2014, \aap, 563, A49

\bibitem[{{Sanders} {et~al.}(2003){Sanders}, {Mazzarella}, {Kim}, {Surace}, \&
  {Soifer}}]{SANDERS03}
{Sanders}, D.~B., {Mazzarella}, J.~M., {Kim}, D., {Surace}, J.~A., \& {Soifer},
  B.~T. 2003, \aj, 126, 1607

\bibitem[{{Sanders} \& {Mirabel}(1996)}]{SANDERS96}
{Sanders}, D.~B., \& {Mirabel}, I.~F. 1996, \araa, 34, 749

\bibitem[{{Schinnerer} {et~al.}(2013){Schinnerer}, {Meidt}, {Pety}, {Hughes},
  {Colombo}, {Garc{\'{\i}}a-Burillo}, {Schuster}, {Dumas}, {Dobbs}, {Leroy},
  {Kramer}, {Thompson}, \& {Regan}}]{SCHINNERER13}
{Schinnerer}, E., {Meidt}, S.~E., {Pety}, J., {et~al.} 2013, \apj, 779, 42

\bibitem[{{Schinnerer} {et~al.}(2017){Schinnerer}, {Meidt}, {Colombo},
  {Chandar}, {Dobbs}, {Garc{\'{\i}}a-Burillo}, {Hughes}, {Leroy}, {Pety},
  {Querejeta}, {Kramer}, \& {Schuster}}]{SCHINNERER17}
{Schinnerer}, E., {Meidt}, S.~E., {Colombo}, D., {et~al.} 2017, \apj, 836, 62

\bibitem[{{Schinnerer} {et~al.}(2019){Schinnerer}, {Hughes}, {Leroy}, {Groves},
  {Blanc}, {Kreckel}, {Bigiel}, {Chevance}, {Dale}, {Emsellem}, {Faesi},
  {Glover}, {Grasha}, {Henshaw}, {Hygate}, {Kruijssen}, {Meidt}, {Pety},
  {Querejeta}, {Rosolowsky}, {Saito}, {Schruba}, {Sun}, \&
  {Utomo}}]{SCHINNERER19}
{Schinnerer}, E., {Hughes}, A., {Leroy}, A., {et~al.} 2019, \apj, 887, 49

\bibitem[{{Schmidt} {et~al.}(2016){Schmidt}, {Bigiel}, {Klessen}, \& {de
  Blok}}]{SCHMIDT16}
{Schmidt}, T.~M., {Bigiel}, F., {Klessen}, R.~S., \& {de Blok}, W.~J.~G. 2016,
  \mnras, 457, 2642

\bibitem[{{Schruba} {et~al.}(2018){Schruba}, {Bialy}, \&
  {Sternberg}}]{SCHRUBA18}
{Schruba}, A., {Bialy}, S., \& {Sternberg}, A. 2018, \apj, 862, 110

\bibitem[{{Schruba} {et~al.}(2019){Schruba}, {Kruijssen}, \&
  {Leroy}}]{SCHRUBA19}
{Schruba}, A., {Kruijssen}, J.~M.~D., \& {Leroy}, A.~K. 2019, \apj, 883, 2

\bibitem[{{Schruba} {et~al.}(2021){Schruba}, {Leroy}, {Bolatto}, {Dalcanton},
  {Sandstrom}, {Scoville}, {Walter}, \& {Weisz}}]{SCHRUBA21}
{Schruba}, A., {Leroy}, A.~K., {Bolatto}, A.~D., {et~al.} 2021, \apj\ to be
  subm

\bibitem[{{Schruba} {et~al.}(2010){Schruba}, {Leroy}, {Walter}, {Sandstrom}, \&
  {Rosolowsky}}]{SCHRUBA10}
{Schruba}, A., {Leroy}, A.~K., {Walter}, F., {Sandstrom}, K., \& {Rosolowsky},
  E. 2010, \apj, 722, 1699

\bibitem[{{Schruba} {et~al.}(2011){Schruba}, {Leroy}, {Walter}, {Bigiel},
  {Brinks}, {de Blok}, {Dumas}, {Kramer}, {Rosolowsky}, {Sandstrom},
  {Schuster}, {Usero}, {Weiss}, \& {Wiesemeyer}}]{SCHRUBA11}
{Schruba}, A., {Leroy}, A.~K., {Walter}, F., {et~al.} 2011, \aj, 142, 37

\bibitem[{{Schruba} {et~al.}(2012){Schruba}, {Leroy}, {Walter}, {Bigiel},
  {Brinks}, {de Blok}, {Kramer}, {Rosolowsky}, {Sandstrom}, {Schuster},
  {Usero}, {Weiss}, \& {Wiesemeyer}}]{SCHRUBA12}
---. 2012, \aj, 143, 138

\bibitem[{{Schruba} {et~al.}(2017){Schruba}, {Leroy}, {Kruijssen}, {Bigiel},
  {Bolatto}, {de Blok}, {Tacconi}, {van Dishoeck}, \& {Walter}}]{SCHRUBA17}
{Schruba}, A., {Leroy}, A.~K., {Kruijssen}, J.~M.~D., {et~al.} 2017, \apj, 835,
  278

\bibitem[{{Schuster} {et~al.}(2007){Schuster}, {Kramer}, {Hitschfeld},
  {Gar\'c{\i}a-Burillo}, \& {Mookerjea}}]{SCHUSTER07}
{Schuster}, K.~F., {Kramer}, C., {Hitschfeld}, M., {Gar\'c{\i}a-Burillo}, S.,
  \& {Mookerjea}, B. 2007, \aap, 461, 143

\bibitem[{{Semenov} {et~al.}(2018){Semenov}, {Kravtsov}, \&
  {Gnedin}}]{SEMENOV18}
{Semenov}, V.~A., {Kravtsov}, A.~V., \& {Gnedin}, N.~Y. 2018, \apj, 861, 4

\bibitem[{{Shaya} {et~al.}(2017){Shaya}, {Tully}, {Hoffman}, \&
  {Pomar{\`e}de}}]{SHAYA17}
{Shaya}, E.~J., {Tully}, R.~B., {Hoffman}, Y., \& {Pomar{\`e}de}, D. 2017,
  \apj, 850, 207

\bibitem[{{Shen} {et~al.}(2003){Shen}, {Mo}, {White}, {Blanton}, {Kauffmann},
  {Voges}, {Brinkmann}, \& {Csabai}}]{SHEN03}
{Shen}, S., {Mo}, H.~J., {White}, S. D.~M., {et~al.} 2003, \mnras, 343, 978

\bibitem[{{Sheth} {et~al.}(2005){Sheth}, {Vogel}, {Regan}, {Thornley}, \&
  {Teuben}}]{SHETH05}
{Sheth}, K., {Vogel}, S.~N., {Regan}, M.~W., {Thornley}, M.~D., \& {Teuben},
  P.~J. 2005, \apj, 632, 217

\bibitem[{{Sheth} {et~al.}(2010){Sheth}, {Regan}, {Hinz}, {Gil de Paz},
  {Men{\'e}ndez-Delmestre}, {Mu{\~n}oz-Mateos}, {Seibert}, {Kim},
  {Laurikainen}, {Salo}, {Gadotti}, {Laine}, {Mizusawa}, {Armus},
  {Athanassoula}, {Bosma}, {Buta}, {Capak}, {Jarrett}, {Elmegreen},
  {Elmegreen}, {Knapen}, {Koda}, {Helou}, {Ho}, {Madore}, {Masters},
  {Mobasher}, {Ogle}, {Peng}, {Schinnerer}, {Surace}, {Zaritsky},
  {Comer{\'o}n}, {de Swardt}, {Meidt}, {Kasliwal}, \& {Aravena}}]{SHETH10}
{Sheth}, K., {Regan}, M., {Hinz}, J.~L., {et~al.} 2010, \pasp, 122, 1397

\bibitem[{{Silk}(1997)}]{SILK97}
{Silk}, J. 1997, \apj, 481, 703

\bibitem[{{Simonian} \& {Martini}(2017)}]{SIMONIAN17}
{Simonian}, G.~V., \& {Martini}, P. 2017, \mnras, 464, 3920

\bibitem[{{Solomon} {et~al.}(1987){Solomon}, {Rivolo}, {Barrett}, \&
  {Yahil}}]{SOLOMON87}
{Solomon}, P.~M., {Rivolo}, A.~R., {Barrett}, J., \& {Yahil}, A. 1987, \apj,
  319, 730

\bibitem[{{Somerville} \& {Dav{\'e}}(2015)}]{SOMERVILLE15}
{Somerville}, R.~S., \& {Dav{\'e}}, R. 2015, \araa, 53, 51

\bibitem[{{Sorai} {et~al.}(2019){Sorai}, {Kuno}, {Muraoka}, {Miyamoto},
  {Kaneko}, {Nakanishi}, {Nakai}, {Yanagitani}, {Tanaka}, {Sato}, {Salak},
  {Umei}, {Morokuma-Matsui}, {Matsumoto}, {Ueno}, {Pan}, {Noma}, {Takeuchi},
  {Yoda}, {Kuroda}, {Yasuda}, {Yajima}, {Oi}, {Shibata}, {Seta}, {Watanabe},
  {Kita}, {Komatsuzaki}, {Kajikawa}, {Yashima}, {Cooray}, {Baji}, {Segawa},
  {Tashiro}, {Takeda}, {Kishida}, {Hatakeyama}, {Tomiyasu}, \&
  {Saita}}]{SORAI19}
{Sorai}, K., {Kuno}, N., {Muraoka}, K., {et~al.} 2019, \pasj, 125

\bibitem[{{Sun} {et~al.}(2018){Sun}, {Leroy}, {Schruba}, {Rosolowsky},
  {Hughes}, {Kruijssen}, {Meidt}, {Schinnerer}, {Blanc}, {Bigiel}, {Bolatto},
  {Chevance}, {Groves}, {Herrera}, {Hygate}, {Pety}, {Querejeta}, {Usero}, \&
  {Utomo}}]{SUN18}
{Sun}, J., {Leroy}, A.~K., {Schruba}, A., {et~al.} 2018, \apj, 860, 172

\bibitem[{{Sun} {et~al.}(2020{\natexlab{a}}){Sun}, {Leroy}, {Ostriker},
  {Hughes}, {Rosolowsky}, {Schruba}, {Schinnerer}, {Blanc}, {Faesi},
  {Kruijssen}, {Meidt}, {Utomo}, {Bigiel}, {Bolatto}, {Chevance}, {Chiang},
  {Dale}, {Emsellem}, {Glover}, {Grasha}, {Henshaw}, {Herrera},
  {Jimenez-Donaire}, {Lee}, {Pety}, {Querejeta}, {Saito}, {Sandstrom}, \&
  {Usero}}]{SUN20}
{Sun}, J., {Leroy}, A.~K., {Ostriker}, E.~C., {et~al.} 2020{\natexlab{a}},
  \apj, 892, 148

\bibitem[{{Sun} {et~al.}(2020{\natexlab{b}}){Sun}, {Leroy}, {Schinnerer},
  {Hughes}, {Rosolowsky}, {Querejeta}, {Schruba}, {Liu}, {Saito}, {Herrera},
  {Faesi}, {Usero}, {Pety}, {Kruijssen}, {Ostriker}, {Bigiel}, {Blanc},
  {Bolatto}, {Boquien}, {Chevance}, {Dale}, {Deger}, {Emsellem}, {Glover},
  {Grasha}, {Groves}, {Henshaw}, {Jimenez-Donaire}, {Kim}, {Klessen},
  {Kreckel}, {Lee}, {Meidt}, {Sandstrom}, {Sardone}, {Utomo}, \&
  {Williams}}]{SUN20B}
{Sun}, J., {Leroy}, A.~K., {Schinnerer}, E., {et~al.} 2020{\natexlab{b}},
  \apjl, 901, L8

\bibitem[{{Tan}(2000)}]{TAN00}
{Tan}, J.~C. 2000, \apj, 536, 173

\bibitem[{{Telford} {et~al.}(2020){Telford}, {Dalcanton}, {Williams}, {Bell},
  {Dolphin}, {Durbin}, \& {Choi}}]{TELFORD20}
{Telford}, O.~G., {Dalcanton}, J.~J., {Williams}, B.~F., {et~al.} 2020, \apj,
  891, 32

\bibitem[{{Thompson} \& {Krumholz}(2016)}]{THOMPSON16}
{Thompson}, T.~A., \& {Krumholz}, M.~R. 2016, \mnras, 455, 334

\bibitem[{{Thompson} {et~al.}(2005){Thompson}, {Quataert}, \&
  {Murray}}]{THOMPSON05}
{Thompson}, T.~A., {Quataert}, E., \& {Murray}, N. 2005, \apj, 630, 167

\bibitem[{{Tonry} {et~al.}(2001){Tonry}, {Dressler}, {Blakeslee}, {Ajhar},
  {Fletcher}, {Luppino}, {Metzger}, \& {Moore}}]{TONRY01}
{Tonry}, J.~L., {Dressler}, A., {Blakeslee}, J.~P., {et~al.} 2001, \apj, 546,
  681

\bibitem[{{Tremonti} {et~al.}(2004){Tremonti}, {Heckman}, {Kauffmann},
  {Brinchmann}, {Charlot}, {White}, {Seibert}, {Peng}, {Schlegel}, {Uomoto},
  {Fukugita}, \& {Brinkmann}}]{TREMONTI04}
{Tremonti}, C.~A., {Heckman}, T.~M., {Kauffmann}, G., {et~al.} 2004, \apj, 613,
  898

\bibitem[{{Tress} {et~al.}(2020{\natexlab{a}}){Tress}, {Smith}, {Sormani},
  {Glover}, {Klessen}, {Mac Low}, \& {Clark}}]{TRESS20}
{Tress}, R.~G., {Smith}, R.~J., {Sormani}, M.~C., {et~al.} 2020{\natexlab{a}},
  \mnras, 492, 2973

\bibitem[{{Tress} {et~al.}(2020{\natexlab{b}}){Tress}, {Sormani}, {Smith},
  {Glover}, {Klessen}, {Mac Low}, {Clark}, \& {Duarte-Cabral}}]{TRESS20b}
{Tress}, R.~G., {Sormani}, M.~C., {Smith}, R.~J., {et~al.} 2020{\natexlab{b}},
  arXiv e-prints, arXiv:2012.05919

\bibitem[{{Tully} {et~al.}(2016){Tully}, {Courtois}, \& {Sorce}}]{TULLY16}
{Tully}, R.~B., {Courtois}, H.~M., \& {Sorce}, J.~G. 2016, \aj, 152, 50

\bibitem[{{Tully} {et~al.}(2009){Tully}, {Rizzi}, {Shaya}, {Courtois},
  {Makarov}, \& {Jacobs}}]{TULLY09}
{Tully}, R.~B., {Rizzi}, L., {Shaya}, E.~J., {et~al.} 2009, \aj, 138, 323

\bibitem[{{Utomo} {et~al.}(2017){Utomo}, {Bolatto}, {Wong}, {Ostriker},
  {Blitz}, {Sanchez}, {Colombo}, {Leroy}, {Cao}, {Dannerbauer},
  {Garcia-Benito}, {Husemann}, {Kalinova}, {Levy}, {Mast}, {Rosolowsky}, \&
  {Vogel}}]{UTOMO17}
{Utomo}, D., {Bolatto}, A.~D., {Wong}, T., {et~al.} 2017, \apj, 849, 26

\bibitem[{{Utomo} {et~al.}(2018){Utomo}, {Sun}, {Leroy}, {Kruijssen},
  {Schinnerer}, {Schruba}, {Bigiel}, {Blanc}, {Chevance}, {Emsellem},
  {Herrera}, {Hygate}, {Kreckel}, {Ostriker}, {Pety}, {Querejeta},
  {Rosolowsky}, {Sandstrom}, \& {Usero}}]{UTOMO18}
{Utomo}, D., {Sun}, J., {Leroy}, A.~K., {et~al.} 2018, \apjl, 861, L18

\bibitem[{{Vazdekis} {et~al.}(2016){Vazdekis}, {Koleva}, {Ricciardelli},
  {R{\"o}ck}, \& {Falc{\'o}n-Barroso}}]{VAZDEKIS16}
{Vazdekis}, A., {Koleva}, M., {Ricciardelli}, E., {R{\"o}ck}, B., \&
  {Falc{\'o}n-Barroso}, J. 2016, \mnras, 463, 3409

\bibitem[{{Vazquez-Semadeni}(1994)}]{VAZQUEZSEMADENI94}
{Vazquez-Semadeni}, E. 1994, \apj, 423, 681

\bibitem[{{Virtanen} {et~al.}(2020){Virtanen}, {Gommers}, {Oliphant},
  {Haberland}, {Reddy}, {Cournapeau}, {Burovski}, {Peterson}, {Weckesser},
  {Bright}, {van der Walt}, {Brett}, {Wilson}, {Jarrod Millman}, {Mayorov},
  {Nelson}, {Jones}, {Kern}, {Larson}, {Carey}, {Polat}, {Feng}, {Moore}, {Vand
  erPlas}, {Laxalde}, {Perktold}, {Cimrman}, {Henriksen}, {Quintero}, {Harris},
  {Archibald}, {Ribeiro}, {Pedregosa}, {van Mulbregt}, \&
  {Contributors}}]{SCIPY2020}
{Virtanen}, P., {Gommers}, R., {Oliphant}, T.~E., {et~al.} 2020, Nature
  Methods, 17, 261

\bibitem[{{Vutisalchavakul} {et~al.}(2016){Vutisalchavakul}, {Evans}, \&
  {Heyer}}]{VUTISALCHAVUAKUL16}
{Vutisalchavakul}, N., {Evans}, II, N.~J., \& {Heyer}, M. 2016, \apj, 831, 73

\bibitem[{{Walch} {et~al.}(2015){Walch}, {Girichidis}, {Naab}, {Gatto},
  {Glover}, {W{\"u}nsch}, {Klessen}, {Clark}, {Peters}, {Derigs}, \&
  {Baczynski}}]{WALCH15}
{Walch}, S., {Girichidis}, P., {Naab}, T., {et~al.} 2015, \mnras, 454, 238

\bibitem[{{Watkins} {et~al.}(2016){Watkins}, {Mihos}, \& {Harding}}]{WATKINS16}
{Watkins}, A.~E., {Mihos}, J.~C., \& {Harding}, P. 2016, \apj, 826, 59

\bibitem[{{Weigel} {et~al.}(2016){Weigel}, {Schawinski}, \&
  {Bruderer}}]{WEIGEL16}
{Weigel}, A.~K., {Schawinski}, K., \& {Bruderer}, C. 2016, \mnras, 459, 2150

\bibitem[{{Wilson} {et~al.}(2012){Wilson}, {Warren}, {Israel}, {Serjeant},
  {Attewell}, {Bendo}, {Butner}, {Chanial}, {Clements}, {Golding}, {Heesen},
  {Irwin}, {Leech}, {Matthews}, {M{\"u}hle}, {Mortier}, {Petitpas},
  {S{\'a}nchez-Gallego}, {Sinukoff}, {Shorten}, {Tan}, {Tilanus}, {Usero},
  {Vaccari}, {Wiegert}, {Zhu}, {Alexander}, {Alexander}, {Azimlu}, {Barmby},
  {Brar}, {Bridge}, {Brinks}, {Brooks}, {Coppin}, {C{\^o}t{\'e}},
  {C{\^o}t{\'e}}, {Courteau}, {Davies}, {Eales}, {Fich}, {Hudson}, {Hughes},
  {Ivison}, {Knapen}, {Page}, {Parkin}, {Rigopoulou}, {Rosolowsky}, {Seaquist},
  {Spekkens}, {Tanvir}, {van der Hulst}, {van der Werf}, {Vlahakis}, {Webb},
  {Weferling}, \& {White}}]{WILSON12}
{Wilson}, C.~D., {Warren}, B.~E., {Israel}, F.~P., {et~al.} 2012, \mnras, 424,
  3050

\bibitem[{{Wilson} \& {Rood}(1994{\natexlab{a}})}]{WILSON94}
{Wilson}, T.~L., \& {Rood}, R. 1994{\natexlab{a}}, \araa, 32, 191

\bibitem[{{Wilson} \& {Rood}(1994{\natexlab{b}})}]{WILSONROOD94}
---. 1994{\natexlab{b}}, \araa, 32, 191

\bibitem[{{Wong} \& {Blitz}(2002)}]{WONG02}
{Wong}, T., \& {Blitz}, L. 2002, \apj, 569, 157

\bibitem[{{Wong} {et~al.}(2011){Wong}, {Hughes}, {Ott}, {Muller}, {Pineda},
  {Bernard}, {Chu}, {Fukui}, {Gruendl}, {Henkel}, {Kawamura}, {Klein},
  {Looney}, {Maddison}, {Mizuno}, {Paradis}, {Seale}, \& {Welty}}]{WONG11}
{Wong}, T., {Hughes}, A., {Ott}, J., {et~al.} 2011, \apjs, 197, 16

\bibitem[{{Wong} {et~al.}(2013){Wong}, {Xue}, {Bolatto}, {Leroy}, {Blitz},
  {Rosolowsky}, {Bigiel}, {Fisher}, {Ott}, {Rahman}, {Vogel}, \&
  {Walter}}]{WONG13}
{Wong}, T., {Xue}, R., {Bolatto}, A.~D., {et~al.} 2013, \apjl, 777, L4

\bibitem[{{Wong} {et~al.}(2019){Wong}, {Hughes}, {Tokuda}, {Indebetouw},
  {Onishi}, {Bandurski}, {Chen}, {Fukui}, {Glover}, {Klessen}, {Pineda},
  {Roman-Duval}, {Sewi{\l}o}, {Wojciechowski}, \& {Zahorecz}}]{WONG19}
{Wong}, T., {Hughes}, A., {Tokuda}, K., {et~al.} 2019, \apj, 885, 50

\bibitem[{{Wright} {et~al.}(2010){Wright}, {Eisenhardt}, {Mainzer}, {Ressler},
  {Cutri}, {Jarrett}, {Kirkpatrick}, {Padgett}, {McMillan}, {Skrutskie},
  {Stanford}, {Cohen}, {Walker}, {Mather}, {Leisawitz}, {Gautier}, {McLean},
  {Benford}, {Lonsdale}, {Blain}, {Mendez}, {Irace}, {Duval}, {Liu}, {Royer},
  {Heinrichsen}, {Howard}, {Shannon}, {Kendall}, {Walsh}, {Larsen}, {Cardon},
  {Schick}, {Schwalm}, {Abid}, {Fabinsky}, {Naes}, \& {Tsai}}]{WRIGHT10}
{Wright}, E.~L., {Eisenhardt}, P.~R.~M., {Mainzer}, A.~K., {et~al.} 2010, \aj,
  140, 1868

\bibitem[{{Yim} {et~al.}(2020){Yim}, {Wong}, {Rand}, \& {Schinnerer}}]{YIM20}
{Yim}, K., {Wong}, T., {Rand}, R.~J., \& {Schinnerer}, E. 2020, \mnras, 494,
  4558

\bibitem[{{Young} {et~al.}(1996){Young}, {Allen}, {Kenney}, {Lesser}, \&
  {Rownd}}]{YOUNG96}
{Young}, J.~S., {Allen}, L., {Kenney}, J.~D.~P., {Lesser}, A., \& {Rownd}, B.
  1996, \aj, 112, 1903

\bibitem[{{Young} \& {Scoville}(1991)}]{YOUNG91}
{Young}, J.~S., \& {Scoville}, N.~Z. 1991, \araa, 29, 581

\bibitem[{{Young} {et~al.}(1995){Young}, {Xie}, {Tacconi}, {Knezek}, {Viscuso},
  {Tacconi-Garman}, {Scoville}, {Schneider}, {Schloerb}, {Lord}, {Lesser},
  {Kenney}, {Huang}, {Devereux}, {Claussen}, {Case}, {Carpenter}, {Berry}, \&
  {Allen}}]{YOUNG95}
{Young}, J.~S., {Xie}, S., {Tacconi}, L., {et~al.} 1995, \apjs, 98, 219

\bibitem[{{Young} {et~al.}(2011){Young}, {Bureau}, {Davis}, {Combes},
  {McDermid}, {Alatalo}, {Blitz}, {Bois}, {Bournaud}, {Cappellari}, {Davies},
  {de Zeeuw}, {Emsellem}, {Khochfar}, {Krajnovi{\'c}}, {Kuntschner},
  {Lablanche}, {Morganti}, {Naab}, {Oosterloo}, {Sarzi}, {Scott}, {Serra}, \&
  {Weijmans}}]{YOUNG11}
{Young}, L.~M., {Bureau}, M., {Davis}, T.~A., {et~al.} 2011, \mnras, 414, 940

\bibitem[{{Zucker} {et~al.}(2018){Zucker}, {Battersby}, \&
  {Goodman}}]{ZUCKER18}
{Zucker}, C., {Battersby}, C., \& {Goodman}, A. 2018, \apj, 864, 153

\bibitem[{{Zuckerman} \& {Evans}(1974)}]{ZUCKERMAN74}
{Zuckerman}, B., \& {Evans}, N.~J., I. 1974, \apjl, 192, L149

\end{thebibliography}

\begin{appendix}
\section{Contributions}
\label{sec:contrib}

The design, execution, processing, and scientific exploitation of PHANGS--ALMA was a team effort, with major contributions from many people and input from the entire team. This paper also reflects major direct and indirect contributions from many people. We summarize some of the key contributions here.

\smallskip

\noindent \textbf{Observation Design, Data Processing, and Quality Assurance of the ALMA Data:} Since 2016, observation design, quality assurance, pipeline and algorithm development, and data processing have all been organized through the ``PHANGS ALMA Data Reduction'' (ADR) working group. J.~Pety has led this group since the beginning of the PHANGS collaboration, and the active members participating in almost all key activities and doing the core work on observations and data processing have been C.~Faesi, C.~Herrera, A.~Hughes,  D.~Liu, A.~Leroy, T.~Saito, E.~Rosolowsky, E.~Schinnerer, A.~Schruba, and A.~Usero. The imaging and postprocessing pipeline for the interferometric data and the software to create derived products has been mostly developed by by A.~Leroy, D.~Liu, E.~Rosolowsky, and T.~Saito. For the total power data, the processing and software development have been led by C.~Faesi, C.~Herrera, J.~Pety, and A.~Usero. A.~Hughes led quality assurance efforts and software development. A.~Hygate and K.~Sliwa also made early key contributions to a wide range of ADR efforts. A broad cross-section of the PHANGS--ALMA team also contributed to quality assurance efforts and gave excellent, frequent feedback. These included: A.~Barnes, I.~Beslic, M.~Chevance, J.~den Brok, C.~Eibensteiner, C.~Faesi, A.~Garc\'{i}a-Rodr\'{i}guez, C.~Herrera, A.~Hygate, M.~Jimenez~Donaire, J.~Kim, A.~Leroy, D.~Liu, J.~Pety, J.~Puschnig, M.~Querejeta, E.~Rosolowsky, T.~Saito, A.~Sardone, E.~Schinnerer, A.~Schruba, J.~Sun, A.~Usero, D.~Utomo, T.~Williams. Please also note that \citet{LEROY21a} includes a more detailed description of contributions to the PHANGS--ALMA data processing and product generation pipeline.

\smallskip

\noindent \textbf{Management of the PHANGS Collaboration:} E.~Schinnerer has served as the leader of the PHANGS collaboration since 2015. G.~Blanc, E.~Emsellem, A.~Leroy, and E.~Rosolowsky have acted as the PHANGS steering committee. E.~Rosolowsky has served as team manager since 2018. The entire PHANGS core team provides key input and oversight to all major collaboration decisions. The core team includes: F.~Bigiel, G.~Blanc, E.~Emsellem, A.~Escala, B.~Groves, A.~Hughes, K.~Kreckel, J.M.D.~Kruijssen, J.~Lee, A.~Leroy, S.~Meidt, M.~Querejeta, J.~Pety, E.~Rosolowsky, P.~Sanchez-Blazquez, K.~Sandstrom, E.~Schinnerer, A.~Schruba, and A.~Usero. Scientific exploitation of PHANGS--ALMA has taken place largely in the context of the ``Cold ISM Structure and Its Relation to Star Formation'' (Cold ISM) and ``Large Scale Dynamical Processes'' (Dynamics) science working groups. The ``Cold ISM'' group was led by A.~Hughes and K.~Kreckel in 2019 and C.~Faesi and A.~Hughes since 2020. The ``Dynamics'' group has been led by S.~Meidt and M.~Querejeta since 2019.

\smallskip

\noindent \textbf{Calculation of Galaxy Properties:} G.~Anand, J.~Lee, and the PHANGS--HST team led compilation of distances for the sample. A.~Leroy, S.~Meidt, M.~Querejeta, K.~Sandstrom, J.~Sun, and T.~Williams played active roles in the discussion and determination of stellar masses and sizes for the sample. I.~Ho, A.~Leroy, and T.~Williams have curated or developed software for curating the master table of PHANGS target properties. E.~Behrens and A.~Leroy created by-hand masks for artifacts and stars in the multiwavelength data used to estimate the SFR and stellar masses. M.~Querejeta played a key role in many aspects of obtaining and processing the IRAC data used for stellar mass estimation. Benchmarks against the MUSE-based galaxy property estimates relied on the hard work of the PHANGS--MUSE team, led by E.~Emsellem with major contributions to SFR and M$_\star$ estimates by F.~Belfiore, I.~Pessa, and P.~Sanchez Blazquez. The PHANGS sample definition and the SFR and M$_\star$ estimation built on work on this topic by A.~Leroy, K.~Sandstrom, J. Chastenet, and I. Chiang, heavily leveraging previous work by S.~Salim, M.~Boquien, and J.~Lee.

\smallskip

\noindent \textbf{Preparation of this Paper:} A.~Leroy led preparation of the text and figures in close collaboration with E.~Schinnerer. Frequent heavy edits and early, repeated high level scientific input throughout were contributed by A.~Hughes, E.~Rosolowsky, K.~Sandstrom, and A.~Schruba. The survey motivation in Section \ref{sec:motivation} distills many years of collective scientific discussions and proposals by the team. S.~Meidt and M.~Querejeta provided major input on the section presenting the estimation of galaxy properties. The entire team provided multiple rounds of careful vetting. C.~Faesi and A.~Usero compiled the detailed log of observations for the total power data.

\smallskip

\noindent \textbf{Observatory and Community Support:} Across multiple large and small projects, PHANGS--ALMA has benefited from outstanding support from the Joint ALMA Observatory, the North American ALMA Science Center (NAASC) at the National Radio Astronomy Observatory (NRAO), and the European Southern Observatory (ESO). The Large Program was carried out with the NAASC as supporting ALMA Regional Center, and the NAASC staff, including ARC manager A.~Remijan, have been incredibly responsive, helpful, and supportive with a wide range of technical issues. The ESO ALMA ARC has been similarly supportive whenever issues arose. PHANGS--ALMA built heavily on the high quality ALMA calibration pipeline (E.~Humphreys et al.\ in preparation) and the \texttt{CASA} software package. We also heavily leveraged the work by the \texttt{astropy} collaboration and the broader scientific \texttt{python} community. We  also acknowledge the astronomical IDL community, which was instrumental in many early parts of this work.

\end{appendix}

\end{document}